\documentclass[aps,prb, twocolumn, superscriptaddress, amsmath, amsfonts, amssymb, amsthm, floatfix]{revtex4-2}
\usepackage{graphicx}
\usepackage{dcolumn}
\usepackage[unicode=true,
 bookmarks=false,
 breaklinks=true,pdfborder={0 0 1},backref=false,colorlinks=true]
 {hyperref}
\hypersetup{pdfcreator={},pdfproducer={LaTeX with hyperref},linkcolor=blue,anchorcolor=blue,citecolor=blue,filecolor=red,menucolor=red,urlcolor=blue,pdfstartview=FitV,pdfhighlight=/I,hypertexnames=true}
\makeatletter
\usepackage{lmodern}
\usepackage{mathptmx, newtxtext, newtxmath, xspace, xfrac}
\usepackage{colortbl}
\usepackage{bbm}
\usepackage{siunitx}
\usepackage{comment}
\usepackage[table, dvipsnames]{xcolor}

\usepackage{esint}

\usepackage{array}[=2016-10-06]
\usepackage{booktabs}
\usepackage{multirow}
\usepackage{varwidth}

\usepackage{bm}
\usepackage{braket}

\usepackage{ragged2e}

\usepackage{natbib}
\setcitestyle{numbers}

\renewcommand{\eqref}[1]{Eq. (\ref{#1})}

\providecommand\@nil{} 
\makeatother

\begin{document}
\title{Theory of Electronic Nematic Criticality Constrained by Elastic Compatibility}
\author{W. Joe Meese}
\affiliation{Department of Physics, The Grainger College of Engineering, University of Illinois Urbana-Champaign, Urbana, IL 61801, USA}
\affiliation{Anthony J. Leggett Institute for Condensed Matter Theory, The Grainger College of Engineering, University of Illinois Urbana-Champaign, Urbana, IL 61801, USA}
\author{Rafael M. Fernandes}
\affiliation{Department of Physics, The Grainger College of Engineering, University of Illinois Urbana-Champaign, Urbana, IL 61801, USA}
\affiliation{Anthony J. Leggett Institute for Condensed Matter Theory, The Grainger College of Engineering, University of Illinois Urbana-Champaign, Urbana, IL 61801, USA}
\date{February 18, 2026}
\begin{abstract}
The defining property of electronic nematicity -- the spontaneous breaking of rotational symmetry -- implies an unavoidable coupling between the nematic order parameter and elastic strain fields, known as nemato-elasticity. While both quantities are rank-2 tensors, the strain tensor is constrained through the Saint Venant compatibility relations. These constraints, which take the form of three coupled second-order partial differential equations, arise from the lattice displacement vector's role as the potential field. The compatibility relations reflect the underlying gauge invariance of geometric deformations, which are violated only in the presence of crystalline defects. In this work, we develop a theory of nemato-elasticity that incorporates elastic compatibility explicitly through a co-rotating helical basis. Within our formalism, we show that elasticity bestows tensor compatibility upon the electronic nematic order parameter by suppressing incompatible electronic nematic fluctuations. As a result, nemato-elasticity is markedly different from bare nematicity. In ideal media devoid of defects, we show that the suppression of incompatible nematicity underlies direction-selective criticality, even in the absence of crystalline anisotropy. In realistic systems with defects, meanwhile, we show that elastic pinning fields emanate from distributions of quenched defects, generating random longitudinal and transverse conjugate fields experienced by the local nematic order parameter. The coexistence of direction-selective nematic criticality with pinning effects from random fields is naturally explained within our theory from the transformation to the helical basis, implying that local experimental probes of nematicity will be influenced by a linear -- but nonlocal -- combination of long-ranged and short-ranged helical nematic modes. Because the compatibility relations are gauge constraints endowed in the isotropic medium, our results constitute universal features of nemato-elastic criticality present in all crystalline systems.
\end{abstract}
\maketitle

\section{Introduction}

Initially proposed for cuprates and quantum Hall systems \cite{Kivelson98, Eisenstein99_nem, Oganesyan01, kivelsonHowDetectFluctuating2003}, the phenomenon of electronic nematicity has quickly expanded, since the discovery of iron-based superconductors \cite{paglioneHightemperatureSuperconductivityIronbased2010, Chu2010_nem,Chuang2010_nem,Chu12_nem,Fernandes12_nematic,Fernandes2014, Bohmer_Meingast_review,bohmerNematicityNematicFluctuations2022, fernandesIronPnictidesChalcogenides2022}, to an ever growing list of quantum materials with rather dissimilar properties \cite{Tamegai2019_nem,Cho2020_nem,Drucker2024_nem,RMF08_Potts_nematic_strain,Hwangbo2024_nem,Sun2024_nem,Tan2024_nem,Geim11_nem,Little2020_nem,Hamill2021_nem,Cho2022_nem,Silber2024_nem,Beaudin2022_nem,Mackenzie07_nem,Jiang2019_nem,Cao2021_nem,RubioVerdu2022_nem,Zhang2022_nem,Siddiquee2022_nem,Venkatesan2024_nem,Hossain2024_nem}. On the one hand, this rich landscape of materials in which electronic degrees of freedom can drive the breaking of a rotational symmetry point to a broad range of possible microscopic mechanisms, from spin fluctuations \cite{Xu2008,Fang2008,Fernandes12_nematic, Fernandes2014} to multi-component pairing excitations \cite{Hecker2018,Fernandes2019}. On the other hand, they also raise questions about universal properties of electronic nematic phenomena, particularly close to critical points, either thermal or quantum \cite{Oganesyan01,Paul17}. Moreover, besides their own interest as a new form of quantum many-body state, electronic nematicity is often intertwined with other degrees of freedom and quantum phases of interest, most notably unconventional superconductivity \cite{Lederer2015,Lederer2017,Klein2018}.

Despite their microscopic dissimilarities, electronic nematic  phases all share a common feature which distinguishes them from classical nematic liquids: they are crystalline solids. Indeed, while the elastic properties of nematic liquids are contained in the nematic degrees of freedom, the electronic nematic order parameter inherits the elastic properties of the underlying lattice \cite{qiGlobalPhaseDiagram2009, Fernandes10_nematic}. This fact introduces two important effects for the phenomenological description of electronic nematicity. First, in isotropic liquid crystals, the nematic director can condense in any direction in space, meaning its order parameter is a rank-2 traceless symmetric quadrupolar tensor with five degenerate components in three spatial dimensions \cite{chaikinPrinciplesCondensedMatter1995}. In electronic nematics, meanwhile, the discrete rotational symmetry of the solid splits the fivefold degeneracy encoded in the tensor into different irreducible representations of the quadrupole within the corresponding point group of the lattice \cite{Fradkin_review,RMF14_classification_nematics}. The crystal anisotropy can reduce the five degenerate components down to a single ``Ising'' nematic order parameter, as is the case in the tetragonal-to-orthorhombic phase transitions of the iron-based superconductors \cite{fernandesIronPnictidesChalcogenides2022}, 2DEGs \cite{Fradkin_review}, cuprates \cite{Davis10_nem} or collective Jahn-Teller systems \cite{maharajTransverseFieldsTune2017, RMF13_disorder_TmVO4}. Two-dimensional hexagonal systems, meanwhile, realize the 3-state Potts nematic universality class, since the fivefold degeneracy of $\text{SO}(3)$ is reduced to only twofold in the crystal \cite{Hecker2018, Cho2020_nem, Fernandes2020,RMF06_Potts_nematic_strain, Little2020_nem, RMF08_Potts_nematic_strain, Hwangbo2024_nem, hattoriOrbitalMoireQuadrupolar2024}. Whereas these materials are effectively two-dimensional (or actually two-dimensional), there are also reports of electronic nematicity exhibiting a higher threefold degeneracy within cubic systems, returning electronic nematic phenomenology back to the fully three-dimensional regime \cite{wangVerweyTransitionEvolution2023, AnantPaper}. Regardless of the point group, however, each of these nematic irreducible representations are all subsets of the original five basis functions from $\text{SO}(3)$.

The second major effect is \textit{nemato-elasticity} -- the bilinear coupling between electronic nematicity and symmetry-breaking elastic strains -- and is the subject of this work and an accompanying Letter \cite{ShortPaper}. At first sight, such a bilinear coupling may seem trivial, since the structural degrees of freedom are not critical on their own, but only via their coupling to the electronic nematic degrees of freedom. However, elastic degrees of freedom are subjected to constraints that will be inevitably inherited by the critical nematicity. Indeed, it has been shown that nemato-elasticity has profound effects near the electronic nematic critical point that are not present in liquid crystals. Some of these effects include the development of a spontaneous symmetry-breaking ferroelastic strain \cite{Chu12_nem, Fernandes12_nematic, sanchezQuantitativeRelationshipStructural2022, KaanPaper}, complete softening of the appropriate elastic modulus by nematic fluctuations \cite{Fernandes10_nematic, bohmerNematicSusceptibilityHoledoped2014}, and direction-selective criticality \cite{Karahasanovic16, Paul17, Fernandes2020, Hecker2022}. The first two effects reflect how electronic correlations can strongly alter the structural properties, while the latter effect arises from elastostatic fluctuations renormalizing the bare electronic nematic critical behavior. Originally associated with acoustic-phonon-driven ferroelastic instabilities \cite{Cowley76, Folk76, folkCriticalDynamicsElastic1979}, direction-selective criticality is an effect where thermal fluctuations of the order parameter soften anisotropically in a crystalline environment, i.e., the nematic susceptibility only diverges along specific momentum-space directions. This ultimately leads to only a sub-dimensional manifold of critical fluctuations with specific high-symmetry momenta, while suppressing most fluctuations at other momenta.  In electronic nematics, this effect is believed to be responsible for the essentially mean-field critical behavior observed for Ising-nematics in low-dimensional systems \cite{Chu12_nem, Karahasanovic16, Paul17, sanchezQuantitativeRelationshipStructural2022}.

Despite the theoretical progress in nematic criticality gained from phonon-mediated direction-selective criticality, there remain important discrepancies between theoretical predictions and experimental observations, particularly at mesoscopic length scales. There has long been experimental evidence in samples that the electronic nematic order parameter condenses into domains \cite{tanatarDirectImagingStructural2009, Buchner2010, Ran2011,Niedziela2012, Rosenthal2014, Dioguardi2015, Dioguardi2016, Forrest2016, Ren2021, shimojimaDiscoveryMesoscopicNematicity2021,curroNematicityGlassyBehavior2022, RMF08_Potts_nematic_strain, RMF13_disorder_TmVO4} with random field disorder being a likely culprit for domain formation \citep{Imry_Ma, Binder1983, nattermannInstabilitiesIsingSystems1983, grinsteinSurfaceTensionRoughening1983, Carlson2006, Phillabaum2012, vojtaPhasesPhaseTransitions2013, vojtaDisorderQuantumManyBody2019, mirandaPhaseDiagramFrustrated2021, yeStripeOrderImpurities2022, RMF01_RFBM, RMF08_Potts_nematic_strain, yangCoarseningDynamicsIsingnematic2025}. If indeed pinning effects from random fields are responsible for the domains, then standard arguments from statistical mechanics would contend that the sharp nematic phase transition associated with long-range order should be heavily suppressed, if not lost altogether, in low-dimensional systems \cite{Imry_Ma, Binder1983, grinsteinSurfaceTensionRoughening1983}.  Moreover, while there are many sources of local random fields for the nematic order parameter, particularly in doped materials \cite{Fradkin_review}, one of the most obvious arises in the form of random \textit{strain} fields \cite{RMF03_smectic_defect, RMF01_RFBM, RMF13_disorder_TmVO4}. These naturally arise as a consequence of nemato-elasticity in materials endowed with unavoidable structural disorder. Such random strains, particularly those emanating from  crystalline defects,  should be long-ranged and anisotropic \cite{eshelbyContinuumTheoryLattice1956, kronerekkehartContinuumTheoryDefects1981, muratoshioMicromechanicsDefectsSolids1987}, only amplifying the strength of random strain disorder \cite{weinribCriticalPhenomenaSystems1983, vojtaPhasesPhaseTransitions2013, vojtaDisorderQuantumManyBody2019}. 

From the perspective of elasticity, however, the same elastic fluctuations that give rise to to the sharp mean-field phase transition also mediate the long-range random pinning fields from structural defects. This presents a seeming paradox which we seek to resolve with our proposed theory of nemato-elastic criticality. How can elastic fluctuations that mediate interactions favoring spontaneous symmetry breaking simultaneously also mediate random pinning fields which eliminate nematic criticality? Moreover, at the microscopic level, detailed experiments have observed a puzzling decoupling between the local strain and the electronic nematic order parameter, seemingly at odds with the expectation that these two fields are bilinearly coupled \cite{Ren2021}.

The focus of our theory is on a crucial feature of elasticity theory that, while important for structural disorder, has been overlooked in the electronic nematic literature: the Saint Venant compatibility relations \cite{muratoshioMicromechanicsDefectsSolids1987, kleinertGaugeFieldsSolids1989,Lookman2003,Littlewood2014, beekmanDualGaugeField2017}. These are a set of second-order partial differential equations that act as constraints between the various components of the spatially modulated strain tensor. The continuous deformation of a medium -- one that displaces every unit cell from its equilibrium position at $\boldsymbol{x}$ to its displaced position at $\boldsymbol{x} + \boldsymbol{u}(\boldsymbol{x})$ -- requires that the information contained within the strain tensor must reflect the \textit{vector} displacement field, $\boldsymbol{u}(\boldsymbol{x})$, rather than being a truly tensorial quantity like the electronic nematic order parameter, $\varphi_{ij}(\boldsymbol{x})$. Formally, these constraints arise because elasticity is a $\text{U}(1)$ tensor gauge theory \cite{kleinertGaugeFieldsSolids1989, chaikinPrinciplesCondensedMatter1995, beekmanDualGaugeField2017, Pretko2018_PRL, Pretko2019, gaaFractonelasticityDualityTwisted2021}. Using Kr\"oner's incompatibility operator \citep{kronerekkehartContinuumTheoryDefects1981}, one writes them as 
\begin{equation}
    \text{inc}(\varepsilon)_{ij} \equiv \epsilon_{ikl}\epsilon_{jmn}\partial_k\partial_m\varepsilon_{ln} = 0,\label{eq:kroner_inc_real-space}
\end{equation}
where $\epsilon_{ijk}$ is the Levi-Civita symbol. The expressions above are analogous to the curl-free constraints imposed on the electrostatic field, $(\boldsymbol{\nabla}\times \boldsymbol{E})_i = \epsilon_{ijk}\partial_jE_k = 0$, which guarantees the existence of a scalar-valued electrostatic potential field that satisfies $\boldsymbol{E} = -\boldsymbol{\nabla}\phi$. It is straightforward to show that the infinitesimal strain tensor of elastostatics, 
\begin{equation}
    \varepsilon_{ij}(\boldsymbol{x}) = \tfrac{1}{2}\left[\partial_j u_i(\boldsymbol{x}) + \partial_i u_j(\boldsymbol{x}) \right], \label{eq:real_space_strain_tensor_definition}
\end{equation}
satisfies \eqref{eq:kroner_inc_real-space}. Thus, the displacement vector plays the part of a \textit{vector-valued} potential field for the strain \textit{tensor}. Any other tensor that satisfies \eqref{eq:kroner_inc_real-space} is called \textit{compatible}, whereas any tensor that does not is called \textit{incompatible} \cite{kronerekkehartContinuumTheoryDefects1981, kleinertGaugeFieldsSolids1989}. These terms for tensors are analogous to the terms ``conservative'' or ``irrotational'', and ``nonconservative'' or ``rotational'', used for vector fields, respectively.

We emphasize that the compatibility relations are \textit{geometrical} constraints -- they must be obeyed if the deformation arises from a differentiable, \textit{single-valued}, displacement vector field, $\boldsymbol{u}(\boldsymbol{x})$. When the compatibility relations are violated, then the deformation may be constructed from a multi-valued displacement vector field, attributed to structural defects which, in equilibrium, are responsible for plastic deformation \cite{eshelbyContinuumTheoryLattice1956, dewitLinearTheoryStatic1970, Dewit1973, dewitTheoryDisclinationsIII1973, dewitTheoryDisclinationsIV1973, kronerekkehartContinuumTheoryDefects1981, kleinertDoubleGaugeTheory1983, kleinertGaugeTheoryDefect1984, muratoshioMicromechanicsDefectsSolids1987, kleinertGaugeFieldsSolids1989, beekmanDualGaugeField2017}. In contrast to the elastic regime, in which a solid returns to its equilibrium position after the applied external stress is removed, in the plastic regime the structural defects are impacted by the external stress, such that the solid cannot return to its equilibrium position.

From the perspective of electronic nematicity, a phenomenon which is most conveniently expressed in terms of the group theoretic language of condensed matter physics, the compatibility relations impose non-trivial relationships between the irreducible representations of the strain tensor. Importantly, as we will show, for every nonzero momentum infinitesimally close to the Brillouin zone center, the compatibility relations relate strain components that transform trivially at the zone center with those that transform nontrivially. This fundamentally changes how one understands the symmetry analysis -- and subsequent consequences -- of \textit{inhomogeneous} nemato-elasticity compared with the \textit{homogeneous} case.

Given that the electronic nematic order parameter lowers a system's rotational symmetry, it can only couple directly to the strain components that transform as the same irreducible representation. One would expect from the symmetry analysis, therefore, that there is always at least one component of the strain tensor that will never be bilinearly coupled to electronic nematicity: the symmetry-preserving, volume-changing \textit{dilatation} strain, $\text{Tr}(\varepsilon) = \varepsilon_{xx} + \varepsilon_{yy} + \varepsilon_{zz} = \boldsymbol{\nabla} \cdot \boldsymbol{u}(\boldsymbol{x})$, given its invariance under rotations and reflections \footnote{In crystals with low-symmetries, there may be other trivially transforming components that one would expect to be independent of electronic nematicity -- for example, in axial crystals, the $\varepsilon_{zz}$, component transforms trivially.}.  However, we will show that this expectation is not upheld for the vast majority of electronic nematic fluctuations because they inherit the constraints of the compatibility relations.
\begin{figure}
    \centering
    \includegraphics[width=\columnwidth]{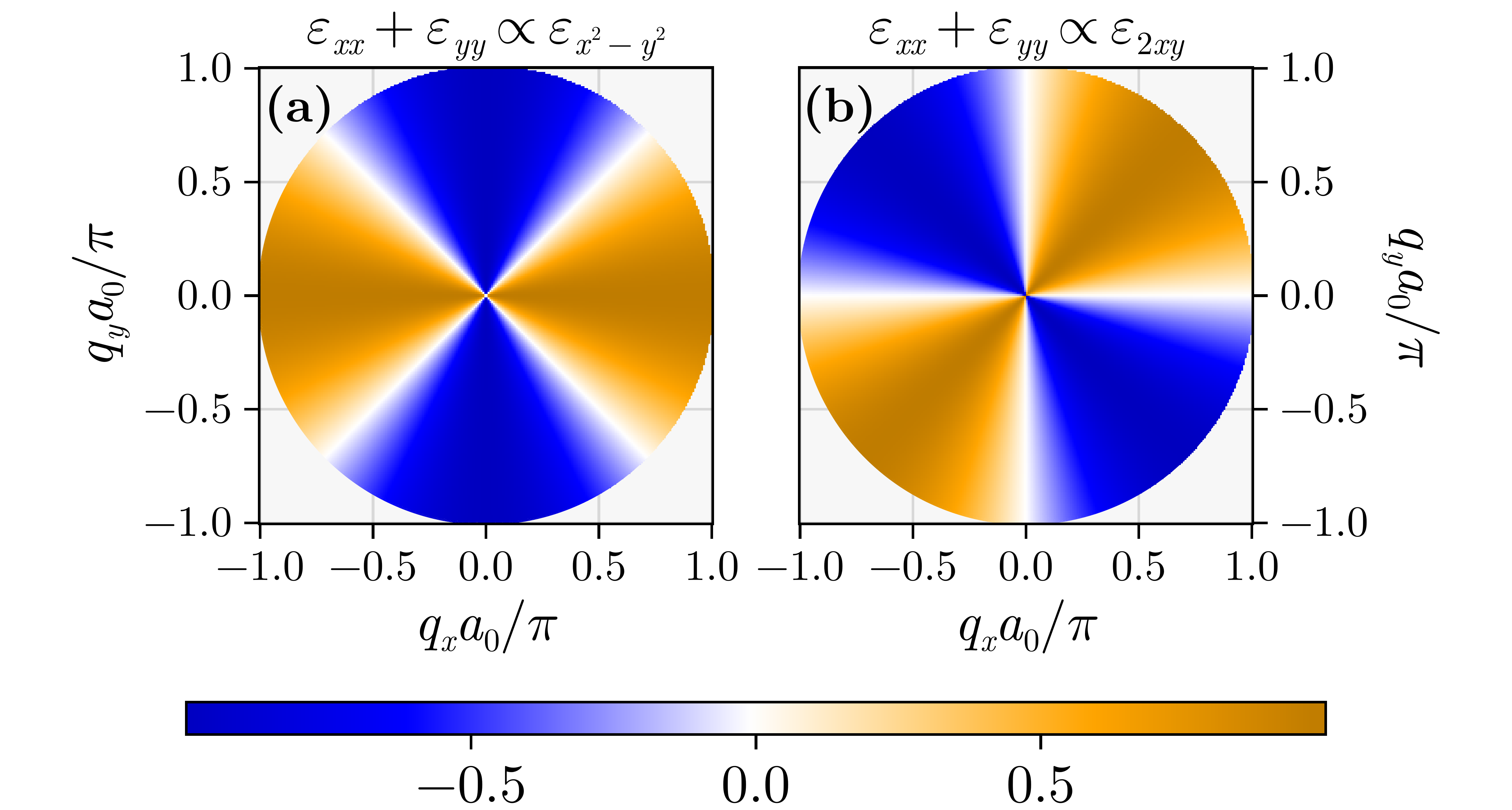}
    \caption{Two-dimensional elastic compatibility relation enforced in momentum space. Each panel shows the dilatation, $\varepsilon_{xx} + \varepsilon_{yy} = \partial_xu_x + \partial_yu_y$ induced by a symmetry-breaking static strain wave with well-defined wavevector, $\boldsymbol{q}$. The direction-dependence of these constraints are present in systems of any crystal symmetry. The colorbar represents the form factors that relate the symmetry-breaking \textbf{(a)} deviatoric, $\varepsilon_{x^2 - y^2}=\varepsilon_{xx} - \varepsilon_{yy}$, and the \textbf{(b)} shear, $\varepsilon_{2xy} = 2\varepsilon_{xy}$, strains to the symmetry-preserving dilatations, as enforced by the single compatibility condition in 2D. }
    \label{fig:SVCR_2D}
\end{figure}

It is simplest to understand how electronic nematic fluctuations couple to symmetry-preserving dilatations by considering the two-dimensional limit first. Throughout the rest of the paper, we will work in the fully-three dimensional elastic medium. For planar deformations, the so-called ``plane strain'' problems of elasticity require that $u_z(\boldsymbol{x}) \equiv 0$  and $\partial_z u_i(\boldsymbol{x}) \equiv 0$ \cite{landauTheoryElasticity1970, muratoshioMicromechanicsDefectsSolids1987, kleinertGaugeFieldsSolids1989}. Thus, there are only three nonzero strain components, $\varepsilon_{xx}$, $\varepsilon_{yy}$, and $\varepsilon_{xy}$, while there are two components in the displacement vector: $\boldsymbol{u} = (u_x, u_y)$. There is only one compatibility relation needed as a constraint to match the number of independent components of these two quantities: $\text{inc}(\varepsilon)_{33} = 0$. Written out in its usual form, it is
\begin{equation}
    \text{inc}(\varepsilon)_{33} = \partial_x^2 \varepsilon_{yy}  +\partial_y^2 \varepsilon_{xx} - 2\partial_{x}\partial_{y}\varepsilon_{xy} = 0,
\end{equation}
given that $\varepsilon_{xy}=\varepsilon_{yx}$. In symmetry-reduced form, meanwhile, the above becomes 
\begin{equation}
    \nabla^2 (\varepsilon_{xx} + \varepsilon_{yy}) = (\partial_x^2 - \partial_y^2)(\varepsilon_{xx} - \varepsilon_{yy}) + (2\partial_x\partial_y)(2\varepsilon_{xy}). \label{eq:intro1}
\end{equation}

The left-hand side of the equation above, being the trace of the elastic strain tensor in two-dimensions, represents local volume-changing dilatation strain $\varepsilon_{x^2 +y^2}\equiv \varepsilon_{xx} + \varepsilon_{yy}$. The right-hand side, meanwhile, contains only the traceless volume-preserving deviatoric $\varepsilon_{x^2 -y^2} \equiv \varepsilon_{xx} - \varepsilon_{yy}$ and shear $\varepsilon_{2xy} \equiv 2\varepsilon_{xy}$ strain components. The combinations of derivatives acting on each strain component transform the same way as the strain component itself, e.g. $\partial_x\partial_y$ has the same symmetry as the shear $\varepsilon_{2xy}$. Consequently, the two-dimensional Saint-Venant compatibility condition in \eqref{eq:intro1} -- a constraint that enforces the tensor-gauge theoretic structure of elasticity -- mixes all the strain components when $q\neq0$ \cite{ShortPaper, KaanPaper}. Importantly, in the long-wavelength but non-uniform limit, the implication is that the strain components with distinct symmetry characters at $q=0$ are \textit{not} independent of one another for finite $q$. Because of this, non-uniform ($q\neq0$) and uniform ($q=0)$ elastic strain must be treated separately.

In momentum space, \eqref{eq:intro1} is written more succinctly as 
\begin{equation}
    \varepsilon_{x^2 + y^2,\boldsymbol{q}} = \cos\left( 2\phi \right) \varepsilon_{x^2 - y^2,\boldsymbol{q}} + \sin\left( 2\phi \right)\varepsilon_{2xy,\boldsymbol{q}}, \label{eq:2D_SVCR_polar}
\end{equation}
where the wavevector is parameterized in the plane as $\boldsymbol{q}\equiv q(\cos\phi,\sin\phi)$ \cite{ShortPaper, KaanPaper}. In this form, the non-analytic relationship between the dilatation, deviatoric, and shear strains is clearer. Since this relationship is exact for all Fourier modes with $\boldsymbol{q}\neq\boldsymbol{0}$, but it does \textit{not} depend on the magnitude of the momentum $q \equiv |\boldsymbol{q}|$, the compatibility relations are intrinsically non-analytic as $q\rightarrow 0$. Thus, when the electronic nematic order parameter, or any other order parameter for that matter, couples to the traceless part of the strain tensor, the specific \textit{direction} associated with the strain's momentum is of extreme importance. Because of this non-analyticity, the effects of nemato-elasticity for $q = 0$ and \textit{in the limit} $q \rightarrow 0$ are completely different. When we consider the \textit{limit} $q \rightarrow 0$, the role of the momentum's \textit{direction} is essential, even in a medium with continuous rotational symmetry.

These non-analytic trigonometric form factors are plotted in momentum space in Fig.~\ref{fig:SVCR_2D}. The momenta in either panel where the form factors are nonzero force non-uniform deviatoric and shear strains to induce non-uniform dilatations. Clearly this is the vast majority of momentum directions, and they convert either $\varepsilon_{x^2 -y^2}$ or $\varepsilon_{2xy}$  strains into symmetry-preserving, volume-changing strain. The relationship between these irreducible representations is universal and must occur in all crystals. Note that the compatibility condition in Eq. (\ref{eq:intro1}) is perfectly allowed from a symmetry perspective,  since all terms transform trivially under any action from the orthogonal group $\text{O}(3)$, and thus under all point-group symmetry operations of any crystal. The key point is that the elastic compatibility relations are not symmetry-enforced constraints, but geometric constraints enforced by the gauge-field nature of elasticity theory.

In the special case of a $\varphi_{2xy}$-nematic order parameter, such as that in the iron-based superconductors \citep{Fernandes2014, Bohmer_Meingast_review, fernandesIronPnictidesChalcogenides2022, bohmerNematicityNematicFluctuations2022, KaanPaper}, the nematic fluctuations will naturally induce $\varepsilon_{2xy}$ strain fluctuations through nemato-elasticity. However, \eqref{eq:2D_SVCR_polar} and Fig.~\ref{fig:SVCR_2D}(b) show that nematic fluctuations will also necessarily induce dilatations \textit{unless} $\sin(2\phi) = 0$. These induced dilatations incur a higher energy cost associated with the bulk modulus of the material, which is often the largest elastic constant of the materials of interest. Similarly, for $\varphi_{x^2-y^2}$-nematics, dilatations are generated \textit{unless} $\cos(2\phi) = 0$. These are the critical momentum directions discussed before in the context of  ``direction-selective criticality'' in Ising electronic nematics \cite{Karahasanovic16, Paul17}. Analogously, in 3-state Potts nematics of 2D hexagonal systems, where the order parameter is the $(\varphi_{x^2 -y^2},\varphi_{2xy})$ doublet, then \eqref{eq:2D_SVCR_polar} implies that dilatations are generated \textit{unless} the nematic director is orthogonal to the momentum quadrupole, $(q_x^2 -q_y^2, 2q_xq_y)$, an effect predicted previously named ``nemato-orbital coupling'' \cite{Fernandes2020}. In those works, direction-selective criticality  was obtained by considering the coupling between the electronic nematic order parameter and the acoustic phonons -- the Goldstone modes that emerge in elasticity theory -- which usually have anisotropic velocities. We argue here that this effect is actually a direct consequence of the compatibility relations. Although they are indirectly encoded in the phonons, the latter are not the microscopic mechanism responsible for this effect. The fact that for a crystalline point group the five-component nematic $d$-orbital order parameters is split into the irreducible representations of that group, combined with the single compatibility relation in \eqref{eq:intro1} and \eqref{eq:2D_SVCR_polar}, is what \textit{selects} the soft momentum directions as those that decouple the symmetry-breaking strain from the symmetry-preserving dilatation strain. 

The elastic compatibility's role in direction-selective criticality is  reflected by the fact that the use of the compatibility relations in \eqref{eq:2D_SVCR_polar} in deriving direction-selective criticality circumvents knowledge of the crystalline point group. We reiterate that these constraints exist even in the 2D isotropic continuum because they arise from the geometrical relationship between $\boldsymbol{u}$ and $\varepsilon_{ij}$. Because of this, direction-selective criticality must also be a fundamental and universal consequence of compatibility, rather than being an effect of crystalline anisotropy. While these arguments completely account for direction-selective criticality for two-dimensional planar nematics, elastic deformation of real materials is fundamentally three-dimensional. Thus, there are two additional independent compatibility relations in \eqref{eq:kroner_inc_real-space} that determine the interdependence of the six components of the strain tensor. In turn, this means that in-plane nematic fluctuations can induce out-of-plane deformations, and \textit{vice versa}. 

In this paper we develop a fully three-dimensional theory of nemato-elasticity. The key ingredient of this framework is to write down the electronic nematic order parameter not in the usual $d$-orbital basis of irreducible representations, but in a basis we dub ``helical'', which automatically accounts for the presence (or absence) of tensor compatibility. We focus on the isotropic limit, since it is clear from the 2D limit that direction-selectivity in crystals is precipitated down from geometrical constraints on deformation to strains within crystals. Within our theory we find that nemato-elastic criticality can always be thought of as a ``planar'' instability because the five degenerate nematic basis functions of $\text{SO}(3)$ split at each momentum into a critical doublet and a non-critical singlet and doublet. The critical components are identified with the compatible sector of the electronic nematic order parameter, and can be written as the symmetric gradient of a purely transverse displacement vector field.  The non-critical components, meanwhile, are identified with incompatible electronic nematicity, and likewise couple to crystalline defects because of their mutual orthogonality to the compatible sector of nematicity. We discuss the several implications of our results for the general phenomenology of electronic nematics, resolving the aforementioned apparent contradictions. 

The structure of the paper is as follows. In Section \ref{sec:homogeneous_nematoelasticity}, we establish the nemato-elastic bilinear in an isotropic medium, and elucidate how it renormalizes the bare electronic nematic free energy in the case of homogeneous deformation. In Section \ref{sec:inhomogeneous_elasticity_and_compatibility}, we contrast homogeneous with inhomogeneous deformations, and develop our theory of nemato-elasticity while paying special attention to the Saint Venant compatibility relations. In this section we define our ``helical basis'' for nemato-elasticity, and contrast it with the ``$d$-orbital'' and Cartesian bases conventionally used in nematicity and elasticity problems, respectively. Starting in Section \ref{sec:compatible_restrictions_on_nematicity}, we apply our helical basis to the problem of nemato-elasticity, and find an anisotropic renormalization of the bare electronic mass in the isotropic elastic continuum. In Section \ref{subsec:suppression_of_incompatible_nematicity}, we use the compatibility relations to prove that the anisotropic nematic mass suppresses incompatible electronic nematicity. From there, in Section \ref{sec:polar_and_planar_nematicity}, we use this principle of incompatible nematic suppression to isolate the character of critical electronic nematic fluctuations in a 3D medium with only  ``polar'' and ``planar'' momenta, and derive direction-selective criticality in the isotropic continuum explicitly. In Section \ref{sec:elastic_incompatibility_and_nematoplasticity}, we incorporate elastic incompatibility, in the form of infinitesimal continuum plasticity, into our theory of nemato-elasticity for applications in structurally disordered electronic nematics. We show that ``nemato-plasticity'' can only couple to the non-critical components of the helical nematic order parameter. The results indicate the protection of long-range nematic order despite short-range electronic nematic pinning by long-range structural disorder. We develop an analytically soluble model of disorder using an ensemble of uncorrelated straight edge dislocations and compute the disorder-averaged real-space correlation functions inevitably present for the local electronic nematic order parameters. We discuss our theory in the greater context of electronic nematic quantum materials in Section \ref{sec:discussion}. Additionally, we provide various appendices (Appendices \ref{app:Gell-Mann-Decomp-Tensors}-\ref{app:harmonic_Greens_functions}) to supplement some calculations in the main text, including a discussion of the symmetry-allowed cubic invariant in the bare electronic nematic free energy (Appendix \ref{app:anharmonic_terms_free_energy}), connections between our formalism based on elastic compatibility and conventional acoustic phonons (Appendix \ref{app:helical_dynamical_matrix_crystals}), the solution for the plasticity problem within the helical strain formalism for an elastic crystal of arbitrary symmetry (Appendix \ref{app:elastostatic_plastic_solution}) and the expansion of the elastic strain from a distribution of multipolar crystalline defects (Appendix \ref{app:plastic_strain_defect_density_tensor}).

\section{Homogeneous Nemato-elasticity \label{sec:homogeneous_nematoelasticity}}
In this section, we begin the discussion by reviewing the simplest case of fluctuating \textit{homogeneous} strains in an isotropic medium \cite{qiGlobalPhaseDiagram2009, canoInterplayMagneticStructural2010,Karahasanovic16,Paul17,Carvalho2019,Fernandes2020}. The symmetry analysis serves as the foundation for the theory that incorporates spatially inhomogeneous strains throughout the rest of the paper. We show that fluctuating homogeneous strains renormalize the bare electronic nematic mass by introducing long-ranged, all-to-all, interactions with strength inversely proportional to the elastic shear modulus.  We show that these interactions do not couple the nematic order parameters in the conventional $d$-orbital basis.

The electronic nematic order parameter in three spatial dimensions is defined as the local electronic quadrupolar density,
\begin{equation}
    \varphi_{ij}\left(\boldsymbol{x}\right)=-\left\langle \hat{\psi}^{\dagger}\left(\boldsymbol{x}\right)\left(\partial_{i}\partial_{j}-\tfrac{1}{3}\nabla^{2}\delta_{ij}\right)\hat{\psi}\left(\boldsymbol{x}\right)\right\rangle ,\label{eq:definition_nematic_tensor_expectations}
\end{equation}
where $\hat{\psi}(\boldsymbol{x})$ is the many-body electronic wavefunction \citep{Oganesyan01, Fradkin_review, Fernandes2014, RMF14_classification_nematics}. Note that while the focus here in on the charge degrees of freedom, the expression above can be generalized in a straightforward way to include also spin, orbital, or superconducting degrees of freedom \cite{chandraIsingTransitionFrustrated1990,wuFermiLiquidInstabilities2007,Xu2008,Fang2008, zachariasMultiscaleQuantumCriticality2009a,valenzuelaPomeranchukInstabilityDoped2008,kontaniOriginOrthorhombicTransition2011, fanfarilloSpinorbitalInterplayTopology2015,parkElectronicInstabilitiesKagome2021,Fernandes12_nematic,Hecker2018,Fernandes2019, RMF08_Potts_nematic_strain,galiCriticalNematicPhase2024, grandiTheoriesChargedrivenNematicity2024, kieselUnconventionalFermiSurface2013, hattoriOrbitalMoireQuadrupolar2024}. In isotropic media,  the components of $\varphi_{ij}$  transform under rotations as the fivefold-degenerate $\ell = 2$  irreducible representation of $\text{SO}(3)$. The five nematic order parameters can be chosen to transform as the $d$-orbitals, $\{d_{z^2},d_{x^2-y^2},d_{2yz},d_{2zx},d_{2xy}\}$, and are likewise labeled with similar subscripts: $\varphi_{z^2}$, $\varphi_{x^2-y^2}$, \textit{etc}., as is done in \citep{RMF14_classification_nematics}. We construct these order parameters from the traceless $\varphi_{ij}$ tensor according to the following convention:
\begin{equation}
    \begin{array}{ccccc}
\varphi_{z^{2}}\equiv\frac{\varphi_{xx}+\varphi_{yy}-2\varphi_{zz}}{\sqrt{3}}, &  & \varphi_{x^{2}-y^{2}}\equiv\varphi_{xx}-\varphi_{yy},\\
\varphi_{2yz}\equiv2\varphi_{yz}, &  & \varphi_{2zx}\equiv2\varphi_{zx}, &  & \varphi_{2xy}\equiv2\varphi_{xy}.
\end{array}
\end{equation}
These expressions define the \textit{$d$-orbital basis} used in this paper and in Ref. \cite{ShortPaper}. In two spatial dimensions, the fivefold degeneracy of $\text{SO}(3)$ is split in $\text{SO}(2)$ into one $\ell = 0$ singlet, $\varphi_{z^2}$,  an $\ell = 1$ doublet, $(\varphi_{2yz},\varphi_{2zx})$,  and an $\ell = 2$  doublet, $(\varphi_{x^2 -y^2},\varphi_{2xy})$. Given that in 2D, all derivatives with respect to $z$  vanish, this implies that the $\ell = 1$  doublet vanishes identically: $\varphi_{2yz} = \varphi_{2zx} = 0$. The in-plane quadrupolar doublet, $(\varphi_{x^2 -y^2},\varphi_{2xy})$, accounts for  the planar electronic nematic order parameters conventionally studied in quantum materials, despite most of them being 3D crystals.

The elastic strain tensor, $\varepsilon_{ij}$, being a rank-2 symmetric tensor itself, can be decomposed into symmetry-preserving and symmetry-breaking contributions that transform as $\ell=0$  and $\ell=2$ irreducible representations, respectively. Whereas the nematic order parameter is traceless by definition, the strain tensor has a nonzero trace, which quantifies the symmetry-preserving volume changes due to some distortion. In the elasticity literature, it is known as the \textit{dilatation} \cite{eshelbyContinuumTheoryLattice1956,Eshelby1957,dewitTheoryDisclinationsIII1973, dewitTheoryDisclinationsIV1973}. Given that the dilatation does not break any symmetries, it cannot couple bilinearly to the electronic nematic order parameter. This contrasts with the symmetry-breaking \textit{deviatoric}  strains in the $\ell = 2$ irreducible representation. Strictly speaking, ``deviatoric'' refers to the traceless part of any tensor which encapsulates both the shear (off-diagonal) strains as well as the symmetry-breaking diagonal contributions \cite{Eshelby1957, Lookman2003}. For brevity, we refer to both the diagonal and off-diagonal symmetry-breaking contributions as ``deviatoric'' strains. The strain tensor is then written as 
\begin{equation}
    \varepsilon_{ij}=\tfrac{1}{3}\varepsilon_{0}\delta_{ij}+\left(\varepsilon_{ij}-\tfrac{1}{3}\varepsilon_{0}\delta_{ij}\right), \label{eq:strain_separation_dilatation_deviatoric}
\end{equation}
where the first term contains the scalar dilatation strain, $\varepsilon_0 \equiv \text{Tr}(\varepsilon)$, and the second represents the symmetry-breaking deviatoric strain. The deviatoric strain, being a five-component $\ell = 2$ irreducible representation, can also be decomposed in the $d$-orbital basis as
\begin{equation}
    \begin{array}{ccccc}
\varepsilon_{z^{2}}\equiv\frac{\varepsilon_{xx}+\varepsilon_{yy}-2\varepsilon_{zz}}{\sqrt{3}}, &  & \varepsilon_{x^{2}-y^{2}}\equiv\varepsilon_{xx}-\varepsilon_{yy},\\
\varepsilon_{2yz}\equiv2\varepsilon_{yz}, &  & \varepsilon_{2yz}\equiv2\varepsilon_{zx}, &  & \varepsilon_{2xy}\equiv2\varepsilon_{xy}.
\end{array}\label{eq:deviatoric_strain_d_orbitals}
\end{equation}

The nemato-elastic contribution to the free energy, in an isotropic medium, takes the form 
\begin{equation}
    \Delta\mathcal{F}\left[\varphi_{ij},\varepsilon_{ij}\right]\equiv-\lambda_{\text{0}}\int_{x}\boldsymbol{\varphi}\left(\boldsymbol{x}\right)\cdot\boldsymbol{\varepsilon}\left(\boldsymbol{x}\right),\label{eq:nematoelasticity_isotropic_medium_written_out}
\end{equation}
where $\lambda_0$  is the nemato-elastic coupling constant and $\int_x \equiv \int \text{d}^3x$. In the above, we vectorized the $d$-orbital basis as
\begin{equation} \begin{aligned}\boldsymbol{\varphi} & \equiv\left(\varphi_{z^{2}},\varphi_{x^{2}-y^{2}},\varphi_{2yz},\varphi_{2zx},\varphi_{2xy}\right)^{\text{T}},\\
\boldsymbol{\varepsilon} & \equiv\left(\varepsilon_{z^{2}},\varepsilon_{x^{2}-y^{2}},\varepsilon_{2yz},\varepsilon_{2zx},\varepsilon_{2xy}\right)^{\text{T}},
\end{aligned}
\label{eq:definition_of_nematic_order_deviatoric_strain_orbital_basis}
\end{equation}
according to the Gell-Mann decomposition in Appendix \ref{app:Gell-Mann-Decomp-Tensors}. The isotropic elastic free energy is given by 
\begin{equation}
    \mathcal{F}_{\text{elas}}=\frac{1}{2}\int_{x}\left\{ \lambda\varepsilon_{0}^{2}\left(\boldsymbol{x}\right)+2\mu\varepsilon_{ij}\left(\boldsymbol{x}\right)\varepsilon_{ji}\left(\boldsymbol{x}\right)\right\} ,\label{eq:isotropic_bare_elastic_free_energy}
\end{equation}
where $\lambda$  is the Lam\'e constant and $\mu$  is the shear modulus \cite{landauTheoryElasticity1970, chaikinPrinciplesCondensedMatter1995}. 
Omitting the position dependence momentarily, the second term evaluates to $\varepsilon_{ij}\varepsilon_{ji}=\frac{1}{3}\varepsilon_{0}^{2}+\frac{1}{2}\left(\boldsymbol{\varepsilon}\cdot\boldsymbol{\varepsilon}\right)$ using \eqref{eq:trace_AA_GellMann_representation}.
Under homogeneous distortions, the total elastic part of the free energy, $\mathcal{F}=\mathcal{F}_{\text{elas}}+\Delta\mathcal{F}$, becomes 
\begin{align}
    \mathcal{F}	=V\left\{ \tfrac{1}{2}\left(\lambda+\tfrac{2\mu}{3}\right)\varepsilon_{0}^{2}+\tfrac{1}{2}\mu\left(\boldsymbol{\varepsilon}\cdot\boldsymbol{\varepsilon}\right)-\lambda_{0}\left(\boldsymbol{\overline{\varphi}}\cdot\boldsymbol{\varepsilon}\right)\right\} ,  \label{eq:eq:total_elastic_free_energy_homogeneous_strain}
\end{align}
where $V$  is the volume and $\overline{\varphi}_a\equiv \frac{1}{V}\int_x\varphi_a(\boldsymbol{x})$. The spatial average, $\overline{\varphi}_a$, is proportional to the zero-momentum nematic order parameter, $\overline{\varphi}_a = \frac{1}{V}\varphi_{a,\boldsymbol{q}=\boldsymbol{0}}$.  Thermodynamic stability forces both the bulk modulus, $B \equiv \lambda+2\mu/3$, and the shear modulus, $\mu$, to be nonzero. Materials with positive Poisson ratios obey $B>\mu$, whereas so-called auxetic metamaterials can satisfy $B<\mu$ \cite{landauTheoryElasticity1970, chaikinPrinciplesCondensedMatter1995}. In either case, a useful strict inequality is that 
\begin{equation}
    0 < \tfrac{4\mu}{3} <  B + \tfrac{4\mu}{3} = \lambda + 2\mu. \label{eq:varrho_less_than_1}
\end{equation}
Ferroelastic instabilities drive the renormalized shear modulus $\tilde{\mu}$ to zero, whether they be structurally driven \cite{Cowley76, Folk76, folkCriticalDynamicsElastic1979} (proper ferroelastics) or electronically driven as in the case of electronic nematics \cite{Fernandes12_nematic, Bohmer_Meingast_review, fernandesIronPnictidesChalcogenides2022} (pseudoproper ferroelastics). In the latter case, the soft electronic nematic fluctuations drive $\tilde{\mu}\rightarrow0$, whereas the renormalized bulk modulus $\tilde{B}$ remains nonzero.  While that means the elastic medium is unstable to symmetry-breaking shear distortions, since $\tilde{B} > 0$, it remains stable against dilatation strain, indicating that these strains remain energetically expensive at the nemato-elastic transition.

Proceeding under the assumption that the ferroelastic transition is concomitant, then the bare elastic moduli are nonzero and the bare elastic free energy is stable at the harmonic level. One can then consider the correction to the nematic free energy due to fluctuating \textit{homogeneous} strains. Minimizing with respect to the homogeneous strain yields an effective long-ranged nematic interaction of the form 
\begin{equation}
    \Delta\mathcal{F}_{\text{eff}}\left[\varphi\right]=-\frac{\lambda_{0}^{2}}{2\mu V}(\boldsymbol{\varphi}_{\boldsymbol{q}=\boldsymbol{0}}\cdot\boldsymbol{\varphi}_{\boldsymbol{q}=\boldsymbol{0}})=-\frac{\lambda_{0}^{2}}{2\mu V}\int_{xx^{\prime}}\boldsymbol{\varphi}\left(\boldsymbol{x}\right)\cdot\boldsymbol{\varphi}\left(\boldsymbol{x}^{\prime}\right),\label{eq:effective_long-range_interaction_homogeneous_distortion}
\end{equation}
consistent with previous derivations within elastic crystals \cite{qiGlobalPhaseDiagram2009, canoInterplayMagneticStructural2010, Karahasanovic16, Paul17, Fernandes2020, Hecker2022}. In this derivation, however, the renormalization of the nematic mass is by fluctuating homogeneous strain in an \textit{isotropic} medium. The only difference between this case and the crystalline cases is the number of components in the critical nematic irreducible representation. The bare shear modulus, $\mu$, must also be adapted for the symmetry-breaking deviatoric strain in the same irreducible representation. The essential physics, regardless of the form of the correction, is clear even in the isotropic medium: homogeneous strains generate long-ranged, all-to-all, interactions for the electronic nematic order parameter which do not couple different nematic symmetry channels. While the last point regarding symmetry may seem obvious, it is not the case for fluctuating \textit{inhomogeneous} strains. Indeed, as we show in this paper, the compatibility relations couple different nematic symmetry channels, \textit{even in isotropic media}, with profound effects on nematic criticality.

\section{
Inhomogeneous Elasticity and Tensor Compatibility \label{sec:inhomogeneous_elasticity_and_compatibility}}
We show in this work that the simple, all-to-all symmetry-preserving, effective nematic interactions of the homogeneous deformation are replaced by more complex expressions when we consider inhomogeneous strains. In fact, we will obtain a more general form that contains the all-to-all interactions built in by taking the appropriate zero-momentum limit. The compatibility relations couple different nematic basis functions because they can couple the different strain components provided that the momentum transfer is not identically zero. Given that the degree of interdependence between the components is controlled by the momentum direction, in Subsection~\ref{subsec:helical_strain_tensor} we use helicity to develop a basis for elasticity that is invariant with respect to changes in the momentum direction. We then apply the same ``helical'' basis for generic rank-2 symmetric tensors such as electronic nematicity in Subsection~\ref{subsec:helical_gell-mann_basis}.

\subsection{Helical Representation of the Strain Tensor \label{subsec:helical_strain_tensor}}
In an isotropic system, the wavevector $\boldsymbol{q}$ provides an orthornormal basis $\{\hat{e}_1, \hat{e}_2,\hat{e}_3\}$ that can be obtained through a rotation $\mathcal{R}(\hat{q})\equiv (\hat{e}_1\,\vert\,\hat{e}_2\,\vert\,\hat{e}_3)$ from any of the coordinate axes onto  $\hat{q}$. For concreteness, we set $\hat{e}_1=\hat{q}$, rendering the two other basis vectors as the azimuthal and spherical polar unit vectors, $\hat{e}_2=\hat{e}_\phi$ and $\hat{e}_3=-\hat{e}_\theta$, respectively. This leads to the rotation matrix
\begin{align}
  \mathcal{R}(\hat{q})  =\begin{pmatrix}\cos\phi\sin\theta & -\sin\phi & -\cos\phi\cos\theta\\
\sin\phi\sin\theta & \phantom{-}\cos\phi & -\sin\phi\cos\theta\\
\cos\theta & 0 & \sin\theta
\end{pmatrix}. \label{eq:R(q)_rotation_matrix}
\end{align}
The helicity operator, $\mathfrak{h}(\hat{q})$, is defined in terms of the generators of rotations, $\{S_x, S_y, S_z\}$, as the scalar product  \citep{kleinertGaugeFieldsSolids1989, beekmanDualGaugeField2017}:
\begin{equation}
    \mathfrak{h}(\hat{q}) \equiv \text{i}\hat{q} \cdot \boldsymbol{S}. \label{eq:helicity_operator}
\end{equation}
When $\hat{q}=\hat{x}$, the helicity operator is simply $\text{i}S_x$, given as 
\begin{equation}
    \mathfrak{h}(\hat{q}=\hat{x}) = \begin{pmatrix}
        0 & 0 & 0
        \\
        0 & 0 & -\text{i}
        \\
        0 & \text{i} & 0
    \end{pmatrix},
\end{equation}
which has the normalized eigenvectors
\begin{equation}
    \begin{array}{ccc}
\hat{h}_{0}(\hat{q} = \hat{x})=\hat{x}, &  & \hat{h}_{\pm}(\hat{q} = \hat{x})=\frac{1}{\sqrt{2}}\left(\hat{y}\pm\text{i}\hat{z}\right)\end{array},
\end{equation}
with helical eigenvalues of $\{0, \pm 1\}$. These eigenvalues are the angular momenta with respect to the wavevector. Using the rotation matrix $\mathcal{R}(\hat{q})$, one can find the helical eigenvectors with respect to any other $\hat{q}$  through
\begin{equation}
    \begin{array}{ccc}
\hat{h}_{0}\left(\hat{q}\right)=\hat{e}_{1}, &  & \hat{h}_{\pm}\left(\hat{q}\right)=\frac{1}{\sqrt{2}}\left(\hat{e}_{2}\pm \text{i}\hat{e}_{3}\right).\end{array}
\end{equation}

Inverting these expressions, the spherical basis vectors and helical basis vectors for arbitrary $\hat{q}$ are related via 
\begin{equation}
\begin{aligned}\hat{e}_{1} & =\hat{q}=\hat{h}_{0}\left(\hat{q}\right),\\
\hat{e}_{2} & =\tfrac{1}{\sqrt{2}}\left[\hat{h}_{+}\left(\hat{q}\right)+\hat{h}_{-}\left(\hat{q}\right)\right],\\
\hat{e}_{3} & =\tfrac{1}{\text{i}\sqrt{2}}\left[\hat{h}_{+}\left(\hat{q}\right)-\hat{h}_{-}\left(\hat{q}\right)\right].
\end{aligned}
\end{equation}
It is clear that the transverse unit vectors, $\hat{e}_{2,3}$, correspond to a linearly polarized helical basis, whereas $\hat{h}_\pm$ are circularly polarized. However, in the isotropic medium, it is important to emphasize that any choice of two orthonormal vectors in the transverse plane to $\hat{q}$ are degenerate with any other, reflecting a residual $\text{SO}(2)$ symmetry about the $\hat{q}$ axis, associated with different basis vectors in the transverse plane.

We now represent the (linear) strain tensor in this helical basis. In real-space, the strain tensor is given by \eqref{eq:real_space_strain_tensor_definition}. In Fourier space, the relationship between the strain and displacement vector amplitudes follows as 
\begin{equation}
    \varepsilon_{ij} \equiv \tfrac{\text{i}}{2}\left( q_j u_i + q_i u_j \right),\label{eq:strain_qiuj}
\end{equation}
where we have omitted the explicit wavevector dependence of the displacement vector $\boldsymbol{u}$. In the linearly polarized helical basis, we can expand the displacement vector into longitudinal and transverse components as
\begin{equation}
    \begin{array}{ccccc}
\varepsilon_{1}^{h}\equiv\text{i}q\left(\hat{e}_{1}\cdot\boldsymbol{u}\right), &  & \varepsilon_{2}^{h}\equiv\text{i}q\left(\hat{e}_{2}\cdot\boldsymbol{u}\right), &  & \varepsilon_{3}^{h}\equiv\text{i}q\left(\hat{e}_{3}\cdot\boldsymbol{u}\right).
\end{array} \label{eq:definition_of_helical_strains}
\end{equation}
Using the transformation matrix in \eqref{eq:R(q)_rotation_matrix}, and exploiting its orthogonality, it follows that 
\begin{equation}
   \boldsymbol{\varepsilon}_h =   \text{i}q\mathcal{R}^{\text{ T}}(\hat{q}) \cdot \boldsymbol{u}, \label{eq:helical_strain_from_displacement}
\end{equation}
where $\boldsymbol{\varepsilon}_h=\{\varepsilon_1^h,\varepsilon^h_2,\varepsilon_3^h \}$ are the set of ``helical strain'' amplitudes. By inverting this expression, which holds at any $q>0$, then the strain tensor can be written as
\begin{equation}
    \varepsilon_{ij}	=\tfrac{1}{2}\left(\hat{e}_{1,i}\hat{e}_{\beta,j}+\hat{e}_{1,j}\hat{e}_{\beta,j}\right)\varepsilon_{\beta}^{h}, \label{eq:helical_strain_tensor_e1iekj}
\end{equation}
where we have used a Greek index to differentiate the helical indices $1,2,3$ from the Cartesian components $x,y,z$.

\eqref{eq:helical_strain_tensor_e1iekj} represents the helical representation of the strain tensor. It provides a map to the ``Cartesian'' representation of the strain amplitudes (left-hand side) from the one longitudinal and two transverse strains contained in $\boldsymbol{\varepsilon}_h$ -- the linearly polarized ``helical strains.''  It is important to establish that the amplitudes at opposite momenta are related through 
\begin{equation}
    \begin{aligned}
        \hat{e}_\alpha(-\hat{q}) &= \left(-1\right)^{1+\delta_{\alpha, 3}} \hat{e}_\alpha(\hat{q}),
        \\
       \varepsilon^h_\alpha(-\hat{q}) &= \left(-1\right)^{\delta_{\alpha, 3}} \varepsilon^h_\alpha(\hat{q})^*,
    \end{aligned}\label{eq:ehat_helical_strain_negative_qhat}
\end{equation}
where the asterisk denotes complex conjugation. The second equality follows from \eqref{eq:definition_of_helical_strains} and the fact that $u_i(-\boldsymbol{q})=u_i(\boldsymbol{q})^*$, given the displacement vector is a real vector field in real-space. As a result, while $\varepsilon^h_{1,2}(\boldsymbol{x})$  are real in real-space, $\varepsilon_3^h(\boldsymbol{x})$ is purely imaginary. The strain tensor, meanwhile, must be real in real-space. Indeed, it follows from \eqref{eq:helical_strain_tensor_e1iekj} and \eqref{eq:ehat_helical_strain_negative_qhat} that $\varepsilon_{ij}(-\hat{q}) = \varepsilon_{ij}(\hat{q})^*$.

From a symmetric perspective, \eqref{eq:helical_strain_tensor_e1iekj} couples different irreducible representations for the strain at finite momentum. There is no contradiction with group theory, since the momentum-dependent pre-factors ensure that all symmetry transformation properties are preserved. Moreover, from \eqref{eq:definition_of_helical_strains}, the longitudinal strain, $\varepsilon_1^h$, always transforms as a scalar whereas the two transverse strains, $\varepsilon^h_{2,3}$, transform together as a vectorial doublet under rotations in the transverse plane to $\hat{q}$. This contrasts with the Cartesian components of $\varepsilon_{ij}$ which transform as various linear combinations of $d$-wave spherical harmonics. This coupling occurs because the wavevector breaks the $\text{SO}(3)$  symmetry down to $\text{SO}(2)$ through the formation of its little group, where the $C_\infty$ axis of $\text{SO}(2)$ is along $\hat{q}$.  It is stressed that this happens at any nonzero $q$, and therefore only occurs for spatially modulated strains, regardless of how small $q\equiv|\boldsymbol{q}|>0$ is. One contrasts this situation with the case of homogeneous strains where rotational isotropy is not broken by a wavevector.

\subsection{Helical Gell-Mann Basis for Rank-2 Symmetric Tensors \label{subsec:helical_gell-mann_basis}}
The quadrupolar form factors in \eqref{eq:helical_strain_tensor_e1iekj} couples the scalar dilatation strain, $\varepsilon_1^h$, with the transverse helical strains, $\varepsilon_{2,3}^h$, when mapping them onto the usual Cartesian strain components, $\varepsilon_{ij}$. While these form factors completely cover all of linear elasticity, they do not form a complete basis of quadrupolar form factors in a five-dimensional vector space which describe arbitrary quadrupolar tensors, including electronic nematicity. In this subsection, we will define a complete basis for these form factors and use them in our discussion of nemato-elasticity in Section~\ref{sec:compatible_restrictions_on_nematicity}.

We define a symmetric quadrupolar tensor in the helical basis, $\mathcal{Q}^{\alpha\beta}$, through its components as
\begin{equation}
    \mathcal{Q}^{\alpha\beta}_{ij}(\hat{q})\equiv\frac{1}{2}\left(\hat{e}_{\alpha,j} \hat{e}_{\beta,i} + \hat{e}_{\alpha,i} \hat{e}_{\beta,j} \right).\label{eq:Qabij}
\end{equation}
The elastic form factors in \eqref{eq:helical_strain_tensor_e1iekj} correspond to the case with $\alpha=1$, allowing a concise way of writing the Cartesian components of the strain tensor as
\begin{equation}
    \varepsilon_{ij} = \mathcal{Q}_{ij}^{1\alpha}(\hat{q})\,\varepsilon_{\alpha}^h, \label{eq:strain_from_Q1alpha_helical}
\end{equation}
where  summation over the repeated helical Greek indices is implied. Inverting the momentum direction, in accordance with \eqref{eq:ehat_helical_strain_negative_qhat}, shows that this $\mathcal{Q}^{\alpha\beta}$ tensor satisfies 
\begin{equation}
    \mathcal{Q}^{\alpha\beta}_{ij}(-\hat{q}) = \left(-1\right)^{\delta_{\alpha, 3} + \delta_{\beta, 3}}\mathcal{Q}^{\alpha\beta}_{ij}(\hat{q}). \label{eq:Qab_under_qhat_to_minus_qhat}
\end{equation}
For brevity, we will omit the explicit momentum dependence in the following, unless its role in the expressions is not clear. This tensor, $\mathcal{Q}^{\alpha\beta}_{ij}$,  is symmetric in both its Latin and Greek indices. The trace over the Latin indices follows as 
\begin{equation}
    \text{Tr}\,(\mathcal{Q}^{\alpha\beta})=\hat{e}_\alpha \cdot \hat{e}_\beta = \delta_{\alpha\beta}.\label{eq:trace_Qab}
\end{equation}
Likewise the trace over the product is
\begin{equation}
    \text{Tr}\,\left( \mathcal{Q}^{\alpha\beta}\mathcal{Q}^{\gamma\delta} \right) = \frac{1}{2}\left( \delta_{\alpha\gamma}\delta_{\beta\delta} + \delta_{\alpha\delta}\delta_{\beta\gamma} \right).\label{eq:traceQabQgd}
\end{equation}
From these expressions, we find that the three tensors relevant for elasticity, $\{\mathcal{Q}^{11}, \mathcal{Q}^{12}, \mathcal{Q}^{13}\}$, are trace-orthogonal (see \eqref{eq:trace_properties_Q1a}).

Given that \eqref{eq:Qabij} is symmetric in its Latin indices, it may be transformed into a vectorial representation with \eqref{eq:GellMann_decomp_tensor}:
\begin{equation}
\begin{aligned}\mathcal{Q}^{\alpha\beta}_{ij} & =\frac{1}{3}Q_{0}^{\alpha\beta}\lambda^{0}_{ij}+\frac{1}{2}\boldsymbol{Q}^{\alpha\beta}\cdot\boldsymbol{\lambda}_{ij},\\
Q_{0}^{\alpha\beta} & =\text{Tr}\,\left(\mathcal{Q}^{\alpha\beta}\right)=\delta_{\alpha\beta},\\
\boldsymbol{Q}^{\alpha\beta} & =\text{Tr}\,\left(\boldsymbol{\lambda}\mathcal{Q}^{\alpha\beta}\right)=\hat{e}_{\alpha}^{\text{T}}\cdot\boldsymbol{\lambda}\cdot\hat{e}_{\beta},
\end{aligned}\label{eq:Qab_tensor_GellMann_Decomp}
\end{equation}
Likewise, from \eqref{eq:GellMann_Trace_AB} and \eqref{eq:traceQabQgd}, it follows that the Gell-Mann ``quadrupolar vectors,'' $\boldsymbol{Q}^{\alpha\beta}$, satisfy the following scalar product at every momentum, $\hat{q}$: 
\begin{equation}
    \boldsymbol{Q}^{\alpha\beta}\cdot\boldsymbol{Q}^{\gamma\delta}=\delta_{\alpha\gamma}\delta_{\beta\delta}+\delta_{\alpha\delta}\delta_{\beta\gamma}-\frac{2}{3}\delta_{\alpha\beta}\delta_{\gamma\delta}.\label{eq:innerproduct_QabQgd_vectors}
\end{equation}
We recover a set of orthogonality conditions within the vectorial representation of these quadrupolar tensors.  In particular, $\boldsymbol{Q}^{11}$, $\boldsymbol{Q}^{12}$, and $\boldsymbol{Q}^{13}$ are mutually orthogonal (see \eqref{eq:dot_product_Q1alpha}). From these vectors, one can write the strain as 
\begin{equation}
    \varepsilon_{ij}=\frac{1}{3}\varepsilon_{1}^{h}\lambda_{ij}^{0}+\frac{1}{2}\left(\varepsilon_{1}^{h}\boldsymbol{Q}^{11}+\varepsilon_{2}^{h}\boldsymbol{Q}^{12}+\varepsilon_{3}^{h}\boldsymbol{Q}^{13}\right)\cdot\boldsymbol{\lambda}_{ij}. \label{eq:strain_tensor_GellMann_decomp_Qabvecs}
\end{equation}
We emphasize that there are \textit{six} Cartesian components of the strain tensor, but they clearly only depend on \textit{three} helical strains: one longitudinal and two transverse. This is reflected by the fact that, when written in the helical basis, the strain tensor automatically satisfies the Saint Venant compatibility relations. Indeed, an explicit calculation gives (see Appendix \ref{app:quadrupolar_form_factors}):

\begin{equation}
    \text{inc}\,\left(\mathcal{Q}^{\alpha\beta}\right)_{ij}=-q^{2}\epsilon_{1\alpha\gamma}\epsilon_{1\beta\gamma}\mathcal{Q}_{ij}^{\gamma\delta}.\label{eq:inc_of_Qab_simplified_early}
\end{equation}
Therefore, using Eq. (\ref{eq:strain_from_Q1alpha_helical}) and the anti-symmetric properties of the Levi-Civita symbol, it follows that $ \text{inc}\,\left( \varepsilon \right)_{ij}=0$. This result is of course the same as what one would obtain using the conventional basis for strain comprised of the displacement vector, \eqref{eq:strain_qiuj}. The utility of the helical basis is, however, that the helical strains are defined specifically in a coordinate system that co-rotates with the momentum. We can therefore always track the symmetry-preserving dilatation strain $\varepsilon_1^h$ involved in nemato-elasticity through the momentum-dependence in \eqref{eq:strain_tensor_GellMann_decomp_Qabvecs}. This is the main motivation to introduce the helical basis.

From Eq. (\ref{eq:strain_tensor_GellMann_decomp_Qabvecs}) , we see that, whereas the trace of the strain tensor only depends on the longitudinal strain, $\varepsilon_1^h$, the deviatoric part of the strain tensor, $\boldsymbol{\varepsilon}$, depends on \textit{all three} helical strain amplitudes, including the dilatation. 
The deviatoric strain, in the co-rotating vectorial representation, assumes the form
\begin{equation}
    \boldsymbol{\varepsilon}=\varepsilon_{1}^{h}\boldsymbol{Q}^{11}+\varepsilon_{2}^{h}\boldsymbol{Q}^{12}+\varepsilon_{3}^{h}\boldsymbol{Q}^{13}.\label{eq:deviatoric_strain_vector_helical_decomposition}
\end{equation}

While these three quadrupolar vectors account for all of linear elasticity, they \textit{do not} constitute a complete basis for electronic nematicity. Since the nematic vector in \eqref{eq:definition_of_nematic_order_deviatoric_strain_orbital_basis} is a five-component object, we define the nematic order parameter in the $d$-orbital basis in terms of projectors on the five orthonormal vectors, $\{ \boldsymbol{\hat{Q}}_a \}$ that  span the five-dimensional ``nematic'' vector space. 
\begin{equation}
    \boldsymbol{\varphi}=\sum_{a=1}^{5}\Phi_{a}\boldsymbol{\hat{Q}}_{a},\quad\Phi_{a}\equiv\boldsymbol{\hat{Q}}_{a}\cdot\boldsymbol{\varphi},\label{eq:nematic_Qvec_decomposition}
\end{equation}
These five orthonormal vectors are constructed in terms of the six components of  $\boldsymbol{Q}^{\alpha\beta}$ as shown in Appendix \ref{app:quadrupolar_form_factors} and summarized in \eqref{eq:vectorial_Qhatvec_basis}.  Because each of these five vectors co-rotate with the wavevector, their character and orthornormality is maintained for all $\boldsymbol{q} \neq \boldsymbol{0}$. Moreover,
being orthonormal, their outer products satisfy the completeness relation:
\begin{equation}
    \text{I}=\sum_{a=1}^{5}\boldsymbol{\hat{Q}}_{a}^{\phantom{\text{T}}}\boldsymbol{\hat{Q}}_{a}^{\text{T}},\label{eq:Qvec_completeness}
\end{equation}
where $\text{I}$ is the $5\times 5$ identity matrix. 

The  helical decomposition of the deviatoric strain tensor in this five-dimensional basis follows from \eqref{eq:deviatoric_strain_vector_helical_decomposition} as 
\begin{equation}
    \boldsymbol{\varepsilon}\equiv\sum_{a=1}^{5}\varepsilon_{a}\boldsymbol{\hat{Q}}_{a}=\left(\tfrac{2}{\sqrt{3}}\varepsilon_{1}^{h}\right)\boldsymbol{\hat{Q}}_{1}+\varepsilon_{2}^{h}\boldsymbol{\hat{Q}}_{2}+\varepsilon_{3}^{h}\boldsymbol{\hat{Q}}_{3}, \label{eq:deviatoric_strain_Qvec_decomp}
\end{equation}
The difference between \eqref{eq:deviatoric_strain_vector_helical_decomposition} and \eqref{eq:deviatoric_strain_Qvec_decomp} is entirely within the contribution from the dilatation, given that $\boldsymbol{Q}^{11}$ is not normalized, but $\boldsymbol{\hat{Q}}_1$ is. Thus, $\varepsilon_1 = \tfrac{2}{\sqrt{3}}\varepsilon^h_1$, $\varepsilon_{2,3} = \varepsilon^h_{2,3}$, and $\varepsilon_{4,5}=0$. The five-component nematic order parameter, $\boldsymbol{\varphi}$, meanwhile, will be generally projected along all \textit{five} $\boldsymbol{\hat{Q}}_a$ basis vectors. Even in an isotropic medium, this reduces the phase space of nematicity that benefits from nemato-elasticity. To see this, we write the nemato-elastic energy density from \eqref{eq:nematoelasticity_isotropic_medium_written_out} in Fourier space. Given that both $\boldsymbol{\varphi}(\boldsymbol{x})$ and $\boldsymbol{\varepsilon}(\boldsymbol{x})$ are real fields, then for each $d$-orbital basis function, $\varphi_a(-\boldsymbol{q})=\varphi_a(\boldsymbol{q})^*$ and $\varepsilon_a(-\boldsymbol{q})=\varepsilon_a(\boldsymbol{q})^*$. Using the orthonormality of the helical $\boldsymbol{\hat{Q}}$ basis vectors, the nemato-elastic bilinear becomes
\begin{align}
\lambda_0\text{Re}\left(\boldsymbol{\varphi}^{\dagger}\cdot\boldsymbol{\varepsilon}\right)	&=\lambda_0\text{Re}\left(\sum_{a=1}^{5}\Phi_{a}^{*}\varepsilon_{a}^{\phantom{*}}\right)\nonumber\\
	&=\lambda_0\text{Re}\left(\tfrac{2}{\sqrt{3}}\Phi_{1}^{*}\varepsilon_{1}^{h}+\Phi_{2}^{*}\varepsilon_{2}^{h}+\Phi_{3}^{*}\varepsilon_{3}^{h}\right),\label{eq:deviatoric_strain_vec_nematic_vec}
\end{align}
for each nonzero wavevector, $\boldsymbol{q}$. The utility of the helical basis is observed in the nemato-elastic coupling above, valid for each momentum direction $\hat{q}$. In the conventional $d$-orbital basis, $\operatorname{SO}(3)$ symmetry enforces that all five nematic order parameters are explicitly coupling to the five deviatoric strain components. However, the compatibility relations mix the six components of the strain tensor for each momentum $\boldsymbol{q}$ in a non-analytic manner, reducing the independent strain degrees of freedom from six to three. The fact that there are only three independent strain modes -- one longitudinal and two transverse -- is manifest in the helical basis, where it is clear that the compatibility relations project the electronic nematic order parameter only into a three-dimensional subspace. This breaks the isotropic fivefold degeneracy at the \textit{bilinear} level, reflecting the fact that, for a fixed non-zero momentum, the electronic nematic order parameter no longer has the rotational invariance it has at zero-momentum.

At each nonzero wave-vector, the compatibility relations generate distinct effective ``helical'' nemato-elastic couplings for the helical nematic order parameters. 
For $\Phi_{2,3}$, the coupling is unchanged between the $d$-orbital representation and the helical, whereas for $\Phi_1$, the effective coupling is $\tfrac{2}{\sqrt{3}}\lambda_0$, and it vanishes entirely for $\Phi_{4,5}$. In the case of $\Phi_{4,5}$, this means these nematic order parameters do not participate in nemato-elasticity at all, and they must reside in a nematic subspace that exists outside the domain of linear elasticity theory.

Before concluding this section, we point out that the co-rotating \textit{vectorial} representation is easily transformed into a co-rotating \textit{tensorial} representation using a basis of trace-orthonormal basis tensors, $\{\vartheta^a,\;a=1,\dots,5\}$. Indeed, it follows that \eqref{eq:nematic_Qvec_decomposition}
has an analogous tensorial decomposition. This tensorial representation will be much more convenient in discussing the role of nematic incompatibility, particularly in three spatial dimensions, as discussed in Section \ref{sec:elastic_incompatibility_and_nematoplasticity}. The tensorial representation of \eqref{eq:nematic_Qvec_decomposition} is given by 
\begin{equation}
    \begin{aligned}&\varphi_{ij}  =\frac{1}{\sqrt{2}}\sum_{a=1}^{5}\Phi_{a}\vartheta_{ij}^{a}, & &\Phi_{a}  =\sqrt{2}\text{Tr}\left(\vartheta^{a}\varphi\right),\\
&\text{Tr}\left(\vartheta^{a}\right)  =0, & &\text{Tr}\left(\vartheta^{a}\vartheta^{b}\right)  =\delta_{ab}.
\end{aligned}\label{eq:nematic_varThetaTensor_decomposition}
\end{equation}
This definition preserves the normalization of the nematic order parameter, $\boldsymbol{\varphi}^\dagger\cdot\boldsymbol{\varphi}=2\text{Tr}\left(\varphi^\dagger\varphi\right)$ from \eqref{eq:GellMann_Trace_AB}, using the same coefficients, $\{\Phi_a,\;a=1,\dots,5\}$, from the vectorial decomposition in \eqref{eq:nematic_Qvec_decomposition}. The basis tensors $\{\vartheta_a\}$ are related to the basis vectors $\{\boldsymbol{\hat{Q}}_a\}$ by substituting \eqref{eq:nematic_Qvec_decomposition} into \eqref{eq:nematic_and_strain_GellMann_decomp} to obtain
\begin{equation}
    \vartheta_{ij}^{a}=\tfrac{1}{\sqrt{2}}\left(\boldsymbol{\hat{Q}}_{a}\cdot\boldsymbol{\lambda}_{ij}\right).\label{eq:def_varTheta_GellMann_of_Qhat}
\end{equation}
For completeness, we explicitly list these basis tensors in Appendix \ref{app:quadrupolar_form_factors} in \eqref{eq:varThetaTensor_to_Qvechat_to_Qabij}. Importantly, these co-rotating tensors constitute a complete basis for rank-2 symmetric tensors at every momentum $\boldsymbol{q}$. Thus, if $A_{ij}=A_{ij}(\boldsymbol{q})$ is a rank-2 symmetric tensor at momentum, $\boldsymbol{q}$, then it can be written as 
\begin{equation}
   \begin{aligned}A_{ij} & =\frac{1}{3}\left(\text{Tr}\,A\right)\delta_{ij}+\frac{1}{\sqrt{2}}\sum_{a=1}^{5}A_{a}\vartheta_{ij}^{a},\\
A_{a} & =\sqrt{2}\text{Tr}\left(\vartheta^{a}A\right)=\boldsymbol{\hat{Q}}_{a}\cdot\boldsymbol{A}.
\end{aligned}\label{eq:tensor_varThetaTensor_decomposition}
\end{equation}
This decomposition can be viewed as a co-rotating generalization of \eqref{eq:GellMann_decomp_tensor}. Importantly, the characters of the coefficients, $\{\text{Tr}\,A, A_a\}$, remain invariant as the wavevector changes its direction in space. 

\section{Compatible Restrictions on Electronic Nematic Fluctuations \label{sec:compatible_restrictions_on_nematicity}}
With the complete helical basis for electronic nematicity constructed, we apply it to the problem of nemato-elasticity in the isotropic continuum. Importantly, the character of the helical nematic order parameters, $\{\Phi_a\}$, will remain the same regardless of the momentum direction of the nematic fluctuations. Similarly, the helical nematic order parameters will only couple to helical strain amplitudes with the same character. This contrasts with the $d$-orbital basis, $\{\varphi_a\}$, where the compatibility relations couple the symmetry-breaking strains generated by electronic nematic fluctuations with symmetry-preserving dilatation strains from \eqref{eq:deviatoric_strain_vector_helical_decomposition}. 

In Subsection \ref{subsec:renormalization_of_electronic_nematic_mass}, we find the effective nematic free energy renormalized by thermally fluctuating elastostatic strain fields and show that even in the isotropic medium, one still recovers a set of constraints which splits the five-dimensional critical nematic manifold down to a two-dimensional one. In Subsection \ref{subsec:suppression_of_incompatible_nematicity}, we show that these constraints are an expression of the geometry of elasticity theory through the suppression of \textit{incompatible} electronic nematicity.

\subsection{Renormalization of Electronic Nematic Mass in Isotropic Media \label{subsec:renormalization_of_electronic_nematic_mass}}

Before we begin, it is important to emphasize that an instability in a given nematic channel corresponds to the condensation of a subset of order parameters condensing in the $d$-orbital basis $\{\varphi_a\}$, whereas the nemato-elastic coupling is more clearly represented in the helical basis of nematic order parameters  $\{\Phi_a\}$, as it accounts for the elastic compatibility relations in a consistent, $\hat{q}$-independent manner. Thus, our strategy is to integrate out the strain components in the helical basis first.

In an isotropic system, the nemato-elastic free energy in \eqref{eq:isotropic_bare_elastic_free_energy} diagonalizes within the helical basis, and assumes the form
\begin{align}
   \mathcal{F}_{\text{elas}}	&=\frac{1}{2V}\sum_{\boldsymbol{q}}\left[\left(\lambda+2\mu\right)\left|\varepsilon_{1}^{h}\right|^{2}+\mu\sum_{\alpha=2,3}\left|\varepsilon_{\alpha}^{h}\right|^{2}\right]\nonumber
   \\
	&\phantom{=}-\frac{\lambda_{0}}{2V}\sum_{\boldsymbol{q}}\left(\frac{2}{\sqrt{3}}\Phi_{1}^{*}\varepsilon_{1}^{h}+\sum_{\alpha=2,3}\Phi_{\alpha}^{*}\varepsilon_{\alpha}^{h}+\text{c.c.}\right).
\end{align}
Minimizing the free energy with respect to the helical strain amplitudes yields the following equations of state
\begin{equation}
    \begin{array}{ccc}
\varepsilon_{1}^{h}=\frac{2}{\sqrt{3}}\frac{\lambda_{0}}{\lambda+2\mu}\,\Phi_{1}, &  & \varepsilon_{2,3}^{h}=\frac{\lambda_{0}}{\mu}\,\Phi_{2,3}\end{array}.\label{eq:elastic_equations_of_state}
\end{equation}
The minimization thus produces an effective correction to the electronic nematic free energy given by 
\begin{equation}
    \Delta\mathcal{F}_{\text{eff}}\left[\varphi\right]=-\frac{\lambda_{0}^{2}}{2V}\sum_{\boldsymbol{q}}\left\{ \frac{4\left|\Phi_{1}\right|^{2}}{3\left(\lambda+2\mu\right)}+\frac{1}{\mu}\sum_{\alpha=2,3}\left|\Phi_{\alpha}\right|^{2}\right\} .\label{eq:effective_nematic_free_energy_Phi123}
\end{equation}
This term renormalizes the quadratic term of the bare isotropic nematic free energy in the $d$-orbital basis, yielding an effective nematic free energy of the form
\begin{align}
    \mathcal{F}_{\text{eff}}\left[\boldsymbol{\varphi}\right]	&=\mathcal{F}_{\text{nem}}\left[\boldsymbol{\varphi}\right]+\Delta\mathcal{F}_{\text{eff}}\left[\boldsymbol{\varphi}\right]\nonumber
    \\
	&=\frac{1}{2V}\sum_{\boldsymbol{q}}\boldsymbol{\varphi}^{\dagger}\cdot\left\{ \left(r+q^{2}\right)\text{I}-\mathcal{M}\left(\hat{q}\right)\right\} \cdot\boldsymbol{\varphi}+\mathcal{O}\left(\varphi^{3}\right).
\end{align}
In the bare free energy, $r>0$  is the tuning parameter that triggers a continuous bare nematic transition at $r=0$ (assuming that no higher order terms render the transition first-order).  To obtain this expression, we simply used the definition in \eqref{eq:nematic_Qvec_decomposition} to obtain the $5\times 5$ direction-dependent nematic-mass-correction matrix
\begin{equation}
   \mathcal{M}\left(\hat{q}\right)=\tfrac{\lambda_{0}^{2}}{\mu}\left( \boldsymbol{\hat{Q}}_{2}^{\phantom{\text{T}}}\boldsymbol{\hat{Q}}_{2}^{\text{T}}+\boldsymbol{\hat{Q}}_{3}^{\phantom{\text{T}}}\boldsymbol{\hat{Q}}_{3}^{\text{T}}+\varrho\boldsymbol{\hat{Q}}_{1}^{\phantom{\text{T}}}\boldsymbol{\hat{Q}}_{1}^{\text{T}}\right) . \label{eq:effective_electronic_nematic_mass_M_compatible_sector}
\end{equation}
The material-dependent nonuniversal ratio of the bare elastic constants, $\varrho$, is defined as 
\begin{equation}
    \varrho\equiv\frac{4\mu}{3\left(\lambda+2\mu\right)}\in\left(0,1\right), \label{eq:varrho_definition}
\end{equation}
with the bounds obtained from the inequality in \eqref{eq:varrho_less_than_1}. We simplify this expression with \eqref{eq:Qvec_completeness}, and obtain
\begin{equation}
   \mathcal{M}\left(\hat{q}\right)=\tfrac{\lambda_{0}^{2}}{\mu}\left\{ \text{I}-\left[\left(1-\varrho\right)\boldsymbol{\hat{Q}}_{1}^{\phantom{\text{T}}}\boldsymbol{\hat{Q}}_{1}^{\text{T}}+\hat{\boldsymbol{Q}}_{4}^{\phantom{\text{T}}}\boldsymbol{\hat{Q}}_{4}^{\text{T}}+\boldsymbol{\hat{Q}}_{5}^{\phantom{\text{T}}}\boldsymbol{\hat{Q}}_{5}^{\text{T}}\right]\right\} . \label{eq:effective_electronic_nematic_mass_M}
\end{equation}
This mass correction has two contributions: one isotropic in momentum and one anisotropic. Indeed, one observes that, upon substitution, the nematic free energy becomes 
\begin{align}
    \mathcal{F}_{\text{eff}}^{(2)}\left[\boldsymbol{\varphi}\right]	&=\frac{1}{2V}\sum_{\boldsymbol{q}}\left(r-\tfrac{\lambda_{0}^{2}}{\mu}+q^{2}\right)\left(\boldsymbol{\varphi}^{\dagger}\cdot\boldsymbol{\varphi}\right)\nonumber
\\
	&\phantom{=}+\frac{1}{2V}\sum_{\boldsymbol{q}} \frac{\lambda_0^2}{\mu} \left\{ \left(1-\varrho\right)\left|\Phi_{1}\right|^{2}+\sum_{\alpha=4,5}\left|\Phi_{\alpha}\right|^{2}\right\},
\label{eq:effective_quadratic_term_nematicity}
\end{align}
truncated at second order. The isotropic part of the free energy, proportional to $\boldsymbol{\varphi}^{\dagger}\cdot\boldsymbol{\varphi} = \sum_{a=1}^5 |\Phi_a|^2$, enhances the bare nematic transition from $r = 0$ to $r_c\equiv \lambda_0^2/\mu$. This is consistent with the self-energy correction due to homogeneous distortions in \eqref{eq:effective_long-range_interaction_homogeneous_distortion}. Being isotropic, this term does not couple the nematic basis functions in either the $d$-orbital or helical representations. The anisotropic part, however, \textit{does} couple the nematic order parameter written in the $d$-orbital basis. Being the sum of nonnegative numbers, the anisotropic term increases the energy of the nematic fluctuations. The \textit{minimum} of the electronic nematic free energy  thus \textit{selects} the momentum directions that satisfy the following three constraints
\begin{equation}
    \begin{aligned}\Phi_{1} & =\boldsymbol{\hat{Q}}_{1}\cdot\boldsymbol{\varphi}=\sqrt{3}\text{Tr}\left(\mathcal{Q}^{11}\varphi\right)=0,\\
\Phi_{4} & =\boldsymbol{\hat{Q}}_{4}\cdot\boldsymbol{\varphi}=2\text{Tr}\left(\mathcal{Q}^{23}\varphi\right)=0,\\
\Phi_{5} & =\boldsymbol{\hat{Q}}_{5}\cdot\boldsymbol{\varphi}=\text{Tr}\left[\left(\mathcal{Q}^{22}-\mathcal{Q}^{33}\right)\varphi\right]=0.
\end{aligned}\label{eq:Phi1=Phi4=Phi5=0}
\end{equation}
These three constraints reduce the number of degenerate electronic nematic components from five to two, \textit{even in an isotropic medium}. However, as long as these constraints are met, then the phase transition for the inhomogeneous strain case with $q>0$ appears at the same value $r_c$ as that computed for the homogeneous case with $q = 0$ (see \eqref{eq:effective_long-range_interaction_homogeneous_distortion}). Unlike the case of an order parameter that only couples to the fluctuating dilatation strain (see Appendix~\ref{app:bilinear_coupling_to_dilatation}), this mass renormalization is continuous at the origin, provided the three conditions in \eqref{eq:Phi1=Phi4=Phi5=0} are met. Thus, the homogeneous case is ``built-in'' to this more general form for inhomogeneous strain. The major physical difference between the inhomogeneous renormalization and the homogeneous renormalization, however, is the splitting of the fivefold nematic degeneracy clearly seen in the helical basis through the three conditions in \eqref{eq:Phi1=Phi4=Phi5=0}.

Remarkably, the splitting of the helical order parameters in \eqref{eq:Phi1=Phi4=Phi5=0} is produced by the same symmetry-allowed nemato-elastic coupling as in the case of homogeneous deformations. Moreover, the splitting occurs for nemato-elasticity within an isotropic medium with full $\text{SO}(3)$ rotational symmetry. The crucial difference between the homogeneous and inhomogeneous cases, however, is the presence of the wavevector in the latter, enforced by the compatible strain coupling to the nematic fluctuations. Without the displacement vector's role as the vector-valued potential for the strain, and the subsequent compatibility relations it imposes on elasticity, one would expect the fivefold nematic degeneracy of an isotropic system to persist, as is the case for homogeneous deformations.

To understand how the wavevector changes the rotational symmetry of the isotropic medium, we consider its (rotational) little group -- defined by the rotations that leave the wavevector invariant \cite{hamermeshGroupTheory1989,dresselhausGroupTheoryApplication2008}. This corresponds to $\text{SO}(2)$ with the rotational axis being parallel to $\hat{q} = \hat{e}_1$. Within the little group, the fivefold degeneracy of the helical order parameters is split. This is evident from the distinct nemato-elastic couplings in the helical representation, written in \eqref{eq:deviatoric_strain_vec_nematic_vec}. The $(\Phi_2,\Phi_3)$ doublet retains the $\lambda_0$ coupling constant experienced by \textit{all} five $d$-orbital order parameters, whereas the $\Phi_1$ component couples to the dilatation strain with strength $\tfrac{2}{\sqrt{3}}\lambda_0$, and the $(\Phi_4,\Phi_5)$ doublet loses its coupling to the strains outright. This hierarchy arises from the little group's $\text{SO}(2)$ symmetry, where the equivalence between the $\hat{e}_1, \hat{e}_2,\hat{e}_3$ directions is clearly broken given the $\hat{e}_1$-vector's invariance. Because of this, the helical nematic order parameters are also inequivalent. For example, because the quantity $\Phi_1$ only depends on $\hat{e}_1$, it transforms as a ``scalar'', i.e., an angular-momentum-projection zero ($m=0$ irreducible representation of  $\text{SO}(2)$). In contrast, the transverse vectors $\hat{e}_2$ and $\hat{e}_3$ together constitute the $m=1$ irreducible representation of $\text{SO}(2)$ \textit{at each momentum}. The doublet $\Phi_{2,3}$ thus transforms as a ``vector'' within the little group of the wavevector, given its dependence on the tensor product of $\hat{e}_{1}$ with either $\hat{e}_{2}$ or $\hat{e}_3$. Finally, the remaining $(\Phi_4,\Phi_5)$ doublet transforms as a ``quadrupole'' ($m=2$ irreducible representation), since it depends on tensor products of $\hat{e}_2$ and $\hat{e}_3$. 

Given that the helical nematic order parameters are split into inequivalent irreducible representations, it is instructive to understand how this inequivalence affects the low-energy degrees of freedom at the nematic critical point. This is the subject of the remainder of this paper. Up to now, what we have been established is that the fivefold degeneracy of the nematic order parameters is lost. Of the original five nematic degrees of freedom, \eqref{eq:Phi1=Phi4=Phi5=0} establishes that fluctuations of the $(\Phi_2, \Phi_3)$ ``vectorial doublet'' cause an instability. Meanwhile, the $\Phi_{1,4,5}$ order parameters remain gapped, because of their coupling with dilatation strain ($\Phi_1$) or their oblivion to compatible strains ($\Phi_{4,5}$). However, from the definition of the helical order parameters in \eqref{eq:nematic_Qvec_decomposition}, the $d$-orbital content of the non-critical helical order parameters change as the momentum changes. This strongly suggests that whether a given $d$-orbital nematic order parameter, such as $\varphi_{2xy}$, softens at the nematic critical point depends on the direction of its momentum. This restriction will occur even in the absence of crystalline anisotropy. This is the basis of the direction-selective criticality discussed in more details in the next section. 

Before moving onto the next subsection, we comment on nemato-elastic criticality in the $d$-orbital basis.  By \textit{criticality}, here we are specifically referring to a second-order thermal phase transition with a divergent correlation length as $r \rightarrow r_c$ from above. As a result, we only need to consider the structure of the quadratic coefficient in the field theory. While a non-elastic three-dimensional isotropic system allows for a cubic invariant in the free energy that drives the bare nematic transition first-order \cite{chaikinPrinciplesCondensedMatter1995, RMF14_classification_nematics}, nemato-elasticity can change the nematic transition to second-order in an isotropic solid. Details are provided in Appendix \ref{app:anharmonic_terms_free_energy}. The remainder of our discussion about what we dub \textit{compatible electronic nematic criticality} will focus on the region of the phase diagram where the transition remains continuous despite the presence of a sufficiently small cubic invariant. 

To conclude this section, we compute the static nematic susceptibilities for each order parameter in the $d$-orbital basis, 
\begin{align}
    \chi_{aa}(\boldsymbol{q})&\equiv \frac{1}{T}\langle \varphi_a(\boldsymbol{q})\varphi_a(-\boldsymbol{q})\rangle \nonumber
    \\
    &= \frac{1}{T}\sum_{b,c=1}^5\hat{Q}_{b,a}(\hat{q})\hat{Q}_{c,a}(\hat{q})\langle \Phi_b(\boldsymbol{q})\Phi_c(\boldsymbol{q})^*\rangle. \label{eq:orbital_nematic_susceptibility}
\end{align}
where $T$ is the temperature.  Writing the effective nematic free energy of \eqref{eq:effective_quadratic_term_nematicity} first in the helical basis, we find   
\begin{align}
    \mathcal{F}_{\text{eff}}^{(2)}\left[\boldsymbol{\varphi}\right]	&=\frac{1}{2V}\sum_{\boldsymbol{q}} \tilde{r}_c(\boldsymbol{q}) \sum_{a=2,3} |\Phi_a|^2 \nonumber
\\
	&\phantom{=}+\frac{1}{2V}\sum_{\boldsymbol{q}}\left\{ \tilde{r}_1(\boldsymbol{q})\left|\Phi_{1}\right|^{2}+r(\boldsymbol{q})\sum_{\alpha=4,5}\left|\Phi_{\alpha}\right|^{2} \right\},
\label{eq:effective_quadratic_term_helical_nematicity}
\end{align}
where $r(\boldsymbol{q})\equiv r + q^2$, $\tilde{r}_1(\boldsymbol{q}) \equiv r(\boldsymbol{q}) - \lambda_0^2\varrho/\mu$, and $\tilde{r}_c(\boldsymbol{q}) \equiv r(\boldsymbol{q}) - \lambda_0^2/\mu$. We note that these inverse susceptibilities obey a strict hierarchy: $\tilde{r}_c(\boldsymbol{q}) < \tilde{r}_1(\boldsymbol{q}) < r(\boldsymbol{q})$ for all $\boldsymbol{q}$ (since $\varrho <1$ in \eqref{eq:varrho_definition}). In the helical representation, we observe that the effective nematic free energy is diagonal, and therefore
\begin{equation}
    \langle \Phi_b(\boldsymbol{q})\Phi_c(\boldsymbol{q})^*\rangle \propto \delta_{bc} \langle |\Phi_b(\boldsymbol{q})|^2\rangle.
\end{equation}
Substituting this result into \eqref{eq:orbital_nematic_susceptibility}, the static susceptibilities in the $d$-orbital basis follow as 
\begin{align}
    \chi_{aa}(\boldsymbol{q})&\equiv \frac{\hat{Q}^2_{2,a}(\hat{q}) + \hat{Q}^2_{3,a}(\hat{q})}{r(\boldsymbol{q}) - \lambda_0^2/\mu} \nonumber
    \\
    &\phantom{=} + \frac{\hat{Q}^2_{1,a}(\hat{q})}{r(\boldsymbol{q}) - \lambda_0^2\varrho/\mu} + \frac{\hat{Q}^2_{4,a}(\hat{q}) + \hat{Q}^2_{5,a}(\hat{q})}{r(\boldsymbol{q})}. \label{eq:angle_dependent_orbital_electronic_nematic_susceptibility}
\end{align}
As the system is tuned from the isotropic phase across the nematic transition, the leading divergence will occur in the first term, arising from the compatible helical nematic fluctuations. This term is maximized for momentum directions where the form factors $\hat{Q}^2_{2,a}(\hat{q})$ and $\hat{Q}^2_{3,a}(\hat{q})$ are maximal. Thus, at the transition, the fluctuations in the $d$-orbital basis are dominated by the critical fluctuations of the helical amplitudes $\Phi_{2,3}$. However, it is also clear, meanwhile, that this equivalence only exists at the largest length scales: when $q\rightarrow 0$ along these selected momentum directions. Generally, however, $d$-orbital nematic fluctuations at any length scale will be a linear combination of all of the helical nematic fluctuations. This behavior distinguishes macroscopic long-wavelength electronic nematicity from mesoscopic short-wavelength electronic nematicity. 

\subsection{Suppression of Incompatible Electronic Nematicity \label{subsec:suppression_of_incompatible_nematicity}}

The nematic propagator will become critical first only along momentum directions where $\Phi_1=\Phi_4=\Phi_5=0$, from \eqref{eq:Phi1=Phi4=Phi5=0}. It therefore stands to question exactly what each of these projections in the co-rotating helical basis represent physically. The first, $\Phi_1$, follows from \eqref{eq:elastic_equations_of_state} as a symmetry-preserving dilatation stress that, in the long-wavelength limit, corresponds to energetically expensive volume changes of the medium associated with the bulk modulus.

However, there are no equations of state to define the amplitudes $\Phi_{4,5}$, both of which remain un-condensed at nemato-elastic criticality. To gain insight, we use \eqref{eq:nematic_varThetaTensor_decomposition}. By Fourier transforming the incompatibility operator from \eqref{eq:kroner_inc_real-space}, we obtain
\begin{equation}
    \text{inc}\,\left(A\right)_{ij}=-\epsilon_{ikl}\epsilon_{jmn}q_{k}q_{m}A_{ln},\label{eq:KronerInc_Fourier_Space}
\end{equation}
where $A_{ij}$ is an arbitrary second-rank tensor. 

When the incompatibility operator acts on $\vartheta_{ij}^a$, it will act on the various components of $\mathcal{Q}^{\alpha\beta}_{ij}$, through \eqref{eq:varThetaTensor_to_Qvechat_to_Qabij}. Applying \eqref{eq:inc_of_Qab_simplified_early} to the helical basis tensors establishes that $\{\mathcal{Q}^{1\alpha},\,\alpha=1,2,3\}$, $\mathcal{Q}^{23}$,  and $\mathcal{Q}^{22}-\mathcal{Q}^{33}$  are eigentensors of the incompatibility operator (see \eqref{eq:inc_Qab_5_tensors}). The set  $\{\mathcal{Q}^{1\alpha},\,\alpha=1,2,3\}$ has vanishing incompatiblity, whereas the latter two tensors each have eigenvalue $q^2>0$. Additionally, we observe that the incompatibility of the identity tensor at $\hat{q}$ is
\begin{equation}
    \text{inc}\left(\text{I}\right)_{ij}=q^{2}\left(\mathcal{Q}_{ij}^{11}-\delta_{ij}\right).\label{eq:incI_fourier_space}
\end{equation}
Unlike the $\mathcal{Q}^{\alpha\beta}$ tensors, the identity is not an eigentensor of the incompatibility operator.

Using these identities, the incompatibility of the $\{\vartheta^a\}$ tensors follows straightforwardly (see \eqref{eq:inc_varTheta_tensors}). Thus, the incompatibility of a generic rank-2 symmetric tensor, $A$, is 
\begin{align}
    \text{inc}\,A	=\frac{1}{3}\left(\text{Tr}\,A-\tfrac{\sqrt{3}}{2}A_{1}\right)\text{inc}\left(\text{I}\right)+\frac{q^{2}}{\sqrt{2}}\left(A_{4}\vartheta^{4}+A_{5}\vartheta^{5}\right).\label{eq:incompatibility_of_A_varTheta_decomp}
\end{align}
Applying the incompatibility operator to the electronic nematic order parameter, we find that 
\begin{equation}
    \text{inc}\left(\varphi\right)= -\frac{1}{2\sqrt{3}}\Phi_{1}\text{inc}\left(\text{I}\right)+\frac{q^{2}}{\sqrt{2}}\left(\Phi_{4}\vartheta^{4}+\Phi_{5}\vartheta^{5}\right) .
\end{equation}
The conclusion is, therefore, that arbitrary electronic nematic fluctuations will have \textit{incompatible } contributions -- those that \textit{cannot} be written as a symmetric derivative of some vector field. These are clearly the eigentensor components $\Phi_{4,5}$ which, along with the volume-changing dilatations $\Phi_1$, are suppressed when the system undergoes nemato-elastic criticality. The $\Phi_1$ amplitude also reflects a degree of \textit{incompatible} electronic nematicity. This is despite $\boldsymbol{Q}^{11}$ appearing in the strain tensor (\eqref{eq:strain_tensor_GellMann_decomp_Qabvecs}), which is, by definition, \textit{compatible}. The reason for this behavior is because the nematic order parameter is strictly traceless, whereas the strain tensor has a trace: the dilatation. Indeed, using \eqref{eq:incompatibility_of_A_varTheta_decomp} for the strain tensor, we immediately see that $\text{inc}(\varepsilon) = 0$, since 
\begin{equation}
    \text{Tr}\,\varepsilon=\varepsilon_{1}^{h}=\frac{\sqrt{3}}{2}\varepsilon_{1},\quad\varepsilon_{4}=\varepsilon_{5}=0.\label{eq:condition_on_Trace_for_compatibility}
\end{equation}
Thus, it is the dilatation that renders the strain compatible, whereas the $\Phi_1$  contribution of nematicity reflects its intrinsic incompatibility as a traceless tensor. 

Likewise, it is important to note that if a rank-2 symmetric tensor, $B$, is \textit{completely} \textit{incompatible}, then it must be true that $B_1 = 0$, just as $B_2 = B_3 = 0$. Therefore, the symmetric tensor
\begin{equation}
    B_{ij}=\frac{1}{3}\left(\text{Tr}\,B\right)\delta_{ij}+\frac{1}{\sqrt{2}}\left(B_{4}\vartheta_{ij}^{4}+B_{5}\vartheta_{ij}^{5}\right),\label{eq:totally_incompatible_tensor_varTheta_decomp}
\end{equation}
is a tensor with zero compatibility. Note that, just as a completely compatible tensor has only three independent components, so too does a completely incompatible tensor.

To contextualize this tensor analysis within electronic nematics, the completely incompatible contributions to $\varphi_{ij}$, $\Phi_{4,5}$, are the most suppressed because they do not participate in nemato-elasticity. The helical ``scalar'' component, $\Phi_1$, is partially incompatible, and is therefore also suppressed because its compatible contribution induces costly dilatation strain, as is evident from the appearance of $\varrho$ in \eqref{eq:effective_quadratic_term_nematicity}. The critical nemato-elastic order parameters are only the completely compatible contributions to nematicity: $\Phi_{2,3}$. It is only the condensation of these fields that will drive the nemato-elastic instability in the long-wavelength limit. This can be understood in the following way. Because electronic nematic fluctuations induce compatible strains, they induce the vector-valued displacement field that serves as the vector-valued potential for the strain tensor. In turn, the critical tensor-valued electronic nematicity must itself be compatible. Hence, the critical nematic tensor itself must be proportional to the symmetric gradient of that same vector-valued potential field. Therefore, the planar constraints on critical nematicity in \eqref{eq:Phi1=Phi4=Phi5=0} arise to suppress \textit{incompatible} electronic nematicity.  We have thus established that electronic nematicity inherits the compatibility relations of elasticity theory even in the isotropic continuum. It thereby follows that these same restrictions are enforced within the crystalline point groups that comprise actual quantum materials.

\section{Direction-Selective Nematic Criticality \label{sec:polar_and_planar_nematicity}}

\begin{figure*}[t!]
    \centering
    \includegraphics[width=\linewidth]{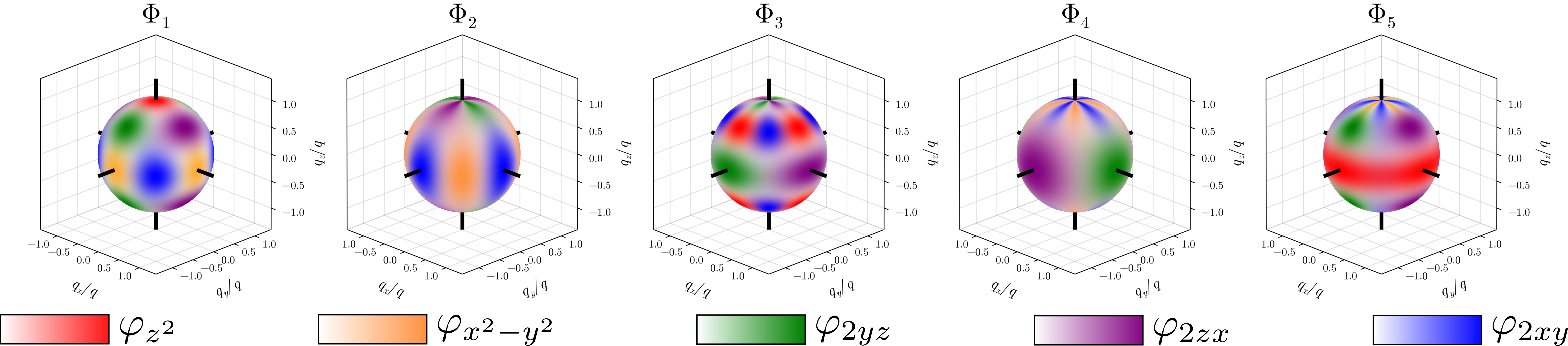}
    \caption{\justifying Linear transformation from the $d$-orbital nematic basis, $\boldsymbol{\varphi}$, into the helical basis, $\boldsymbol{\Phi}$, as functions of the wavevector, $\boldsymbol{q}$. The $\Phi_{1,2,4}$  panels are also shown in Ref.~\citep{ShortPaper}. Each panel visualizes the square-amplitude $(\hat{Q}_{a,b})^2$ to determine the contribution to the $\Phi_a$  helical order parameter from the $\varphi_b$  $d$-orbital order parameter.  For $\Phi_3$, the $\varphi_{z^2}$  and $\varphi_{x^2 - y^2}$ ($\varphi_{2xy}$) amplitudes have coincident maximum magnitude along, for example, the  $[101]$ ($[111]$)  directions.}
    \label{fig:supp_varphi_to_Phi_mapping}
\end{figure*}

Integrating out the compatible elastic modes revealed that the effect of nemato-elastic coupling is to suppress \textit{incompatible} nematic modes at the phase transition. In the helical basis, this ``incompatible sector'' is clearly identified as the amplitudes $\Phi_1$, $\Phi_4$, and $\Phi_5$. The $d$-orbital content of these modes, however, is momentum-direction-dependent. This means that only a subset of the nematic order parameters in the $d$-orbital basis are critical along a given momentum direction. Because the mapping between the helical and $d$-orbital bases is non-analytic at $q=0$, one \textit{must} consider the $q\rightarrow 0, \,q \neq 0$ limit to identify which $d$-orbital basis functions are compatible -- and therefore critical -- along a given momentum-direction. This effect, which is already present in the isotropic lattice, is what enables the direction-selective nematic criticality obtained previously in crystalline lattices \citep{Karahasanovic16, Paul17, Hecker2018, Fernandes2020, Hecker2022, stewardElasticQuantumCriticality2025}. The connections between the isotropic lattice and these previously studied anisotropic crystals is reviewed in Appendix \ref{app:helical_dynamical_matrix_crystals}. 

We emphasize that, in the fully isotropic lattice treated here, all momentum directions are by definition degenerate. Therefore, one can always find a specific directions where one of the five $d$-orbital order parameters is unstable. This is of course expected from the Cowley description of ferroelastic phase transitions \cite{Cowley76}, since there exists an orthogonal plane to each momentum direction that separates the transverse acoustic modes from the longitudinal modes. Once crystalline anisotropies are included, the degeneracy between all momentum directions is lifted, and the phonon velocity only vanishes along specific directions, implying direction-selective ferroelastic criticality. The situation is analogous in the case of electronic nematics: in the isotropic lattice, the elastic compatibility conditions induce a coupling between the momentum direction and the nematic director (which defines which of the five nematic components condenses). In the isotropic lattice, this coupling cannot, by construction, break the rotational symmetry of the lattice. But, because of this coupling between momentum direction and nematic director, any infinitesimal crystalline anisotropy will result in direction-selective nematic transition. This is the reason why we attribute the phenomenon of direction-selective criticality to the elastic compatibility conditions.

To demonstrate how the coupling between momentum and nematic order-parameter components restrict the allowed nematic soft modes along a given direction, in this section we will consider two orientations of the momentum, and assume that the compatible nematic sector becomes unstable along these directions in momentum space. We will show how one then maps from the helical basis back to the $d$-orbital basis to identify which of these more conventional nematic order parameters are compatible and will generate symmetry-breaking deviatoric strain through long-wavelength ($q\rightarrow 0$) transverse elastic displacements. The first is the ``polar'' limit where the critical wavevector is confined to a single direction  -- the $z$-axis. The second is the ``planar'' limit where the critical wavevector is free to rotate within a plane. This second limit corresponds to the case when the wavevector lies entirely within the $xy$-plane. 

By considering these two cases, we will establish direction-selective electronic nematic criticality within an isotropic elastic medium. We will show that because the $d$-orbital content of the helical nematic order parameters explicitly depends on the momentum, and that the critical manifold is spanned only by $(\Phi_2,\Phi_3)$, then the orientation of the critical nematic director relative to its momentum will be constrained.

Generally speaking, we write the transformation between the helical and $d$-orbital nematic order parameters from \eqref{eq:nematic_Qvec_decomposition} as $\boldsymbol{\varphi}\equiv \mathbb{D}(\hat{q}) \cdot \boldsymbol{\Phi}$, where the columns of the $5\times 5$  orthogonal matrix, $\mathbb{D}(\hat{q})$, are the co-rotating $\{\boldsymbol{\hat{Q}}_a,\;a=1,\dots,5\}$  basis vectors. Written explicitly, where $\hat{q} = (\cos\phi\sin\theta,\sin\phi\sin\theta,\cos\theta)$, the matrix is

\begin{widetext}
\begin{align}
    \mathbb{D}\left(\hat{q}\right)	&=\left(\boldsymbol{\hat{Q}}_{1}\,|\,\hat{\boldsymbol{Q}}_{2}\,|\,\boldsymbol{\hat{Q}}_{3}\,|\,\boldsymbol{\hat{Q}}_{4}\,|\,\boldsymbol{\hat{Q}}_{5}\right)\nonumber
    \\
	&=\begin{bmatrix}-\frac{1+3\cos2\theta}{4} & 0 & -\frac{\sqrt{3}}{2}\sin2\theta & 0 & \frac{\sqrt{3}}{2}\sin^{2}\theta\\
\frac{\sqrt{3}}{2}\cos2\phi\sin^{2}\theta & -\sin2\phi\sin\theta & -\frac{1}{2}\cos2\phi\sin2\theta & \sin2\phi\cos\theta & -\frac{1}{4}\cos2\phi\left(3+\cos2\theta\right)\\
\frac{\sqrt{3}}{2}\sin\phi\sin2\theta & \cos\phi\cos\theta & -\sin\phi\cos2\theta & \cos\phi\sin\theta & \frac{1}{2}\sin\phi\sin2\theta\\
\frac{\sqrt{3}}{2}\cos\phi\sin2\theta & -\sin\phi\cos\theta & -\cos\phi\cos2\theta & -\sin\phi\sin\theta & \frac{1}{2}\cos\phi\sin2\theta\\
\frac{\sqrt{3}}{2}\sin2\phi\sin^{2}\theta & \cos2\phi\sin\theta & -\frac{1}{2}\sin2\phi\sin2\theta & -\cos2\phi\cos\theta & -\frac{1}{4}\sin2\phi\left(3+\cos2\theta\right)
\end{bmatrix}.\label{eq:Dmatrix}
\end{align}
\end{widetext}
The inverse mapping from the usual $d$-orbital basis into the co-rotating helical basis for nematicity is given concisely as $\boldsymbol{\Phi} = \mathbb{D}^{\text{T}}(\hat{q})\cdot \boldsymbol{\varphi}$. The individual elements of $\mathbb{D}$  can therefore be interpreted as an ``overlap amplitude'' identifying how strongly each $d$-orbital order parameter, $\varphi_{b}$, contributes to the helical order parameter, $\Phi_a$, at each wavevector, $\hat{q}$. In Fig.~\ref{fig:supp_varphi_to_Phi_mapping} we visualize this mapping by fixing the $d$-orbital nematic order parameter and then plotting the square amplitude for each of the helical basis functions. These square amplitudes are the same as the form factors that appear in the $d$-orbital nematic susceptibilities from \eqref{eq:angle_dependent_orbital_electronic_nematic_susceptibility}, though in that case we used the elements of the $\mathbb{D}(\hat{q})$ matrix, rather than $\mathbb{D}^{\text{T}}(\hat{q})$ matrix. For instance, to show the contribution of $\varphi_{z^2}$ to each of the helical order parameters $\Phi_{1,\dots,5}$, in Fig.~\ref{fig:supp_varphi_to_Phi_mapping}, we fix $\boldsymbol{\varphi} = (1,0,0,0,0)^{\text{T}}$, and then compute $\boldsymbol{\Phi} = \mathbb{D}^{\text{T}}(\hat{q}) \cdot \boldsymbol{\varphi} = \boldsymbol{\hat{Q}}_1$. The resulting colormap shows $(\hat{Q}_{1,a})^2$  on the sphere for each $\Phi_{a}$. The same procedure follows for the other four $d$-orbital nematic basis functions in the figure.

The phenomenon of direction-selective nematic criticality follows directly from Fig.~\ref{fig:supp_varphi_to_Phi_mapping}. As explained above, the nematic instability is described by the condensation of a symmetry-breaking order parameter $\varphi_a$ in the $d$-orbital basis. Because of the nemato-elastic coupling, this instability is achieved along certain directions for which the overlap between $\varphi_a$ and the three incompatible nematic order parameters in the helical basis $\Phi_{1,4,5}$ vanish. In other words, the instability is set by the momenta for which  $(\hat{Q}_{1,a})^2=(\hat{Q}_{3,a})^2=(\hat{Q}_{5,a})^2=0$. Take, for instance, the case of Ising-nematicity $\varphi_{2xy}$ on a tetragonal lattice. Clearly, from Fig.~\ref{fig:supp_varphi_to_Phi_mapping}, these conditions are satisfied when the momentum direction $\hat{q}$ lies along the $[100]$ and $[010]$ directions. Note also that, along these directions, there is a perfect overlap between $\varphi_{2xy}$ and $\Phi_2$. Furthermore, because of the mutual orthogonality of the helical basis vectors, $\{ \boldsymbol{\hat{Q}}_a \}$, if there is perfect overlap between $\varphi_{2xy}$ and $\Phi_2$, that means that $\varphi_{2xy}$ is orthogonal to \textit{all other} helical order parameters: $\Phi_1$, $\Phi_3$, $\Phi_4$ and $\Phi_5$.

To make broader statements about 3D nemato-elastic criticality, one must compute the volume-changing and incompatible contributions to the helical electronic nematic order parameter arising from the $d$-orbital order parameters. We obtain:

\begin{widetext}
\begin{align}
    \Phi_{1}	&=-\tfrac{1}{4}\varphi_{z^{2}}\left(1+3\cos2\theta\right)+\tfrac{\sqrt{3}}{2}\sin\theta\left\{ \sin\theta\left(\varphi_{x^{2}-y^{2}}\cos2\phi+\varphi_{2xy}\sin2\phi\right)+2\cos\theta\left(\phi_{2yz}\sin\phi+\phi_{2zx}\cos\phi\right)\right\} ,\label{eq:Phi1_explicit}
    \\
\Phi_{4}	&=\sin\theta\left(\varphi_{2yz}\cos\phi-\varphi_{2zx}\sin\phi\right)+\cos\theta\left(\varphi_{x^{2}-y^{2}}\sin2\phi-\varphi_{2xy}\cos2\phi\right),\label{eq:Phi4_explicit}
\\
\Phi_{5}	&=\sin\theta\left\{ \tfrac{\sqrt{3}}{2}\varphi_{z^{2}}\sin\theta+\cos\theta\left(\varphi_{2yz}\sin\phi+\varphi_{2zx}\cos\phi\right)\right\} -\tfrac{3+\cos2\theta}{4}\left(\varphi_{x^{2}-y^{2}}\cos2\phi+\varphi_{2xy}\sin2\phi\right).\label{eq:Phi5_explicit}
\end{align}
\end{widetext}
To proceed, it is convenient to treat separately two different limits, which we dub ``polar'' and ``planar''. In the polar limit, we fix the wavevector to be nearly coincident with the $z$-axis. Then, using  $\theta=\delta\theta\ll 1$  with $\phi\in[0,2\pi)$, the expressions for the incompatible contributions to nematicity follow as 
\begin{equation}
\begin{aligned}
    \Phi_{1}\left(\theta=\delta\theta\right)	&=-\varphi_{z^{2}}+\mathcal{O}\left(\delta\theta\right),
    \\
\Phi_{4}\left(\theta=\delta\theta\right)	&=\varphi_{x^{2}-y^{2}}\sin2\phi-\varphi_{2xy}\cos2\phi+\mathcal{O}\left(\delta\theta\right),
\\
\Phi_{5}\left(\theta=\delta\theta\right)	&=-\left(\varphi_{x^{2}-y^{2}}\cos2\phi+\varphi_{2xy}\sin2\phi\right)+\mathcal{O}\left(\delta\theta\right).
\end{aligned}
\end{equation}
Since the nemato-elastic contribution to the free energy, \eqref{eq:effective_quadratic_term_nematicity}, enforces that these three components vanish, we conclude that $\varphi_{z^{2}}=0$, $\varphi_{x^2-y^2}=0$, and $\varphi_{2xy}=0$ in the polar limit. In other words, an electronic nematic instability with wavevector close to the $q_z$-axis can only take place in the channels $\varphi_{2xz}$ and $\varphi_{2yz}$. Thus,
writing the polar momentum as $\hat{q}_{\perp}\equiv\delta\theta\left(\hat{x}\cos\phi+\hat{y}\sin\phi\right)+\hat{z}+\mathcal{O}\left(\delta\theta^{2}\right)$, the effective free energy in \eqref{eq:effective_quadratic_term_nematicity}  reduces to 
\begin{align}
   \mathcal{F}_{\text{eff}}^{\left(2\right)}\left[\boldsymbol{\varphi}\right]\rightarrow\frac{1}{2V}\sum_{\boldsymbol{q}_{\perp}}\left(r-\tfrac{\lambda_{0}^{2}}{\mu}+q^{2}\right)\left(\left|\varphi_{2yz}\right|^{2}+\left|\varphi_{2zx}\right|^{2}\right), \label{eq:effective_free_energy_q=zhat}
\end{align}
where the summation over wavevectors  is constrained to the infinitesimal spherical cap centered on $\hat{z}$. The direction selectivity of the nematic instability is clear. When the momentum is constrained to the $\hat{z}$-axis, the fivefold degeneracy of the electronic nematic order parameter is broken down to twofold. In turn, the resulting critical nematic fluctuations, through \eqref{eq:elastic_equations_of_state} and \eqref{eq:helical_strain_tensor_e1iekj}, induce only the strain tensor components $\varepsilon_{yz}\propto\varphi_{2yz}\neq0$  and $\varepsilon_{zx}\propto\varphi_{2zx}\neq0$. Along these critical momentum directions, one can take the limit $q\rightarrow 0$  to obtain the macroscopic shear distortion of the lattice induced by nematicity and recover the same strained medium as in the homogeneous limit \eqref{eq:effective_long-range_interaction_homogeneous_distortion}. Thus the limiting case of homogeneous deformation is contained within the more general inhomogeneous case.

These conditions show that if the critical momentum direction is along $\hat{z}$, then the critical nematic manifold in the $d$-orbital basis is spanned by the out-of-plane doublet $\boldsymbol{\varphi}_\perp^{\text{T}} \equiv (\varphi_{2zx}, \varphi_{2yz})$. The unit director associated with these nematic components, $\boldsymbol{\mathcal{N}}$, can be written in terms of the spherical polar angles $\alpha$ and $\beta$ in the Cartesian coordinates as 
\begin{equation}
    \boldsymbol{\mathcal{N}} \equiv \left( \cos\alpha \sin\beta, \sin\alpha \sin\beta, \cos\beta \right)^{\text{T}}.
\end{equation}
Identifying the electronic nematic tensor in Cartesian coordinates as $\varphi_{ij} \equiv \varphi_\perp (\mathcal{N}_i \mathcal{N}_j - \tfrac{1}{3}\delta_{ij})$ \cite{RMF14_classification_nematics} shows that the critical doublet can be written as
\begin{equation}
    \boldsymbol{\varphi}_\perp = \varphi_\perp \sin(2\beta) \left( \cos\alpha, \sin\alpha  \right)^{\text{T}}.
\end{equation}
Substituting this expression into \eqref{eq:effective_free_energy_q=zhat} yields a free energy density given by 
\begin{equation}
    f_{\text{eff}}^{(2)}[\boldsymbol{\varphi}_\perp] = \left(r - \tfrac{\lambda_0^2}{\mu} + q^2 \right)\sin^2(2\beta) |\varphi_\perp|^2.
\end{equation}
Above the transition, $r > \lambda_0^2/\mu$, $\langle \varphi_\perp \rangle = 0$. However, below the transition, $r < \lambda_0^2/\mu$, and the electronic nematic condenses with $\langle \varphi_\perp \rangle \neq 0$, allowing for a well-defined director angle. At the mean-field level, the director will lie along $\beta = \pi/4$ such that $\sin^2(2\beta)$ is maximized. This defines the nematic director to be at $45^{\text{o}}$ relative to the momentum direction, $\hat{q} = \hat{z}$. This example illustrates  the emergence of direction-selective criticality in an isotropic medium. We note that the lack of an $\alpha$-dependence in the free energy reflects the residual $\text{SO}(2)$ symmetry in the isotropic medium associated with rotations about $\hat{z}$.

 Expressing the two helical order parameters in terms of the $d$-orbital basis and the director, with \eqref{eq:Dmatrix}, one obtains
\begin{equation}
    \begin{aligned}
        \Phi_2 &= \varphi_{2yz} \cos\phi - \varphi_{2zx}\sin\phi = \varphi_\perp\sin(\alpha - \phi),
        \\
        \Phi_3 &= -\varphi_{2yz} \sin\phi - \varphi_{2zx}\cos\phi = -\varphi_\perp \cos( \alpha - \phi),
    \end{aligned}
\end{equation}
where we used the fact that $\beta = \pi/4$ and $\hat{q} = \hat{z}$. It follows then that certain momentum directions, encoded in $\phi$, map the helical order parameters uniquely onto the individual $d$-orbital order parameters -- even on the spherical cap in the polar limit that we are considering. For example, taking the momentum to lie along $[\delta\theta,0,1]$, i.e., $\phi=0$ and infinitesimal $\delta\theta$, yields $\Phi_2 \propto \varphi_{2yz}$ and $\Phi_3 \propto \varphi_{2zx}$. 

By rotating the momentum away from the polar limit, one further expects the critical nematic director to remain oriented at $45^{\text{o}}$ relative to the critical momentum directions. We show that this remains true in the more familiar planar limit, with critical momenta in the $(q_x,q_y)$ plane, $\hat{q}=\hat{x}\cos\phi + \hat{y}\sin\phi$. \eqref{eq:Phi1_explicit}, \eqref{eq:Phi4_explicit}, and \eqref{eq:Phi5_explicit} now become
\begin{equation}
   \begin{aligned}\Phi_{1}\left(\theta=\tfrac{\pi}{2}\right) & =\tfrac{1}{2}\varphi_{z^{2}}+\tfrac{\sqrt{3}}{2}\boldsymbol{\mathfrak{D}}^{\text{T}}\left(\phi\right)\cdot\boldsymbol{\varphi}_{\parallel},\\
\Phi_{4}\left(\theta=\tfrac{\pi}{2}\right) & =(\hat{z}\times\hat{e}_{1})\cdot\boldsymbol{\varphi}_{\perp},\\
\Phi_{5}\left(\theta=\tfrac{\pi}{2}\right) & =\tfrac{\sqrt{3}}{2}\varphi_{z^{2}}-\tfrac{1}{2}\boldsymbol{\mathfrak{D}}^{\text{T}}\left(\phi\right)\cdot\boldsymbol{\varphi}_{\parallel},
\end{aligned}
\end{equation}
where $\boldsymbol{\mathfrak{D}}^{\text{T}}(\phi)\equiv(\cos2\phi,\sin2\phi)$, $\boldsymbol{\varphi}_\parallel^{\text{T}}\equiv (\varphi_{x^2-y^2},\varphi_{2xy})$, and $\boldsymbol{\varphi}_\perp$ is the same out-of-plane doublet used in the polar limit. The effective nematic free energy in \eqref{eq:effective_quadratic_term_nematicity}  takes the form
\begin{align}
    \mathcal{F}_{\text{eff}}^{\left(2\right)}\left[\boldsymbol{\varphi}\right]	&\rightarrow\frac{1}{2V}\sum_{\boldsymbol{q}_{\parallel}}\bigg\lbrace\left(r-\tfrac{\lambda_{0}^{2}}{\mu}+q^{2}\right)\left(\boldsymbol{\varphi}^{\dagger}\cdot\boldsymbol{\varphi}\right) \nonumber
    \\
	&\phantom{\rightarrow\frac{1}{2V}\sum_{\boldsymbol{q}}}+\tfrac{\lambda_{0}^{2}}{\mu}\left(1-\tfrac{3\varrho}{4}\right)\left|\boldsymbol{\mathfrak{D}}^{\text{T}}\left(\phi\right)\cdot\boldsymbol{\varphi}_{\parallel}\right|^{2} \nonumber
    \\
	&\phantom{\rightarrow\frac{1}{2V}\sum_{\boldsymbol{q}}}+\tfrac{\lambda_{0}^{2}}{\mu}\left[\left(1-\varrho\right)\left|\varphi_{z^{2}}\right|^{2}+\left|\hat{e}_{2}\cdot\boldsymbol{\varphi}_{\perp}\right|^{2}\right] \nonumber
    \\
	&\phantom{\rightarrow\frac{1}{2V}\sum_{\boldsymbol{q}}}-\tfrac{\varrho\sqrt{3}}{4}\left[\varphi_{z^{2}}^{*}\boldsymbol{\mathfrak{D}}^{\text{T}}\left(\phi\right)\cdot\boldsymbol{\varphi}_{\parallel}+\text{c.c.}\right]\bigg\rbrace. \label{eq:planar_free_energy_intermediate}
\end{align}
In the third line above, we simplified the condition that $\Phi_4 = 0$ since  $\hat{z}\times\hat{e}_{1} = \hat{e}_{2}$ when $\hat{q} = \hat{e}_1$ is in the $xy$-plane. From these expressions, we see that nemato-elastic criticality will occur only when $\varphi_{z^2} = 0$,  $\boldsymbol{\mathfrak{D}}^{\text{T}}\left(\phi\right)\cdot\boldsymbol{\varphi}_{\parallel}=0$, and $\hat{e}_{2}\cdot\boldsymbol{\varphi}_{\perp} = 0$. Clearly, the planar limit selects only nematic instabilities with $\varphi_{z^2}=0$.  Meanwhile, the second condition is known from two-dimensional hexagonal systems as ``nemato-orbital'' coupling that renormalizes the 3-state Potts nematic director \citep{Fernandes2020}. It is identical to the 2D compatibility relation in \eqref{eq:2D_SVCR_polar} with the deviatoric and shear strain amplitudes replaced by the $d$-orbital $\varphi_{x^2-y^2}$ and $\varphi_{2xy}$ amplitudes, respectively. In a similar spirit to the polar limit, we recast the two-component in-plane nematic order parameter in terms of an amplitude and the director $\boldsymbol{\mathcal{N}}$ as 
\begin{equation}
    \boldsymbol{\varphi}_\parallel \equiv \varphi_\parallel \sin^2(\beta) \left( \cos2\alpha, \sin2\alpha  \right)^{\text{T}}.
\end{equation}
The second constraint becomes 
\begin{equation}\boldsymbol{\mathfrak{D}}^{\text{T}}\left(\phi\right)\cdot\boldsymbol{\varphi}_{\parallel} = \varphi_\parallel \sin^2(\beta) \cos\left(2\phi - 2\alpha\right) =0,
\end{equation}
restricting the nematic director's orientation in terms of the momentum direction, namely, $\alpha = \phi \pm \pi/4$. 

While the method for determining the direction-selective criticality in this work is based on helical-basis elastostatics, in Appendix \ref{app:helical_dynamical_matrix_crystals}, we show that the usual method employed in previous works -- which involves diagonalizing a dynamical matrix to determine anisotropic phonon velocities \cite{Cowley76, Folk76, Karahasanovic16, Paul17, Hecker2018, Fernandes2020, Hecker2022, stewardElasticQuantumCriticality2025} -- can be circumvented by using the helical basis in crystals. In particular, in Appendix \ref{app:helical_dynamical_matrix_crystals}, we show that the elastic compatibility relations determine exactly the directions along which a phonon velocity can vanish, which in turn are associated with the directions along which the longitudinal and transverse acoustic phonons are decoupled eigenmodes of the crystal. This follows from two results of Appendix \ref{app-subsec:helical_anisotropic_elasticity}: (i) The frequency-dependent phonons do not determine the critical directions, instead they inherit them from the constraints of elastic compatibility already present in the isotropic elastostatic regime. (ii) The helical form factors in \eqref{eq:Dmatrix} can be used to determine exactly where a symmetry-breaking strain component is associated with a purely transverse elastic displacement vector, whereas the same strain component for any other momentum direction will necessarily have some dilatational character associated with a longitudinal displacement vector. In short, direction-selective criticality in crystals is therefore a consequence of the non-analytic elastic compatibility conditions, a property that is inherited by the acoustic phonons.

It is unsurprising, therefore, that in nemato-elastic problems, the same constraints are projected onto the electronic nematic order parameter that generates elastic strain. In Appendix \ref{app-subsec:helical_anisotropic_nemato-elasticity}, we show again that one can consider only the helical form factors for the various $d$-orbitals to determine the direction-selective criticality in anisotropic crystals as a consequence of elastic compatibility. The latter, of course, is already present even in isotropic lattices. In crystals, the crystalline point-group symmetry will determine which of the five $d$-orbitals transform as non-trivial irreducible representations \cite{RMF14_classification_nematics}. The nemato-elastic action first projects out the purely incompatible components of the electronic nematic order parameter, already constraining the phase space of nematic fluctuations. Using the helical decomposition, one then finds that the momentum directions where the conjugate strain component corresponds to a pure transverse elastic mode will be exactly the same momentum directions where the electronic nematic $d$-orbital mode ceases to generate dilatation strain, which imposes an additional energy cost from the bulk modulus. 

The results discussed in this section in the isotropic solid, and shown in Fig. \ref{fig:supp_varphi_to_Phi_mapping}, therefore represent the maximum set of critical momentum directions associated with the full set of $d$-orbital basis functions. To apply this method to crystals, as shown in Appendix \ref{app-subsec:helical_anisotropic_nemato-elasticity}, one uses the point-group symmetry to select the subset of critical $d$-orbital basis functions, corresponding to components that transform as non-trivial irreducible representations of the point group. Then the manifold of critical momenta follows from the set of directions for which these nematic components project exclusively onto the compatible doublet $(\Phi_2, \Phi_3)$ and not on the incompatible and dilatational sectors.  

\begin{figure}
     \centering
     \includegraphics[width=\columnwidth]{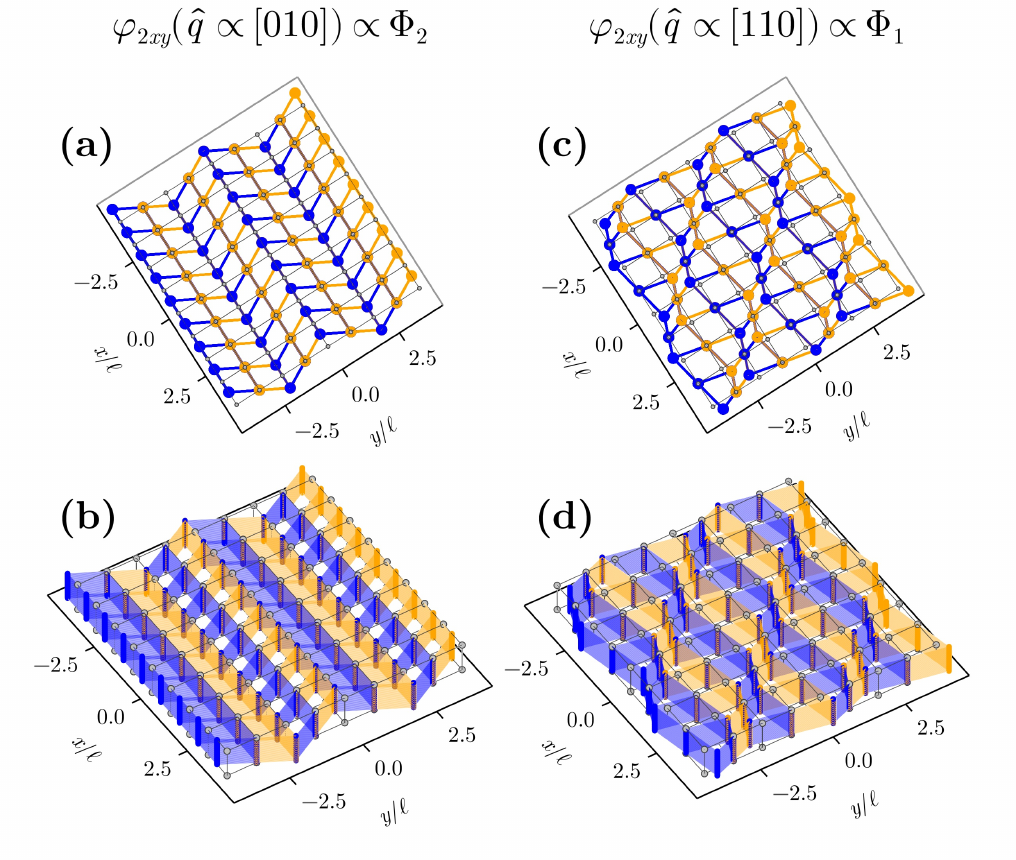}
     \caption{\justifying Elastic deformations of a lattice in the presence of a compatible planar $\varphi_{2xy}$ nematic order parameter with well-defined momentum, $\boldsymbol{q}$. In all figures, the modulation is sinusoidal, with identical amplitudes that are greatly exaggerated for visualization purposes. The equilibrium positions are indicated by the gray points, and the displaced locations are shown in blue and orange, obtained from solving the nemato-elastic equations of state, \eqref{eq:elastic_equations_of_state}. The contrast of blue and orange is solely to emphasize displacements of volume elements of linear size $\ell$. Top-down \textbf{(a) }and side \textbf{(b)} views of the strained three-dimensional isotropic medium undergoing a critical $\varphi_{2xy}$ fluctuation mode, proportional to the helical $\Phi_2$ nematic amplitude with $\hat{q} \propto [010]$. Top-down \textbf{(c)} and side \textbf{(d)} views of a non-critical $\varphi_{2xy}$ fluctuation mode, proportional to the helical $\Phi_1$ nematic amplitude with $\hat{q}\propto [110]$. The suppression of the $\varphi_{2xy}$ wave in \textbf{(c,d)} is apparent from the accompanying dilatation strain, whereas the critical wave in \textbf{(a,b)} does not induce local-volume changes. }
     \label{fig:compatible_planar_deformations}
 \end{figure}

Returning to the planar limit of helical elastostatics, we demonstrate the deformation generated by a static planar nematic wave in a 3D isotropic medium in Fig.~\ref{fig:compatible_planar_deformations} by showing the simplified case of a tetragonal-to-orthorhombic $\varphi_{2xy}$-nematic, corresponding to $\alpha = \pi/4$. There is a clear difference between the nemato-elastic strain induced by planar modulations along critical directions ($\varphi_{2xy} \propto \Phi_2$) with that induced by modulation along non-critical directions ($\varphi_{2xy} \propto \Phi_1$). In the latter case, $\boldsymbol{\mathfrak{D}}(\pi/4)\cdot\boldsymbol{\varphi}_\parallel \neq 0$ because local dilatation strains accompany the shear strain, as is required by the compatibility relations and the helical equations of state in \eqref{eq:elastic_equations_of_state}. The lattice displacement vector in the figure is obtained by directly integrating the sinusoidal strain wave, as was done for a purely 2D system in Ref. \cite{KaanPaper}.

Returning to \eqref{eq:planar_free_energy_intermediate}, the third condition, $\hat{e}_{2}\cdot\boldsymbol{\varphi}_{\perp} = 0$, is a new type of nemato-orbital coupling for out-of-plane nematics, which is absent in the two-dimensional theory. Returning to the out-of-plane director, $\boldsymbol{\varphi}_\perp$, expressed in terms of the director $\boldsymbol{\mathcal{N}}$ , this third condition yields
\begin{equation}
    \hat{e}_{2}\cdot\boldsymbol{\varphi}_{\perp} = \varphi_\perp \sin(2\beta)\sin\left( \alpha - \phi \right)
\end{equation}
with the solution being that $\alpha = \phi$, showing that the out-of-plane director lies in the $(\hat{e}_1,\hat{z})$-plane.
 \begin{figure*}
    \centering
    \includegraphics[width=1\linewidth]{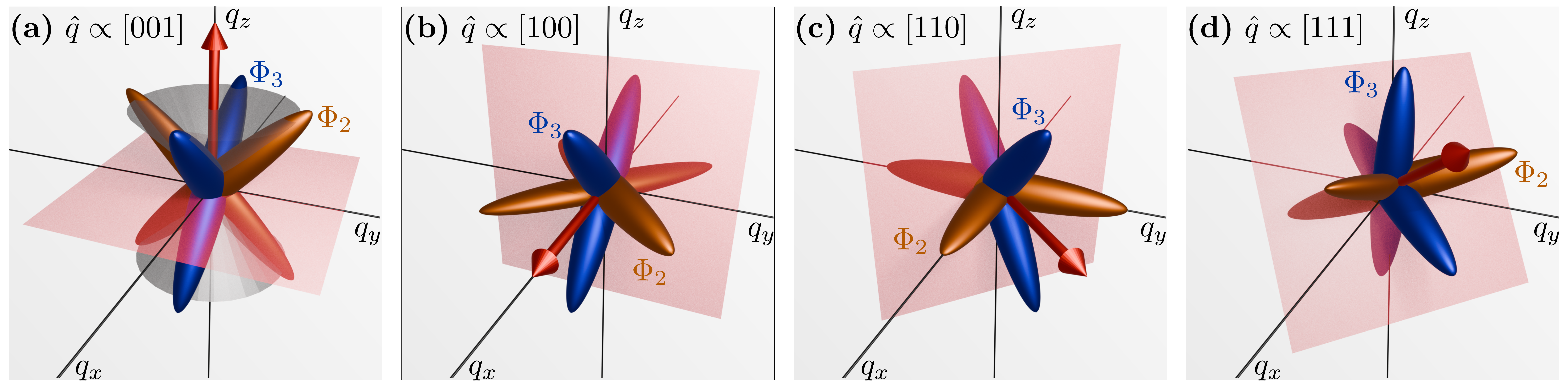}
    \caption{\justifying Schematic of universal direction-selective nematic criticality induced by elastic compatibility. Each panel shows different nematic momenta directions, $\hat{q}$, as the red arrow. The orange and blue ellipsoids represent the possible critical $(\Phi_2, \Phi_3)$ helical nematic order parameters for a given momentum direction. The semi-major axes of the ellipsoids are oriented parallel to the critical directors in the $d$-orbital basis selected by the specified momenta. Note that the critical director can point in any direction along the cone spanned by the blue and orange ellipsoids, shown explicitly in \textbf{(a)}. Nematic directors that lie in the plane transverse to the momentum (red) are incompatible $(\Phi_4, \Phi_5)$. \textbf{(a)} When the critical momentum is polar, $\hat{q} \propto [001]$, the $(\varphi_{2yz},\varphi_{2zx})$ $d$-orbital order parameters span the critical nematic manifold, with directors aligned along the $[011]$ and $[101]$ axes, respectively. \textbf{(b,c)} When the critical momentum is in the plane, there is an in-plane director and an out-of-plane director. \textbf{(b)} For $\hat{q}\propto [100]$ (or $[010]$), the in-plane critical nematic order parameter is $\varphi_{2xy}$. \textbf{(c)} When $\hat{q}\propto [110]$ (or $[1\bar{1}0]$), the critical in-plane nematic order parameter is $\varphi_{x^2 - y^2}$. \textbf{(d)} By rotating the critical momentum in the full 3D space, while the corresponding nematic orientations change in the $d$-orbital basis, it is always the $(\Phi_2, \Phi_3)$ helical doublet which spans the critical manifold.
    \label{fig:dsc_qhat_directors}
    }
\end{figure*}

Enforcing $\varphi_{z^2} = 0$, these expressions can be reduced to an effective free energy for a 3D electronic nematic solely experiencing in-plane critical momenta:
\begin{align}
    \mathcal{F}_{\text{eff}}^{\left(2\right)}\left[\boldsymbol{\varphi}\right]	&\rightarrow\frac{1}{2V}\sum_{\boldsymbol{q}_\parallel}\bigg\lbrace\left(r-\tfrac{\lambda_{0}^{2}}{\mu}+q^{2}\right)\left(\boldsymbol{\varphi}_{\parallel}^{\dagger}\cdot\boldsymbol{\varphi}_{\parallel}^{\phantom{\dagger}}\right)\nonumber
    \\
    &\phantom{\rightarrow\frac{1}{2V}\sum_{\boldsymbol{q}}}+\tfrac{\lambda_{0}^{2}}{\mu}\left(1-\tfrac{3\varrho}{4}\right)\left|\boldsymbol{\mathfrak{D}}^{\text{T}}\left(\phi\right)\cdot\boldsymbol{\varphi}_{\parallel}\right|^{2}\nonumber
    \\
	&\phantom{\rightarrow\frac{1}{2V}\sum_{\boldsymbol{q}}}+\left(r-\tfrac{\lambda_{0}^{2}}{\mu}+q^{2}\right)\left(\boldsymbol{\varphi}_{\perp}^{\dagger}\cdot\boldsymbol{\varphi}_{\perp}^{\phantom{\dagger}}\right)\nonumber
    \\
    &\phantom{\rightarrow\frac{1}{2V}\sum_{\boldsymbol{q}}}+\tfrac{\lambda_{0}^{2}}{\mu}\left|\hat{e}_{2}\cdot\boldsymbol{\varphi}_{\perp}\right|^{2}\bigg\rbrace.
\end{align}
Thus, the nematic order parameter, which is fivefold degenerate in the bare theory, splits into two twofold degenerate irreducible representations. The first one deals with the familiar planar nematic order parameters in $\boldsymbol{\varphi}_\parallel$, whereas the second one corresponds to the out-of-plane nematics $\boldsymbol{\varphi}_\perp$ with in-plane spatial modulations. Each of these irreducible representations is then further constrained by the derivative-dependent nemato-orbital couplings, $\boldsymbol{\mathfrak{D}}^{\text{T}}\left(\phi\right)\cdot\boldsymbol{\varphi}_{\parallel} = 0$ and $\hat{e}_{2}\cdot\boldsymbol{\varphi}_{\perp} = 0$. This ultimately yields two separate critical compatible nematic directors, both lying in planes that contain $\hat{q}$. Indeed, treating these directors separately, the in-plane director maximizes the free energy below the transition for $\beta = \pi/2$, while the out-of-plane director does so for $\beta = \pi/4$. Thus, the directors can be expressed in terms of the in-plane momentum azimuthal angle $\phi$  as  $\boldsymbol{\mathcal{N}}_\parallel = (\cos(\phi\pm \pi/4),\sin(\phi\pm \pi/4), 0)$ and $\boldsymbol{\mathcal{N}}_\perp = \frac{1}{\sqrt{2}}(\cos\phi,\sin\phi,1)$. Calculating the angle between these directors and the in-plane momentum $\hat{q}_\parallel = (\cos\phi, \sin\phi,0)$ gives:
\begin{equation}
    \begin{aligned}
        \hat{q}_\parallel \cdot \boldsymbol{\mathcal{N}}_\parallel &= \hat{q}_\parallel \cdot \boldsymbol{\mathcal{N}}_\perp = \tfrac{1}{\sqrt{2}},
    \end{aligned}
\end{equation}
establishing that both directors are individually constrained at an angle of $45^{\text{o}}$ relative to the momentum. Neither lie in the transverse plane to $\hat{q}$, as these directors would generate nematic incompatibility, which must be suppressed across the phase transition.

Fig.~\ref{fig:dsc_qhat_directors} combines the polar and planar limits discussed  in this section to illustrate schematically how the critical nematic directors $\boldsymbol{\mathcal{N}}$ are fixed for a given momentum  $\hat{q}$. The director orientations are shown as the semi-major axes of the ellipsoids. The orientation of the electronic nematic director  $\boldsymbol{\mathcal{N}}$ is fixed to be on a cone that makes an angle of $45^\circ$ relative to the momentum. In this regard, one can interpret this effect as a ``precession'' of the nematic director around the momentum  $\hat{q}$  induced by the elastic compatibility, resembling how a magnetic moment precesses around a magnetic field. Note that this critical cone rotates about the momentum  $\hat{q}$. Regardless of the momentum, however, the critical modes are \textit{always} spanned by the helical doublet $(\Phi_2,\Phi_3)$. Note that a director lying in the transverse plane to the momentum induces incompatible nematicity, encoded in the doublet $(\Phi_4,\Phi_5)$, which cost energy according to Eq. (\ref{eq:effective_quadratic_term_nematicity}). Conversely, a director  $\boldsymbol{\mathcal{N}}$ that is parallel to the momentum $\hat{q}$  induces dilatation, encoded by $\Phi_1$, which also incurs an energy cost. The fact that the critical nematic manifold is restricted by the wavevector, even in the isotropic continuum, is a consequence of the elastic compatibility relations, and will be inevitable in crystalline systems. In those systems, while the discrete rotational symmetry further splits the degeneracy between the $d$-orbital nematic order parameters, it will not change the meaning of the helical order parameters relative to the wavevector.

\vspace{1em}
\section{Elastic Incompatibility and Emergent Nemato-Plasticity \label{sec:elastic_incompatibility_and_nematoplasticity}}
Our investigation of the role of tensor compatibility in nemato-elasticity showed that the electronic nematic order parameter inherits the constraints imposed by the compatibility relations. Moreover, we found that the requirement to suppress incompatible electronic nematic fluctuations led to the universal phenomenon of direction-selective criticality. In elasticity theory, it is well-known how to incorporate incompatible strains, which arise in the regime of plastic deformation by crystalline defects \cite{eshelbyContinuumTheoryLattice1956, dewitLinearTheoryStatic1970, Dewit1973, dewitTheoryDisclinationsIII1973, dewitTheoryDisclinationsIV1973, kronerekkehartContinuumTheoryDefects1981, kleinertDoubleGaugeTheory1983, kleinertGaugeTheoryDefect1984, muratoshioMicromechanicsDefectsSolids1987, kleinertGaugeFieldsSolids1989, beekmanDualGaugeField2017, Pretko2018_PRL,Pretko2019, gaaFractonelasticityDualityTwisted2021}. The goal of this section is to incorporate incompatible elastic strains into our theory of nemato-elasticity.

We first provide a brief review in Subsection \ref{subsec:reveiw_incompatible_elasticity} of the details of plasticity, viewed as a ``defect gauge theory'' \cite{kleinertDoubleGaugeTheory1983, kleinertTwoGaugeFields1985}. Our review will follow the notation given in both Mura's and Kleinert's textbooks, Refs. \cite{muratoshioMicromechanicsDefectsSolids1987} and \cite{kleinertGaugeFieldsSolids1989}, respectively, and occasionally draw upon the approaches to continuum plasticity in \citep{eshelbyContinuumTheoryLattice1956, dewitLinearTheoryStatic1970, Dewit1973, dewitTheoryDisclinationsIII1973, dewitTheoryDisclinationsIV1973,kronerekkehartContinuumTheoryDefects1981,valsakumarGaugeTheoryDefects1988}.

In subsection \ref{subsec:incompatible_nematicity_defects}, we demonstrate how to incorporate defect gauge theory into the helical strain formalism developed for nemato-elasticity. We show that the anisotropic contribution to the electronic nematic mass renormalization in \eqref{eq:effective_electronic_nematic_mass_M} specifically projects out the incompatible part of the nematic tensor. Moreover, we show that the non-critical incompatible electronic nematic order parameters experience plastic strains as a conjugate field, whereas the compatible critical electronic nematic order parameters are insensitive to plastic defects. This further distinguishes mesoscopic electronic nematicity in the $d$-orbital representation from the macroscopic critical helical electronic nematic orders. Finally, we study an ensemble of straight edge dislocations as an analytically tractable example of a plastic strains in Subsection \ref{subsec:dislocations}, and use the helical formalism to compute their impact on the $d$-orbital electronic nematic order parameters as a random conjugate field.

\subsection{Incompatible Elasticity and the Defect-Density Tensor \label{subsec:reveiw_incompatible_elasticity}}
Even in the limit of infinitesimal deformations, not all strains in a crystal are compatible. In particular, in the presence of crystalline defects, the elastic strain tensor, now denoted by the script $\mathscr{E}_{ij}$, is incompatible \citep{eshelbyContinuumTheoryLattice1956, dewitLinearTheoryStatic1970, Dewit1973, dewitTheoryDisclinationsIII1973, dewitTheoryDisclinationsIV1973, muratoshioMicromechanicsDefectsSolids1987, kleinertGaugeFieldsSolids1989}.  Incompatible strains emanate from defects such as disclinations, dislocations, vacancies, and intersititials \citep{eshelbyContinuumTheoryLattice1956, kronerekkehartContinuumTheoryDefects1981, Sahoo1984, muratoshioMicromechanicsDefectsSolids1987}. Whatever the defects are, they define a new equilibrium state for the medium that has been plastically deformed away from the ideal case. In this new equilibrium state, these defects are ``stress-free,'' and any stresses within the material further deform the medium away from this new ``defective equilibrium'' \citep{eshelbyContinuumTheoryLattice1956, muratoshioMicromechanicsDefectsSolids1987}. 

Quantitatively, one measures the deformation relative to this new defective equilibrium. Thus, the elastic strain is modified to 
\begin{equation}
 \mathscr{E}_{ij}\left(\boldsymbol{x}\right)\equiv\varepsilon_{ij}\left(\boldsymbol{x}\right)-\varepsilon_{ij}^{p}\left(\boldsymbol{x}\right)=\tfrac{1}{2}\left[\partial_{j}u_{i}\left(\boldsymbol{x}\right)+\partial_{i}u_{j}\left(\boldsymbol{x}\right)\right]-\varepsilon_{ij}^{p}\left(\boldsymbol{x}\right), \label{eq:definition_of_elastic_strain}
\end{equation}
where $\varepsilon_{ij}$  denotes the compatible ``total strain'' tensor and $\varepsilon^p_{ij}$  is the ``plastic strain'' tensor generated by the defects. Note that $\varepsilon^p_{ij}$  is the strain measured with respect to the ideal, defect-free, limit. Hooke's Law for stress is also expressed relative to the new defective equilibrium and takes the form
\begin{equation}
    \sigma_{ij}\left(\boldsymbol{x}\right)=C_{ijkl}\mathscr{E}_{kl}\left(\boldsymbol{x}\right)=C_{ijkl}\left[\varepsilon_{kl}\left(\boldsymbol{x}\right)-\varepsilon_{kl}^{p}\left(\boldsymbol{x}\right)\right],
\end{equation}
where $\sigma_{ij}$  is the stress tensor \citep{eshelbyContinuumTheoryLattice1956, muratoshioMicromechanicsDefectsSolids1987, kleinertGaugeFieldsSolids1989}. Since the total strain is compatible, it follows that the incompatibility of the elastic strain, $\eta_{ij}$, is given by
\begin{equation}
    \eta_{ij}\left(\boldsymbol{x}\right)\equiv\text{inc}\left[\mathscr{E}\left(\boldsymbol{x}\right)\right]_{ij}=-\text{inc}\left[\varepsilon^{p}\left(\boldsymbol{x}\right)\right]_{ij}. \label{eq:eta_tensor_incompatibility_of_E}
\end{equation}
This tensor, $\eta_{ij}$, is known as the defect density tensor, and is related to the disclination density, $\Theta_{ij}$, and the dislocation density, $\alpha_{ij}$, as
 described in detail in references \citep{dewitLinearTheoryStatic1970, Dewit1973, dewitTheoryDisclinationsIII1973, dewitTheoryDisclinationsIV1973, muratoshioMicromechanicsDefectsSolids1987, kleinertGaugeFieldsSolids1989,beekmanDualGaugeField2017}.

The presence of incompatible strains generated by defects leads to an additional gauge redundancy in elasticity \citep{kleinertDoubleGaugeTheory1983, valsakumarGaugeTheoryDefects1988, kleinertGaugeFieldsSolids1989}. This becomes apparent by considering that the physical content encoded by the plastic strain tensor is contained in \eqref{eq:eta_tensor_incompatibility_of_E}. In words, the incompatible contribution to the observable elastic strain, $\mathscr{E}$, defines the incompatible contribution to the plastic strain tensor. Therefore, one can add any compatible tensor to the plastic strain and obtain the same defect density. This is analogous to adding an arbitrary gradient to the vector potential of magnetostatics, $\boldsymbol{A}(\boldsymbol{x})$, and obtaining the exact same static magnetic field, $\boldsymbol{B}(\boldsymbol{x}) = \boldsymbol{\nabla}\times \boldsymbol{A}(\boldsymbol{x})$, indicating that only the curl of the vector potential is physical. Since this gauge redundancy in plasticity occurs because of the inclusion of the plastic strain tensor, the resulting theory is called the ``defect gauge theory'' of plasticity \citep{kleinertDoubleGaugeTheory1983, kleinertTwoGaugeFields1985, valsakumarGaugeTheoryDefects1988, kleinertGaugeFieldsSolids1989, valsakumarForceTwoParallel1996,beekmanDualGaugeField2017}. 

However, adding an arbitrary compatible tensor to $\varepsilon^p(\boldsymbol{x})$  in \eqref{eq:definition_of_elastic_strain}  will clearly change the observable elastic strain, unless the total strain is modified in a gauge-covariant manner. That is to say, the following local gauge transformations
\begin{equation}
    \begin{aligned}\varepsilon_{ij}^{p}\left(\boldsymbol{x}\right) & \rightarrow\left[\varepsilon_{ij}^{p}\left(\boldsymbol{x}\right)\right]^{\prime}=\varepsilon_{ij}^{p}\left(\boldsymbol{x}\right)-\tfrac{1}{2}\left[\partial_{j}\Lambda_{i}\left(\boldsymbol{x}\right)+\partial_{i}\Lambda_{j}\left(\boldsymbol{x}\right)\right],\\
\varepsilon_{ij}\left(\boldsymbol{x}\right) & \rightarrow\left[\varepsilon_{ij}\left(\boldsymbol{x}\right)\right]^{\prime}=\varepsilon_{ij}\left(\boldsymbol{x}\right)-\tfrac{1}{2}\left[\partial_{j}\Lambda_{i}\left(\boldsymbol{x}\right)+\partial_{i}\Lambda_{j}\left(\boldsymbol{x}\right)\right],\\
\mathscr{E}_{ij}\left(\boldsymbol{x}\right) & \rightarrow\left[\mathscr{E}_{ij}\left(\boldsymbol{x}\right)\right]^{\prime}=\mathscr{E}_{ij}\left(\boldsymbol{x}\right),
\end{aligned}
\end{equation}
leave the stress state of the system invariant. In the above, we have used the position-dependent vector, $\boldsymbol{\Lambda}(\boldsymbol{x})$, to denote the potential for the compatible tensor in the defect gauge transformation. Comparing with \eqref{eq:definition_of_elastic_strain}, one can write an equivalent gauge-covariant derivative for the displacement vector, $\boldsymbol{u}$, as 
\begin{equation}
    \partial_{j}u_{i}\left(\boldsymbol{x}\right)\rightarrow\left[\partial_{j}u_{i}\left(\boldsymbol{x}\right)\right]^{\prime}=\partial_{j}u_{i}\left(\boldsymbol{x}\right)-\partial_{j}\Lambda_{i}\left(\boldsymbol{x}\right), \label{eq:distortion_tensor_gauge_transformation}
\end{equation}
where the defect-gauge covariant derivative, $\mathcal{D}_i$, is defined by $\mathcal{D}_{j}u_{i}\left(\boldsymbol{x}\right)\equiv\partial_{j}u_{i}\left(\boldsymbol{x}\right)-\partial_{j}\Lambda_{i}\left(\boldsymbol{x}\right)$ \citep{valsakumarGaugeTheoryDefects1988, kleinertGaugeFieldsSolids1989}. The second term $\partial_{j}\Lambda_{i}$ that acts as the defect-gauge field is known as the ``plastic distortion'' tensor,  $\beta^p_{ij}$, and its symmetric part is the plastic strain tensor, $\varepsilon^p_{ij}$.

These gauge transformations reflect the fact that, in order to describe an incompatible tensor in terms of a displacement vector, the latter needs to be multivalued. In other words, if we obtain a vector field, $\boldsymbol{v}$, such that $2\mathscr{E}_{ij}(\boldsymbol{x})=\partial_jv_i(\boldsymbol{x}) + \partial_i v_j(\boldsymbol{x})$, then one can always shift it by a single-valued vector field $\boldsymbol{\Lambda}(\boldsymbol{x})$ and obtain the same defect density. This problem is endemic in chemically pure media, whether isotropic or crystalline, since every constituent element, or unit cell, is indistinguishable from any other. In this regard, the multivaluedness reflects an ambiguity in uniquely assigning an equilibrium position of such constituents in a defective medium.  The gauge field formalism simply avoids the need for multivalued functions \citep{kleinertGaugeFieldsSolids1989}. 

\begin{figure}
    \centering
    \includegraphics[width=\columnwidth]{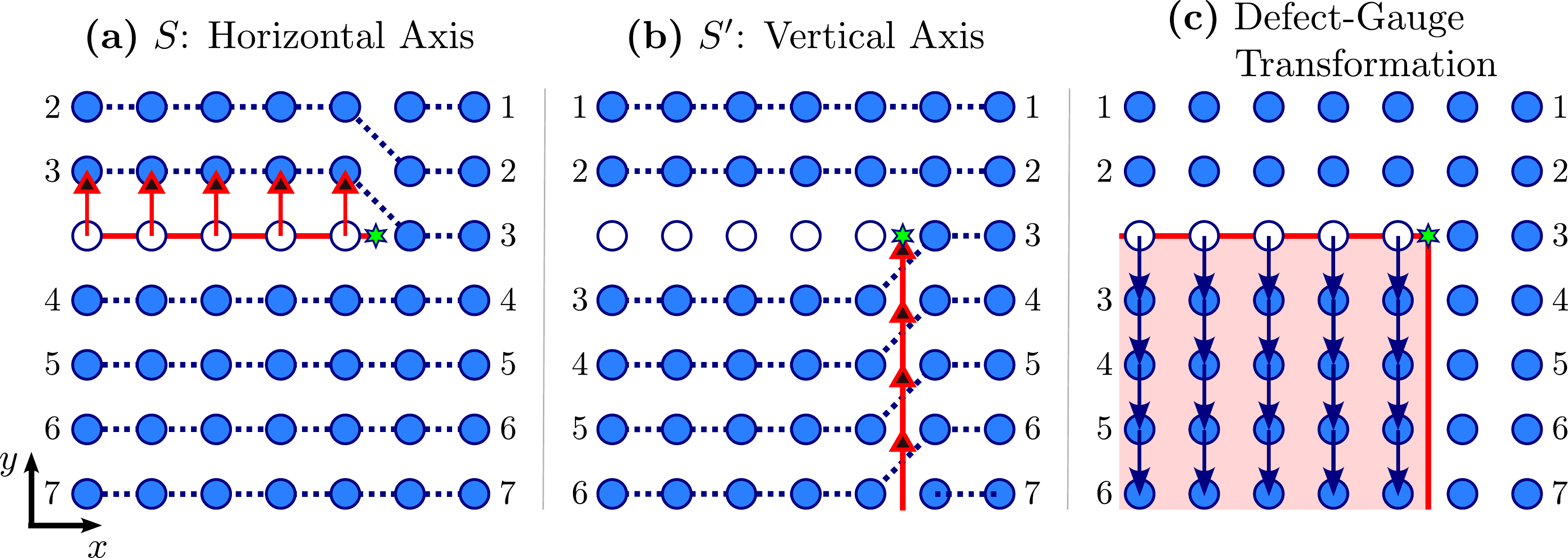}
    \caption{Defect-gauge transformation acting on a straight edge dislocation. In each figure, the dislocation core is the green star. The Burgers vector is represented by the red vertical arrows and the crystal planes in the dislocated medium are numbered. \textbf{(a)} The Volterra surface is taken along the leftward horizontal axis, $S$, creating a discontinuous upward jump in the 3, 2, and 1, crystal planes, as read from right-to-left. \textbf{(b)} An equivalent Volterra surface is drawn vertically, $S^\prime$, now leading to a downward jump of planes $3$-$7$, read from right-to-left. Both $S$ and $S^\prime$  are anchored to the dislocation core (green star) and have the same Burgers vector. \textbf{(c)} The appropriate defect-gauge transformation, as described in the text, amounts to a redefinition of the crystal planes around the defect core, effectively displacing all sites within the volume enclosed by $S$ and $S^\prime$. This physically equivalent redefinition of the displacement vector exemplifies defect gauge redundancy. }
    \label{fig:slip_gauge_transformation}
\end{figure}

From the perspective of the defects, this gauge ambiguity can be understood with the example of Volterra cutting surfaces in the elastic continuum that define dislocations and disclinations. In particular, if one keeps the defect lines (or cores) fixed, these cutting surfaces can be translated and rotated without affecting the defect density tensor \citep{dewitLinearTheoryStatic1970, Dewit1973, dewitTheoryDisclinationsIII1973, dewitTheoryDisclinationsIV1973, kleinertDoubleGaugeTheory1983, kleinertGaugeFieldsSolids1989}, as shown with the example case of a straight edge dislocation in Fig.~\ref{fig:slip_gauge_transformation}. In this light, these defect surfaces are unphysical, but the defect lines are physically observable objects \citep{chaikinPrinciplesCondensedMatter1995}. Within the context of geometry, the Volterra process of removing (or adding) material to a simply-connected domain renders it multiply-connected.   An application of Weingarten's Theorem to the Volterra surfaces in an elastic continuum establishes that, since the compatiblity conditions are satisfied everywhere other than the Volterra surface, the displacement vector can discontinuously change across that surface by no more than a translation and rigid rotation \cite{dewitLinearTheoryStatic1970, dewitTheoryDisclinationsIV1973, kleinertGaugeFieldsSolids1989}. These define the Burgers vector, $\boldsymbol{b}$, and Frank vector, $\boldsymbol{\Omega}$, of the dislocation and disclination, respectively. This plastic jump in the displacement vector takes the form 
\begin{equation}
    \Delta u^{p}_{i}\left(\boldsymbol{x} \in S\right)=b_{i}+\epsilon_{ijk}\Omega_{j}x_{k},
\end{equation}
where $S$ is the Volterra cutting surface. In the above, we have also assumed that the defect line intercepts the origin. By integrating this field (or its curl) over any closed circuit around the defect line, one obtains the Burgers  (or Frank) vector, establishing these as the topological defects of elasticity \cite{kleinertGaugeFieldsSolids1989,chaikinPrinciplesCondensedMatter1995,Pretko2018_PRL, Pretko2019}. In Fig.~\ref{fig:slip_gauge_transformation}(a), we show how an straight edge dislocation with a core aligned along the $\hat{z}$-axis is formed from a Volterra surface $S$, taken to be the $xz$ plane. The third crystal plane of atoms -- those that lie along the Volterra surface -- is ``removed'' in order to plastically deform the defect-free medium into the dislocated medium. The Burgers vector is given by $\boldsymbol{b} = \hat{y}$, indicating that in the deformed medium, the third crystal plane is discontinuously translated vertically in the figure.

While the Volterra surface in Fig.~\ref{fig:slip_gauge_transformation}(a) is sufficient, it is not unique. The same topological vectors, $\boldsymbol{b}$ and $\boldsymbol{\Omega}$, can be obtained with a different Volterra surface, $S^\prime$, provided that is also pinned to the same defect line. One can show that this change in surface amounts to a redefinition of the plastic displacement by a single-valued vector field  \citep{kleinertGaugeFieldsSolids1989}. We illustrate a special case of this transformation in Fig.~\ref{fig:slip_gauge_transformation} for the case of the straight edge dislocation. Comparing Figs.~\ref{fig:slip_gauge_transformation}(a) and (b), the dislocation core is fixed, as indicated by the green star, but the Volterra surface is changed from the $xz$ ($S$) plane to the $yz$ plane ($S^\prime$). As shown in the figure, while the result is a relabeling of which atoms in the dislocated medium correspond to the third through seventh crystal planes, the actual physical state of the dislocated medium is identical to that shown in Fig.~\ref{fig:slip_gauge_transformation}(a). The only difference is a relabeling of the atomic positions. Given that this change in Volterra surface only changes the plastic displacement, $u^p$, by a single-valued vector field, it amounts to a \textit{compatible}  change in plastic strain. Thus, it is not a physical change of the system because it does not change the \textit{incompatibility} of the elastic strain. The single-valued vector field constituting the unphysical, \textit{compatible}, change to the plastic strain is shown in Fig.~\ref{fig:slip_gauge_transformation}(c) everywhere within the red highlighted region.

This concludes our brief review of the gauge theory of defects in the limit of infinitesimal continuum elasticity. For completeness, we show how to use the helical formalism to obtain the elastic strain fields, $\mathscr{E}_{ij}(\boldsymbol{x})$, and total strain fields, $\varepsilon_{ij}(\boldsymbol{x})$, generated by the plastic strain tensor, $\varepsilon^p_{ij}(\boldsymbol{x})$, for a medium of arbitrary symmetry in Appendix \ref{app:elastostatic_plastic_solution}. The helical elastic strain Fourier amplitudes sourced by the plastic strain follows from \eqref{eq:plastic_elastostatics_solution_intermediaries} as 
\begin{equation}
    \mathscr{E}_{ij}=\left[\mathcal{Q}_{ij}^{1\alpha}\mathscr{S}_{\alpha\beta}\left(\hat{q}\right)\mathcal{Q}_{lk}^{1\beta}C_{klmn}-\tfrac{1}{2}\left(\delta_{im}\delta_{jn}+\delta_{in}\delta_{jm}\right)\right]\varepsilon_{mn}^{p},\label{eq:elastic_strain_solution_anisotropic_medium}
\end{equation}
in a medium with elastic constants $C_{klmn}$. As shown in Appendix \ref{app:elastostatic_plastic_solution}, the quantity $\mathscr{S}_{\alpha\beta}$  is the ``helical compliance matrix,'' defined by 
\begin{equation}
\mathcal{S}_{\alpha\beta}\left(\hat{q}\right)\left(\mathcal{Q}_{ji}^{1\beta}C_{ijkl}\mathcal{Q}_{kl}^{1\gamma}\right)\equiv\delta_{\alpha\gamma}.
\end{equation}
As proven in Appendix \ref{app:elastostatic_plastic_solution}, the elastic strain tensor $\mathscr{E}_{ij}$ identically vanishes if the incompatibility of the plastic strain vanishes, $\text{inc}(\varepsilon^p)_{ij} = 0$, regardless of the elastic medium's symmetry. Therefore, the defect gauge formalism isolates only the physical incompatibility contained within the plastic strain, $\varepsilon^p_{ij}$.
In the special case of an isotropic medium, we find 
\begin{equation}
    \mathscr{E}_{ij}=\frac{1}{3}\varrho\Delta_{p}\delta_{ij}-\frac{1}{\sqrt{2}}\left[\left(1-\varrho\right)\Delta_{p}\vartheta_{ij}^{1}+\sum_{a=4,5}\varepsilon_{a}^{p}\vartheta_{ij}^{a}\right], \label{eq:elastic_strain_solution_isotropic_medium}
\end{equation}
where $\varrho$  is defined in \eqref{eq:varrho_definition} and  $\Delta_p \equiv \frac{\sqrt{3}}{2}\varepsilon_1^p - \varepsilon_0^p$ quantifies the incompatible part of the plastic dilatation, $\varepsilon^p_0 \equiv \text{Tr}(\varepsilon^p)$ (see \eqref{eq:incompatibility_of_A_varTheta_decomp}).  As before, $\varepsilon^p_a \equiv \boldsymbol{\hat{Q}}_a \cdot \boldsymbol{\varepsilon}^p = \sqrt{2}\,\text{Tr}(\vartheta^a\varepsilon^p)$ are the components of the plastic strain in the helical basis.  

\subsection{Incompatible Electronic Nematicity and Defects \label{subsec:incompatible_nematicity_defects}}
One of the central results of the previous section was that, in the presence of defects and plastic strains, only the combination $\mathscr{E}_{ij}\left(\boldsymbol{x}\right)\equiv\varepsilon_{ij}\left(\boldsymbol{x}\right)-\varepsilon_{ij}^{p}\left(\boldsymbol{x}\right)$ is gauge invariant and therefore observable. Returning to nemato-elasticity, given that electronic nematicity is observable, it can only couple to the components of the \textit{elastic} strain tensor $\mathscr{E}_{ij}$. Hence, in materials with structural defects, one must generalize the bilinear $\boldsymbol{\varphi}(\boldsymbol{x})\cdot \boldsymbol{{\varepsilon}}(\boldsymbol{x})$  to  $\boldsymbol{\varphi}(\boldsymbol{x})\cdot \boldsymbol{\mathscr{E}}(\boldsymbol{x})$. Using the Gell-Mann decomposition of a tensor, \eqref{eq:GellMann_decomp_tensor}, one finds that the nemato-elastic bilinear assumes the form: $\boldsymbol{\varphi}\left(\boldsymbol{x}\right)\cdot\boldsymbol{\mathscr{E}}\left(\boldsymbol{x}\right)=\boldsymbol{\varphi}\left(\boldsymbol{x}\right)\cdot\boldsymbol{\mathscr{\varepsilon}}\left(\boldsymbol{x}\right)-\boldsymbol{\varphi}\left(\boldsymbol{x}\right)\cdot\boldsymbol{\varepsilon}^{p}\left(\boldsymbol{x}\right)$. The elastic free energy density, in real-space, is then 
\begin{align}
   f_{\text{elas}}\left(\varphi,\mathscr{E}\right)	&=\frac{1}{2}\left\{ \left(\lambda+\tfrac{2\mu}{3}\right)\left[\text{Tr}\,\mathscr{E}\left(\boldsymbol{x}\right)\right]^{2}+\mu\boldsymbol{\mathscr{E}}\left(\boldsymbol{x}\right)\cdot\boldsymbol{\mathscr{E}}\left(\boldsymbol{x}\right)\right\} \nonumber
	\\
    &\phantom{=}-\lambda_{0}\boldsymbol{\varphi}\left(\boldsymbol{x}\right)\cdot\boldsymbol{\mathscr{E}}\left(\boldsymbol{x}\right).
\end{align}
The difference between the defect-free case is the presence of two additional sources of total strain: plastic deviatoric \textit{and} plastic dilatation strain. 

After expanding the plastic strain tensor in the helical representation from \eqref{eq:plastic_strain_varTheta_decomp}  and \eqref{eq:Deltap_measuring_incompatible_dilatation}, the nemato-elastic bilinear becomes
\begin{align}
    \boldsymbol{\varphi}^{\dagger}\cdot\boldsymbol{\mathscr{E}}	&=\sum_{a=1}^{5}\Phi_{a}^{*}\mathscr{E}_{a}^{\phantom{*}} \nonumber
    \\
	&=\tfrac{2}{\sqrt{3}}\Phi_{1}^{*}\left(\varepsilon_{1}^{h}-\varepsilon_{0}^{p}-\Delta_{p}\right)\nonumber
    \\
	&\phantom{=}+\sum_{\alpha=2,3}\Phi_{\alpha}^{*}\left(\varepsilon_{\alpha}^{h}-\varepsilon_{\alpha}^{p}\right)-\sum_{\alpha=4,5}\Phi_{\alpha}^{*}\varepsilon_{\alpha}^{p}.
\end{align}
In the above, the differences $\varepsilon^h_1 - \varepsilon_0^p$ and $\varepsilon_\alpha^h - \varepsilon_\alpha^p$ for $\alpha = 2,3$, reflect the gauge redundancy in the plastic strain tensor associated with the addition of an arbitrary compatible tensor. By coupling the helical nematic order parameter to these differences, nemato-elasticity remains defect-gauge invariant. The elastic free energy density in the presence of defects is 
\begin{align}
    f_{\text{elas}}\left(\varphi,\varepsilon,\varepsilon^{p}\right)	&=\frac{1}{2}\left\{ \left(\lambda+2\mu\right)\left|\varepsilon_{1}^{h}-\varepsilon_{0}^{p}\right|^{2}+\mu\sum_{a=2,3}\left|\varepsilon_{a}^{h}-\varepsilon_{a}^{p}\right|^{2}\right\} \nonumber
    \\
	&\phantom{=}-\tfrac{\lambda_{0}}{2}\bigg\lbrace\tfrac{2}{\sqrt{3}}\left(\Phi_{1}^{*}+\tfrac{2}{\sqrt{3}}\tfrac{\mu}{\lambda_{0}}\Delta_{p}^{*}\right)\left(\varepsilon_{1}^{h}-\varepsilon_{0}^{p}\right) \nonumber
    \\
	&\phantom{=-\frac{\lambda_{0}}{2}\bigg\lbrace}+\sum_{a=2,3}\Phi_{a}^{*}\left(\varepsilon_{a}^{h}-\varepsilon_{0}^{p}\right)+\text{c.c.}\bigg\rbrace \nonumber
    \\
	&\phantom{=}+\frac{1}{2}\mu\left(\frac{4}{3}\left|\Delta_{p}\right|^{2}+\sum_{a=4,5}\left|\varepsilon_{a}^{p}\right|^{2}\right)\nonumber
    \\
	&\phantom{=}+\frac{\lambda_{0}}{2}\left(\frac{2}{\sqrt{3}}\Phi_{1}^{*}\Delta_{p}+\sum_{a=4,5}\Phi_{a}^{*}\varepsilon_{a}^{p}+\text{c.c.}\right).
\end{align}
Minimizing with respect to the compatible helical strain yields the following equations of state: 
\begin{equation}
    \begin{aligned}\varepsilon_{1}^{h} & =\varepsilon_{0}^{p}+\frac{2\lambda_{0}}{\sqrt{3}\left(\lambda+2\mu\right)}\left(\Phi_{1}+\frac{2\mu}{\sqrt{3}\lambda_{0}}\Delta_{p}\right),\\
\varepsilon_{2,3}^{h} & =\varepsilon_{2,3}^{p}+\frac{\lambda_{0}}{\mu}\Phi_{2,3}.
\end{aligned}
\label{eq:disordered_elastic_equations_of_state}
\end{equation}
Comparing these equations of state with the defect-free case in \eqref{eq:elastic_equations_of_state} allows one to quickly compute the effective correction to the nematic free energy
\begin{align}
    \Delta f_{\text{eff}}\left(\Phi,\varepsilon^{p}\right)	&=-\frac{\lambda_{0}^{2}}{2\mu}\left\{ \sum_{a=2,3}\left|\Phi_{a}\right|^{2}+\varrho\left|\Phi_{1}+\frac{2\mu}{\sqrt{3}\lambda_{0}}\Delta_{p}\right|^{2}\right\} \nonumber
    \\
	&\phantom{=}+\frac{1}{2}\mu\left(\frac{4}{3}\left|\Delta_{p}\right|^{2}+\sum_{a=4,5}\left|\varepsilon_{a}^{p}\right|^{2}\right) \nonumber
    \\
	&\phantom{=}+\frac{\lambda_{0}}{2}\left(\frac{2}{\sqrt{3}}\Phi_{1}^{*}\Delta_{p}+\sum_{a=4,5}\Phi_{a}^{*}\varepsilon_{a}^{p}+\text{c.c.}\right).
\end{align}
where $\varrho$ is defined in \eqref{eq:varrho_definition}.  The first line has the same form as in the defect-free case, albeit now there is a renormalized coupling between $\Phi_1$  and the incompatible component of the plastic dilatation, $\Delta_p$, since both fields couple to the elastic dilatation, $\varepsilon^h_1 - \varepsilon^p_0$.  The total effective free energy can be separated into three parts 
\begin{align}
    \mathcal{F}_{\text{eff}}^{\left(2\right)}\left[\varphi,\varepsilon^{p}\right]	&=\mathcal{F}_{\text{eff}}^{\left(2\right)}\left[\varphi,0\right] \nonumber
    \\
	&\phantom{=}+\frac{\lambda_{0}}{2V}\sum_{\boldsymbol{q}}\left\{ \frac{2\left(1-\varrho\right)}{\sqrt{3}}\Phi_{1}^{*}\Delta_{p}+\sum_{a=4,5}\Phi_{a}^{*}\varepsilon_{a}^{p}+\text{c.c.}\right\} \nonumber
    \\
	&\phantom{=}+\frac{\mu}{2V}\sum_{\boldsymbol{q}}\left\{ \frac{4}{3}(1-\varrho)\left|\Delta_{p}\right|^{2}+\sum_{a=4,5}\left|\varepsilon_{a}^{p}\right|^{2}\right\}.\label{eq:effective_nematoplastic_free_energy}
\end{align}
The first line corresponds to the usual renormalization of the electronic nematic free energy in an ideal, defect-free crystal, coinciding with \eqref{eq:effective_quadratic_term_nematicity}. The last line is the elastostatic self-energy of the defects, consistent with \eqref{eq:plastic_free_energy_iso_minimum} in the limit of $\varphi \rightarrow 0$, and it clearly only depends on the physical, incompatible, contributions to the plastic strain tensor, $\varepsilon^p_{ij}$.  The second line is new, and represents nonlocal, momentum-dependent, bilinear interactions between the electronic nematic order parameter and the plastic strain. We therefore call it the \textit{nemato-plastic}  free energy, and we write it as 
\begin{equation}
    \mathcal{F}_{\text{np}}\left[\varphi,\varepsilon^{p}\right]\equiv\frac{\lambda_{0}}{2V}\sum_{\boldsymbol{q}}\left\{ \frac{2}{\sqrt{3}}\left(1-\varrho\right)\Phi_{1}^{*}\Delta_{p}+\sum_{\alpha=4,5}\Phi_{\alpha}^{*}\varepsilon_{\alpha}^{p}+\text{c.c.}\right\} \label{eq:nematoplastic_free_energy_Deltap}.
\end{equation}
Thus, the incompatible part of the plastic strain only couples to the electronic nematic order parameters that are already suppressed by direction-selective criticality.  From \eqref{eq:effective_nematoplastic_free_energy}, the quantities $\Phi_{2,3}$  do not experience structural disorder through a bilinear coupling. Indeed, their transition is enhanced by $\lambda_0^2/\mu$, like the case of homogeneous and inhomogeneous deformations in defect-free media (Sections ~\ref{sec:homogeneous_nematoelasticity} and ~\ref{sec:compatible_restrictions_on_nematicity}, respectively). On the other hand, the other three components $\Phi_{1,4,5}$,  which are suppressed in defect-free media, experience the structural disorder. Thus, the compatibility relations of elasticity protect the critical order parameters $\Phi_{2,3}$ from experiencing local defect strains through a nemato-plastic bilinear coupling.

In real-space, however, these distinct effects of defects on the nematic order parameter are less obvious, particularly when one uses the conventional, $d$-orbital basis for electronic nematicity. \eqref{eq:nematoplastic_free_energy_Deltap} establishes a plastic conjugate field for the $d$-orbital electronic nematic order parameter, $\boldsymbol{h}$, defined in momentum space via
\begin{align}
    \boldsymbol{h}_{\boldsymbol{q}}&\equiv-\left\{ \boldsymbol{\Gamma}_{0}\left(\hat{q}\right)\varepsilon_{0,\boldsymbol{q}}^{p}+\Gamma\left(\hat{q}\right)\cdot\boldsymbol{\varepsilon}_{\boldsymbol{q}}^{p}\right\}, \label{eq:conjugate_h_field_momentum_space}
    \\
\boldsymbol{\Gamma}_{0}\left(\hat{q}\right)&\equiv-\tfrac{2}{\sqrt{3}}\left(1-\varrho\right)\boldsymbol{\hat{Q}}_{1}, \label{eq:Gamma0_propagator}
\\
\Gamma\left(\hat{q}\right)&\equiv\left(1-\varrho\right)\boldsymbol{\hat{Q}}_{1}^{\phantom{\text{T}}}\boldsymbol{\hat{Q}}_{1}^{\text{T}}+\boldsymbol{\hat{Q}}_{4}^{\phantom{\text{T}}}\boldsymbol{\hat{Q}}_{4}^{\text{T}}+\boldsymbol{\hat{Q}}_{5}^{\phantom{\text{T}}}\boldsymbol{\hat{Q}}_{5}^{\text{T}}. \label{eq:Gamma_propagator}
\end{align}

In the expression above, we made the wavevector dependence explicit for clarity (though it is implied by definition in $\{\boldsymbol{\hat{Q}}_a\}$). The form factors in the expression above are both even under inversion, $\boldsymbol{\Gamma}_0(-\hat{q}) = \boldsymbol{\Gamma}_0(\hat{q})$ and $\Gamma(-\hat{q}) = \Gamma(\hat{q})$, from \eqref{eq:Qhat_under_qhat_to_minus_qhat}. Importantly, we note that this conjugate field is exactly the elastic strain generated by plasticity -- given in \eqref{eq:elastic_deviatoric_strain_solution_isotropy} and \eqref{eq:elastic_strain_iso_no_nematicity} -- \textit{in the absence of electronic nematicity}. Thus, $\boldsymbol{h}_{\boldsymbol{q}} = \boldsymbol{\mathscr{E}}_{\boldsymbol{q}} \vert_{\varphi = 0}$, implying that one can obtain the defect-generated conjugate field for the nematic order parameter by solving for the physical elastic strain separately from the nematic problem, despite both the nematic and plastic fields being coupled to the same fluctuating total strain. Meanwhile, it is insightful to note that $\Gamma(\hat{q})$ is exactly the anisotropic contribution to the renormalized $5\times 5$ electronic nematic mass matrix in \eqref{eq:effective_electronic_nematic_mass_M}, which can be written as
\begin{equation}
    \mathcal{M}(\hat{q}) = \tfrac{\lambda_0^2}{\mu}\left\{ \mathrm{I} - \Gamma(\hat{q}) \right\},
\end{equation}
exemplifying that direction-selective criticality occurs in the isotropic medium precisely to project out incompatible electronic nematic fluctuations.

In real-space, the nematoplastic free energy as
\begin{align}
\mathcal{F}_{\text{np}}\left[\boldsymbol{\varphi}\left(\boldsymbol{x}\right),\boldsymbol{h}\left[\varepsilon_{ij}^{p}\left(\boldsymbol{x}\right)\right]\right]	&=-\lambda_{0}\int_{x}\boldsymbol{h}\left(\boldsymbol{x}\right)\cdot\boldsymbol{\varphi}\left(\boldsymbol{x}\right),
\\
\boldsymbol{h}\left(\boldsymbol{x}\right)	&=-\int_{x^{\prime}}\bigg\lbrace\boldsymbol{\Gamma}_{0}\left(\boldsymbol{x}-\boldsymbol{x}^{\prime}\right)\varepsilon_{0}^{p}\left(\boldsymbol{x}^{\prime}\right)\nonumber
\\
	&\phantom{=\int_{x^{\prime}}\bigg\lbrace}+\Gamma\left(\boldsymbol{x}-\boldsymbol{x}^{\prime}\right)\cdot\boldsymbol{\varepsilon}^{p}\left(\boldsymbol{x}^{\prime}\right)\bigg\rbrace.\label{eq:h_real_space}
\end{align}
At this level, the defects can generate long-range conjugate fields for all five nematic $d$-orbital order parameters. Furthermore, we note that the character of the defect source can change as the strain is propagated through the medium by both the $5\times 1$ quadrupolar vector $\boldsymbol{\Gamma}_0(\boldsymbol{x})$ and $5\times 5$ matrix $\Gamma(\boldsymbol{x})$. 

Elastic fluctuations mediate long-range and anisotropic interactions in real-space between the physical plastic strain and the electronic nematic order parameter. However, as is already apparent in \eqref{eq:effective_nematoplastic_free_energy}, these interactions are between those components of the electronic nematic order parameter that remain massive at the nemato-elastic critical point. Treating the defects as quenched, then the implication is that the electronic nematic order parameter, in the $d$-orbital basis, suffers from long-range, anisotropic, and correlated random field disorder. Despite this random pinning effect on the $d$-orbital basis, the nemato-elastic criticality survives since the critical nemato-elastic order parameters, $\Phi_{2,3}$, do not experience the structural disorder. Higher order or dynamical effects may become important if the nemato-elastic criticality deviates from the Gaussian fixed point, but such a discussion lies outside the scope of this work.

Yet, as is the case with any bilinear, there is an inverse, or reciprocal, effect as well. In this theory, the reciprocal effect would be plasticity generated by electronic nematicity. If it is the case that $\lambda_0^2/\mu \ll 1$, then one expects that the internal fivefold degeneracy of the bare electronic nematic order parameter, $\boldsymbol{\varphi}$, is approximately held. The effect on the structure can be quite devastating then, since the elastically suppressed components, $\Phi_{1,4,5}$, are almost as soft as the compatible nematic order parameters, $\Phi_{2,3}$. If this happens, then one expects that the observable plastic amplitudes, $\Delta_p$  and $\varepsilon^p_{4,5}$, will be softened considerably by the nematic fluctuations. Such effects may lead to a proliferation of plastic defects within an electronic nematic, although more quantitative statements can only be made by considering higher-order effects of plasticity which lie outside the scope of the present work.

\subsection{Random Nematic Fields Generated by an Ensemble of Straight Edge Dislocations \label{subsec:dislocations}}
In this section, we provide a concrete, analytically tractable example of random conjugate elastic strain fields for the electronic nematic order parameter generated by a distribution of quenched random plastic defects. We focus on straight edge dislocations with Burgers vectors oriented in the $xy$-plane. Once the elastic strains are obtained, we calculate the spatial correlations for these fields, treating the sources -- the defects -- as random uncorrelated variables. This contrasts with more conventional random field approaches where the strain fields themselves are treated as random uncorrelated variables \cite{Imry_Ma, Binder1983, vojtaPhasesPhaseTransitions2013, vojtaDisorderQuantumManyBody2019, RMF01_RFBM}. As explained above, however, a random distribution of strains does not satisfy the elastic compatibility relations. We show that the conjugate elastic strains have long-range autocorrelations, which are reminiscent of correlated random fields \cite{weinribCriticalPhenomenaSystems1983, nattermannInstabilitiesIsingSystems1983, vojtaPhasesPhaseTransitions2013}. However, an important difference is that these autocorrelations are anisotropic and the random fields for the different orbital basis functions are highly correlated. Since the different orbital basis functions generally split due to the discrete rotational symmetry in crystals, random fields in orthogonal symmetry channels act as random \textit{transverse} fields rather than random \textit{longitudinal} ones \citep{maharajTransverseFieldsTune2017, RMF13_disorder_TmVO4}. The statistical-dependence of these random longitudinal and transverse fields that act on orthogonal symmetry channels may pull the electronic nematic criticality back into the quantum regime, even at nonzero temperature where thermal fluctuations dominate the critical behavior in the defect-free case. This suggests that structurally disordered electronic nematics may motivate a new class of random field problem hitherto unexplored in statistical physics. The role of dynamics in structurally disordered electronic nematics is however outside the scope of the current work where we only consider classical critical behavior.

\begin{figure}
    \centering
    \includegraphics[width=\columnwidth]{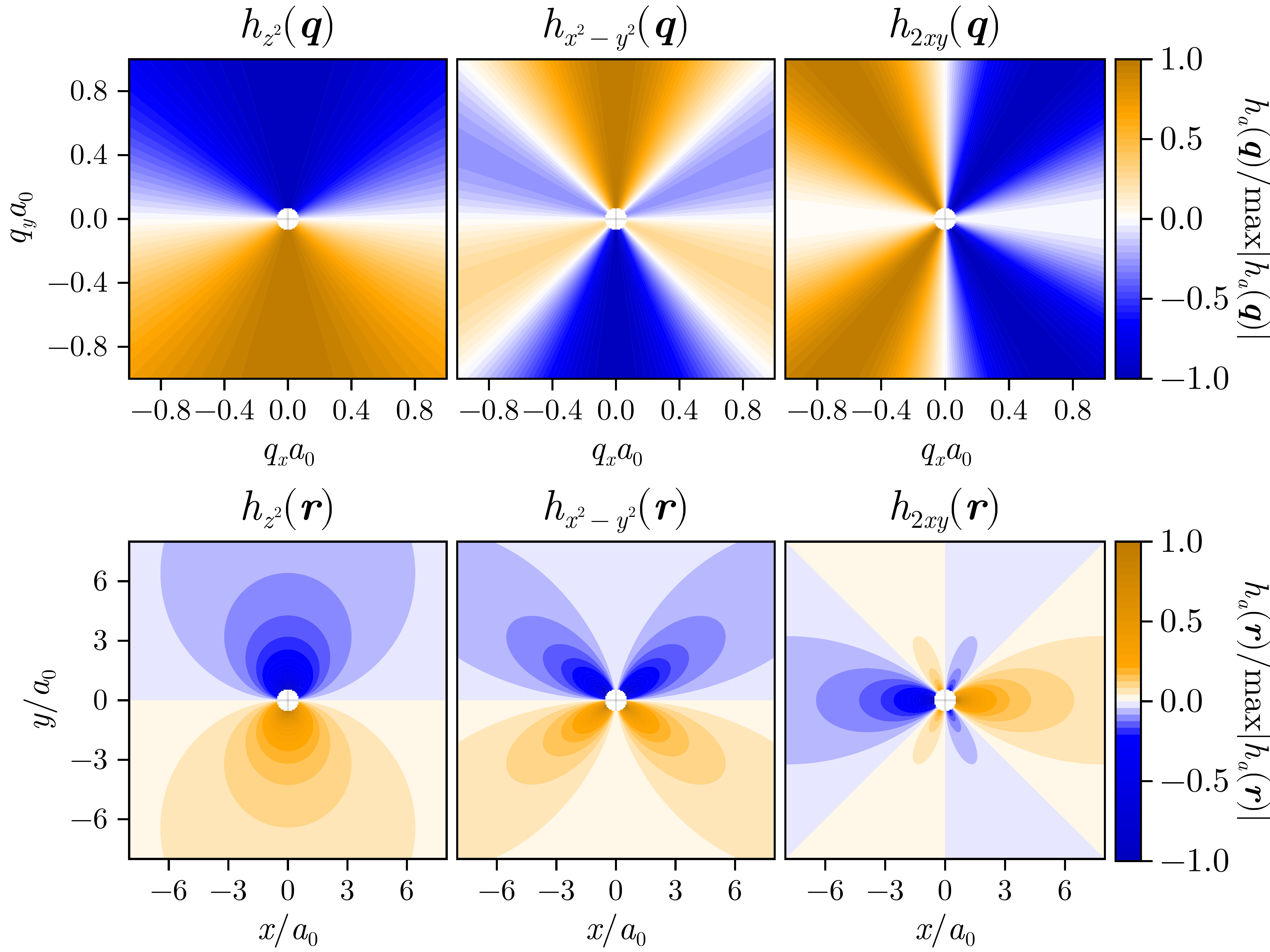}
    \vspace{-0.25in}
    \caption{Fourier- and real-space elastic strain fields emanating from a single straight edge dislocation in an isotropic elastic medium. Each column is one of the three nonzero elastic strain fields in the $d$-orbital basis, and are each normalized by their maximum magnitude. The positive (negative) values correspond to tensile (compressive) strain. The Burgers vector is chosen as $\boldsymbol{b} = a_0\hat{x}$, where $a_0$ is the lattice constant. Top row: The Fourier space fields are shown, from \eqref{eq:hq_disloc_expanded}, as a function of momentum $\boldsymbol{q}$. Bottom row: Real-space elastic strain fields for each nonzero elastic strain in the $d$-orbital basis from \eqref{eq:elastic_dislocation_fields_real-space_polar}. The minimum radius is taken to be $a_0/2$. }
    \label{fig:single_dislocation_field}
\end{figure}

We employ the nemato-plastic theory to compute the elastic strain fields conjugate to the nematic order parameter in the $d$-orbital basis, $\boldsymbol{\varphi}$, created by a single straight edge dislocation. The plastic distortion tensor for the dislocation anchored to the $z$-axis and with Burgers vector $\boldsymbol{b}=(b_x,b_y)$  in the convenient ``slip gauge'' is given by 
\begin{equation}
    \beta_{ij}^{p}\left(\boldsymbol{x}\right)=b_{x}\delta_{i,x}\delta_{j,y}\,\delta\left(y\right)\Theta\left(-x\right)-b_{y}\delta_{i,y}\delta_{j,x}\,\delta\left(x\right)\Theta\left(-y\right),
\end{equation}
as worked out in Appendix~\ref{app:elastic_strains_from_dislocations}.
The plastic strain tensor is its symmetric part. Since $\lambda^{1}_{ij} = \delta_{i1}\delta_{j2} + \delta_{i2}\delta_{j1}$  from \eqref{eq:GellMann_Matrices}, then the plastic strain is 
\begin{equation}
    \varepsilon_{ij}^{p}\left(\boldsymbol{x}\right)=\frac{1}{2}\left\{ b_{x}\delta\left(y\right)\Theta\left(-x\right)-b_{y}\delta\left(x\right)\Theta\left(-y\right)\right\} \lambda_{ij}^{1},
\end{equation}
from which we readily obtain its decomposition in the $d$-orbital basis 
\begin{equation}
    \boldsymbol{\varepsilon}^{p}\left(\boldsymbol{x}\right)=\left(0,0,0,0,b_{x}\delta\left(y\right)\Theta\left(-x\right)-b_{y}\delta\left(x\right)\Theta\left(-y\right)\right)^{\text{T}},
\end{equation}
and $\varepsilon^p_0(\boldsymbol{x}) = 0$. Clearly, only $\varepsilon^p_{2xy}$  is nonzero for this defect-gauge choice. We obtain the conjugate strain, $\boldsymbol{h}(\boldsymbol{x})$, via Fourier transform, using \eqref{eq:conjugate_h_field_momentum_space}. The Fourier transform of $\varepsilon^p_{2xy}(\boldsymbol{x})$ gives
\begin{align}
    \varepsilon_{2xy}^{p}	&=\int_{x}\text{e}^{-\text{i}\boldsymbol{q}\cdot\boldsymbol{x}}\left\{ b_{x}\delta\left(y\right)\Theta\left(-x\right)-b_{y}\delta\left(x\right)\Theta\left(-y\right)\right\} 
    \\
	&=2\pi\delta\left(q_{z}\right)\,\lim_{\eta\rightarrow0^{+}}\left(\frac{\text{i}b_{x}}{q_{x}+\text{i}\eta}-\frac{\text{i}b_{y}}{q_{y}+\text{i}\eta}\right),
\end{align}
where we have taken the infinite-volume limit. Simplifying and taking the principal part, we have 
\begin{equation}
    \varepsilon_{2xy}^{p}\left(\boldsymbol{q}\right)=4\pi\text{i}\,\delta\left(q_{z}\right)\left(\frac{b_{x}\sin\phi-b_{y}\cos\phi}{q\sin2\phi}\right),\label{eq:2xy_plastic_strain_source_FS}
\end{equation}
where $\phi$ is the polar angle of  the momentum $\boldsymbol{q}$. The conjugate field, $\boldsymbol{h}$, then follows as $\boldsymbol{h} =- \Gamma(\hat{q}) \cdot \boldsymbol{\varepsilon}^p$, where $\Gamma(\hat{q})$  is given by  \eqref{eq:Gamma_propagator}. The detailed steps are performed in Appendix \ref{app:elastic_strains_from_dislocations}. Expanding and simplifying \eqref{eq:hq_disloc_compact}, we obtain
\begin{align}
    \boldsymbol{h}	=\delta\left(q_{z}\right)\, \tfrac{2\pi\text{i}}{1-\nu_3}\tfrac{b_{x}\sin\phi-b_{y}\cos\phi}{q}\left(\tfrac{1-2\nu_{3}}{\sqrt{3}},-\cos2\phi,0,0,-\sin2\phi\right)^{\text{T}}.\label{eq:hq_disloc_expanded}
\end{align}
As can be seen from these expressions, and from the top row of Fig.~\ref{fig:single_dislocation_field}, all field amplitudes vanish for momenta parallel to the Burgers vector. Moreover, the two in-plane $h_{x^2- y^2}$ and $h_{2xy}$ fields have additional zeroes due to the incompatible projector $\Gamma(\hat{q})$. Indeed, these extra directions show that the elastic strain fields specifically lack momenta along the critical, and therefore compatible, directions for the electronic nematic orders. This is verified by comparing directly with \eqref{eq:2D_SVCR_polar}, Fig.~\ref{fig:SVCR_2D}, and Fig.~\ref{fig:supp_varphi_to_Phi_mapping}, which show that $\varphi_{x^2-y^2}$ ($\varphi_{2xy}$) is critical when $\cos(2\phi) = 0$ ($\sin(2\phi) = 0$). Performing an inverse Fourier transformation, the elastic strain field can be written in real-space as 
\begin{equation}
    \begin{aligned}h_{z^{2}}\left(\boldsymbol{r}_{\parallel}\right) & =\frac{1-2\nu_{3}}{2\pi\left(1-\nu_{3}\right)\sqrt{3}}\,\frac{\left(-\hat{z}\right)\cdot\left(\boldsymbol{b}\times\hat{r}_{\parallel}\right)}{r_{\parallel}},\\
h_{x^{2}-y^{2}}\left(\boldsymbol{r}_{\parallel}\right) & =-\frac{\boldsymbol{b}\cdot\hat{r}_{\parallel}}{2\pi\left(1-\nu_{3}\right)}\,\frac{\sin2\phi_{r}}{r_{\parallel}},\\
h_{2xy}\left(\boldsymbol{r}_{\parallel}\right) & =\frac{\boldsymbol{b}\cdot\hat{r}_{\parallel}}{2\pi\left(1-\nu_{3}\right)}\,\frac{\cos2\phi_{r}}{r_{\parallel}},
\end{aligned}\label{eq:elastic_dislocation_fields_real-space_polar}
\end{equation}
with $\boldsymbol{r}_\parallel \equiv (x,y,0) = r_\parallel(\cos\phi_r,\,\sin\phi_r,0)$. 

We have recovered the well-known expressions for the elastic strain fields around a single straight edge dislocation \citep{landauTheoryElasticity1970, dewitTheoryDisclinationsIV1973, Hameed2022, RMF13_disorder_TmVO4},  written in the more convenient $d$-orbital basis of electronic nematicity. These fields are plotted on a continuous mesh for $r_\parallel > a_0/2$ in Fig. 
\ref{fig:single_dislocation_field}, where $a_0$ is taken to be the lattice constant. One clearly observes long-range behavior emanating from a single defect, as well as an anisotropy developing with respect to the coordinate axes. We emphasize that the defect source, $\varepsilon^p_{ij}$, is only of $2xy$ character, whereas the field that couples to the electronic nematic order parameter has additional $z^2$ and $x^2-y^2$ character due to the nonlocality of the helical transformation in real-space.

\begin{figure*}
    \centering
    \includegraphics[width=\linewidth]{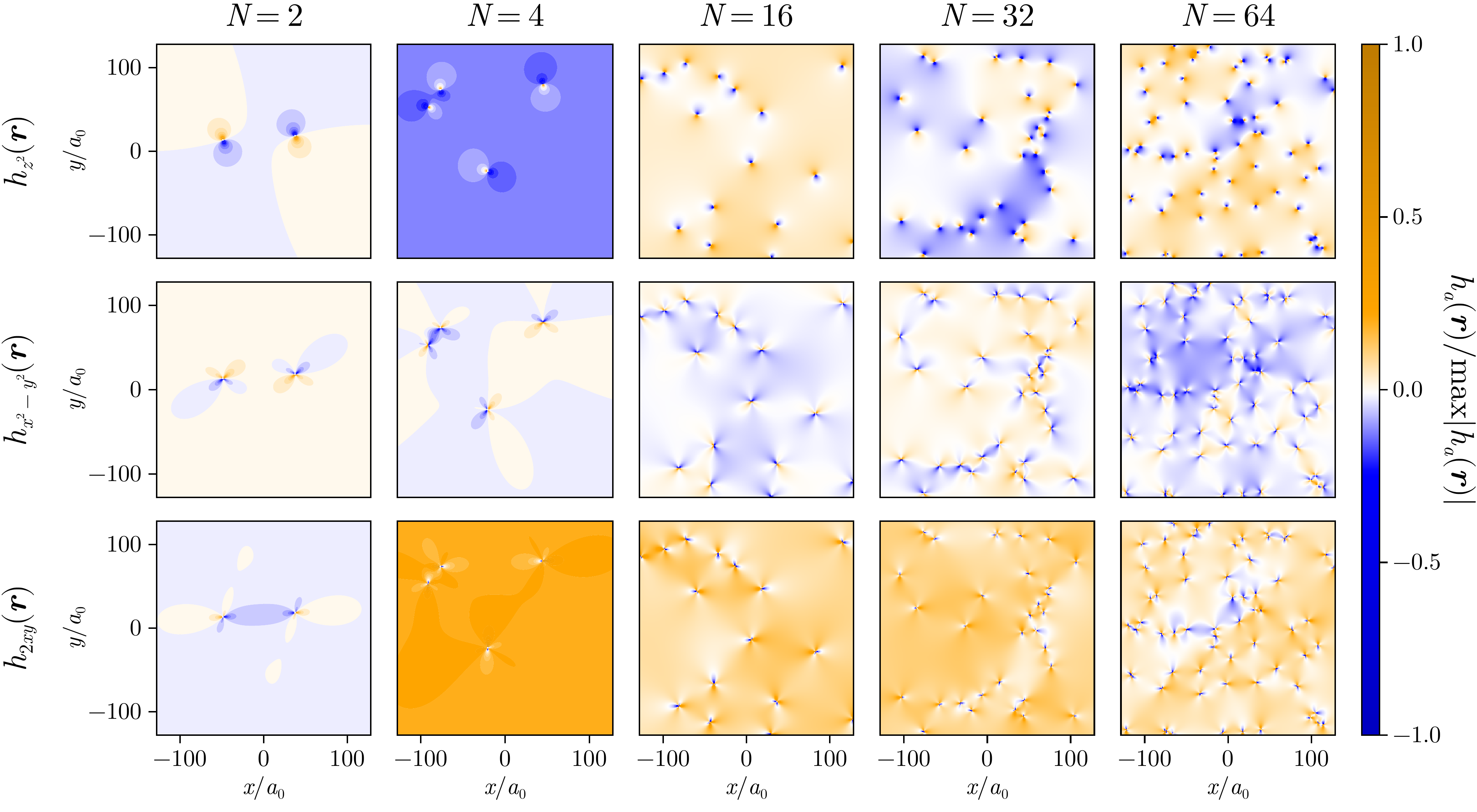}
    \vspace{-0.1in}
    \caption{Elastic strain field realizations of $N$ uncorrelated random straight edge dislocations. The rows show the superimposed strain fields for each dislocation, calculated from \eqref{eq:elastic_dislocation_fields_real-space_polar}. Each field is normalized by its maximum value, and the minimum radial distance from each dislocation core is taken to be $a_0/2$.  The columns sweep various concentrations of dislocations by increasing $N$ from left to right. The strains within each column are computed from the same set of random dislocations. The $N = 2,4$ columns are contour plots to highlight the different regions of compressive and tensile strain, whereas $N=16,32,64$ all show a continuous color gradient.  In each realization, the dislocation cores are randomly chosen uniformly. The length of each Burgers vectors is $a_0$, and half of their orientations are sampled uniformly within the $xy$-plane. The other half are opposite the first to enforce a vanishing total Burgers vector in each realization. }
    \label{fig:dislocation_realizations}
\end{figure*}
We now consider an ensemble of $N$  uncorrelated, noninteracting, quenched dislocations as a model for random elastic strain disorder for the electronic nematic order parameter within a structurally disordered isotropic medium. We denote the elastic strain from the $\alpha^{\text{th}}$ edge dislocation with Burgers vector $\boldsymbol{b}^\alpha$  and core location $\boldsymbol{x}^\alpha$  by $\boldsymbol{h}\left(\boldsymbol{x}-\boldsymbol{x}^{\alpha};\boldsymbol{b}^{\alpha}\right)$. The elastic strain from an ensemble of $N$ dislocations follows in real and momentum space as
\begin{align}
    \boldsymbol{h}\left(\boldsymbol{x}\right)&=\sum_{\alpha=1}^{N}\boldsymbol{h}\left(\boldsymbol{x}-\boldsymbol{x}^{\alpha};\boldsymbol{b}^{\alpha}\right),
    \\
    \boldsymbol{h}_{\boldsymbol{q}}&=-\sum_{\alpha=1}^{N}\text{e}^{\text{i}\boldsymbol{q}\cdot\boldsymbol{x}^{\alpha}}\Gamma\left(\hat{q}\right)\cdot\boldsymbol{\varepsilon}^{p}\left(\boldsymbol{q};\boldsymbol{b}^{\alpha}\right).
\end{align}
Example realizations for various dislocation concentrations in an isotropic elastic medium are shown in Fig.~\ref{fig:dislocation_realizations}, where the strain fields are plotted on a continuous square mesh within a system of linear size $L = 128a_0$, with $a_0$ being the lattice constant. The minimum distance from any dislocation core was taken to be $a_0/2$ to avoid the singularities. These realizations uniformly sample random dislocation cores by choosing any point $(x,y)\in(-L,L)^2$, but they enforce that the total Burgers vector vanish. From Fig.~\ref{fig:dislocation_realizations}, one observes that even at high dislocation concentrations, the total strain field in the orbital basis retains spatial structure, having large regions of space that favor either compressive or tensile strain. The strain fields that decay as $1/r$ in \eqref{eq:elastic_dislocation_fields_real-space_polar} indicate that the entire dislocated medium experiences local symmetry breaking strains, which would tend to locally pin the electronic nematic order parameters in the $d$-orbital basis. This statement remains true even at the lowest concentrations because the strains are long-ranged, ultimately arising from the displacement vector's role in elasticity as the vector-valued potential for the strain tensor. This feature, in turn, is an expression of elasticity theory's global Euclidean symmetry that leaves the energy invariant under rigid translations and rotations.

To better quantify the spatial structure contained within this defect ensemble, we compute the correlation functions from each realization, and then perform a disorder average over all realizations. The $5\times5$ correlation matrix for the ensemble, $\mathcal{C}_{ab}(\boldsymbol{q})$, is given by
\begin{equation}
\begin{aligned}
    \mathcal{C}\left(\boldsymbol{q}\right)	&\equiv\boldsymbol{h}_{\boldsymbol{q}}^{\phantom{\text{T}}}\boldsymbol{h}_{-\boldsymbol{q}}^{\text{T}}=\sum_{\alpha,\beta=1}^{N}\text{e}^{\text{i}\boldsymbol{q}\cdot\left(\boldsymbol{x}^{\alpha}-\boldsymbol{x}^{\beta}\right)}\mathscr{K}^{\alpha\beta}\left(\boldsymbol{q}\right),
\\
\mathscr{K}^{\alpha\beta}\left(\boldsymbol{q}\right)
&\equiv\Gamma\left(\hat{q}\right)\cdot\boldsymbol{\varepsilon}^{p}(\boldsymbol{q};\boldsymbol{b}^{\alpha})\boldsymbol{\varepsilon}^{p}(-\boldsymbol{q};\boldsymbol{b}^{\beta})^{\text{T}}\cdot\Gamma^{\text{T}}\left(-\hat{q}\right).
\end{aligned}\label{eq:definition_of_correlation_single_realization}
\end{equation}
For simplicity, we assume that the dislocations are uncorrelated between themselves, that they can appear with uniform probability throughout the medium, and that their Burgers vectors are distributed isotropically. From these assumptions, it is possible to derive six nonzero, disorder-averaged, spatial correlation functions for the various components of the elastic strain field, $\boldsymbol{h}$, as shown in Appendix \ref{app:elastic_strains_from_dislocations}. These are given by
\begin{widetext}
\begin{equation}
 \begin{aligned}\left\llbracket\mathcal{C}_{z^2,z^2}\left(\boldsymbol{q}\right)\right\rrbracket & =2\pi^{2}N\xi^{2}\,\frac{(1-2\nu_{3})^{2}}{3}\frac{\delta\left(q_{z}\right)}{q^{2}}, & \left\llbracket \mathcal{C}_{z^2,x^2-y^2}\left(\boldsymbol{q}\right)\right\rrbracket  & =-2\pi^{2}N\xi^{2}\,\frac{1-2\nu_{3}}{\sqrt{3}}\frac{\delta\left(q_{z}\right)\cos\left(2\phi\right)}{q^{2}},\\
\left\llbracket \mathcal{C}_{x^2-y^2,x^2-y^2}\left(\boldsymbol{q}\right)\right\rrbracket  & =2\pi^{2}N\xi^{2}\,\frac{\delta\left(q_{z}\right)\left[1+\cos\left(4\phi\right)\right]}{2q^{2}}, & \left\llbracket \mathcal{C}_{z^2,2xy}\left(\boldsymbol{q}\right)\right\rrbracket  & =-2\pi^{2}N\xi^{2}\,\frac{1-2\nu_{3}}{\sqrt{3}}\frac{\delta\left(q_{z}\right)\sin\left(2\phi\right)}{q^{2}},\\
\left\llbracket \mathcal{C}_{2xy,2xy}\left(\boldsymbol{q}\right)\right\rrbracket  & =2\pi^{2}N\xi^{2}\,\frac{\delta\left(q_{z}\right)\left[1-\cos\left(4\phi\right)\right]}{2q^{2}}, & \left\llbracket \mathcal{C}_{x^2-y^2,2xy}\left(\boldsymbol{q}\right)\right\rrbracket  & =2\pi^{2}N\xi^{2}\,\frac{\delta\left(q_{z}\right)\sin\left(4\phi\right)}{2q^{2}},
\end{aligned}\label{eq:dis_avg_correlation_function_FS}
\end{equation}
\end{widetext}
where we have defined the parameter $\xi^2 \equiv \llbracket b^2 \rrbracket/(1-\nu_3)^2$. With the exception of $\llbracket\mathcal{C}_{z^2,z^2}(\boldsymbol{q})\rrbracket$,  all of these correlations are anisotropic. The cross-correlations exhibit anti-correlations between the different basis functions as well.

\begin{figure}
    \centering
    \includegraphics[width=\columnwidth]{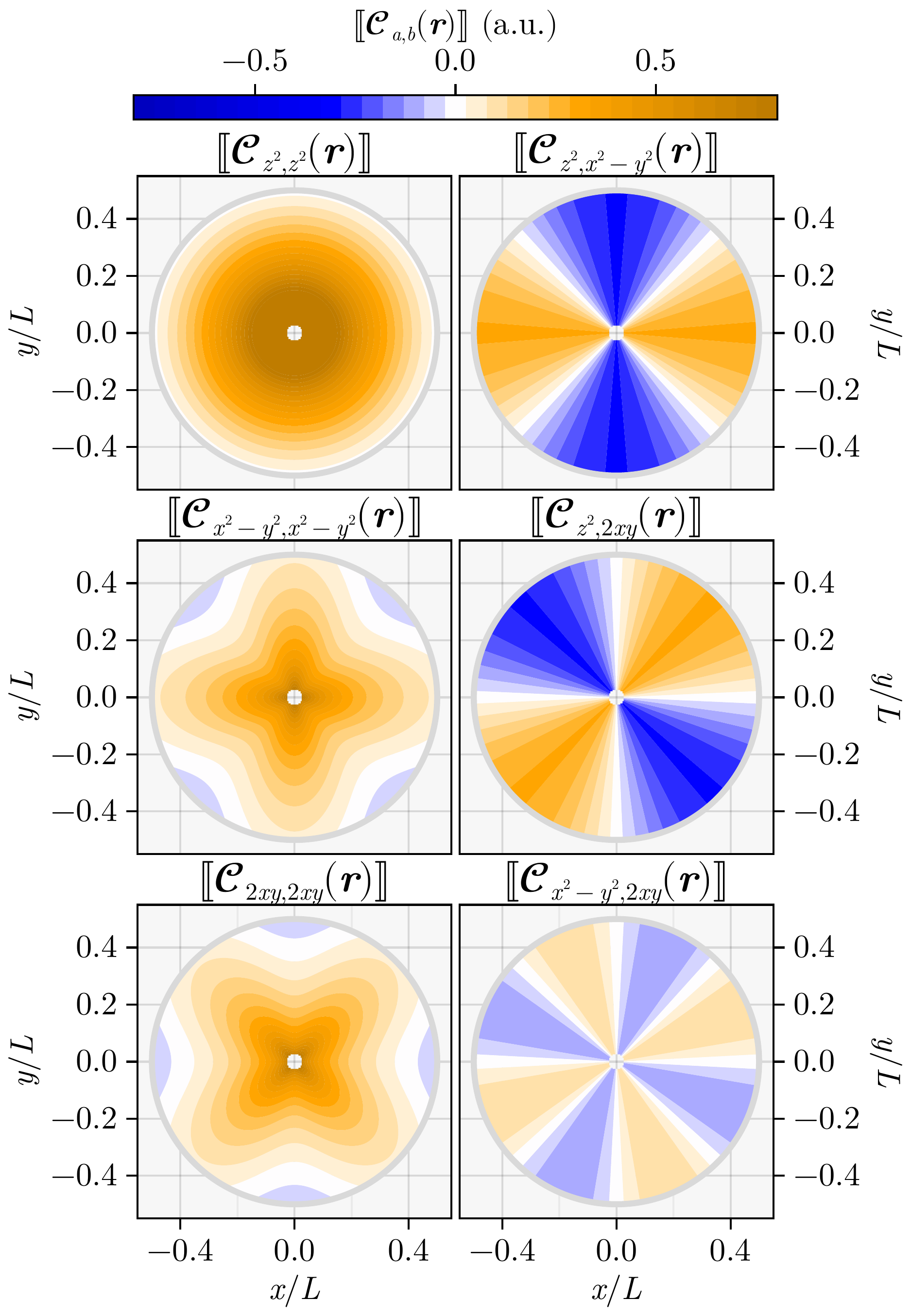}
    \caption{Disorder-averaged real-space elastic strain correlations from an ensemble of uncorrelated straight edge dislocations in a system of size $L$. The expressions for these functions are givenin Eqs. (\ref{eq:dis_avg_correlation_function_RS}) and are all normalized by their maximum magnitudes.  The subscripts label the $d$-orbital basis functions in $\boldsymbol{h}(\boldsymbol{r})$.  While the spatial autocorrelation functions (left-column) display a radial dependence, the cross-correlation functions between $d$-orbital basis functions do not. The planar correlation functions $\llbracket\mathcal{C}_{x^2-y^2,x^2-y^2}\rrbracket$, $\llbracket\mathcal{C}_{2xy,2xy}\rrbracket$, and $\llbracket\mathcal{C}_{x^2-y^2,2xy}\rrbracket = \llbracket\mathcal{C}_{2xy,x^2-y^2}\rrbracket$ appear in Ref.~\citep{ShortPaper} as well. }
    \label{fig:in_and_out_of_plane_dislocs_correlations}
\end{figure}

Finally, we compute these correlation functions in real-space, as shown in Appendix \ref{app:elastic_strains_from_dislocations}. There is a logarithmic divergence that naturally appears in the limit that $r_\parallel \rightarrow 0$ due to the breakdown of continuum elasticity near the dislocation core. After regularizing this integral by the macroscopic but finite size of the system, $L$, then the real-space correlations follow as
\begin{widetext}
\begin{equation}
    \begin{aligned}\left\llbracket \mathcal{C}_{z^2,z^2}\left(\boldsymbol{r}_{\parallel}\right)\right\rrbracket  & =N\xi^{2}\,\tfrac{(1-2\nu_{3})^{2}}{3}\log\left(\tfrac{L}{2r_{\parallel}}\right), & \left\llbracket \mathcal{C}_{z^2,x^2-y^2}\left(\boldsymbol{r}_{\parallel}\right)\right\rrbracket  & =\tfrac{1}{2}N\xi^{2}\,\tfrac{1-2\nu_{3}}{\sqrt{3}}\cos\left(2\phi_{r}\right),\\
\left\llbracket \mathcal{C}_{x^2-y^2,x^2-y^2}\left(\boldsymbol{r}_{\parallel}\right)\right\rrbracket  & =\tfrac{1}{4}N\xi^{2}\left\{ \log\left(\tfrac{L}{2r_{\parallel}}\right)+\tfrac{1}{4}\cos\left(4\phi_{r}\right)\right\} , & \left\llbracket \mathcal{C}_{z^2,2xy}\left(\boldsymbol{r}_{\parallel}\right)\right\rrbracket  & =\tfrac{1}{2}N\xi^{2}\,\tfrac{1-2\nu_{3}}{\sqrt{3}}\sin\left(2\phi_{r}\right),\\
\left\llbracket \mathcal{C}_{2xy,2xy}\left(\boldsymbol{r}_{\parallel}\right)\right\rrbracket  & =\tfrac{1}{4}N\xi^{2}\left\{ \log\left(\tfrac{L}{2r_{\parallel}}\right)-\tfrac{1}{4}\cos\left(4\phi_{r}\right)\right\} , & \left\llbracket \mathcal{C}_{x^2-y^2,2xy}\left(\boldsymbol{r}_{\parallel}\right)\right\rrbracket  & =\tfrac{1}{8}N\xi^{2}\,\sin\left(4\phi_{r}\right),
\end{aligned}
\label{eq:dis_avg_correlation_function_RS}
\end{equation}
\end{widetext}
as shown in the appendix. The spatial dependence of these functions are shown in Fig.~\ref{fig:in_and_out_of_plane_dislocs_correlations}.

This concludes the discussion of straight edge dislocations. To summarize, these correlation functions demonstrate long-range, non-analytic, and statistical interdependence of the random elastic strain fields emanated by an ensemble of independent and identically distributed dislocations. One identifies auto-correlations as well as correlations between different $d$-orbital basis functions, implying that the presence of generic defects will produce conjugate symmetry-breaking fields for all five electronic nematic order parameters in the $d$-orbital basis. These fields, albeit random, can never be assumed independent of one another. Furthermore, the anisotropy embedded within these long-range correlations, to our knowledge, has not been considered before when one discusses long-range correlated random field disorder. Because the anisotropy persists at all length scales, it will likely induce new statistical physics for disordered critical nemato-elasticity.

While here we focused on the case of straight edge dislocations, we remark that these conclusions are generally valid for any distribution of plastic defects. Since the defect density tensor, $\eta_{ij}$, is related to the physical incompatibility contained within the plastic strain, \eqref{eq:eta_tensor_incompatibility_of_E}, one can readily obtain the conjugate elastic strain $\boldsymbol{h} = \boldsymbol{\mathscr{E}}|_{\varphi=0}$ in terms of the defect density $\eta_{ij}$, rather than in terms of the plastic strain $\varepsilon^p_{ij}$. We show how this is done within our helical formalism in Appendix \ref{app:plastic_strain_defect_density_tensor}. Quite generally, $\eta_{ij}$ can be expanded in multipoles, which correspond to different types of defects -- for instance, the dipole term corresponds to dislocations whereas the quadrupole term corresponds to vacancies \cite{RMF03_smectic_defect}. If one considers the defect density as a distribution of quenched multipoles, one still finds a similar anisotropic family of quenched elastic strains which will host long-ranged, anisotropic, correlations. Simply put, when the defects are treated as random, rather than the strains themselves, it is clear that different symmetry channels will strongly depend on one another and cannot be treated as independent disorder fields.

\section{Discussion \label{sec:discussion}}

Electronic nematicity, a rotational symmetry-breaking phase of matter, appears within a wide range of interesting quantum materials and can be induced by a wide range of microscopic mechanisms. Given the appearance of universal electronic nematic signatures across many distinct families of quantum materials, a phenomenological description applicable to all is warranted, and indeed necessary, to better understand electronic nematicity as a quantum phase of matter.

One of the most important effects that electronic nematicity has in crystalline quantum materials is the bilinear coupling between the electronic nematic order parameter and the symmetry-breaking components of the elastic strain tensor. This nemato-elastic coupling is absent in classical nematic liquids, given their lack of shear rigidity. Meanwhile, because it does exist in quantum materials, nemato-elasticity has shed light on many aspects of electronic nematicity that differentiate it further from nematicity in liquid crystals. For example, the reduced rotational symmetry in crystalline environments allows for electronic nematicity to manifest different universality classes \cite{Fernandes2014, Hecker2018, Fernandes2020, RMF14_classification_nematics}. Moreover, elastic fluctuations tend to suppress the phase space for thermal nematic fluctuations, effectively favoring the mean-field nematic criticality \cite{Karahasanovic16, Paul17, Fernandes2020} frequently observed in experiments \cite{Chu2010_nem, Fernandes10_nematic, Chu12_nem, sanchezTransportStructuralCorrespondence2021,sanchezQuantitativeRelationshipStructural2022}. Additionally, coupling the nematic order locally to intrinsic defect strains introduces random field disorder for the nematic order parameter \cite{RMF01_RFBM, RMF13_disorder_TmVO4}. Such disorder may explain the apparent mesoscopic nematic inhomogeneities that have been frequently observed via local probes in actual samples \cite{tanatarDirectImagingStructural2009, Buchner2010, Niedziela2012,Rosenthal2014,Dioguardi2015, Dioguardi2016, Forrest2016, Ren2021, shimojimaDiscoveryMesoscopicNematicity2021,curroNematicityGlassyBehavior2022, RMF08_Potts_nematic_strain, RMF13_disorder_TmVO4}. 

While the qualitative understanding of electronic nematicity gained from understanding these two individual aspects of nemato-elasticity is strong, they together seem contradictory. How is it that the same nemato-elastic interactions favor mean-field criticality while simultaneously propagating random pinning fields from omnipresent crystalline defects? This question is not purely academic -- even in systems where the disorder is specifically tuned, and therefore rotational symmetry is already broken at small distances, one can still observe sharp criticality indicative of a diverging electronic nematic correlation length \cite{Ran2011, jescheCouplingStructuralMagnetic2010, RMF13_disorder_TmVO4}.

We have shown in this paper, and in the accompanying shorter paper, Ref.~\cite{ShortPaper}, that nemato-elasticity quantitatively transforms the behavior of the bare electronic nematic order parameter, and addresses this apparent contradiction. The transformation is initiated by the Saint-Venant compatibility relations, well-known to elasticity theory as a geometrical constraint of continuous deformations. In short, the electronic nematic order parameter inherits the compatibility relations from elasticity theory. By satisfying them, one establishes a nonlocal transformation between the usual ``$d$-orbital basis'' of electronic nematicity and the two critical ``helical'' nematic order parameters. These helical order parameters, being the components of a manifestly compatible tensor, are blind to random incompatible strain fields generated by structural disorder, and thus can remain critical even in a plastically deformed medium. In this sense, our theory provides a framework to understand how the presence of a sharp electronic nematic phase transition is protected in structurally disordered media.

Within the context of electronic nematics overall, our theory of critical nemato-elasticity primarily establishes that electronic nematic fluctuations, which are degenerate in the bare theory, are not in an elastic medium. This is because electronic nematicity is represented by a tensor quantity that can have both compatible and incompatible contributions, whereas the elastic strain tensor in ideal media exclusively contains compatible contributions. In the context of crystalline nematics, the degeneracy of different orbital nematic fluctuations even in the \textit{same} critical irreducible representation may split depending on their momenta -- indeed, this is the phenomenon of direction-selective criticality. While being responsible for the unexpected mean-field behavior of electronic nematic phase transitions in low dimensions, the suppression of incompatible nematicity may also be responsible for an apparent planar universality exhibited across the many electronic nematics families, even the fully three-dimensional ones. By ``planar universality,'' we mean that the critical momenta near electronic nematic phase transitions lies in the basal plane, despite many of these materials having different electronic and structural properties \cite{fernandesIronPnictidesChalcogenides2022}. Within the context of our theory, this planar universality may be accounted for since the electronic nematic directors must be coplanar with their critical momenta. Given that most actual nematics condense the $\varphi_{x^2-y^2}$ or $\varphi_{2xy}$ basis functions due to their uniaxial symmetry, this means that the only critical momenta can lie in the $xy$ plane so as to suppress nematic incompatibility.   While there may be other reasons for this behavior, it would be somewhat coincidental that 3D tetragonal-to-orthorhombic nematics with different electronic properties all exhibit effectively two-dimensional physics governed by in-plane transfer momenta close to criticality. 

From a mesoscopic perspective, meanwhile, the fact that the $d$-orbital and helical bases are related by a local transformation in momentum-space requires that they are related by a nonlocal transformation in real-space. Thus, one cannot associate local probes of electronic anisotropies with the local critical nematic order parameter. In terms of a given $d$-orbital nematic order parameter, such as $\varphi_{2xy}$, the critical macroscopic order parameter is only realized by first softening along the directions where $\varphi_{2xy}$ maps uniquely onto the critical helical doublet, $(\Phi_{2}, \Phi_{3})$, and then taking the infinite wavelength limit. This distinction is not just academic: given the growing experimental capabilities in measuring mesoscopic nematicity \cite{ shimojimaDiscoveryMesoscopicNematicity2021, Ren2021, gursoyDarkfieldXrayMicroscopy2025, KaanPaper}, one must contend with the nonlocality that is endemic to elastic deformation in electronic nematics. For example, there is evidence that nematic domain boundaries may generate mesoscopic strain modulations within the bulk domains, with wavevectors exactly aligned along the critical directions that force the nematic tensor to be compatible \cite{KaanPaper}. 

Additionally, Ref. \cite{Ren2021} reported a striking disparity between microscopic nematicity and the associated shear structural deformations. This is a puzzling observation, since one would naively expect from the bilinear nemato-elastic coupling that the microscopic $\varphi_{2xy}$  nematic order parameter triggers a microscopic shear deformation $\varepsilon_{xy}$. Our results, however, show that this discrepancy is a consequence of the nonlocal transformation between the helical and $d$-orbital representations of the strain tensor. Given the presumptive role of structural disorder in the experiments performed in that work, we would associate the local nematicity in the $d$-orbital basis directly with pinning effects from crystalline defects, which are orthogonal to the critical (helical) nematic order parameters.  Because of the nonlocal transformation between the $d$-orbital and helical representations, one expects the local $\varphi_{2xy}(\boldsymbol{x})$ nematic modulations to be composed of non-critical helical modulations $\Phi_{1,4,5}$. Thus, local probes will be influenced by non-critical anisotropies -- those that always remain massive and short-ranged. 

Our theory establishes a framework to discuss plasticity and nematicity; hence, the term, ``nemato-plasticity.'' Given the reciprocity that exists in nemato-elasticity, it is reasonable to ask whether electronic nematicity, as collective electronic modes, can induce plastic deformation in their host crystal. While this lies outside the scope of the current work, we emphasize that our helical theory of nemato-elasticity provides the first defect-gauge invariant formalism within which one can discuss plasticity generated by electronic nematic fluctuations. Given the strong influence that plasticity and plastic deformation can have on electronic properties, as recently demonstrated experimentally in the quantum paraelectric SrTiO$_3$ \cite{Hameed2022}, the possibility that \textit{intrinsic} electronic nematicity can be a tuning parameter for plastic deformation is an exciting prospect that has not been considered before.

We have established the crucial and unifying role of the Saint Venant compatibility relations in nemato-elasticity. By developing a formalism that is particularly sensitive to them, the compatibility relations provide a resolution for the apparent paradox within nemato-elasticity. Because of the tensor gauge field constraints they impose on the strain tensor, elastic fluctuations indeed enhance the critical electronic correlations that spontaneously drive the system into the nematic phase. However, they do not enhance any nematic fluctuations, rather only those that reside within a critical \textit{compatible} manifold. Because this compatible sector is orthogonal to the non-critical, \textit{incompatible} nemato-plastic phase space, the compatibility relations protect electronic nematic criticality in the presence of structural disorder.  Given the crucial role of electronic nematicity in stabilizing other correlated electronic phases -- such as unconventional superconductivity \cite{Lederer2015,Lederer2017,Klein2018} -- our results will have far reaching impacts within many quantum materials.

\begin{acknowledgments}
The authors would like to thank J. Freedberg, P. Littlewood, J. Vi\~nals, A. Rastogi, A. Chakraborty, and R. Aquino for many insightful discussions. We also thank J. Freedberg for providing helpful feedback on the manuscript.
\end{acknowledgments}


\appendix 

\section{Gell-Mann Decomposition of Rank-2 Tensors \label{app:Gell-Mann-Decomp-Tensors}}
Electronic nematicity is frequently discussed in terms of crystalline point groups, and therefore is usually represented as multi-component \textit{vectorial} irreducible representations within those groups \citep{RMF14_classification_nematics}. This situation contrasts with elasticity, which is most conveniently expressed in its \textit{tensorial} form -- that is, with respect to its Cartesian coordinates \citep{landauTheoryElasticity1970}. This is true not just in more obvious engineering applications, but also in gauge-theoretic formulations \citep{kleinertDoubleGaugeTheory1983, valsakumarGaugeTheoryDefects1988, kleinertGaugeFieldsSolids1989, beekmanDualGaugeField2017}.  The Gell-Mann matrices of $\text{SU}(3)$ provide a convenient machinery to transform rank-2 tensors between their tensorial and vectorial representations. In this appendix, we tabulate the Gell-Mann matrix conventions used and derive some identities used throughout this work.

Let $A$  be a rank-2 symmetric tensor in three spatial dimensions. It can always be decomposed in the following form 
\begin{equation}
    A=\frac{1}{3}\left(\text{Tr}\,A\right)\lambda^{0}+\frac{1}{2}\boldsymbol{A}\cdot\boldsymbol{\lambda},\quad\boldsymbol{A}=\text{Tr}\left(\boldsymbol{\lambda}A\right),\label{eq:GellMann_decomp_tensor}
\end{equation}
where $\lambda^0$  is the identity and where we have defined the 5-component Gell-Mann vector to be $\boldsymbol{\lambda}\equiv(\lambda^8, \lambda^3,\lambda^6, \lambda^4,\lambda^1)^{\text{T}}$. Here use the following definitions of the traceless symmetric Gell-Mann matrices:
\begin{equation}
    \begin{array}{ccccc}
\lambda^{1}=\begin{pmatrix}0 & 1 & 0\\
1 & 0 & 0\\
0 & 0 & 0
\end{pmatrix}, &  & \lambda^{3}=\begin{pmatrix}1 & 0 & 0\\
0 & -1 & 0\\
0 & 0 & 0
\end{pmatrix}, &  & \lambda^{4}=\begin{pmatrix}0 & 0 & 1\\
0 & 0 & 0\\
1 & 0 & 0
\end{pmatrix},\\
\lambda^{6}=\begin{pmatrix}0 & 0 & 0\\
0 & 0 & 1\\
0 & 1 & 0
\end{pmatrix}, &  & \lambda^{8}=\frac{1}{\sqrt{3}}\begin{pmatrix}1 & 0 & 0\\
0 & 1 & 0\\
0 & 0 & -2
\end{pmatrix}.
\end{array}\label{eq:GellMann_Matrices}
\end{equation}
The other three Gell-Mann matrices, $\lambda^{2,5,7}$, are all anti-symmetric, and are not important for this work. The ordering of the Gell-Mann vector, $\boldsymbol{\lambda}$, is chosen to mirror the orbital ordering of the basis functions of the electronic nematic order parameter in \eqref{eq:definition_of_nematic_order_deviatoric_strain_orbital_basis}.  One can show \eqref{eq:GellMann_decomp_tensor} holds even in the case where $A$  has a nonzero anti-symmetric part, using the two trace properties of the Gell-Mann matrices \footnote{While \eqref{eq:GellMann_decomp_tensor} holds for any tensor, $A$, the meaning of its vectorial representation, $\boldsymbol{A}$, will be different if the antisymmetric Gell-Mann matrices are included.}:
\begin{equation}
    \text{Tr}\left(\lambda^{a}\right)=3\delta_{0a},\quad\text{Tr}\left(\lambda^{a}\lambda^{b}\right)=2\delta_{ab}. \label{eq:trace_properties_GellMann}
\end{equation}
Using these properties, it follows that $\boldsymbol{A}$ is in one-to-one correspondence with the deviatoric, traceless, part of the tensor. This vector, $\boldsymbol{A}$, uniquely maps the deviatoric part of the symmetric tensor $A$ in the $d$-orbital basis onto the $A_{ij}$ components in the Cartesian basis.

Let $B$  be another rank-2 tensor (with or without an antisymmetric part). From \eqref{eq:trace_properties_GellMann}, it follows that 
\begin{equation}
    \text{Tr}\left(AB\right)=\frac{1}{3}\left(\text{Tr}\,A\right)\left(\text{Tr}\,B\right)+\frac{1}{2}\left(\boldsymbol{A}\cdot\boldsymbol{B}\right).\label{eq:GellMann_Trace_AB}
\end{equation}
The above expression is a consequence of the Fierz identity for Gell-Mann matrices,
\begin{equation}
    \lambda_{ij}^{a}\lambda_{kl}^{a}=2\left(\delta_{il}\delta_{jk}-\tfrac{1}{3}\delta_{ij}\delta_{kl}\right). \label{eq:FierzIdentity}
\end{equation}
From \eqref{eq:GellMann_Trace_AB}, it follows that 
\begin{equation}
    \text{Tr}\left(AA\right)=\frac{1}{3}\left(\text{Tr}\,A\right)^{2}+\frac{1}{2}\left(\boldsymbol{A}\cdot\boldsymbol{A}\right). \label{eq:trace_AA_GellMann_representation}
\end{equation}

Moving back to nemato-elasticity, both the nematic order parameter from \eqref{eq:definition_nematic_tensor_expectations} and strain tensor in \eqref{eq:strain_separation_dilatation_deviatoric} can be written in vectorial form as 
\begin{equation}
    \varphi_{ij}=\frac{1}{2}\boldsymbol{\varphi}\cdot\boldsymbol{\lambda}_{ij},\quad\varepsilon_{ij}=\frac{1}{3}\left(\text{Tr}\,\varepsilon\right)\lambda_{ij}^{0}+\frac{1}{2}\boldsymbol{\varepsilon}\cdot\boldsymbol{\lambda}_{ij}, \label{eq:nematic_and_strain_GellMann_decomp}
\end{equation}
with $\text{Tr}\,\varphi\equiv 0$. The vector $\boldsymbol{\varepsilon}$ is in one-to-one correspondence with the deviatoric part of the strain tensor. 

\section{Quadrupolar Form Factor Identities\label{app:quadrupolar_form_factors}}
In this appendix, we provide some explicit identities necessary for constructing the co-rotating helical basis of deviatoric tensors. 

We start by calculating the traces and trace-norms of the  $\{\mathcal{Q}^{11}, \mathcal{Q}^{12}, \mathcal{Q}^{13}\}$ tensors of compatible elasticity. From \eqref{eq:trace_Qab} and \eqref{eq:traceQabQgd}, it follows that
\begin{equation}
    \begin{array}{ccccc}
\text{Tr}\,\left(\mathcal{Q}^{11}\right)=1, &  & \text{Tr}\,\left(\mathcal{Q}^{11}\mathcal{Q}^{11}\right)=1, &  & \text{Tr}\,\left(\mathcal{Q}^{11}\mathcal{Q}^{12}\right)=0,\\
\text{Tr}\,\left(\mathcal{Q}^{12}\right)=0, &  & \text{Tr}\,\left(\mathcal{Q}^{12}\mathcal{Q}^{12}\right)=\frac{1}{2}, &  & \text{Tr}\,\left(\mathcal{Q}^{11}\mathcal{Q}^{13}\right)=0,\\
\text{Tr}\,\left(\mathcal{Q}^{13}\right)=0, &  & \text{Tr}\,\left(\mathcal{Q}^{13}\mathcal{Q}^{13}\right)=\frac{1}{2}, &  & \text{Tr}\,\left(\mathcal{Q}^{12}\mathcal{Q}^{13}\right)=0,
\end{array}\label{eq:trace_properties_Q1a}
\end{equation}
establishing them as a set of three trace-orthogonal tensors at each momentum, $\hat{q}$.

Using the Gell-Mann decomposition in \eqref{eq:Qab_tensor_GellMann_Decomp}, we find that
\begin{equation}
    \begin{aligned}\mathcal{Q}^{11} & =\tfrac{1}{3}\lambda^{0}+\tfrac{1}{2}\boldsymbol{Q}^{11}\cdot\boldsymbol{\lambda}=\tfrac{1}{3}\lambda^{0}+\tfrac{1}{2}\left(\hat{e}_{1}^{\text{T}}\cdot\boldsymbol{\lambda}\cdot\hat{e}_{1}\right)\cdot\boldsymbol{\lambda},\\
\mathcal{Q}^{12} & =\tfrac{1}{2}\boldsymbol{Q}^{12}\cdot\boldsymbol{\lambda}=\tfrac{1}{2}\left(\hat{e}_{1}^{\text{T}}\cdot\boldsymbol{\lambda}\cdot\hat{e}_{2}\right)\cdot\boldsymbol{\lambda},\\
\mathcal{Q}^{13} & =\tfrac{1}{2}\boldsymbol{Q}^{13}\cdot\boldsymbol{\lambda}=\tfrac{1}{2}\left(\hat{e}_{1}^{\text{T}}\cdot\boldsymbol{\lambda}\cdot\hat{e}_{3}\right)\cdot\boldsymbol{\lambda}.
\end{aligned}
\end{equation}
The Gell-Mann vectors satisfy:
\begin{equation}
    \begin{array}{ccc}
\boldsymbol{Q}^{11}\cdot\boldsymbol{Q}^{11}=\frac{4}{3}, &  & \boldsymbol{Q}^{11}\cdot\boldsymbol{Q}^{12}=0,\\
\boldsymbol{Q}^{12}\cdot\boldsymbol{Q}^{12}=1, &  & \boldsymbol{Q}^{11}\cdot\boldsymbol{Q}^{13}=0,\\
\boldsymbol{Q}^{12}\cdot\boldsymbol{Q}^{12}=1, &  & \boldsymbol{Q}^{12}\cdot\boldsymbol{Q}^{13}=0.
\end{array}\label{eq:dot_product_Q1alpha}
\end{equation}
To normalize these vectors, and to find two other orthonormal vectors necessary to complete the basis, we appeal to \eqref{eq:innerproduct_QabQgd_vectors}, repeated here for convenience as
\begin{equation}
    \boldsymbol{Q}^{\alpha\beta}\cdot\boldsymbol{Q}^{\gamma\delta}=\delta_{\alpha\gamma}\delta_{\beta\delta}+\delta_{\alpha\delta}\delta_{\beta\gamma}-\frac{2}{3}\delta_{\alpha\beta}\delta_{\gamma\delta}.\tag{\ref{eq:innerproduct_QabQgd_vectors}}
\end{equation}
The remaining two orthogonal vectors can be obtained from the expression above by setting $\alpha = 1$, representing the known vectors from elasticity in \eqref{eq:dot_product_Q1alpha}. Then $\boldsymbol{Q}^{23}$ can be obtained immediately since each term in the inner product vanishes. This is the fourth basis vector. The fifth follows a similar logic, but since all values of the Greek indices $\{1,2,3\}$ have been used by the first four basis vectors, we must appeal to repeated indices, as is the case for $\boldsymbol{Q}^{11}$. Since this will incur a nonzero overlap with $\boldsymbol{Q}^{11}$ from the last term in \eqref{eq:innerproduct_QabQgd_vectors}, by taking the difference between $\boldsymbol{Q}^{22}$ and $\boldsymbol{Q}^{33}$, we eliminate this overlap, and arrive at the last basis vector needed for a complete set. The definitions of the orthonormalized Gell-Mann vectors in the co-rotating helical basis are therefore
\begin{equation}
    \begin{aligned}\boldsymbol{\hat{Q}}_{1} & \equiv\tfrac{\sqrt{3}}{2}\boldsymbol{Q}^{11} = \tfrac{\sqrt{3}}{2}\left(\hat{e}_1^{\mathrm T} \cdot \boldsymbol{\lambda} \cdot \hat{e}_1 \right), \\ \boldsymbol{\hat{Q}}_{2} & \equiv\boldsymbol{Q}^{12}=\hat{e}_1^{\mathrm T} \cdot \boldsymbol{\lambda} \cdot \hat{e}_2,\\
\boldsymbol{\hat{Q}}_{3} & \equiv\boldsymbol{Q}^{13} = \hat{e}_1^{\mathrm T} \cdot \boldsymbol{\lambda} \cdot \hat{e}_3, \\ \boldsymbol{\hat{Q}}_{4} & \equiv\boldsymbol{Q}^{23} =\hat{e}_2^{\mathrm T} \cdot \boldsymbol{\lambda} \cdot \hat{e}_3, \\ \boldsymbol{\hat{Q}}_{5} & \equiv\tfrac{1}{2}\left(\boldsymbol{Q}^{22}-\boldsymbol{Q}^{33}\right) = \tfrac{1}{2}\left(\hat{e}_2^{\mathrm T} \cdot \boldsymbol{\lambda} \cdot \hat{e}_2 - \hat{e}_3^{\mathrm T} \cdot \boldsymbol{\lambda} \cdot \hat{e}_3 \right).
\end{aligned}\label{eq:vectorial_Qhatvec_basis}
\end{equation}
These helical basis vectors each have five components in the $d$-orbital basis and are mututually orthonormal at every momentum $\boldsymbol{q}\neq\boldsymbol{0}$. Their orthonormality can be established using \eqref{eq:innerproduct_QabQgd_vectors} in a similar tabulation as \eqref{eq:dot_product_Q1alpha}. These quadrupolar unit vectors at $\hat{q}$ are directly proportional to those at $-\hat{q}$. From \eqref{eq:ehat_helical_strain_negative_qhat}, it follows that 
\begin{equation}
\boldsymbol{\hat{Q}}_{a}(-\hat{q}) = \left(-1\right)^{\delta_{a,3} + \delta_{a,4}}\boldsymbol{\hat{Q}}_a(\hat{q}). \label{eq:Qhat_under_qhat_to_minus_qhat}
\end{equation}
With these basis vectors established, in a similar spirit to \eqref{eq:GellMann_decomp_tensor}, we can write any rank-2 symmetric tensor $A_{ij}=A_{ji}$ in a helical Gell-Mann decomposition as 
\begin{equation}
    A_{ij} =\frac{1}{3}\left(\text{Tr}\,A\right)\delta_{ij}+\frac{1}{2}\left(\sum_{a=1}^{5}A_{a}\boldsymbol{\hat{Q}}_{a}\right) \cdot \boldsymbol{\lambda}_{ij}, \quad A_a \equiv \boldsymbol{\hat{Q}}_a \cdot \boldsymbol{A}. \label{eq:Aij_helical_GellMann_decomp_tensor_from_vectorial}
\end{equation}

We now show how the $\{\vartheta^a\}$ basis tensors in \eqref{eq:def_varTheta_GellMann_of_Qhat}  are related to the basis vectors in \eqref{eq:vectorial_Qhatvec_basis}  and ultimately the $\mathcal{Q}^{\alpha\beta}_{ij}$  form factors defined in \eqref{eq:Qabij}. In particular, these tensors are related by 
\begin{equation}
    \begin{aligned}\vartheta_{ij}^{1} & =\tfrac{1}{\sqrt{2}}\left(\boldsymbol{\hat{Q}}_{1}\cdot\boldsymbol{\lambda}_{ij}\right)=\sqrt{\tfrac{3}{2}}\left(\mathcal{Q}_{ij}^{11}-\tfrac{1}{3}\delta_{ij}\right)\\
\vartheta_{ij}^{2} & =\tfrac{1}{\sqrt{2}}\left(\boldsymbol{\hat{Q}}_{2}\cdot\boldsymbol{\lambda}_{ij}\right)=\sqrt{2}\mathcal{Q}_{ij}^{12},\\
\vartheta_{ij}^{3} & =\tfrac{1}{\sqrt{2}}\left(\boldsymbol{\hat{Q}}_{3}\cdot\boldsymbol{\lambda}_{ij}\right)=\sqrt{2}\mathcal{Q}_{ij}^{13},\\
\vartheta_{ij}^{4} & =\tfrac{1}{\sqrt{2}}\left(\boldsymbol{\hat{Q}}_{4}\cdot\boldsymbol{\lambda}_{ij}\right)=\sqrt{2}\mathcal{Q}_{ij}^{23},\\
\vartheta_{ij}^{5} & =\tfrac{1}{\sqrt{2}}\left(\boldsymbol{\hat{Q}}_{5}\cdot\boldsymbol{\lambda}_{ij}\right)=\tfrac{1}{\sqrt{2}}\left(\mathcal{Q}_{ij}^{22}-\mathcal{Q}_{ij}^{33}\right).
\end{aligned}\label{eq:varThetaTensor_to_Qvechat_to_Qabij}
\end{equation}
These basis tensors -- being traceless, rank-2, and symmetric -- contain exactly the same information as the \textit{vectorial} basis in \eqref{eq:vectorial_Qhatvec_basis}. However, they are explicitly trace-orthonormal with respect to \textit{Cartesian} coordinates at every momentum $\boldsymbol{q}\neq\boldsymbol{0}$. The 5-component basis vectors $\{\boldsymbol{\hat{Q}}_\alpha(\hat{q})\}$ are useful for calculations specifically in the $d$-orbital space, whereas the basis tensors $\{\vartheta^\alpha(\hat{q})\}$ are conducive for calculations in Cartesian space.

From direct substitution, it follows from \eqref{eq:Qhat_under_qhat_to_minus_qhat} that by reversing the momentum, the $\vartheta$ basis tensors obey
\begin{equation}
    \vartheta_{ij}^a(-\hat{q}) = \left(-1\right)^{\delta_{a,3} + \delta_{a,4}}\vartheta_{ij}^a(\hat{q}).
\end{equation}
From these basis tensors, any rank-2 symmetric tensor, $A_{ij} = A_{ji}$, can be decomposed in momentum space as \eqref{eq:tensor_varThetaTensor_decomposition} in the main text. We repeat it here for completeness as
\begin{equation}
   \begin{aligned}A_{ij} & =\frac{1}{3}\left(\text{Tr}\,A\right)\delta_{ij}+\frac{1}{\sqrt{2}}\sum_{a=1}^{5}A_{a}\vartheta_{ij}^{a},\\
A_{a} & =\sqrt{2}\text{Tr}\left(\vartheta^{a}A\right)=\boldsymbol{\hat{Q}}_{a}\cdot\boldsymbol{A}.
\end{aligned}\tag{\ref{eq:tensor_varThetaTensor_decomposition}}
\end{equation}
This \textit{tensorial} decomposition is completely equivalent to the \textit{vectorial} one in \eqref{eq:Aij_helical_GellMann_decomp_tensor_from_vectorial}, with the exact same coefficients. However, the tensorial one proves more convenient for calculations involving tensor incompatibility.

 To find the incompatibility of a given tensor, it follow from \eqref{eq:tensor_varThetaTensor_decomposition} and \eqref{eq:varThetaTensor_to_Qvechat_to_Qabij} that one must compute $\text{inc}(\text{I})$  and $\text{inc}(\mathcal{Q}^{\alpha\beta})$. Starting with the latter, and applying \eqref{eq:KronerInc_Fourier_Space}, we have
\begin{align}
    \text{inc}\,\left(\mathcal{Q}^{\alpha\beta}\right)_{il}	&=-\tfrac{q^{2}}{2}\epsilon_{ijk}\epsilon_{lmn}\hat{e}_{1,j}\hat{e}_{1,m}\left(\hat{e}_{\alpha,k}\hat{e}_{\beta,n}+\hat{e}_{\alpha,n}\hat{e}_{\beta,k}\right)\nonumber
	\\
    &=-\tfrac{q^{2}}{2}\bigg\lbrace\left(\hat{e}_{1}\times\hat{e}_{\alpha}\right)_{i}\left(\hat{e}_{1}\times\hat{e}_{\beta}\right)_{l}\nonumber
    \\
	&\phantom{=-\frac{q^{2}}{2}\bigg\lbrace}+\left(\hat{e}_{1}\times\hat{e}_{\beta}\right)_{i}\left(\hat{e}_{1}\times\hat{e}_{\alpha}\right)_{l}\bigg\rbrace.
\end{align}
Given that the $\hat{e}_\alpha$  unit vectors are a right-handed orthonormal set at each $\hat{q}$, it follows that $\left(\hat{e}_{\alpha}\times\hat{e}_{\beta}\right)_{i}=\epsilon_{\alpha\beta\gamma}\hat{e}_{\gamma,i}$. Thus, we find that 
\begin{equation}
    \text{inc}\,\left(\mathcal{Q}^{\alpha\beta}\right)_{ij}=-q^{2}\epsilon_{1\alpha\gamma}\epsilon_{1\beta\gamma}\mathcal{Q}_{ij}^{\gamma\delta}.
\end{equation}
Applying this formula to the five orthogonal $\mathcal{Q}^{\alpha\beta}$ tensors yields
\begin{equation}
    \begin{aligned}&\text{inc}\left(\mathcal{Q}^{11}\right)=0,  & &\text{inc}\left(\mathcal{Q}^{12}\right)=0,\\
&\text{inc}\left(\mathcal{Q}^{13}\right)=0,   & &\text{inc}\left(\mathcal{Q}^{23}\right)=q^{2}\mathcal{Q}^{23}, &  & 
\\
&\text{inc}\left(\mathcal{Q}^{22}-\mathcal{Q}^{33}\right)=q^{2}\left(\mathcal{Q}^{22}-\mathcal{Q}^{33}\right),
\end{aligned}\label{eq:inc_Qab_5_tensors}
\end{equation}
which shows that each of these tensors is an eigentensor of the incompatibility operator. The incompatibility of the identity tensor, meanwhile, follows from
\begin{align}
    \text{inc}\left(\text{I}\right)_{il}	&=-q^{2}\epsilon_{ijk}\epsilon_{lmn}\hat{e}_{1,j}\hat{e}_{1,m}\delta_{kn}\nonumber
    \\
	&=-q^{2}\left(\delta_{il}\delta_{jm}-\delta_{im}\delta_{jl}\right)\hat{e}_{1,j}\hat{e}_{1,m}.
\end{align}
Simplifying the last expression then yields \eqref{eq:incI_fourier_space}. From these identities, it follows that
\begin{equation}
    \begin{aligned}\text{inc}\,\vartheta^{1} & =\sqrt{\tfrac{3}{2}}\text{inc}\left(\mathcal{Q}^{11}-\frac{1}{3}\text{I}\right)=-\tfrac{q^{2}}{\sqrt{6}}\left(\mathcal{Q}^{11}-\text{I}\right),\\
\text{inc}\,\vartheta^{2} & =\sqrt{2}\text{inc}\,\mathcal{Q}^{12}=0,\\
\text{inc}\,\vartheta^{3} & =\sqrt{2}\text{inc}\,\mathcal{Q}^{12}=0,\\
\text{inc}\,\vartheta^{4} & =\sqrt{2}\text{inc}\,\mathcal{Q}^{23}=q^{2}\vartheta^{4}\\
\text{inc}\,\vartheta^{5} & =\tfrac{1}{\sqrt{2}}\text{inc}\left(\mathcal{Q}^{22}-\mathcal{Q}^{33}\right)=q^{2}\vartheta^{5}.
\end{aligned}. \label{eq:inc_varTheta_tensors}
\end{equation}
From these expressions, the incompatibility of any rank-2 symmetric tensor follows as \eqref{eq:incompatibility_of_A_varTheta_decomp} in the main text and repeated here for convenience:
\begin{align}
    \text{inc}\,A	=\frac{1}{3}\left(\text{Tr}\,A-\tfrac{\sqrt{3}}{2}A_{1}\right)\text{inc}\left(\text{I}\right)+\frac{q^{2}}{\sqrt{2}}\left(A_{4}\vartheta^{4}+A_{5}\vartheta^{5}\right).\tag{\ref{eq:incompatibility_of_A_varTheta_decomp}}
\end{align}

\section{Bilinear Order Parameter Coupling to Dilatation\label{app:bilinear_coupling_to_dilatation}}
In this appendix, we show that a bilinear coupling to dilatation leads to a discontinuous mass renormalization for an scalar order parameter, $\mathcal{A}$, which is invariant under spatial rotations and inversion \cite{villainSELFCONSISTENCYLANDAUS1970,Nelson1996,dzeroQuantumCriticalEnd2006,hacklKondoVolumeCollapse2008, zachariasMottMetalInsulatorTransition2012, Guzman2019}. Unlike the bilinear coupling to deviatoric strain, such as nemato-elasticity, coupling to dilatation does not lead to direction-selective criticality since the renormalization is discontinuous at $\boldsymbol{q}=\boldsymbol{0}$.

We assume that the bilinear coupling is of the form 
\begin{equation}
    \Delta\mathcal{F} = -\gamma\int_x \mathcal{A}(\boldsymbol{x})\varepsilon_0(\boldsymbol{x}),
\end{equation}
with $\gamma$ being the coupling constant for the theory. Writing the dilatation in terms of the homogeneous and inhomogeneous contributions yields
\begin{equation}
    \varepsilon_0(\boldsymbol{x}) = \overline{\varepsilon}_0 + \frac{1}{V}\sum_{q>0} \text{e}^{\text{i}\boldsymbol{q}\cdot \boldsymbol{x}}\varepsilon_{0,\boldsymbol{q}},
\end{equation}
with a bilinear coupling given by
\begin{equation}
    \Delta\mathcal{F} = -\frac{\gamma\, V}{2}\;\overline{\mathcal{A}}\,\overline{\varepsilon_0}-\frac{\gamma}{2V}\sum_{q>0}\left( \mathcal{A}^* \varepsilon_1^h + \text{c.c.}\right),
\end{equation}
in the helical representation with $\varepsilon_{0,\boldsymbol{q}} = \varepsilon^h_{1,\boldsymbol{q}}$. Here $\overline{A} \equiv \frac{1}{V}\int_x \mathcal{A}(\boldsymbol{x})$ and $\overline{\varepsilon_0}\equiv \frac{1}{V}\int_x \varepsilon_0(\boldsymbol{x})$. Notice how the $\varepsilon_{2,3}^h$ helical strains are not involved at all in this problem. The total elastic free energy is then
\begin{align}
    \mathcal{F} &= V\left\{ \frac{1}{2} B\, \overline{\varepsilon_0}^2 - \gamma\, \overline{\mathcal{A}}\,\overline{\varepsilon_0}\right\} \nonumber
    \\
    &\phantom{=} + \frac{1}{2V} \sum_{q>0} \left\{ (\lambda + 2\mu) |\varepsilon^h_1|^2 + \mu\sum_{a=2,3} |\varepsilon^h_a|^2 - \gamma(\mathcal{A}^*\varepsilon_1^h + \text{c.c.})\right\},  
\end{align}
where the bulk modulus is $B = \lambda + \tfrac{2\mu}{3}$. We have already taken the homogeneous deviatoric strains to be zero in the above given the lack of homogeneous deviatoric stress. The equations of state are obtained as 
\begin{equation}
    \begin{aligned}
        \overline{\varepsilon}_0 &= \tfrac{\gamma}{B}\,\overline{\mathcal{A}},
        \\
        \varepsilon^h_1 &= \tfrac{\gamma}{\lambda + 2\mu}\,\mathcal{A},
        \\
        \varepsilon^h_{2,3} &= 0.
    \end{aligned}
\end{equation}
The effective correction to the free energy for $\mathcal{A}(\boldsymbol{x})$ then assumes the form
\begin{equation}
    \mathcal{F}_{\text{eff}} = -\frac{\gamma^2}{2V}\left\{\frac{1}{B} \mathcal{A}_{\boldsymbol{q}=\boldsymbol{0}}^2 + \frac{1}{B + \tfrac{4\mu}{3}}\sum_{q>0} \left\vert \mathcal{A}\right\vert^2\right\}, 
\end{equation}
where $\mathcal{A}_{\boldsymbol{q}=\boldsymbol{0}} = \int_x \mathcal{A}(\boldsymbol{x}) = V\,\overline{\mathcal{A}}$. Note that 
\begin{equation}
    \frac{1}{B}-\frac{1}{B+\tfrac{4\mu}{3}} = \frac{\varrho}{B} > 0,
\end{equation}
since both $B>0$ and $\mu>0$ are required by thermodynamic stability. The quantity $\varrho$ in the expression above is defined in \eqref{eq:varrho_definition}. Thus, the mass renormalization is \textit{discontinuous} as $q\rightarrow 0 $ for \textit{all} momentum directions. Furthermore, the largest enhancement comes from the homogeneous component, rather than the inhomogeneous component. As a result, the $\mathcal{A}_{\boldsymbol{q}=\boldsymbol{0}}$ order parameter is critical before any of the inhomogeneous ones with nonzero momentum. This contrasts with the case of direction-selective nematic criticality in \eqref{eq:effective_quadratic_term_nematicity}, where if the three direction-dependent constraints in \eqref{eq:Phi1=Phi4=Phi5=0} are met, then the inhomogeneous renormalization is identical to the homogeneous renormalization in \eqref{eq:effective_long-range_interaction_homogeneous_distortion}.
We also note the lack of directionality present in these effective interactions. This is because the order parameter couples directly to the dilatation without any direction-dependent form factors. Again, this is distinct from nemato-elastic criticality. We note that direction-selectivity through nemato-elasticity is also distinct from the so-called ``Larkin-Pikin'' mechanism that drives second-order transitions to weakly first-order transitions. In this case, because the square of the order parameter couples to the dilatation, fluctuations of the latter renormalize the order parameter's quartic coefficient rather than its mass \cite{larkinPHASETRANSITIONSFIRST1969, khmelnitskiiPhaseTransitionsCompressible1975,qiGlobalPhaseDiagram2009, chandraQuantumAnnealedCriticality2020}.

\section{Role of the Cubic Term in the Isotropic Electronic Nematic Free Energy \label{app:anharmonic_terms_free_energy}}

In this appendix, we consider the role that the symmetry-allowed cubic term in the bare electronic nematic free energy has on the compatible electronic nematic criticality discussed in Section \ref{sec:compatible_restrictions_on_nematicity}. In an isotropic three-dimensional system, such a cubic invariant is allowed, and a quartic term is needed for stability \cite{chaikinPrinciplesCondensedMatter1995}. Since the cubic invariant in the bare nematic theory immediately drives the transition first-order, one needs to address the fate of compatible electronic nematic criticality in the presence of the cubic term. The fact that nemato-elasticity causes only two linearly independent combinations of nematic order parameters to be soft across the phase transition, while the other three remain gapped, suggests that the first-order transition driven by the cubic term can be circumvented.  Here we show that this is precisely the case, and that the continuous nematic transition can survive in a solid when the nemato-elastic coupling dominates over the cubic invariant coefficient. This distinguishes electronic nematic phenomenology in solids from nematicity in classical liquid crystals, since compatible electronic nematics can undergo second-order phase transitions despite the presence of the cubic invariant.

\subsubsection{Origin of the Cubic Term in Bare Nematicity}

We first discuss the role of the higher-order (i.e., beyond quadratic) terms in the free energy in the $d$-orbital basis and then proceed to discuss its form in the helical basis. Recall that we write the electronic nematic order parameter in tensorial form as  
\begin{equation}
    \varphi_{ij}(\boldsymbol{x}) = \frac{1}{2}\boldsymbol{\varphi}(\boldsymbol{x}) \cdot \boldsymbol{\lambda}_{ij},
\end{equation}
where $\boldsymbol{\varphi} \equiv (\varphi_{z^2}, \varphi_{x^2 -y^2}, \varphi_{2yz}, \varphi_{2zx}, \varphi_{2xy})^{\operatorname{T}}$ is the five-component nematic vector written in the conventional $d$-orbital basis and $\boldsymbol{\lambda}\equiv (\lambda^8, \lambda^3, \lambda^6, \lambda^4, \lambda^1)^{\operatorname{T}}$ is the corresponding vector of symmetric Gell-Mann matrices. The quadratic, cubic, and quartic invariants are simply $\operatorname{Tr}(\varphi^2)$, $\operatorname{Tr}(\varphi^3)$, $\operatorname{Tr}(\varphi^4)$, respectively \cite{RMF14_classification_nematics}. 

The quadratic invariant in the $d$-orbital basis is
\begin{equation}
    \operatorname{Tr}(\varphi^2) = \frac{1}{2} \boldsymbol{\varphi}\cdot\boldsymbol{\varphi},
\end{equation}
as is given in the main text. Because the electronic nematic order parameter is a rank-2 symmetric traceless tensor, one can exploit its spectral decomposition in the $d$-orbital basis to obtain the quartic invariant in terms of the quadratic invariant. It follows then that 
\begin{equation}
    \operatorname{Tr}(\varphi^4) = \frac{1}{2}\left[ \operatorname{Tr}(\varphi^2)\right]^2 = \frac{1}{8}(\boldsymbol{\varphi}\cdot\boldsymbol{\varphi})^2. \label{eq:quartic_invariant_quadratic_invariant}
\end{equation}
The cubic invariant, meanwhile, is written in terms of the $d$-orbital basis functions as
\begin{align}
    \operatorname{Tr}(\varphi^3) &= \frac{3}{8}\left[ (\varphi_{2zx}^2 - \varphi_{2yz}^2)\varphi_{x^2 - y^2} +  (2\varphi_{2yz}\varphi_{2zx})\varphi_{2xy} \right] \nonumber
    \\
    &\phantom{=} + \frac{\sqrt{3}}{8}\left[ 2\left( \varphi_{x^2 -y^2}^2 + \varphi_{2xy}^2 \right) -  \left( \varphi_{2yz}^2 + \varphi_{2zx}^2 + \frac{2}{3} \varphi_{z^2}^2 \right)\right]\varphi_{z^2}.\label{eq:trace_phitensor^3_expanded}
\end{align}
The expression above also shows that in the purely 2D limit where $\varphi_{z^2} = \varphi_{2yz} = \varphi_{2zx} = 0$, then $\operatorname{Tr}(\varphi^3) = 0$, as is expected for 2D electronic nematics \cite{Oganesyan01}.

The bare nematic free energy can then be expanded to fourth-order as
\begin{align}
    \mathcal{F}_{\text{nem}}^{(0)}[\varphi] &=  \frac{1}{2V}\sum_{\boldsymbol{q}} (r + q^2) |\boldsymbol{\varphi}_{\boldsymbol{q}}|^2 \nonumber
    \\ 
    &\phantom{=}+  \int_x\; \left(\frac{1}{4}\, u|\boldsymbol{\varphi}(\boldsymbol{x})|^4 - \frac{4}{\sqrt{3}}\,g\,\varphi_{ij}(\boldsymbol{x}) \varphi_{jk}(\boldsymbol{x}) \varphi_{ki}(\boldsymbol{x}) \right),
\end{align}
in a system of volume $V$, where $r$ is the thermal tuning parameter, $u>0$ is required for stability, and $g>0$ is the coefficient of the cubic invariant. This expression generally leads to a first-order transition in both classical liquid crystals \cite{chaikinPrinciplesCondensedMatter1995} and in bare electronic nematics \cite{RMF14_classification_nematics}. In order to capture the fivefold degeneracy of the $d$-orbital basis functions within $\operatorname{SO}(3)$, one can employ the so-called ``$(\boldsymbol{nml})$-decomposition'' of the nematic order parameter, which can be written in terms of an amplitude, $|\boldsymbol{\varphi}|\equiv \sqrt{\boldsymbol{\varphi}\cdot \boldsymbol{\varphi}}$, an angle tuning between uniaxiality and biaxiality of the order parameter, and three mutually orthogonal unit vectors $\boldsymbol{n}$, $\boldsymbol{m}$, and $\boldsymbol{l}$ \cite{RMF14_classification_nematics}. At the mean-field level, this free energy undergoes a bare discontinuous phase transition into a uniaxial nematic state, allowing us to write the nematic order parameter in terms of a single unit vector, $\boldsymbol{n}$, as 
\begin{equation}
    \varphi_{ij} = \tfrac{\sqrt{3}}{2} |\boldsymbol{\varphi}| \left(n_i n_j - \tfrac{1}{3}\delta_{ij}\right).
\end{equation}
The cubic invariant follows simply in this parameterization as $\operatorname{Tr}(\varphi^3) = |\boldsymbol{\varphi}|^3/4\sqrt{3}$. The bare mean-field free energy then becomes 
\begin{equation}
    f_{\text{nem}}^{(0)}[\varphi] = \frac{1}{2}r|\boldsymbol{\varphi}|^2 + \frac{1}{4}|\boldsymbol{\varphi}|^4 - \frac{1}{3}g|\boldsymbol{\varphi}|^3, \quad g>0.
\end{equation}

The discontinuous phase transition occurs with $r>0$ when the following two equations are simultaneously satisfied for $\varphi\neq0$:
\begin{align}
  f_{\text{nem}}^{(0)}(\varphi) =    \frac{1}{2}r - \frac{1}{3}g|\boldsymbol{\varphi}| + \frac{1}{4}u|\boldsymbol{\varphi}|^2  &= 0,
    \\
  \partial{ f_{\text{nem}}^{(0)}}/\partial{\varphi} =  r-g|\boldsymbol{\varphi}|+u|\boldsymbol{\varphi}|^2 &= 0.
\end{align}
Together, these conditions imply that the first-order transition occurs at 
\begin{align}
    r_{\text{FO}} = \frac{2g^2}{9u}, \quad |\boldsymbol{\varphi}_{\text{FO}}| = \frac{2g}{3u}, \label{eq:r_firstorder_d_discontinuous}
\end{align}
for $g>0$. In the 2D case, $g = 0$ by symmetry, and the bare mean-field phase transition is continuous at $r=0$. 

\subsubsection{Impact of the Cubic Term on Compatible Nematic Criticality}

The important conclusion from \eqref{eq:r_firstorder_d_discontinuous} is that the energy scale of the cubic invariant, $g$, controls how closely the system is to nematic \textit{criticality} -- defined in this work by a divergent correlation length or nematic susceptibility. When we incorporate nemato-elasticity and integrate over the elastic strains, the bare electronic nematic free energy is renormalized to 
\begin{align}
    \mathcal{F}^{\text{eff}}_{\text{nem}}[\varphi] &= \frac{1}{2V}\sum_{\boldsymbol{q}} \left[ (r + q^2)|\boldsymbol{\varphi}_q|^2 - \boldsymbol{\varphi}^\dagger_{\boldsymbol{q}} \cdot \mathcal{M}(\hat{q}) \cdot \boldsymbol{\varphi}_{\boldsymbol{q}} \right] + \nonumber
    \\
    &\phantom{=}+ \int_x\; \left(\frac{1}{4}\, u|\boldsymbol{\varphi}(\boldsymbol{x})|^4 - \frac{4}{\sqrt 3}\,g\,\varphi_{ij}(\boldsymbol{x}) \varphi_{jk}(\boldsymbol{x}) \varphi_{ki}(\boldsymbol{x}) \right).
\end{align}
The quantity $\mathcal{M}(\hat{q})$ is the $5\times 5$ direction-dependent nematic mass renormalization that projects the electronic nematic fluctuations in the $d$-orbital basis onto the compatible sector. Rewriting the harmonic terms in the helical basis $\{ \Phi_{a}\}$ yields
\begin{align}
    \mathcal{F}^{\text{eff}}_{\text{nem}}[\varphi] &= \frac{1}{2V}\sum_{\boldsymbol{q}} \left(r -\tfrac{\lambda_0^2}{\mu} + q^2\right)\sum_{\alpha=2,3}\left|\Phi_{\alpha,\boldsymbol{q}}\right|^2 \nonumber
    \\
    &\phantom{=} + \frac{1}{2V}\sum_{\boldsymbol{q}}\left(r -\tfrac{\lambda_0^2}{\mu}\varrho + q^2\right)\left|\Phi_{1,\boldsymbol{q}}\right|^2 \nonumber
    \\
    &\phantom{=}+ \frac{1}{2V}\sum_{\boldsymbol{q}}\left(r + q^2\right)\sum_{\alpha=4,5}\left|\Phi_{\alpha,\boldsymbol{q}}\right|^2 \nonumber
    \\
    &\phantom{=}+ \int_x\; \left(\frac{1}{4}\, u|\boldsymbol{\varphi}(\boldsymbol{x})|^4 - \frac{4}{\sqrt 3}\,g\,\varphi_{ij}(\boldsymbol{x}) \varphi_{jk}(\boldsymbol{x}) \varphi_{ki}(\boldsymbol{x}) \right), \label{eq:helical_effective_nematic_free_energy_cubic_quartic}
\end{align}
where $\lambda_0$ is the nemato-elastic coupling, $\mu$ is the bare shear modulus, and $\varrho\in(0,1)$ is a dimensionless ratio of elastic constants. The purely compatible sector corresponds to the $(\Phi_2,\Phi_3)$ doublet, and without the cubic invariant, it is the critical sector of the theory with a renormalized transition temperature given by  $r_c\equiv \lambda_0^2/\mu$.

To understand the impact of the cubic term, we perform a mean-field analysis. Because of the isotropy of the system, we can assume that the nemato-elastic criticality occurs for $q \rightarrow 0$ with $\hat{q} \equiv \hat{z}$. As shown in Section \ref{sec:polar_and_planar_nematicity}, this corresponds to the ``polar limit'' where the $(\Phi_2,\Phi_3)$ doublet in the helical basis maps directly onto the $(\varphi_{2yz},\varphi_{2zx})$ doublet in the $d$-orbital basis. The nematic component that generates dilatation is given by $\Phi_1 = -\varphi_{z^2}$ along this direction, while the incompatible doublet is given by $(\Phi_4, \Phi_5) = (\varphi_{x^2 -y^2},\varphi_{2xy})$. Measuring temperature with respect to $r_c$ as $\tilde{r} \equiv r - r_c$ and taking the $q\rightarrow 0$ limit leads to the following mean-field nematic free energy
\begin{align}
    f^{\text{MF}}_{\text{eff.}}(\hat{q}=\hat{z}) &= \frac{1}{2}\tilde{r}\left( \varphi_{2yz}^2 + \varphi_{2zx}^2\right) + \frac{1}{2}(\tilde{r} + M_1^2)\varphi_{z^2}^2 \nonumber
    \\
    &\phantom{=} + \frac{1}{2}(\tilde{r} + M_{\operatorname{inc.}}^2)\left( \varphi_{x^2-y^2}^2 + \varphi_{2xy}^2\right) +\frac{1}{4}u (\boldsymbol{\varphi}\cdot \boldsymbol{\varphi})^2 \nonumber
    \\
    &\phantom{=} - \frac{\sqrt 3}{2}g\bigg\lbrace (\varphi_{2zx}^2 - \varphi_{2yz}^2)\varphi_{x^2 - y^2} +  (2\varphi_{2yz}\varphi_{2zx})\varphi_{2xy} \nonumber
    \\
    &\phantom{=- \frac{1}{2}g\bigg\{}+ \frac{1}{\sqrt{3}} \bigg[ 2\left( \varphi_{x^2 -y^2}^2 + \varphi_{2xy}^2 \right) \nonumber 
    \\
    &\phantom{=- \frac{1}{2}g\bigg\{+ \frac{1}{\sqrt{3}}}-  \left( \varphi_{2yz}^2 + \varphi_{2zx}^2 + \frac{2}{3} \varphi_{z^2}^2 \right) \bigg] \varphi_{z^2}\bigg\rbrace,
\end{align}
where $M_1^2 \equiv \lambda_0^2(1-\varrho)/\mu$ and $M_{\operatorname{inc.}}^2 \equiv \lambda_0^2/\mu$ are positive. Compared to the bare nematic theory, the quadratic term is explicitly seen to break the fivefold-degeneracy of the $d$-orbital basis functions. This is the internal fivefold degeneracy in nematic director space, which reflects the $\operatorname{SO}(3)$ rotational symmetry in real-space. It is broken by the quadratic term because of the coupling between the chosen momentum direction $\hat{q}$ and the nematic components enforced by the compatibility conditions, as discussed in Section \ref{sec:polar_and_planar_nematicity}. The quadratic term here splits the fivefold internal symmetry down to two doublets and one singlet. This splitting will therefore frustrate the cubic term that breaks the fivefold degeneracy down to threefold, reflecting the tendency for the cubic term to favor uniaxial nematicity over biaxial nematicity \cite{RMF14_classification_nematics}. For an arbitrary momentum direction, we use the helical decomposition in the $q\rightarrow0$ limit to find
\begin{align}
    f^{\text{MF}}_{\text{eff.}}(\hat{q}) &= \frac{1}{2}\tilde{r}\sum_{\alpha=2,3} \Phi_\alpha^2 + \frac{1}{2}(\tilde{r} + M_1^2)\Phi_{1}^2 \nonumber
    \\
    &\phantom{=}+ \frac{1}{2}(\tilde{r} + M_{\operatorname{inc.}}^2)\sum_{\alpha=4,5}\Phi_\alpha^2  \nonumber
    \\
    &\phantom{=} +\frac{1}{4}u (\boldsymbol{\Phi}\cdot \boldsymbol{\Phi})^2 -\frac{1}{3}\,g\,\Phi_1^3 \nonumber
    \\
    &\phantom{=} -\frac{1}{2}\,g\,\left[ \Phi_2^2 + \Phi_3^2 - 2\left( \Phi_4^2 + \Phi_5^2\right) \right]\Phi_1 \nonumber
    \\
    &\phantom{=}- \frac{\sqrt{3}}{2}\,g\,\left[ \left( 2\Phi_2\Phi_3 \right)\Phi_4 + \left( \Phi_2^2 -\Phi_3^2 \right)\Phi_5\right] .
\end{align}

At the mean-field level, it is possible to establish a consistency condition for when our theory of compatible electronic nematic criticality survives the presence of the cubic invariant. If it does survive, only the compatible doublet $(\Phi_2,\Phi_3)$ is critical, and we can neglect the terms anharmonic in $\Phi_{\operatorname{inc.}} \in \{ \Phi_1,\Phi_4, \Phi_5\}$, since these components remain gapped at the critical point. The free energy then becomes
\begin{align}
    f^{\text{MF}}_{\text{eff.}}(\hat{q}) &= \frac{1}{2}\tilde{r}\sum_{\alpha=2,3} \Phi_\alpha^2 + \frac{1}{2}(\tilde{r} + M_1^2)\Phi_{1}^2 \nonumber
    \\
    &\phantom{=}+ \frac{1}{2}(\tilde{r} + M_{\operatorname{inc.}}^2)\sum_{\alpha=4,5}\Phi_\alpha^2  \nonumber
    \\
    &\phantom{=} +\frac{1}{4}u (\Phi_2^2 + \Phi_3^2)^2 -\frac{1}{2}\,g\,\left( \Phi_2^2 + \Phi_3^2 \right)\Phi_1 \nonumber
    \\
    &\phantom{=}- \frac{\sqrt{3}}{2}\,g\,\left[ \left( 2\Phi_2\Phi_3 \right)\Phi_4 + \left( \Phi_2^2 -\Phi_3^2 \right)\Phi_5\right]  + \mathcal{O}(\Phi_{\operatorname{inc.}}^3).
\end{align}
Minimizing with respect to these variables yields the following equations of state:
\begin{align}
    \Phi_1 &= \frac{g}{2(\tilde{r} + M_1^2)}\left(\Phi_2^2 + \Phi_3^2\right),
    \\
    \Phi_4 &= \frac{g\sqrt{3}}{2(\tilde{r} + M_{\operatorname{inc.}}^2)}\,\left(2\Phi_2\Phi_3\right),
    \\
    \Phi_5 &= \frac{g\sqrt{3}}{2(\tilde{r} + M_{\operatorname{inc.}}^2)}\,\left(\Phi_2^2 - \Phi_3^2\right),
\end{align}
and the following effective free energy for the compatible nematic sector
\begin{align}
    f^{\text{MF}}_{\text{comp.}}(\hat{q},\,q\rightarrow0) &= \frac{1}{2}\,\tilde{r}\,(\Phi_2^2 + \Phi_3^2) + \frac{1}{4}\tilde{u}(\tilde{r})(\Phi_2^2 + \Phi_3^2)^2.
\end{align}
This free energy is unstable towards \textit{compatible} electronic nematic criticality when $\tilde{r}\rightarrow 0^+$. Importantly, this mean-field free energy is isomorphic to the two-dimensional isotropic electronic nematic free energy that exhibits continuous criticality in the absence of the lattice \cite{Oganesyan01}. For completeness, returning to the ``polar limit'' with $\hat{q} =\hat{z}$  and repeating these steps in the $d$-orbital basis amounts to replacing the $\Phi_2^2 + \Phi_3^2$  with $\varphi_{2yz}^2 + \varphi_{2zx}^2$ in $f^{\text{MF}}_{\text{comp.}}$. The free energy with $\tilde{r}>0$ for this polar limit is discussed explicitly in Section \ref{sec:polar_and_planar_nematicity}. 

\begin{figure}
    \centering
    \includegraphics[width=\columnwidth]{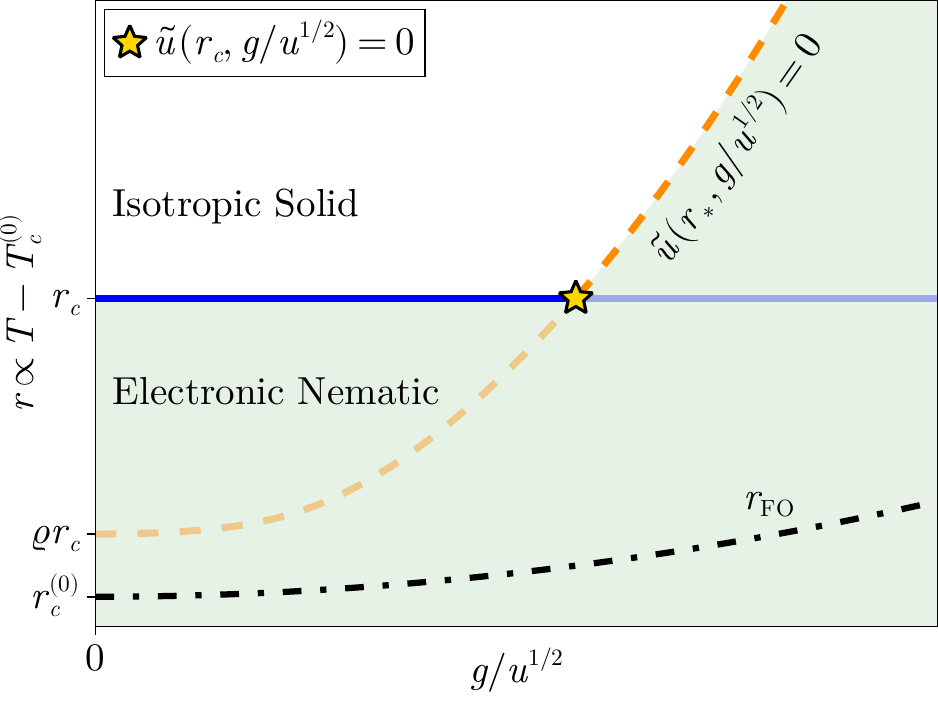}
    \caption{ Mean-field phase diagram of the isotropic-solid to electronic-nematic phase transition. The isotropic solid phase is the white region while the  electronic nematic phase is shaded in green. Unlike liquid crystals where the $d$-orbital basis condenses, due to the elastic compatibility relations, the \textit{compatible} nematic doublet $(\Phi_2,\Phi_3)$ of the helical basis condenses in isotropic solids. The phase boundary is denoted by the solid blue line and the dashed orange line, denoting the critical second-order and incompatible-fluctuation-induced first-order phase transitions, respectively. These curves meet at a tricritical point, indicated by the star. The vertical axis is the thermal tuning parameter $r \propto T - T_c^{(0)}$, where $r_c^{(0)} \propto T_c^{(0)}$ is the temperature of the bare electronic nematic instability. The nemato-elastic coupling enhances this instability to $r_c$ (see Section \ref{sec:compatible_restrictions_on_nematicity}). The horizontal axis tunes the strength of the cubic and quartic invariants allowed in the bare nematic free energy, represented by the phenomenological parameters $g > 0$ and $u > 0$ respectively. The bare nematic criticality is unstable towards a first-order transition at $r_{\operatorname{FO}}$ for any $g>0$. For small enough $g/\sqrt{u}$, the compatible electronic nematic transition remains second-order in a three dimensional isotropic solid. For large $g/\sqrt{u}$, the fluctuations of the incompatible nematic sector renormalize the quartic coefficient from the bare nematic theory to zero, $\tilde{u}(r_\ast, g/\sqrt{u}) = 0$, as shown by \eqref{eq:utilde_renormalized}. }
    \label{fig:cubic_invariant_phase_diagram}
\end{figure}

The main effect of the incompatible nematic fluctuations in the presence of the cubic invariant is to renormalize the quartic coefficient $u$ to $\tilde{u}$:
\begin{equation}
    \tilde{u}\left(\tilde{r},g/u^{1/2}\right)\equiv u \left[1 - \frac{3g^2}{2u}\left( \frac{1}{\tilde{r} + M_{\operatorname{inc.}}^2} + \frac{1}{3}\frac{1}{\tilde{r} + M_{1}^2}\right) \right]. \label{eq:utilde_renormalized}
\end{equation}
A second-order phase transition requires $\tilde{u}>0$ at the phase transition. The minimum value of $\tilde{u}(\tilde{r})$ above the transition occurs for $\tilde{r}\rightarrow 0^+$, from which it is clear that continuous compatible electronic nematic criticality survives provided that 
\begin{equation}
    \frac{3}{2} \left(\frac{\sfrac{g^2}{u}}{\sfrac{\lambda_0^2}{\mu}}\right)\left[ 1 + \frac{1}{3(1 - \varrho)}\right] <  1.
\end{equation}
Given that the renormalized transition temperature in the compatible sector is $r_c = \lambda_0^2/\mu$, the condition above can be written in terms of $r_c$ and the bare first-order transition $r_{\operatorname{FO}}$:
\begin{equation}
    r_c > \frac{27}{4}\left[ 1 + \frac{1}{3(1 - \varrho)}\right] r_{\operatorname{FO}} > r_{\operatorname{FO}}.
\end{equation}
If this inequality holds, then the compatible sector will undergo a second-order nematic transition as described in the main text. When the first inequality is saturated, an increasing cubic invariant drives the compatible critical transition to a tricritical point, wherein incompatible nematic fluctuations renormalize the bare quartic invariant to zero. While the transition changes at this point from second-order to first-order, the specific mechanism is \textit{not} what is found in isotropic-to-nematic first-order transitions in liquid crystals \cite{chaikinPrinciplesCondensedMatter1995}. Instead, because the nemato-elastic coupling bifurcates the nematic fluctuations into the critical compatible sector and the gapped incompatible sector, \textit{fluctuations} of the latter is what destabilizes the effective \textit{quartic} coefficient of the compatible sector. This situation is similar to the Larkin-Pikin mechanism, by which second-order phase transitions can become weakly first-order in compressible lattices \cite{larkinPHASETRANSITIONSFIRST1969}.

We complete this section by obtaining the first-order phase boundary, and sketching the mean-field phase diagram shown in Figure \ref{fig:cubic_invariant_phase_diagram}. Rewriting \eqref{eq:utilde_renormalized} in terms of the temperature, $r = \tilde{r} + r_c \propto T - T_c^{(0)}$, allows one to obtain the first-order phase boundary, $r_\ast(g/u^{1/2})$, defined implicitly through $\tilde{u}(r_\ast, g/u^{1/2}) = 0$. Given that $M_{\operatorname{inc.}}^2 = r_c$ and $M_1^2 = r_c(1-\varrho) < M_{\operatorname{inc.}}^2$, we find an implicit expression for $r_\ast$ given by
\begin{equation}
    1 = \frac{3g^2}{2u}\left[ \frac{1}{r_\ast}+\frac{1}{3(r_\ast-r_c\varrho)} \right].
\end{equation}
Thus, the first-order phase boundary, approached from above, is obtained as
\begin{equation}
    r_\ast\left( g/u^{1/2} \right) = \frac{\varrho r_c}{2}\left( 1 + \frac{2g^2/u}{\varrho r_c}\right)\left[ 1 + \sqrt{1 - \frac{6g^2/u }{\varrho r_c \left(1 + \frac{2g^2/u}{\varrho r_c}\right)^2}} \right].
\end{equation}

We emphasize the fact that elastic fluctuations produce an explicit criterion separating these second-order and first-order phase transitions distinguishes electronic nematicity in \textit{solids} from nematicity in classical \textit{liquid} crystals which lack shear rigidity. In the latter, any nonzero cubic term immediately renders the isotropic-to-nematic transition first-order. In a solid, however, the cubic invariant competes with the nemato-elastic coupling which acts to suppress incompatible nematic order. Thus, the low-energy phenomenology of nematicity in solids is distinct from that in classical liquid crystals, even if the unbroken symmetry group is $\operatorname{SO}(3)$ in both cases. In the case of electronic nematicity in solids, the non-analytic, momentum-direction-dependent, elastic compatibility relations change the nature of the mean-field phase boundary in the isotropic lattice.

\section{Impact of Elastic Compatibility in Anisotropic Nemato-Elastic Crystals \label{app:helical_dynamical_matrix_crystals}}

While our work on direction-selective criticality focuses on elastostatics explicitly, the majority of the ferroelastic and electronic nematic literature deals with the static limit of the full elastodynamic theory \cite{Cowley76, Folk76, folkCriticalDynamicsElastic1979, qiGlobalPhaseDiagram2009, Karahasanovic16, Paul17, Hecker2018, Fernandes2020, Hecker2022, stewardElasticQuantumCriticality2025}. To make connections between the helical representation and elastic compatibility to these earlier works, we show in this appendix how to apply our helical representation to the elastodynamic modes of an anisotropic (i.e., crystalline) medium. 

We will show that the elastic compatibility relations, when applied to ferroelastic and nematic transitions, controls their direction-selective criticality in crystals, without the need to obtain anisotropic acoustic mode velocities. This occurs for two reasons. The first reason, shown in Appendix \ref{app-subsec:helical_anisotropic_elasticity}, is because the elastic compatibility relations force the two transverse elastic displacements to generate all five components of the deviatoric strain tensor for generic momenta $\boldsymbol{q}\neq\boldsymbol{0}$. In lower-symmetry crystals, these deviatoric strain modes may transform as trivial irreducible representations of the crystalline point group, which therefore couple them bilinearly to the longitudinal sector. Thus, as $q\rightarrow 0$, these transverse modes will end up inducing simultaneous volume-changing strain at an additional energy cost. Meanwhile, along high-symmetry directions $\hat{q}_c$, individual deviatoric strain components completely decouple from the longitudinal displacements, and therefore will be the lowest energy strain modes as $q\rightarrow0$. As shown in Appendix \ref{app-subsec:helical_anisotropic_nemato-elasticity}, the second reason is that the nemato-elastic bilinear universally projects out the purely incompatible electronic nematic fluctuations in a momentum-direction-selective manner, already restricting the full phase space of nematic fluctuations in a three-dimensional crystal. We subsequently show that the critical momenta are further selected from this compatible manifold when the compatible nematic fluctuations do not generate dilatation strain. By eliminating dilatation strain in the $q\rightarrow0$ limit, we find the same set of critical directions $\hat{q}_c$ from Appendix \ref{app-subsec:helical_anisotropic_elasticity}. 

The conclusion is therefore that the origin of direction-selectivity in anisotropic crystals is the tensor gauge constraints of the elastic compatibility relations imposed on the electronic nematic order parameter. 
Throughout this appendix, we must explicitly convert between the conventional Cartesian basis of elastodynamics, the standard $d$-orbital basis of electronic nematicity, and the helical basis to manifestly satisfy the elastic compatibility relations. To disambiguate the various indices and notations for the vectors and tensors, we refer to Table \ref{tab:notation_glossary} as a glossary for each basis. The strict distinction of the index labels of Cartesian, $d$-orbital, and helical bases as the sets $\{i,j,k,\dots\}$, $\{a,b,c,\dots\}$, and $\{\alpha,\beta,\gamma,\dots\}$, respectively, is implemented in this appendix. When unambiguous, Einstein summation will be used within each basis over the relevant indices. Otherwise, explicit summations will be written.

\begin{table*}

\caption{Glossary of indices and notation for the vectors and traceless rank-2 symmetric tensors within the three bases used in nemato-elasticity in three-dimensional systems. For crystalline materials, it is crucial to move between these bases, and so the index labeling scheme is strictly enforced in Appendix \ref{app:helical_dynamical_matrix_crystals} for clarity. Cartesian coordinates are denoted by indices $i,j,k,\dots\in\{x,y,z\}$, whereas the $d$-orbital components in the orbital basis are denoted by $a,b,c,\dots\in\{z^2,x^2-y^2,2yz,2zx,2xy\}$. The Cartesian basis is valid for both vectors and tensors, whereas the $d$-orbital basis only applies to the traceless part of rank-2 symmetric tensors. The helical representation applies to vectors and rank-2 symmetric tensors, with Greek indices $\alpha,\beta,\gamma,\dots \in \{ 1,2,3,4,5 \}$. Indices 1, 2, and 3 are restricted to vectors and compatible tensors, while indices 4 and 5 appear in incompatible tensors.  \label{tab:notation_glossary}}
\bigskip{}

\centering{}%
\begin{tabular}{cccccccc}
\toprule
 &  &  & Cartesian $i,j,k,\dots$ &  & $d$-Orbital $a,b,c,\dots$ &  &
\shortstack[c]{Helical $\alpha,\beta,\gamma,\dots$\\(Momentum-Dependent)}\\
\midrule\midrule

 & Indices &  & $x,y,z$ &  & -- &  & $1,2,3$\\
\cmidrule{2-8}
Vector & Basis &  & $\left\{ \hat{x},\hat{y},\hat{z}\right\}$ &  & -- &  &
$\left\{ \hat{e}_{1},\hat{e}_{2},\hat{e}_{3}\right\}$\\
\cmidrule{2-8}
 & Notation &  & $v_{i}$ &  & -- &  &
$v_{\alpha}^{h}\in\left\{ v_{1}^{h},\,v_{2}^{h},\,v_{3}^{h}\right\}$\\
\midrule

 & Indices &  & $xx,yy,zz,yz,zx,xy$ &  &
$z^{2},x^{2}-y^{2},2yz,2zx,2xy$ &  & $1,2,3,4,5$\\
\cmidrule{2-8}
\shortstack[c]{Traceless Rank-2\\Symmetric Tensor}
 & Basis &  &
$\left\{ \hat{x}\otimes\hat{x},\hat{y}\otimes\hat{y},\dots\right\}$ &  &
$\boldsymbol{\lambda}=\left( \lambda_{ij}^{8},\;\lambda_{ij}^{3},\;\lambda_{ij}^{6},\;\lambda_{ij}^{4},\;\lambda_{ij}^{1}\right)$ &  &
$\left\{ \boldsymbol{\hat{Q}}_{1},\boldsymbol{\hat{Q}}_{2},\dots\right\}$
or $\left\{ \vartheta_{ij}^{1},\vartheta_{ij}^{2},\dots\right\}$\\
\cmidrule{2-8}
 & Notation &  & $A_{ij}$ &  &
$A_{ij}=\frac{1}{2}\boldsymbol{A}\cdot\boldsymbol{\lambda}_{ij}
= \frac{1}{2}\sum\limits_{a=1}^{5}A_{a}\lambda^{a}_{ij}$ &  &
$A_{ij}=\frac{1}{2}\sum\limits_{\alpha=1}^{5}A_{\alpha}\boldsymbol{\hat{Q}}_{\alpha}\cdot \boldsymbol{\lambda}_{ij}
=\frac{1}{\sqrt{2}}\sum\limits_{\alpha=1}^{5}A_{\alpha}\vartheta_{ij}^{\alpha}$\\

\bottomrule
\end{tabular}
\end{table*}

\subsection{Helical Elastodynamics in Anisotropic Lattices \label{app-subsec:helical_anisotropic_elasticity}}

In this section, we seek the helical representation of the dynamical matrix governing the elastodynamic waves in anisotropic media. The bare (Minkowski) action in elasticity is obtained by expanding the equilibrium crystal potential to quadratic order in the unit-cell displacements as \cite{qiGlobalPhaseDiagram2009, Karahasanovic16, Paul17, stewardElasticQuantumCriticality2025}:
\begin{equation}
    S_{\text{e}} = \frac{1}{2}\int \operatorname{d}t \int \operatorname{d}^D x\, \left\{ \rho(\partial_t \boldsymbol{u})^2 - C_{ijkl}\partial_i u_j \partial_k u_l\right\},
\end{equation}
where $\rho$ is the mass density of the unit cell. The vector $\boldsymbol{u}=\boldsymbol{u}(t,\boldsymbol{x})$ is the dynamic unit-cell displacement which corresponds to the acoustic phonon branch when the theory is quantized. The action is only a function of gradients of $\boldsymbol{u}(t,\boldsymbol{x})$ to conserve center-of-mass momentum of the entire crystal. The coefficients $C_{ijkl}$ are the rank-4 elastic stiffness constants with the following symmetries: $C_{ijkl}=C_{jikl}=C_{ijlk}=C_{klij}$.

Expanding in Fourier modes, we find that the action is given by
\begin{align}
    S_{\text{e}} &= \frac{1}{2V}\sum_{q>0} \int \frac{\operatorname{d}\omega}{2\pi} \, \bigg[ \rho\,\omega^2 |\boldsymbol{u}(\omega,\boldsymbol{q})|^2 \nonumber
    \\
    &\phantom{= \frac{1}{2V}\sum_{q>0} \int \frac{\operatorname{d}\omega}{2\pi}\, \bigg[}- C_{ijkl}q_i q_k u_j^*(\omega,\boldsymbol{q})\, u_l(\omega,\boldsymbol{q})\bigg],
\end{align}
where the asterisk denotes complex conjugation. The dynamical matrix is therefore given as 
\begin{equation}
    \mathcal{D}_{jl}(\omega, \boldsymbol{q}) = \rho\,\omega^2 \delta_{jl} - C_{ijkl}q_iq_k,
\end{equation}
such that the Lagrangian density is given by
\begin{equation}
    \mathcal{L} = \mathcal{D}_{ij}(\omega,\boldsymbol{q})u^*_i(\omega,\boldsymbol{q})u_j(\omega,\boldsymbol{q}).
\end{equation}
The transformation to the helical modes, being proportional to a rotation, preserves this structure. Inverting our \eqref{eq:helical_strain_from_displacement}, we find that the displacement vector is given in terms of the helical strain amplitudes as
\begin{equation}
    u_i(\omega,\boldsymbol{q}) = -\frac{\operatorname{i}}{|\boldsymbol{q}|} \, \mathcal{R}_{i\alpha}(\hat{q}) \varepsilon_{\alpha}^h(\omega, \boldsymbol{q}).
\end{equation}
Thus, the Lagrangian density is given by 
\begin{equation}
    \mathcal{L} = \mathfrak{D}^{\alpha\beta}(\omega,\boldsymbol{q}) \varepsilon_{\alpha}^{h*}(\omega, \boldsymbol{q})\varepsilon_{\beta}^{h}(\omega, \boldsymbol{q}),
\end{equation}
 with a ``helical'' dynamical matrix given by
\begin{equation}
    \mathfrak{D}^{\alpha\beta}(\omega,\boldsymbol{q})= \rho\, s^2(\omega,q) \delta^{\alpha\beta} -C_{ijkl}\mathcal{Q}_{ij}^{1\alpha}(\hat{q}) \mathcal{Q}_{kl}^{1\beta}(\hat{q}), \label{eq:helical_dynamical_matrix}
\end{equation}
where $s(\omega,q) \equiv \omega/q$ and the tensor $\mathcal{Q}^{\alpha\beta}_{ij}(\hat{q})$ is defined as \eqref{eq:Qabij} in the main text. It is repeated here for convenience as 
\begin{equation}
    \mathcal{Q}^{\alpha\beta}_{ij}(\hat{q})\equiv\frac{1}{2}\left(\hat{e}_{\alpha,j} \hat{e}_{\beta,i} + \hat{e}_{\alpha,i} \hat{e}_{\beta,j} \right).\tag{\ref{eq:Qabij}}
\end{equation}

The helical dynamical matrix in \eqref{eq:helical_dynamical_matrix} describes the coupling between the longitudinal and transverse elastic modes at all frequencies and momenta. As a matrix, it assumes the following structure
\begin{equation}
    \mathfrak{D}\left(\omega,\boldsymbol{q}\right)=\begin{pmatrix}\begin{array}{c|cc}
\mathfrak{D}^{11} & \mathfrak{D}^{12} & \mathfrak{D}^{13}\\[2pt] \hline \\[-6pt] \mathfrak{D}^{12} & \mathfrak{D}^{22} & \mathfrak{D}^{23}\\
\mathfrak{D}^{13} & \mathfrak{D}^{23} & \mathfrak{D}^{33}
\end{array}\end{pmatrix}.\label{eq:helical_dynamical_matrix_matrix_form}
\end{equation}
The horizontal and vertical lines explicitly separate the longitudinal block -- the $\mathfrak{D}^{11}(\omega,\boldsymbol{q})$ element -- from the transverse block: $\mathfrak{D}^{\alpha\beta}$ with $\alpha,\beta\in\{2,3\}$. Importantly, nonzero $\mathfrak{D}^{12}(\omega,\boldsymbol{q})$ and $\mathfrak{D}^{13}(\omega,\boldsymbol{q})$ components bilinearly couple the longitudinal and transverse elastic modes, allowing for transverse displacements to generate longitudinal ones (and \textit{vice versa}).

\subsubsection{Direction-Selective Criticality in the Helical Representation}

Within isotropic elasticity, there are only two elastic stiffness constants, and the longitudinal and transverse helical modes decouple at the harmonic level for all momentum directions. Indeed, the dynamical matrix in \eqref{eq:helical_dynamical_matrix_matrix_form} diagonalizes with $\mathfrak{D}^{12}(\omega,\boldsymbol{q}) = \mathfrak{D}^{13}(\omega,\boldsymbol{q}) = \mathfrak{D}^{23}(\omega,\boldsymbol{q}) = 0$ with $\mathfrak{D}^{22}(\omega,\boldsymbol{q}) = \mathfrak{D}^{33}(\omega,\boldsymbol{q})$ in $\operatorname{SO}(3)$. Following the Cowley description, symmetry-breaking ferroelastic transitions can only occur along momentum directions where the transverse mode decouples from the symmetry-preserving longitudinal mode \citep{Cowley76,Folk76,zachariasQuantumCriticalElasticity2015,Paul17}. In isotropic solids, bare elasticity allows this to occur along any direction in momentum space. In crystals, however, most directions in the Brillouin zone host a symmetry-allowed bilinear coupling between the longitudinal and transverse degrees of freedom. In group theory language, this is simply because they transform as the same (trivial) irreducible representation of the little group of the wavevector. Whereas the longitudinal helical mode, $\varepsilon^h_1 = \mathrm{i}\boldsymbol{q}\cdot \boldsymbol{u}$, always transforms as the trivial irreducible representation, for high-symmetry directions, the transverse helical modes $\{\varepsilon^h_2, \varepsilon^h_3\}$ can transform as a non-trivial irreducible representation. In crystalline ferroelastics -- proper or pseudo-proper -- one must first identify where this longitudinal-transverse coupling happens in the bare elastic action, and only then consider the impact of a bilinear coupling to the critical order parameter \cite{folkCriticalDynamicsElastic1979, Paul17}. While it is more common to remain in the crystalline Cartesian basis, and then diagonalize the entire dynamical matrix, we will show here that the helical representation of elasticity makes the critical directions manifest by only considering the role of elastic compatibility within the static sector, without the need to obtain the phonon dispersions.

Since along the high-symmetry directions -- denoted in momentum space by the unit vectors $\hat{q}_c$ -- the longitudinal and transverse modes transform as separate irreducible representations within the little group of $\hat{q}_c$, a bilinear coupling between them is symmetry forbidden. Thus, the helical dynamical matrix $\mathfrak{D}^{\alpha\beta}(\omega,\boldsymbol{q})$ in \eqref{eq:helical_dynamical_matrix_matrix_form} block-diagonalizes as
\begin{equation}    \mathfrak{D}\left(\omega,\boldsymbol{q}_{c}\right)\equiv\begin{pmatrix}\begin{array}{c|cc}
\mathfrak{D}^{11} & 0 & 0\\
 \hline \\[-6pt] 0 & \mathfrak{D}^{22} & \mathfrak{D}^{23}\\
0 & \mathfrak{D}^{23} & \mathfrak{D}^{33}
\end{array}\end{pmatrix}\label{eq:block_diagonal_helical_dynamical_matrix}
\end{equation}
Clearly, its inverse -- the elastodynamic Green's function -- will then also be block-diagonalized into longitudinal and transverse sectors. This is another advantage of employing the helical representation in crystalline materials. There is no such manifest block-diagonalization in the standard Cartesian coordinates of the crystal. When one instead works within the eigenbasis of the dynamical matrix, while it is diagonal by construction, the longitudinal and transverse modes are generally hybridized, making it difficult to ascertain the critical directions in the purely diagonal representation as well. Thus, the helical representation provides a transparent way to readily obtain the critical directions.

Given that the dynamical inertial term in \eqref{eq:helical_dynamical_matrix} is already diagonal in the helical modes, the block diagonalization in \eqref{eq:block_diagonal_helical_dynamical_matrix} is governed \textit{only} by the \textit{elastostatic} term. It suffices, therefore, to find the critical directions $\hat{q}_c$ where
\begin{equation}
\begin{aligned}
    \mathfrak{D}^{12}(\omega,\boldsymbol{q}_c) &=-C_{ijkl}\mathcal{Q}_{ij}^{11}(\hat{q}_c) \mathcal{Q}_{kl}^{12}(\hat{q}_c) = 0,
    \\
    \mathfrak{D}^{13}(\omega,\boldsymbol{q}_c) &=-C_{ijkl}\mathcal{Q}_{ij}^{11}(\hat{q}_c) \mathcal{Q}_{kl}^{13}(\hat{q}_c) = 0.
\end{aligned}\label{eq:helical_elastic_constants_block_diagonal_generic}
\end{equation}
It is helpful to write these conditions in terms of the helical basis vectors from \eqref{eq:vectorial_Qhatvec_basis}: 
\begin{equation}
\begin{aligned}
    \boldsymbol{\hat{Q}}_1 &= \tfrac{\sqrt{3}}{2}\left(\hat{e}_1^{\mathrm T} \cdot \boldsymbol{\lambda} \cdot \hat{e}_1\right),
    \\
    \boldsymbol{\hat{Q}}_2 &= \hat{e}_1^{\mathrm T} \cdot \boldsymbol{\lambda} \cdot \hat{e}_2,
    \\
    \boldsymbol{\hat{Q}}_3 &= \hat{e}_1^{\mathrm T} \cdot \boldsymbol{\lambda} \cdot \hat{e}_3.
\end{aligned}
\end{equation}
The unit vectors above co-rotate with the momentum $\boldsymbol{q}$, and are defined in Section \ref{sec:inhomogeneous_elasticity_and_compatibility} as either longitudinal to $\boldsymbol{q}$, $\hat{e}_1 \equiv \hat{q}$, or transverse to it $\{\hat{e}_2,\hat{e}_3\}$. Given that 
\begin{equation}
    \mathcal{Q}^{1\alpha}_{ij}(\hat{q}) = \frac{1}{3}\delta^{1\alpha}\delta_{ij} + \frac{1}{2} \boldsymbol{\hat{Q}}_\alpha(\hat{q}) \cdot \boldsymbol{\lambda}_{ij}, \;\; \alpha \in \{1,2,3\}, \label{eq:app_Q1alpha_decomp_GellMann}
\end{equation}
we find that we can write the elastic constants in the helical representation as
\begin{align}
    \mathscr{C}^{\alpha\beta} &\equiv C_{ijkl}^{\phantom{1\alpha}}\mathcal{Q}_{ij}^{1\alpha} \mathcal{Q}_{kl}^{1\beta} \nonumber
    \\
    &= \frac{1}{9} C_{iikk} \delta^{1\alpha}\delta^{1\beta} \nonumber
    \\
    &\phantom{=} + \frac{1}{6}\left(\delta^{1\alpha}\hat{Q}_{\beta, a} + \delta^{1\beta}\hat{Q}_{\alpha, a}\right) C_{iikl}^{\phantom{a}}\lambda^a_{kl} \nonumber
    \\
    &\phantom{=}+\frac{1}{4}\hat{Q}_{\alpha,a}\hat{Q}_{\beta,b}\,\left( C_{ijkl}^{\phantom{b}}\lambda^a_{ij}\lambda_{kl}^b \right), \label{eq:elastic_Cijkl_Qij_Qkl}
\end{align}
where we have dropped the explicit $\hat{q}$-dependence for brevity and have used Einstein summation for the Latin indices,  $a,b\in\{1,2,3,4,5\}$, in the $d$-orbital basis (see Table \ref{tab:notation_glossary}). 

The quantity $C_{iikk}$ is proportional to the crystalline equivalent of the ``bulk modulus'' since it only exists for $\alpha = \beta = 1$. Because of this, the term is already in block-diagonal form. Thus, we only must consider the direction-dependence of the remaining two terms. The term proportional to $C_{iikl}^{\phantom{a}}\lambda^a_{kl}$ couples the dilatation strain to the deviatoric strain. This is only allowed in crystals when the $a^{\text{th}}$ component of the strain tensor in the $d$-orbital basis transforms trivially with respect to the crystalline point group. When it does, one must find directions where $\hat{Q}_{2,a}(\hat{q}_c) = 0$ or $\hat{Q}_{3,a}(\hat{q}_c) = 0$, otherwise transverse elastic modes in the $q\rightarrow0$ limit will generate symmetry-preserving strain, incurring additional energy cost. Finally, the term proportional to $C_{ijkl}^{\phantom{b}}\lambda^a_{ij}\lambda_{kl}^b$ is nonzero whenever the product of the $a^{\text{th}}$ and $b^{\text{th}}$ $d$-orbital basis functions transforms trivially within the crystalline point group. These terms will only block diagonalize for $\hat{Q}_{1,a}(\hat{q}_c)\hat{Q}_{\beta,b}(\hat{q}_c) = 0$. To summarize, block-diagonalization in the helical representation within the bare elastic action occurs if the following conditions are met in the $d$-orbital basis
\begin{align}
    &a\in \text{Trivial\, Irrep.}: &&\hat{Q}_{\beta,a}(\hat{q}_c) = 0, \label{eq:trivial_irrep_decoupling}
    \\
    &a,b\in \text{Non-Trivial\, Irrep.}:&&\hat{Q}_{1,a}(\hat{q}_c)\hat{Q}_{\beta,b}(\hat{q}_c) = 0, \label{eq:nontrivial_irrep_decoupling}
\end{align}
for $\beta \in \{2,3\}$.

\subsubsection{Direction-Selective Criticality in Tetragonal Crystals}

A relevant and instructive example case is the three-dimensional tetragonal system with point group $\text{D}_{4h}$ ($4/mmm$). This point group corresponds to that of Ising tetragonal-to-orthorhombic electronic nematic. Table \ref{tab:tetragonal_deformation_to_other_groups} shows how to alter the tetragonal free energy, by imposing constraints on the independent elastic constants, into the free-energy of higher-symmetry crystals of interest -- with hexagonal $\text{D}_{6h}$ ($6/mmm$) and cubic $\text{O}_h$ ($m\bar{3}m$) point groups  -- as well as how to return to the isotropic lattice with $\operatorname{SO(3)}$ symmetry considered in the main text. In $\text{D}_{4h}$, there are six elastic constants which give rise to the following free-energy density:
\begin{align}
    2f_{\text{elas}} &= C_{A_{1g}^1} ( \varepsilon_{xx} + \varepsilon_{yy} )^2 + C_{A_{1g}^2} \varepsilon_{zz}^2 + C_{A_{1g}^3}(\varepsilon_{xx} + \varepsilon_{yy})\varepsilon_{zz}\nonumber
    \\
    &\phantom{=} + C_{B_{1g}}(\varepsilon_{xx} - \varepsilon_{yy})^2 + C_{B_{2g}}(2\varepsilon_{xy})^2 \nonumber
    \\
    &\phantom{=} + C_{E_{g}}\left[(2\varepsilon_{yz})^2 + (2\varepsilon_{zx})^2\right],
\end{align}
where the subscripts on the elastic constants correspond to the irreducible representations of $\text{D}_{4h}$ in Sch\"onflies notation. 

Transforming into the $d$-orbital basis, we find
\begin{align}
    2f_{\text{elas}} &= C_{0}^{\phantom{2}} \varepsilon_0^2 + C_{z^2}^{\phantom{2}} \varepsilon_{z^2}^2 + C_{0,z^2}\varepsilon_0\varepsilon_{z^2} + C_{B_{1g}}\varepsilon_{x^2-y^2}^2\nonumber
    \\
    &\phantom{=}  + C_{B_{2g}}\varepsilon_{2xy}^2 + C_{E_{g}}\left(\varepsilon_{2yz}^2 + \varepsilon_{2zx}^2\right), \label{eq:tetragonal_free_energy_d-orbitals}
\end{align}
with $\varepsilon_0 \equiv \varepsilon_{xx} + \varepsilon_{yy} + \varepsilon_{zz}$, and where the newly introduced elastic constants follow as
\begin{align}
    C_0 &\equiv \frac{1}{9}\left( 4C_{A_{1g}^1} + C_{A_{1g}^2} + 2C_{A_{1g}^3}\right),
    \\
    C_{z^2} &\equiv \frac{1}{3} \left(C_{A_{1g}^1} + C_{A_{1g}^2} - C_{A_{1g}^3}\right),
    \\
    C_{0,z^2} &\equiv \frac{1}{3\sqrt{3}}\left(-4C_{A_{1g}^1} + 2C_{A_{1g}^2} + C_{A_{1g}^3}\right). \label{eq:C0z2}
\end{align}
In the above, $C_0$ is the bulk modulus.
\begin{table*}
\caption{Recipe to alter the elastic free energy of a tetragonal crystal with point group $\text{D}_{4h}$ into the free energy of hexagonal,
cubic, and isotropic lattices, with symmetry groups $\text{D}_{6h}$ ($6/mmm$), $\text{O}_{h}$ ($m\bar{3}m$), and $\text{SO}(3)$, respectively. The symmetry-reduced tetragonal free energy in \eqref{eq:tetragonal_free_energy_d-orbitals} has six distinct elastic constants, corresponding to four symmetry-breaking $d$-orbital basis functions, and one trivially transforming $d$-orbital ($z^2$). By applying the following constraints to the six tetragonal elastic constants, the symmetry of the elastic action increases to that of hexagonal, cubic, and isotropic lattices. The quantities $\mu^{\text{SO}(3)}$ and $B^{\text{SO}(3)}$ are the shear and bulk moduli defined for isotropic elasticity in the main text (see \eqref{eq:eq:total_elastic_free_energy_homogeneous_strain}). \label{tab:tetragonal_deformation_to_other_groups}}
\vspace{1ex}
\centering
\par\noindent
\begingroup

\begin{tabular}{c@{\hspace{3em}}c@{\hspace{3em}}c}
\toprule
$\text{D}_{4h}\rightarrow\text{D}_{6h}$ &
$\text{D}_{4h}\rightarrow\text{O}_{h}$ &
$\text{D}_{4h}\rightarrow\operatorname{SO}(3)$\\
\midrule\midrule
$C_{B_{1g}}^{(\text{D}_{4h})}=C_{B_{2g}}^{(\text{D}_{4h})}\equiv C_{E_{2g}}^{(\text{D}_{6h})}$ &
$C_{0z^{2}}^{(\text{D}_{4h})}=0$ &
$C_{0z^{2}}^{(\text{D}_{4h})}=0$\\
\midrule
$C_{E_{g}}^{(\text{D}_{4h})}\equiv C_{E_{1g}}^{(\text{D}_{6h})}$ &
$C_{z^{2}}^{(\text{D}_{4h})}=C_{B_{1g}}^{(\text{D}_{4h})}\equiv C_{E_{g}}^{(\text{O}_{h})}$ &
$C_{z^{2}}^{(\text{D}_{4h})}=C_{B_{1g}}^{(\text{D}_{4h})}=C_{B_{2g}}^{(\text{D}_{4h})}=C_{E_{g}}^{(\text{D}_{4h})}\equiv\mu^{(\operatorname{SO}(3))}$\\
\midrule
-- &
$C_{B_{2g}}^{(\text{D}_{4h})}=C_{E_{g}}^{(\text{D}_{4h})}\equiv C_{T_{2g}}^{(\text{O}_{h})}$ &
$C_{0}^{(\text{D}_{4h})}\equiv B^{(\operatorname{SO}(3))}$\\
\midrule
\textbf{Total: }5 Constants &
\textbf{Total: }3 Constants &
\textbf{Total: }2 Constants\\
\bottomrule
\end{tabular}
\endgroup
\end{table*}

For crystals with axial point groups -- such as $\text{D}_{4h}$ and $\text{D}_{6h}$ -- $C_{0,z^2} = C_{iikl}^{\phantom{8}}\lambda^8_{kl} \neq 0$ since $z^2$ transforms as a trivial irreducible representation. Thus, by condition \eqref{eq:trivial_irrep_decoupling}, the transverse helical modes in tetragonal elasticity will generate longitudinal displacements unless $\hat{Q}_{2,z^2}(\hat{q}_c) = 0$ or $\hat{Q}_{3,z^2}(\hat{q}_c) = 0$. Using the helical form factors in \eqref{eq:Dmatrix} and Figure \ref{fig:supp_varphi_to_Phi_mapping}, $\hat{Q}_{2,z^2}(\hat{q}) = 0$ for all $\hat{q}$, whereas $\hat{Q}_{3,z^2}(\hat{q}_c) \propto \sin(2\theta_c) = 0$ only when $\hat{q}_c \cdot \hat{z} = \pm 1 $ or $\hat{q}_c \cdot \hat{z} = 0$. In other words, as discussed in Section \ref{sec:polar_and_planar_nematicity}, transverse elastic modes with momenta outside the ``polar'' and ``planar'' limits will necessarily generate $\varepsilon_{z^2} \propto \varepsilon_{xx} + \varepsilon_{yy} - 2\varepsilon_{zz}$ strain from the elastic compatibility relations. Because this strain mode is a trivial irreducible-representation of the point group, it is not a symmetry-breaking strain, and it will generate an additional energy cost associated with dilatation strain through the bilinear coupling $C_{0,z^2}$.

Restricting our attention to strictly polar or planar momenta, by \eqref{eq:nontrivial_irrep_decoupling} we are left to find directions where
\begin{equation}
    \hat{Q}_{1,a}(\hat{q}_c)\hat{Q}_{\beta,b}(\hat{q}_c) = 0, \;\; \hat{q}_c \in \{ \hat{q}:\;\hat{q}\cdot \hat{z} = \pm 1\; \text{or} \; \hat{q}\cdot \hat{z} = 0\},
\end{equation}
for the reduced set of symmetry-breaking $d$-orbitals in $\text{D}_{4h}$: $a,b \in \{ x^2-y^2, 2yz, 2zx, 2xy \}$. From \eqref{eq:tetragonal_free_energy_d-orbitals}, it is clear that 
\begin{align}
    \hat{Q}_{1,x^2-y^2}(\hat{q}_c)\hat{Q}_{\beta,x^2-y^2}(\hat{q}_c) &= 0,
    \\
    \hat{Q}_{1,2yz}(\hat{q}_c)\hat{Q}_{\beta,2yz}(\hat{q}_c) &= 0,
    \\
    \hat{Q}_{1,2zx}(\hat{q}_c)\hat{Q}_{\beta,2zx}(\hat{q}_c) &= 0,
    \\
    \hat{Q}_{1,2xy}(\hat{q}_c)\hat{Q}_{\beta,2xy}(\hat{q}_c) &= 0.
\end{align}
In order for these symmetry-breaking strains to map onto the transverse modes with $\beta = 2,3$ and still satisfy the block-diagonalization criteria above, we conclude that 
\begin{equation}
    \hat{Q}_{1,a}(\hat{q}_c) = 0,\; a\in \{ x^2-y^2, 2yz, 2zx, 2xy \}.
\end{equation}
Applying this condition the four symmetry-breaking strains identifies the critical directions explicitly from  \eqref{eq:Dmatrix} as
\begin{align}
    \hat{Q}_{1,x^2-y^2}(\hat{q}_c) &= \frac{\sqrt{3}}{2}\cos(2\phi_c)\sin^2(\theta_c) =  0, \label{eq:Qhat1x2y2=0}
    \\
    \hat{Q}_{1,2yz}(\hat{q}_c) &= \frac{\sqrt{3}}{2}\sin(\phi_c)\sin(2\theta_c) = 0,\label{eq:Qhat12yz=0}
    \\
    \hat{Q}_{1,2zx}(\hat{q}_c) &= \frac{\sqrt{3}}{2}\cos(\phi_c)\sin(2\theta_c)= 0, \label{eq:Qhat12zx=0}
    \\
    \hat{Q}_{1,2xy}(\hat{q}_c) &= \frac{\sqrt{3}}{2}\sin(2\phi_c)\sin^2(\theta_c) = 0,\label{eq:Qhat12xy=0}
\end{align}
where $\hat{q}_c = (\cos\phi_c\sin\theta_c,\sin\phi_c\sin\theta_c,\cos\theta_c)$. Since we must have $\sin(2\theta_c) = 0$ to eliminate any generation of dilatation strain from the transverse modes via the $z^2$ strain channel, it is clear that no shear strain in the $E_g$ doublet will generate dilatation. While the transverse modes generate strain for $\hat{q}_c = \pm \hat{z}$, neither the longitudinal nor transverse elastic modes with this momentum direction will generate any in-plane strain with $B_{1g}$ or $B_{2g}$ character within linear elasticity simply because $q_x = q_y = 0$. Meanwhile, the critical directions for $\hat{q}_c \cdot \hat{z} = 0$ for each in-plane irreducible representation are achieved for 
\begin{equation}
    \begin{aligned}
        B_{1g}:&& \cos(2\phi_c) &= 0,
        \\
        B_{2g}:&& \sin(2\phi_c) &= 0,
    \end{aligned}\label{eq:B1gB2gcompatibility=0}
\end{equation}
when $\theta_c = \pi/2$. These are the \textit{exact} same form factors that appear in the two-dimensional elastic compatibility relation in \eqref{eq:2D_SVCR_polar} and Fig. \ref{fig:SVCR_2D}, that show how inhomogeneous deviatoric strain is accompanied by dilatation strain unless these form factors vanish.  

This concludes the discussion of bare elastodynamics in the helical representation. Because elastodynamics is driven by the elastic displacement \textit{vector}, we only considered the helical indices 1, 2, and 3 (see Table \ref{tab:notation_glossary}). To understand the impact of elastic fluctuations on nemato-elasticity in crystals, we must consider again that the electronic nematic order parameter is a traceless rank-2 symmetric \textit{tensor} in a five-dimensional helical representation.

\subsection{Nemato-Elasticity in Anisotropic Crystals \label{app-subsec:helical_anisotropic_nemato-elasticity}}

Including nemato-elasticity within the elastodynamic action is straightforward, but the impact of nemato-elastic coupling results in a highly constrained phase space of electronic nematic fluctuations as a result of elastic compatibility.

In group theory language, only the non-trivial irreducible representations of $d$-orbitals will give rise to electronic nematic order parameters \cite{RMF14_classification_nematics}. For completeness, we will consider the most general symmetry-allowed nemato-elastic bilinear, and restrict our attention to the inhomogeneous components. This gives rise to the following nemato-elastic bilinear in the Minkowski action 
\begin{align}
    S_{\text{ne}} &\equiv   \int \operatorname{d}t \int \operatorname{d}^Dx\, \sum_{a = 1}^5 \lambda_a\varepsilon_a(t,\boldsymbol{x})\varphi_a(t,\boldsymbol{x}) \nonumber
    \\
    &= \frac{1}{2V}\sum_{q>0} \int \frac{\operatorname{d}\omega}{2\pi}\, \boldsymbol{\tilde{\varphi}}^\dagger(\omega,\boldsymbol{q}) \cdot \boldsymbol{\varepsilon}(\omega, \boldsymbol{q}) + \text{h.c.},
\end{align}
which generalizes nemato-elasticity considered previously for individual critical irreducible representations \cite{qiGlobalPhaseDiagram2009, Karahasanovic16, Paul17, Hecker2018, Fernandes2020, Hecker2022, stewardElasticQuantumCriticality2025}. In the bilinear above, we have included all five $d$-orbital electronic nematic order parameters as generic conjugate source fields for the five deviatoric strain components. In moving from the first to the second equality, we rewrote the local coupling in Fourier space and then absorbed the five nemato-elastic couplings $\{\lambda_a\}$ for each $d$-orbital into the nematic vector as 
\begin{equation}
    \boldsymbol{\tilde{\varphi}} = \left( \lambda_{z^2}\varphi_{z^2},  \lambda_{x^2 - y^2}\varphi_{x^2 - y^2}, \lambda_{2yz}\varphi_{2yz} , \lambda_{2zx}\varphi_{2zx}, \lambda_{2xy}\varphi_{2xy} \right)^{\operatorname{T}}.
\end{equation} 
Crystalline anisotropy will dictate that only a subset of these order parameters exist in the critical theory, unless the system is tuned between nematic states of different symmetry character \cite{borisovEvolution$B_2g$Nematics2019, ishidaNovelElectronicNematicity2020}. We will impose these restrictions later in the discussion, but for now we continue to consider all five basis functions.

\subsubsection{Compatible Nemato-Elasticity in Anisotropic Crystals}

Using the helical decomposition of a tensor -- a basis constructed relative to the momentum direction $\hat{q}$ and therefore universally applicable to \textit{all} crystalline point groups -- then we can write the nemato-elastic bilinear as
\begin{align}
    \boldsymbol{\tilde{\varphi}}^\dagger(\omega,\boldsymbol{q})\cdot \boldsymbol{\varepsilon}(\omega,\boldsymbol{q}) &= \frac{2}{\sqrt{3}}\,\tilde{\Phi}_1^*(\omega,\boldsymbol{q}) \varepsilon_1^h(\omega,\boldsymbol{q}) \nonumber
    \\
    &\phantom{=}+ \sum_{\alpha = 2}^3 \tilde{\Phi}_\alpha^*(\omega,\boldsymbol{q}){\varepsilon}_\alpha^{h}(\omega,\boldsymbol{q}),
    \\
    \tilde{\Phi}_\alpha(\omega, \boldsymbol{q}) &\equiv  \boldsymbol{\hat{Q}}_\alpha^{\text{T}}(\hat{q}) \cdot \boldsymbol{\tilde{\varphi}}(\omega, \boldsymbol{q}). 
\end{align}
which generalizes \eqref{eq:deviatoric_strain_vec_nematic_vec} to anisotropic crystals. This immediately shows that the incompatible contributions to the electronic nematic fluctuations are projected out from nemato-elasticity. Indeed, omitting the frequency and momentum dependence for clarity,  we find that the expression above can be written using the completeness relation in \eqref{eq:Qvec_completeness} as
\begin{align}
    \boldsymbol{\tilde{\varphi}}^\dagger\cdot \boldsymbol{\varepsilon} &= \sum_{\alpha = 1}^3 \tilde{\Phi}_\alpha^* \varepsilon_\alpha^{\phantom{*}} \nonumber
    \\
    &= \boldsymbol{\tilde{\varphi}}^\dagger \cdot  \left( \sum_{\alpha = 1}^3  \boldsymbol{\hat{Q}}_{\alpha}^{\phantom{\text{T}}} \boldsymbol{\hat{Q}}_{\alpha}^{\text{T}}\right) \cdot \boldsymbol{\varepsilon} \nonumber
    \\
    &= \boldsymbol{\tilde{\varphi}}^\dagger \cdot  \left( \operatorname{I} - \sum_{\alpha = 4}^5  \boldsymbol{\hat{Q}}_{\alpha}^{\phantom{\text{T}}} \boldsymbol{\hat{Q}}_{\alpha}^{\text{T}}\right) \cdot \boldsymbol{\varepsilon} \nonumber
    \\
    &= \left[ \boldsymbol{\tilde{\varphi}}^\dagger -  \left(\tilde{\Phi}_4^*\boldsymbol{\hat{Q}}_{4}^{\text{T}} + \tilde{\Phi}_5^*\boldsymbol{\hat{Q}}_{5}^{\text{T}}\right)\right] \cdot \boldsymbol{\varepsilon}.\label{eq:nemato-elasticity_projects_out_Phi4Phi5_crystals}
\end{align}
Therefore, the nemato-elastic bilinear itself projects out the purely incompatible components of the electronic nematic order parameter, $\tilde{\Phi}_4$ and $\tilde{\Phi}_5$, restricting the phase space of nematic fluctuations only to the compatible sector. Because nematic \textit{incompatibility} is momentum-direction dependent, then it is clear that the nexus of nematic direction-selectivity in crystals is elastic compatibility. Whereas elasticity itself is anisotropic in crystals, the direction-selectivity originates from the constraints of elastic compatibility inherited by the electronic nematic fluctuations.

As is the case in elastostatics, as shown in Section \ref{sec:compatible_restrictions_on_nematicity}, one can integrate out the helical elastic modes to obtain the renormalization of the electronic nematic mass. At the saddle-point, the helical elastic modes satisfy
\begin{align}
    \varepsilon^h_{\alpha}(\omega,\boldsymbol{q}) &= -\frac{2}{\sqrt{3}}\mathcal{G}^{\alpha1}(\omega,\boldsymbol{q})\;\tilde{\Phi}_1(\omega,\boldsymbol{q}) \nonumber
    \\
    &\phantom{=\;}- \sum_{\beta = 2}^3 \mathcal{G}^{\alpha\beta}(\omega,\boldsymbol{q})\;\tilde{\Phi}_\beta(\omega,\boldsymbol{q}). \label{eq:nemato-elastodynamic_equations_of_state}
\end{align}
In this equation above, we have defined the elastodynamic Green's function as $\mathcal{G}^{\alpha\beta} = (\mathfrak{D}^{-1})^{\alpha\beta}$. Its indices also denote the longitudinal and transverse sectors (see Table \ref{tab:notation_glossary}).

The saddle-point equations importantly reveal that the longitudinal elastodynamic mode, $\varepsilon^h_1(\omega,\boldsymbol{q})$, will generically be generated by the three helical nematic modes, $\tilde{\Phi}_1$, $\tilde{\Phi}_2$, and $\tilde{\Phi}_3$. Considering the $d$-orbital content of $\tilde{\Phi}_1$ reveals that
\begin{equation}
    \tilde{\Phi}_1(\omega,\boldsymbol{q}) = \sum_{a=1}^5 \lambda_a \, \hat{Q}_{1,a}(\hat{q}) \varphi_{a}(\omega,\boldsymbol{q}), \label{eq:Phi1_tilde_longitudinal}
\end{equation}
Focusing on an individual $d$-orbital, say $a = 2xy$, with $\lambda_{a\neq 2xy} = 0$, then the above reduces to
\begin{align}
    \tilde{\Phi}_1(\omega,\boldsymbol{q}) &= \lambda_{2xy} \, \hat{Q}_{1,2xy}(\hat{q}) \varphi_{2xy}(\omega,\boldsymbol{q})\nonumber
    \\
    &= \lambda_{2xy}\frac{\sqrt{3}}{2} \, \sin(2\phi)\sin^2(\theta)\, \varphi_{2xy}(\omega,\boldsymbol{q}),\label{eq:Phi1_tilde_2xy_only_longitudinal}
\end{align}
where $\phi$ and $\theta$ are the azimuthal and polar angles at $\hat{q}$. Thus, it is possible for $\tilde{\Phi}_1 = 0$ while $\varphi_{2xy}\neq0$, provided that one considers the specific direction $\hat{q}_c$ for which $\hat{Q}_{1,a}(\hat{q}_c) = 0$. For the $a=2xy$ $d$-orbital, this occurs for $\hat{q}_c\cdot\hat{z} =\pm 1$ or $\hat{q}_c\cdot\hat{z} = 0$, representing the ``polar'' and ``planar'' limits, respectively,  discussed in Section \ref{sec:polar_and_planar_nematicity}. The polar limit for this specific order parameter, however does not participate in nemato-elasticity since \eqref{eq:nemato-elasticity_projects_out_Phi4Phi5_crystals} projects out the purely incompatible sector and $\tilde{\Phi}_{4,5}(\omega,q\hat{z}) \propto \tilde{\varphi}_{2xy}(\omega,q\hat{z})$ (see Fig. \ref{fig:supp_varphi_to_Phi_mapping}). Thus, only the planar limit with $\hat{q}_c \cdot \hat{z} = 0$ is important for nemato-elasticity of only $2xy$ character. 

Recalling \eqref{eq:nontrivial_irrep_decoupling}, the condition that $\tilde{\Phi}_1 = 0$ in \eqref{eq:Phi1_tilde_2xy_only_longitudinal} vanishes for $\hat{q}_c$ exactly determines where the dynamic matrix $\mathfrak{D}^{\alpha\beta}(\omega,\hat{q})$ -- and its inverse $\mathcal{G}^{\alpha\beta}(\omega,\boldsymbol{q})$ -- block-diagonalize into distinct longitudinal and transverse sectors. Along these directions, transverse elastic displacements can exclusively generate symmetry-breaking $2xy$-deviatoric strain without simultaneously generating dilatation through a coupling to the longitudinal modes. The implication from the saddle-point equations is therefore that 
\begin{equation}
    \varepsilon^h_1(\omega,\boldsymbol{q}) = 0 \;\; \text{while}\;\; \varepsilon^h_\beta(\omega,\boldsymbol{q}) \neq 0 \; \beta \in \{2,3\}.
\end{equation}
Generalizing from only the $a= 2xy$ nematic order parameter in the $d$-orbital basis to any of the other five possible order parameters leads to the condition that \textit{if} $\tilde{\Phi}_1(\omega,\hat{q}_c) = 0$ while the critical order parameter is nonzero, $\tilde{\varphi}_a(\omega,\hat{q}_c) \neq 0$, then $\mathcal{G}^{12}(\omega,\hat{q}_c) = \mathcal{G}^{13}(\omega,\hat{q}_c) = 0$ in \eqref{eq:nemato-elastodynamic_equations_of_state}. In turn, nemato-elastically generated transverse elastic modes decouple from dilatation strain: $\varepsilon_1^h(\omega,\hat{q}_c)$. In the $q\rightarrow0$ limit, these nematic fluctuations completely soften, and induce a pseudo-proper ferroelastic strain. Otherwise, they incur an extra energy cost associated with symmetry-preserving, volume-changing strain in the limit that $q\rightarrow 0$.

To demonstrate how the electronic nematic fluctuations' inheritance of elastic compatibility satisfies the requirements of direction-selectivity critiality in crystals explicitly, we first finish integrating out the elastic modes to obtain the electronic nematic polarization bubble. Its static limit corresponds to the effective correction to the electronic nematic mass. We then focus again on the special case of the three-dimensional tetragonal crystal and obtain the same critical directions obtained in previous works.

Evaluating the elastic action at the saddle-point leads to the following effective correction to the nematic action:
\begin{equation}
    \Delta S_{\text{eff}} = -\frac{1}{2V}\sum_{q>0}\int \frac{\operatorname{d}\omega}{2\pi}\, \sum_{a,b=1}^5 \tilde{\varphi}^*_a(\omega,\boldsymbol{q}) \Pi^{\phantom{\dagger}}_{ab}(\omega,\boldsymbol{q}) \tilde{\varphi}^{\phantom{\dagger}}_b(\omega,\boldsymbol{q}). 
\end{equation}
where the polarization bubble in the $d$-orbital basis is defined by
\begin{equation}
    \Pi_{ab}(\omega,\boldsymbol{q}) = \hat{Q}_{\alpha,a}(\hat{q})\, \mathcal{A}_{\alpha\beta} \mathcal{G}^{\beta\gamma}(\omega,\boldsymbol{q})\mathcal{A}_{\gamma\delta}\;\hat{Q}_{\delta,b}(\hat{q}). \label{eq:full_polarization_bubble}
\end{equation}
In the equation above, we have employed Einstein summation over the three compatible helical indices in Greek $\{1,2,3\}$, and defined the matrix $\mathcal{A}$ as $\mathcal{A} \equiv \operatorname{diag}(2/\sqrt{3}, 1,1 )$. Its static limit corrects the electronic nematic mass, generalizing \eqref{eq:effective_electronic_nematic_mass_M} obtained in the isotropic solid to anisotropic crystals. Taking the static limit of \eqref{eq:helical_dynamical_matrix} and inverting it leads to
\begin{align}
    \mathcal{M}_{ab}(\hat{q}) &\equiv -\Pi_{ab}(0,\boldsymbol{q}) \nonumber
    \\
    &= \sum_{\alpha,\beta,\gamma,\delta =1}^3\hat{Q}_{\alpha,a}(\hat{q})\, \mathcal{A}_{\alpha\beta} \mathcal{S}^{\beta\gamma}(\hat{q})\mathcal{A}_{\gamma\delta}\;\hat{Q}_{\delta,b}(\hat{q}).\label{eq:Mab_within_crystals}
\end{align}
where the ``helical compliance matrix'' is the inverse of $\mathscr{C}^{\alpha\beta}(\hat{q})$:
\begin{equation}
    \mathcal{S}_{\alpha\beta}(\hat{q})\mathscr{C}^{\beta\gamma}(\hat{q}) = \mathcal{S}_{\alpha\beta}(\hat{q})\,C_{ijlk}^{\phantom{1\beta}}\mathcal{Q}_{ij}^{1\beta}(\hat{q}) \mathcal{Q}_{kl}^{1\gamma}(\hat{q}) = \delta^{\alpha\gamma}.
\end{equation}
\eqref{eq:Mab_within_crystals} generalizes the correction to the nematic mass  for an isotropic lattice discussed in the main text, \eqref{eq:effective_electronic_nematic_mass_M_compatible_sector}, to the case of anisotropic crystal lattices. 

\subsubsection{Tetragonal Nemato-Elasticity}

We now return to the example tetragonal crystal from Appendix \ref{app-subsec:helical_anisotropic_elasticity}. There are only four symmetry-breaking $d$-orbital basis functions in point group $\text{D}_{4h}$ that lead to two scalar order parameters in the plane $\varphi_{x^2-y^2} \equiv \varphi_{B_{1g}}$ and $\varphi_{2xy} \equiv \varphi_{B_{2g}}$, as well as an out-of-plane doublet $(\varphi_{2yz},\varphi_{2zx})\equiv\boldsymbol{\varphi}_{E_g}$ \cite{RMF14_classification_nematics}. Omitting the frequency  and momentum dependence, evaluating \eqref{eq:Phi1_tilde_longitudinal} with these four order parameters gives 
\begin{align}
    \tilde{\Phi}_1 &= \lambda_{x^2-y^2} \left( \hat{Q}_{1,x^2-y^2} \varphi_{x^2-y^2}\right) + \lambda_{2xy} \left(\hat{Q}_{1,2xy} \varphi_{2xy} \right) \nonumber
    \\
    &\phantom{=}+ \lambda_{E_g}\left( \hat{Q}_{1,2yz} \varphi_{2yz} + \hat{Q}_{1,2zx} \varphi_{2zx}\right),
\end{align}
where $\lambda_{2yz} = \lambda_{2zx} \equiv \lambda_{E_g}$ by symmetry. If the only critical channel has $B_{2g}$ symmetry (with $\lambda_{x^2-y^2} = \lambda_{E_g} = 0$), we see that $\tilde{\Phi}_1(\omega,\hat{q}_c) = 0$ along directions where $\hat{Q}_{1,2xy}(\hat{q}_c) = 0$, as argued for \eqref{eq:Phi1_tilde_2xy_only_longitudinal}. Similarly, if only the $B_{1g}$ or $E_g$ nematic order parameters are critical, we would instead have $\tilde{\Phi}_1(\omega, \hat{q}_c) = 0$ if $\hat{Q}_{1,a}(\hat{q}_c) = 0$ with $a \in \{x^2-y^2,2yz,2zx\}$. 

In total, these four conditions are given in Eqs. (\ref{eq:Qhat1x2y2=0} - \ref{eq:Qhat12xy=0}) as those that block-diagonalize the bare elastodynamic action in a tetragonal crystal. We now use the helical elastic constants to compute the helical compliance and therefore the correction to the nematic mass in \eqref{eq:Mab_within_crystals}. In $\text{D}_{4h}$, the helical elastic constants follow from \eqref{eq:tetragonal_free_energy_d-orbitals} as 
\begin{align}
    \mathscr{C}^{\alpha\beta} &= C_0 \delta^{1\alpha}\delta^{1\beta} + C_{z^2} \hat{Q}_{\alpha,z^2}\hat{Q}_{\beta,z^2} \nonumber
    \\
    &\phantom{=} + \frac{1}{2}C_{0,z^2}\left(\delta^{1\alpha}\hat{Q}_{\beta,z^2} + \delta^{1\beta}\hat{Q}_{\alpha,z^2}\right)  \nonumber
    \\
    &\phantom{=}+ C_{B_{1g}} \hat{Q}_{\alpha, x^2-y^2}\hat{Q}_{\beta, x^2-y^2} + C_{B_{2g}}\hat{Q}_{\alpha, 2xy}\hat{Q}_{\beta, 2xy} \nonumber
    \\
    &\phantom{=} + C_{E_g} \left( \hat{Q}_{\alpha, 2yz}\hat{Q}_{\beta, 2yz} + \hat{Q}_{\alpha, 2zx}\hat{Q}_{\beta, 2zx} \right), \label{eq:D4h_helical_elastic_constants}
\end{align}
since $\hat{Q}_{\alpha,x^2+y^2+z^2} = \delta^{1\alpha}$. For a critical $E_g$ doublet, $\Phi_1(\omega,\hat{q}_c) = 0$ for $\sin(2\theta_c) = 0$ from Eqs. (\ref{eq:Qhat12yz=0}) and (\ref{eq:Qhat12zx=0}). This occurs in the ``polar'' and ``planar'' limits from Section \ref{sec:polar_and_planar_nematicity} with $\hat{q}_c\cdot\hat{z} = \pm 1$ or $\hat{q}_c\cdot\hat{z} = 0$, respectively. Evaluating \eqref{eq:D4h_helical_elastic_constants} in the polar limit diagonalizes the helical elastic constants as
\begin{equation}
    \mathscr{C}(\hat{q} = \pm \hat{z}) = \begin{pmatrix}
        C_0  - C_{0,z^2} + C_{z^2} & 0 & 0 
        \\
        0 & C_{E_g} & 0
        \\
        0 & 0 & C_{E_g}
    \end{pmatrix},
\end{equation}
The helical compliance matrix $\mathcal{S}^{\alpha\beta}$ is obtained from a trivial inversion. Clearly both transverse elastic modes will only correspond to $2yz$ and $2zx$ shear strains which decouple from dilatation. The correction to the electronic nematic mass then follows from \eqref{eq:Mab_within_crystals} as 
\begin{equation}
    \mathcal{M}_{ab}(\hat{q}_c = \pm \hat{z}) = \frac{1}{C_{E_g}} \left( \delta_{a,2yz}\delta_{b,2yz} + \delta_{a,2zx}\delta_{b,2zx}\right).
\end{equation}

Meanwhile the planar limit with $\hat{q}_c\cdot\hat{z} = 0$ -- which also satisfies \eqref{eq:Qhat1x2y2=0} and \eqref{eq:Qhat12xy=0} for the $B_{1g}$ and $B_{2g}$ nematic order parameters -- we find
\begin{equation}
    \mathscr{C}(\hat{q} \cdot \hat{z} = 0) = \begin{pmatrix}
        \mathscr{C}^{11}(\hat{q} \cdot \hat{z} = 0) & \mathscr{C}^{12}(\hat{q} \cdot \hat{z} = 0) & 0 
        \\
        \mathscr{C}^{12}(\hat{q} \cdot \hat{z} = 0) & \mathscr{C}^{22}(\hat{q} \cdot \hat{z} = 0) & 0
        \\
        0 & 0 & C_{E_g}
    \end{pmatrix}.\label{eq:helical_elastic_constants_tetragonal_planar}
\end{equation}
Only the $\varepsilon^h_3$ transverse modes -- representing out-of-plane displacements -- are decoupled from the longitudinal sector. This transverse mode is exclusively generated by fluctuations of the $\boldsymbol{\varphi}_{E_g}$ doublet. The other coefficients in the equation above are
\begin{align}
    \mathscr{C}^{11}(\hat{q} \cdot \hat{z} = 0) &= C_0 + \frac{1}{2}C_{0,z^2} + \frac{1}{4}C_{z^2} \nonumber
    \\
    &\phantom{=} + \frac{3}{8}\left[ C_{B_{1g}} + C_{B_{2g}} + \left( C_{B_{1g}} - C_{B_{2g}} \right)\cos(4\phi) \right],
    \\
    \mathscr{C}^{12}(\hat{q} \cdot \hat{z} = 0) &= -\frac{\sqrt{3}}{4}\left( C_{B_{1g}} - C_{B_{2g}} \right) \sin(4\phi),
    \\
    \mathscr{C}^{22}(\hat{q} \cdot \hat{z} = 0) &= \frac{1}{2}\left[ C_{B_{1g}} + C_{B_{2g}} - \left( C_{B_{1g}} - C_{B_{2g}} \right)\cos(4\phi) \right].
\end{align}
These coefficients clearly only involve the in-plane deviatoric strains in the $B_{1g}$ and $B_{2g}$ channels. For the $B_{1g}$ and $B_{2g}$ order parameters, \eqref{eq:Phi1_tilde_longitudinal} shows that $\tilde{\Phi}_1(\omega,\hat{q}_c) = 0$ for $\cos(2\phi_c) = 0$ or $\sin(2\phi_c) = 0$, respectively. Notice the bare elastic action again diagonalizes into longitudinal and transverse elastic modes in tetragonal crystals along these directions since $\sin(4\phi_c) = 2\cos(2\phi_c)\sin(2\phi_c) = 0$. Using these directions, the correction to the electronic nematic mass from \eqref{eq:Mab_within_crystals} are
\begin{align}
    \mathcal{M}_{x^2-y^2,x^2-y^2}(\cos(2\phi_c) = 0, \theta_c = \pi/2) &= \frac{1}{C_{B_{1g}}},
    \\
    \mathcal{M}_{2xy,2xy}(\sin(2\phi_c) = 0, \theta_c = \pi/2) &= \frac{1}{C_{B_{2g}}}.
\end{align}

\subsubsection{Hexagonal Nemato-Elasticity in 2D}

Finally, we also consider planar hexagonal crystals with point group $\text{D}_{6h}$.  Within the $xy$-plane, the helical elastic constants are isotropic for all $\phi$ since symmetry enforces $C_{B_{1g}} = C_{B_{2g}}$  (see Table \ref{tab:tetragonal_deformation_to_other_groups}) and $\hat{Q}_{1,z^2}(\hat{q}\cdot\hat{z}=0) = 0$ from \eqref{eq:trivial_irrep_decoupling}. Because $\mathcal{G}^{12}(\omega,\hat{q}_c\cdot \hat{z} = 0)=0$ by symmetry, the saddle-point equations in \eqref{eq:nemato-elastodynamic_equations_of_state} show that dilatation strain is exclusively generated by $\tilde{\Phi}_1$: $\varepsilon^h_1 \propto \tilde{\Phi}_1$. Setting $\lambda_{x^2-y^2} = \lambda_{2xy}$ by symmetry in \eqref{eq:Phi1_tilde_longitudinal}, then direction-selective criticality occurs if 
\begin{equation}
    \lambda_{2xy} \left[ \varphi_{x^2-y^2} \cos(2\phi_c) + \varphi_{2xy} \sin(2\phi_c) \right]= 0, \label{eq:lambda_varphi_planar_Phi1=0}
\end{equation}
representing the director-momentum locking in 3-state Potts nematics originally obtained as in-plane ``nemato-orbital coupling'' in Ref. \cite{Fernandes2020} and discussed in this work in Section \ref{sec:polar_and_planar_nematicity}. When \eqref{eq:lambda_varphi_planar_Phi1=0} is satisfied, then a two-components electronic nematic order parameter can be defined in the plane, up to a negative sign, as 
\begin{equation}
   \boldsymbol{\varphi}_\parallel \equiv \left( \varphi_{x^2-y^2}, \varphi_{2xy} \right) = |\boldsymbol{\varphi}_\parallel|\left[ -\sin(2\phi_c), \cos(2\phi_c)\right]. 
\end{equation}
This order parameter will soften first with a corrected mass proportional to 
\begin{equation}
\mathcal{M}_\parallel = \frac{1}{C_{E_{2g}}},
\end{equation}
where the elastic constant $C_{E_{2g}}$ is given in Table \ref{tab:tetragonal_deformation_to_other_groups}. Note that the mass for the order parameter $|\boldsymbol{\varphi}_\parallel|$ is isotropic in the plane, from \eqref{eq:lambda_varphi_planar_Phi1=0}. It is therefore clear that as $\phi_c$ rotates in the plane, different linear combinations of $\varphi_{x^2-y^2}$ or $\varphi_{2xy}$ fluctuations will be needed to eliminate the dilatation strain generated by $\tilde{\Phi}_1$. Constraints imposed by elastic compatibility are therefore the origin of the nemato-orbital coupling in planar hexagonal crystals.

\section{Helical Elastostatics in Systems with Quenched Plastic Strain \label{app:elastostatic_plastic_solution}}
In this section, we provide a solution to  elastostatics problems involving known quenched plastic strain fields. We solve these problems within the helical strain formalism. The solution is obtained by finding a closed-form expression  for the elastic strain, $\mathscr{E}_{ij}$, given a plastic strain, $\varepsilon^p_{ij}$. With this solution, one can obtain the elastic strain from any distribution of defects by relating $\varepsilon^p_{ij}$ to the defect density tensor, $\eta_{ij}$, as shown in Appendix \ref{app:plastic_strain_defect_density_tensor}. We first construct the solution formally in a system with arbitrary symmetry and then specialize to the case of an isotropic medium.

The equilibrium condition for an isolated medium is that its stress tensor, $\sigma_{ij}(\boldsymbol{x})$, is divergenceless: $\partial_{i}\sigma_{ij}(\boldsymbol{x}) = 0$ \cite{landauTheoryElasticity1970, muratoshioMicromechanicsDefectsSolids1987}. In momentum space, this expression can be written as 
\begin{equation}
    \hat{e}_{\alpha,j}q_{i}\sigma_{ij}=q\hat{e}_{\alpha,j}\hat{e}_{1,i}\sigma_{ij}=q\mathcal{Q}_{ij}^{1\alpha}\sigma_{ij} = 0,
\end{equation}
where we projected the dangling Cartesian index, $j$, into the helical basis. In moving between the second and third equality, we exploited the symmetry of the stress tensor. Through Hooke's Law, one relates the elastic strain to the stress, at which point it follows immediately that equilibrium is maintained in real-space as long as 
\begin{equation}
    C_{ijkl}\partial_{i}\varepsilon_{kl}\left(\boldsymbol{x}\right)=C_{ijkl}\partial_{i}\varepsilon_{kl}^{p}\left(\boldsymbol{x}\right).
\end{equation}
In momentum space, this yields 
\begin{equation}
    \mathcal{Q}_{ji}^{1\alpha}C_{ijkl}\mathcal{Q}_{kl}^{1\beta}\,\varepsilon_{\beta}^{h}=\mathcal{Q}_{ji}^{1\alpha}C_{ijkl}\varepsilon_{kl}^{p}.\label{eq:AppC_1}
\end{equation}
It is useful to now exploit the symmetric ``helical elastic constant'' matrix, $\mathscr{C}^{\alpha\beta}(\hat{q})$,   from \eqref{eq:elastic_Cijkl_Qij_Qkl}, repeated here as
\begin{equation}
    \mathscr{C}^{\alpha\beta}\left(\hat{q}\right)\equiv\mathcal{Q}_{ji}^{1\alpha}(\hat{q})C_{ijkl}\mathcal{Q}_{kl}^{1\beta}(\hat{q}).
\end{equation}
Being real and symmetric, it is diagonalizable, and therefore invertible (provided it is positive-definite). We define its inverse as the ``helical compliance'' matrix,  $\mathscr{S}_{\alpha\beta}(\hat{q}) \equiv [\mathscr{C}^{-1}(\hat{q})]^{\alpha\beta}$. To solve the elastostatics problem given a fixed plastic strain, $\varepsilon^p_{ij}$, one must obtain the helical strain amplitudes, $\varepsilon^h_\alpha$, the total strain amplitudes, $\varepsilon_{ij}$, and the elastic strain amplitudes, $\mathscr{E}_{ij}$. In the helical representation, it is straightforward to invert Eq. (\ref{eq:AppC_1}) and find $\varepsilon^h_\alpha$, from which we can also readily obtain $\varepsilon_{ij}$:
\begin{equation}
    \begin{aligned}\varepsilon_{\alpha}^{h} & =\mathscr{S}_{\alpha\beta}\left(\hat{q}\right)\mathcal{Q}_{ji}^{1\beta}C_{ijkl}\varepsilon_{kl}^{p},\\
\varepsilon_{ij} & =\mathcal{Q}_{ij}^{1\alpha}\varepsilon_{\alpha}^{h}=\mathcal{Q}_{ij}^{1\alpha}\mathscr{S}_{\alpha\beta}\left(\hat{q}\right)\mathcal{Q}_{lk}^{1\beta}C_{klmn}\varepsilon_{mn}^{p}.
\end{aligned}\label{eq:plastic_elastostatics_solution_intermediaries}
\end{equation}
The elastic strain amplitudes follow directly from $\mathscr{E}_{ij} = \varepsilon_{ij} - \varepsilon^p_{ij}$, and are given in the main text as \eqref{eq:elastic_strain_solution_anisotropic_medium}. For convenience, we repeat it here:
\begin{equation}
    \mathscr{E}_{ij}=\left[\mathcal{Q}_{ij}^{1\alpha}\mathscr{S}_{\alpha\beta}\left(\hat{q}\right)\mathcal{Q}_{lk}^{1\beta}C_{klmn}-\tfrac{1}{2}\left(\delta_{im}\delta_{jn}+\delta_{in}\delta_{jm}\right)\right]\varepsilon_{mn}^{p}.\tag{\ref{eq:elastic_strain_solution_anisotropic_medium}}
\end{equation}
The real-space solution follows from the above equation via Fourier transformation over the infinite domain. This constitutes the full solution to the plasticity problem at the level of linear (infinitesimal) strain. While not immediately obvious, if the plastic strain tensor is purely compatible, meaning it has no gauge-invariant physical part, such that $\text{inc}(\varepsilon^p)_{ij} = 0$, then $\mathscr{E}_{ij} = 0$ in \eqref{eq:elastic_strain_solution_anisotropic_medium}. This is true in any medium of any degree of crystalline anisotropy. To prove this, assume that the incompatible part of the plastic strain vanishes, such that the plastic strain can be written in the helical basis as $\varepsilon^p_{ij} = \mathcal{Q}^{1\alpha}_{ij}\varepsilon_\alpha^p$ for $\alpha = 1,2,3$, as we did for the completely compatible strain tensor in \eqref{eq:strain_from_Q1alpha_helical}. Upon substitution into \eqref{eq:elastic_strain_solution_anisotropic_medium}, we will have
\begin{align}
    \mathcal{Q}_{ij}^{1\alpha}\mathscr{S}_{\alpha\beta}\left(\hat{q}\right)\mathcal{Q}_{lk}^{1\beta}C_{klmn}\mathcal{Q}_{mn}^{1\gamma}\varepsilon_\gamma^p &= \mathcal{Q}_{ij}^{1\alpha}\mathscr{S}_{\alpha\beta}\left(\hat{q}\right)\mathscr{C}_{\beta\gamma}(\hat{q})\,\varepsilon_\gamma^p \nonumber
    \\
    &= \mathcal{Q}_{ij}^{1\alpha}\varepsilon_{\alpha}^p.
\end{align}
This establishes the total strain as $\varepsilon_{ij} \equiv \varepsilon_{ij}^p$, rendering the physical elastic strain $\mathscr{E}_{ij} = 0$. In this sense, \eqref{eq:elastic_strain_solution_anisotropic_medium} can be thought of as projecting out only the physical incompatibility contained in the plastic strain $\varepsilon_{ij}^p$. This physical part is generated by a nonzero defect density tensor, shown in Appendix \ref{app:plastic_strain_defect_density_tensor}.

To connect these results with nematicity, it is helpful to unpack the plastic strain, and afterwards, apply it to obtain the elastic strain \eqref{eq:elastic_strain_solution_anisotropic_medium} in an isotropic medium. We thus seek an expression for $C_{ijkl}\varepsilon^p_{kl}$.  Unlike the \textit{compatible} total strain, the \textit{incompatible} plastic strain must be expanded according to \eqref{eq:tensor_varThetaTensor_decomposition} (also Appendix \ref{app:quadrupolar_form_factors})  as 
\begin{equation}
    \varepsilon_{ij}^{p}=\frac{1}{3}\left(\text{Tr}\,\varepsilon^{p}\right)\delta_{ij}+\frac{1}{\sqrt{2}}\sum_{a=1}^{5}\varepsilon_{a}^{p}\vartheta_{ij}^{a}.\label{eq:plastic_strain_varTheta_decomp}
\end{equation}
All six coefficients are required since the plastic strain is not compatible. However, given that the total compatible strain also has a trace and $\vartheta_{ij}^{1,2,3}$  projections, it follows that neither $\text{Tr}\,\varepsilon^p$, nor $\varepsilon^p_{1,2,3}$, are gauge-invariant quantities. This contrasts with the $\varepsilon^p_{4,5}$  components, which clearly are gauge-invariant, since they are the projections along the eigentensors of the incompatibility operator with nonzero eigenvalue. Likewise, from \eqref{eq:condition_on_Trace_for_compatibility}, it follows that one can define the $\varepsilon^p_1$ coefficient as
\begin{equation}
    \varepsilon_{1}^{p}\equiv\frac{2}{\sqrt{3}}\left(\varepsilon_{0}^{p}+\Delta_{p}\right),\label{eq:Deltap_measuring_incompatible_dilatation}
\end{equation}
where $\varepsilon^p_0 \equiv \text{Tr}\,\varepsilon^p$. Then, $\Delta_p$  can be interpreted as the incompatible plastic dilatation. Upon substitution, we find 
\begin{equation}
    C_{ijkl}\varepsilon_{kl}^{p}=\frac{1}{3}\varepsilon_{0}^{p}C_{ijkk}+\frac{1}{\sqrt{2}}\sum_{a=1}^{5}\varepsilon_{a}^{p}\left(C_{ijkl}\vartheta_{kl}^{a}\right).
\end{equation}
In an isotropic medium, $C_{ijkl} = \lambda \delta_{ij}\delta_{kl} + \mu (\delta_{ik}\delta_{jl} + \delta_{il}\delta_{jk})$ \citep{landauTheoryElasticity1970}, from which one obtains $C_{ijkk} = 3(\lambda + 2\mu/3)\delta_{ij}$ and $C_{ijkl}\vartheta^a_{kl} = 2\mu \vartheta^{a}_{ij}$ from \eqref{eq:nematic_varThetaTensor_decomposition} and \eqref{eq:def_varTheta_GellMann_of_Qhat}.  Then,
\begin{equation}
    C_{ijkl}\varepsilon_{kl}^{p}=\left(\lambda+\tfrac{2\mu}{3}\right)\varepsilon_{0}^{p}\delta_{ij}+\sqrt{2}\mu\sum_{a=1}^{5}\varepsilon_{a}^{p}\vartheta_{ij}^{a},
\end{equation}
from which we obtain
\begin{equation}
    \mathcal{Q}_{ji}^{1\beta}C_{ijkl}\varepsilon_{kl}^{p}=\left[\left(\lambda+2\mu\right)\varepsilon_{0}^{p}+\tfrac{4\mu}{3}\Delta_{p}\right]\delta_{1\beta}+\mu\left(\varepsilon_{2}^{p}\delta_{2\beta}+\varepsilon_{3}^{p}\delta_{3\beta}\right),
\end{equation}
using the trace of $\mathcal{Q}^{\alpha\beta}$,  \eqref{eq:varThetaTensor_to_Qvechat_to_Qabij}, and \eqref{eq:Deltap_measuring_incompatible_dilatation}. The helical elastic constants, $\mathscr{C}^{\alpha\beta}$, and the helical compliance, $\mathscr{S}_{\alpha\beta}$, are
\begin{equation}
    \begin{aligned}\mathscr{C}^{\alpha\beta}\left(\hat{q}\right) & =\left(\lambda+2\mu\right)\delta^{1\alpha}\delta^{1\beta}+\mu\sum_{c=2,3}\delta^{c\alpha}\delta^{c\beta},\\
\mathscr{S}_{\alpha\beta}\left(\hat{q}\right) & =\frac{1}{\lambda+2\mu}\delta_{1\alpha}\delta_{1\beta}+\frac{1}{\mu}\sum_{c=2,3}\delta_{c\alpha}\delta_{c\beta}.
\end{aligned}
\end{equation}
The solution in the isotropic medium follows from \eqref{eq:plastic_elastostatics_solution_intermediaries} as
\begin{equation}
 \begin{aligned}\varepsilon_{\alpha}^{h} & =\left(\varepsilon_{0}^{p}+\varrho\Delta_{p}\right)\delta_{1\alpha}+\sum_{b=2,3}\varepsilon_{b}^{p}\delta_{b\alpha},\\
\varepsilon_{ij} & =\frac{1}{3}\left(\varepsilon_{0}^{p}+\varrho\Delta_{p}\right)\delta_{ij}\\
 & \phantom{=}+\frac{1}{\sqrt{2}}\left[\frac{2}{\sqrt{3}}\left(\varepsilon_{0}^{p}+\varrho\Delta_{p}\right)\vartheta_{ij}^{1}+\sum_{a=2,3}\varepsilon_{a}^{p}\vartheta_{ij}^{a}\right]
\end{aligned}\label{eq:helical_and_total_strain_divergenceless_stress}
\end{equation}
with the elastic strain amplitudes given in the main text as \eqref{eq:elastic_strain_solution_isotropic_medium}, repeated here for completeness as
\begin{equation}
    \mathscr{E}_{ij}=\frac{1}{3}\varrho\Delta_{p}\delta_{ij}-\frac{1}{\sqrt{2}}\left[\left(1-\varrho\right)\Delta_{p}\vartheta_{ij}^{1}+\sum_{a=4,5}\varepsilon_{a}^{p}\vartheta_{ij}^{a}\right].\tag{\ref{eq:elastic_strain_solution_isotropic_medium}}
\end{equation}
The quantity $\varrho$ is defined in \eqref{eq:varrho_definition}. Note that only the gauge-invariant contributions to the plasticity tensor, $\{\Delta_p, \varepsilon_{4,5}^p\}$, contribute to the physical elastic strain $\mathscr{E}_{ij}$, as proven in full generality earlier from \eqref{eq:elastic_strain_solution_anisotropic_medium}. The deviatoric part of the elastic strain can be obtained from $\mathscr{E}_{ij}$  using \eqref{eq:GellMann_decomp_tensor} and \eqref{eq:def_varTheta_GellMann_of_Qhat} as
\begin{align}
    \boldsymbol{\mathscr{E}}	&=-\left(1-\varrho\right)\Delta_{p}\boldsymbol{\hat{Q}}_{1} -\sum_{a=4,5}\varepsilon_{a}^{p}\boldsymbol{\hat{Q}}_{a}\nonumber
    \\
	&=\tfrac{2\left(1-\varrho\right)}{\sqrt{3}}\boldsymbol{\hat{Q}}_{1}\varepsilon_{0}^{p}-\left[\left(1-\varrho\right)\boldsymbol{\hat{Q}}_{1}^{\phantom{T}}\boldsymbol{\hat{Q}}_{1}^{\text{T}}+\sum_{a=4,5}\boldsymbol{\hat{Q}}_{a}^{\phantom{\text{T}}}\boldsymbol{\hat{Q}}_{a}^{\text{T}}\right]\cdot\boldsymbol{\varepsilon}^{p}. \label{eq:elastic_deviatoric_strain_solution_isotropy}
\end{align}
In the last expression, we have rewritten the plastic strain in terms of its trace and orbital components. One readily identifies these form factors from \eqref{eq:Gamma0_propagator} and \eqref{eq:Gamma_propagator} in the main text. The elastic strain in the $d$-orbital basis can then be simplified using them as
\begin{equation}
    \boldsymbol{\mathscr{E}} = -\left\{ \boldsymbol{\Gamma}_0(\hat{q})\varepsilon^p_0 + \Gamma(\hat{q}) \cdot \boldsymbol{\varepsilon}^p \right\}. \label{eq:elastic_strain_iso_no_nematicity}
\end{equation}
In the Subsection~\ref{subsec:incompatible_nematicity_defects} we show that this deviatoric elastic strain field, generated by plasticity, couples bilinearly to the electronic nematic order parameter.

Before concluding this appendix, we calculate the free energy minimum for an isotropic system with quenched plastic defect strain. Generalizing the variational free energy in \eqref{eq:isotropic_bare_elastic_free_energy} in the defect gauge theory amounts to
\begin{equation}
    \mathcal{F}_{\text{elas}}\left[\mathscr{E}\right]	=\frac{1}{2V}\sum_{\boldsymbol{q}}\left\{ \left(\lambda+\tfrac{2\mu}{3}\right)\left\vert\text{Tr}\,\mathscr{E}\right\vert^{2}+\mu\boldsymbol{\mathscr{E}}^\dagger\cdot\boldsymbol{\mathscr{E}}\right\}.
\end{equation}
Using \eqref{eq:plastic_strain_varTheta_decomp}, we write the elastic strain as
\begin{align}
    \mathscr{E}_{ij} &= \frac{1}{3}\left( \varepsilon_1^h - \varepsilon_0^p \right)\delta_{ij}+\frac{1}{\sqrt{2}}\left( \tfrac{2}{\sqrt{3}}\varepsilon^h_1 - \varepsilon^p_1 \right)\vartheta_{ij}^1 \nonumber
    \\
    &\phantom{=} + \frac{1}{\sqrt{2}}\left[ \sum_{a=2,3}\left( \varepsilon_a^h - \varepsilon^p_a \right)\vartheta_{ij}^a - \sum_{a=4,5} \varepsilon_a^p\vartheta_{ij}^a \right].
\end{align}
With \eqref{eq:Deltap_measuring_incompatible_dilatation}, it is clear that $\varepsilon_0^p$ and $\varepsilon_1^p$ are not independent. Instead, the physical, incompatible dilatation $\Delta_p$ is. Substituting in \eqref{eq:Deltap_measuring_incompatible_dilatation} for $\varepsilon_1^p$ yields an elastic strain tensor of
\begin{align}
    \mathscr{E}_{ij} &= \frac{1}{3}\left( \varepsilon_1^h - \varepsilon_0^p \right)\delta_{ij}+\frac{1}{\sqrt{2}} \left[\tfrac{2}{\sqrt{3}}\left(\varepsilon^h_1 - \varepsilon^p_0 - \Delta_p \right) \right]\vartheta_{ij}^1 \nonumber
    \\
    &\phantom{=} + \frac{1}{\sqrt{2}}\left[ \sum_{a=2,3}\left( \varepsilon_a^h - \varepsilon^p_a \right)\vartheta_{ij}^a - \sum_{a=4,5} \varepsilon_a^p\vartheta_{ij}^a \right].
\end{align}
Thus, the elastic dilatation is the gauge invariant combination $\varepsilon_1^h - \varepsilon_0^p$, while the elastic Gell-Mann vector is 
\begin{align}
    \boldsymbol{\mathscr{E}} &= \bigg[ \tfrac{2}{\sqrt{3}}\left(\varepsilon^h_1 - \varepsilon^p_0 - \Delta_p \right), \varepsilon_2^h - \varepsilon^p_2, \varepsilon_3^h - \varepsilon^p_3, -\varepsilon_4^p, -\varepsilon^p_5\bigg]^{\text{T}}.
\end{align}
The free energy density follows as
\begin{align}
    f_{\text{elas}}\left(\mathscr{E}\right)	&=\frac{1}{2}\bigg\{ \left(\lambda+\tfrac{2\mu}{3}\right)\left\vert \varepsilon_1^h - \varepsilon_0^p \right\vert^{2}+\tfrac{4}{3}\mu \left\vert \varepsilon^h_1 - \varepsilon^p_0 - \Delta_p\right\vert^2 \nonumber
    \\
    &\phantom{=\frac{1}{2}\bigg\{} + \mu \sum_{a=2,3} \left\vert \varepsilon_a^h - \varepsilon^p_a\right\vert^2 + \mu \sum_{a=4,5} \left\vert \varepsilon^p_a\right\vert^2 \bigg\}\nonumber
    \\
    &=\frac{1}{2}\bigg\{ \left( \lambda + 2\mu \right) \left\vert \varepsilon_1^h - \varepsilon_0^p \right\vert^{2} + \mu \sum_{a=2,3} \left\vert \varepsilon_a^h - \varepsilon^p_a\right\vert^2 \nonumber
    \\
    &\phantom{=\frac{1}{2}} - \tfrac{4}{3}\mu \left[\Delta^*_p\left( \varepsilon_1^h - \varepsilon_0^p\right) + \text{c.c.}\right] \nonumber
    \\
    &\phantom{=\frac{1}{2}} + \tfrac{4}{3}\mu \left\vert\Delta_p\right\vert^2 + \mu \sum_{a=4,5} \left\vert \varepsilon^p_a\right\vert^2 \bigg\}.
\end{align}
In moving from the first to the second equality, we factored out the incompatible dilatation, and reorganized the terms. We now can minimize with respect to the single-valued helical strain fields. Doing so yields
\begin{equation}
    \varepsilon_{\alpha}^{h}  =\left(\varepsilon_{0}^{p}+\varrho\Delta_{p}\right)\delta_{1\alpha}+\sum_{b=2,3}\varepsilon_{b}^{p}\delta_{b\alpha},
\end{equation}
consistent with \eqref{eq:helical_and_total_strain_divergenceless_stress}. This establishes the minimization of the free energy as equivalent to the divergenceless stress condition used in deriving \eqref{eq:helical_and_total_strain_divergenceless_stress}, and it is employed in this work for nemato-elasticity. Substituting this result back into the variational free energy yields
\begin{align}
    \mathcal{F}_{\text{elas}}\left(\varepsilon^p_{ij}\right) = \frac{\mu}{2V}\sum_{\boldsymbol{q}}\left\{ \frac{4}{3}(1-\varrho)\left\vert \Delta_p \right\vert^2 + \sum_{a=4,5}\left\vert \varepsilon_a^p\right\vert^2 \right\}, \label{eq:plastic_free_energy_iso_minimum}
\end{align}
given that $(\lambda +2\mu)\varrho^2 = \tfrac{4}{3}\mu \varrho$ from \eqref{eq:varrho_definition}. We note that the minimum free energy explicitly depends only on the physical incompatible part of the plastic strain tensor generated by defects.

\section{Elastic Strains from Straight Edge Dislocations\label{app:elastic_strains_from_dislocations}}
In this appendix, we discuss the ``slip gauge'' of dislocations to demonstrate the defect-gauge redundancy in plasticity. We then provide the details of how to compute the elastic strain fields from straight edge dislocations within the slip gauge. We also include some relevant intermediate steps in computing their spatial correlation functions. Our derivation of the plastic strain for this type of defect, and the notation used in the derivation, primarily follows Kleinert's book \cite{kleinertGaugeFieldsSolids1989} with influence from de Wit's works \cite{dewitLinearTheoryStatic1970, Dewit1973, dewitTheoryDisclinationsIII1973, dewitTheoryDisclinationsIV1973}.

We begin with a straight edge dislocation described by a Burgers vector $\boldsymbol{b}$ in the $xy$-plane and a dislocation line (or core) oriented along the $z$-axis, as shown in Fig.~\ref{fig:slip_gauge_transformation}. Assuming that the defect is a pure dislocation, then the plastic jump in the displacement vector is given by 
\begin{equation}
    \Delta_{j}u_{i}^{p}\left(\boldsymbol{x}; S\right)=b_{i}\delta_{j}\left(\boldsymbol{x};\, S\right)\equiv b_{i}\int_{S}\text{d}S_j\left(\boldsymbol{x}^{\prime}\right)\,\delta\left(\boldsymbol{x}-\boldsymbol{x}^{\prime}\right),
\end{equation}
where we have adapted the notation for the vector-valued delta-function on the surface $S$ from Refs.~\cite{dewitLinearTheoryStatic1970, Dewit1973, dewitTheoryDisclinationsIII1973, dewitTheoryDisclinationsIV1973, kleinertGaugeFieldsSolids1989} to make the $\boldsymbol{x}$-dependence explicit. The subscript $j$ on the $\Delta$ represents the tensorial nature of the plastic jump. The discontinuity in the displacement vector is measured with respect to the Volterra surface, and therefore the abrupt change can occur in any component $u_{x,y,z}$ across the surface oriented in any direction $x,y,z$. At this point, a defect gauge must be chosen -- that is, one must define the Volterra surface, $S$. We choose the negative-$x$ half of the $xz$-plane, and anchor the dislocation line at the origin, as shown in Fig.~\ref{fig:slip_gauge_transformation}(a). It follows then that $\text{d}\boldsymbol{S} = (0,\text{d}x\,\text{d}z,0)$  and therefore
\begin{align}
   \Delta_{j}u_{i}^{p}\left(\boldsymbol{x}\right)	&=b_{i}\delta_{j,y}\int_{-\infty}^{0}\text{d}x^{\prime}\int_{-\infty}^{\infty}\text{d}z^{\prime}\,\delta\left(x-x^{\prime}\right)\delta\left(y\right)\delta\left(z-z^{\prime}\right)\nonumber
   \\
	&=b_{i}\delta_{j,y}\,\delta\left(y\right)\Theta\left(-x\right),
\end{align}
where $\Theta(x)$  is the usual Heaviside function. This is one gauge choice for the plastic distortion tensor, $\beta^p_{ij}(\boldsymbol{x})$, for a straight edge dislocation. It is consistent with the topology associated with the Burgers vector \cite{landauTheoryElasticity1970, muratoshioMicromechanicsDefectsSolids1987, chaikinPrinciplesCondensedMatter1995}.

This gauge choice turns out to only be convenient when the dislocation is a slip -- meaning that $\boldsymbol{b} \cdot \text{d}\boldsymbol{S}(\boldsymbol{x}) = 0,\; \forall\boldsymbol{x}$, as shown in Fig.~\ref{fig:slip_gauge_transformation}(b). We now demonstrate that a defect gauge transformation can transform any straight edge dislocation with Burgers vector $\boldsymbol{b} = (b_x, b_y)$  into two slip dislocations along either coordinate axis. For $b_x$, the Volterra surface, $S$, suffices. For $b_y$, we rotate the Volterra surface by $90^{\text{o}}$  so that it is now the negative-$y$ half of the $yz$-plane. Then this new $S^\prime$  surface has area element $\text{d}\boldsymbol{S}^\prime = (-\text{d}y\,\text{d}z,0,0)$ and a plastic displacement vector given by 
\begin{equation}
    \Delta_{j}u_{i}^{p}\left(\boldsymbol{x};S^{\prime}\right)=b_{y}\delta_{i,y}\delta_{j}\left(\boldsymbol{x};\, S^{\prime}\right)=-b_{y}\delta_{i,y}\delta_{j,x}\,\delta\left(x\right)\Theta\left(-y\right),
\end{equation}
as illustrated in Fig.~\ref{fig:slip_gauge_transformation}(b). This choice of Volterra surface is also consistent with the topology of the dislocation, and therefore there is an ambiguity that arises between surfaces $S$ and $S^\prime$, as both describe the same topological defect. The difference between the plastic displacements for $b_y$  follows as 
\begin{align}
    \Delta_{j}u_{i}^{p}\left(\boldsymbol{x};S^{\prime}\right)-\Delta_{j}u_{i}^{p}\left(\boldsymbol{x};S\right)	&=-b_{y}\delta_{i,y}\bigg\lbrace\delta_{j,x}\,\delta\left(x\right)\Theta\left(-y\right) \nonumber
    \\
	&\phantom{=-b_{y}\delta_{i,y}\bigg\lbrace}+\delta_{j,y}\,\delta\left(y\right)\Theta\left(-x\right)\bigg\rbrace \nonumber
    \\
	&=-b_{y}\delta_{i,y}\left\{ -\delta_{j}\left(\boldsymbol{x};\,S^{\prime}\right)+\delta_{j}\left(\boldsymbol{x};\,S\right)\right\} \nonumber
    \\
	&=-b_{y}\delta_{i,y}\,\delta_{j}\left(\boldsymbol{x};\,\partial\mathcal{V}\right),
\end{align}
where the boundary of the volume enclosed by the surfaces $S$  and $S^\prime$  is defined as $\partial\mathcal{V}\equiv S - S^\prime$. This nontrivial, position-dependent difference would express itself in the displacement vector, and therefore the total strain tensor, despite the fact that both Volterra surfaces parameterize the same defect. Thus, the edge dislocation exemplifies the defect-gauge redundancy of plasticity.

We proceed with what we dub the ``slip gauge'' for the straight edge dislocation, namely that we can find a new Volterra surface $S^\prime$ that satisfies $\boldsymbol{b}\cdot {\text d}\boldsymbol{S}^\prime(\boldsymbol{x}) = 0$, for all $\boldsymbol{x}\in S^\prime$. To show that it is gauge-equivalent to the dislocation defined by $S$, we demonstrate that the change in Volterra surface amounts only to a compatible change in the plastic strain tensor -- that is, it amounts to a redefinition of the displacement vector. For this case, as shown in Fig.~\ref{fig:slip_gauge_transformation}(c), the Volterra surfaces $S$ and $S^\prime$ encapsulate the rectangular prism, $\mathcal{V}=\left\{ \boldsymbol{x} \in \mathbb{R}^3:\; x < 0 \; \text{and} \; y < 0 \right\}$.  We have defined its surface normals such that they are outer-facing (hence the negative sign on $S^\prime$). We now define a delta-function acting on the volume $\mathcal{V}$  through its sifting property as $\int_{x}f\left(\boldsymbol{x}\right)\delta\left(\boldsymbol{x};\,\mathcal{V}\right)\equiv\int_{\mathcal{V}}\text{d}^{3}x\,f\left(\boldsymbol{x}\right)$. Using the test function $f(\boldsymbol{x}) = \partial_i g(\boldsymbol{x})$ and the divergence theorem, it can be shown that $\partial_i \delta(\boldsymbol{x};\, \mathcal{V}) = -\delta_i(\boldsymbol{x};\, \partial \mathcal{V})$ \citep{Dewit1973}. Thus, the difference between the plastic fields for the $b_y$  component of the Burgers vector are 
\begin{equation}
    \Delta_{j}u_{i}^{p}\left(\boldsymbol{x};S^{\prime}\right)-\Delta_{j}u_{i}^{p}\left(\boldsymbol{x};S\right)=b_{y}\delta_{i,y}\partial_{j}\delta\left(\boldsymbol{x};\,\mathcal{V}\right).
\end{equation}

The defect-gauge transformation from $S\rightarrow S^\prime$ is therefore summarized in the displacement vector as 
\begin{equation}
    \boldsymbol{u}\left(\boldsymbol{x};S\right)\rightarrow\boldsymbol{u}\left(\boldsymbol{x};S^{\prime}\right)=\boldsymbol{u}\left(\boldsymbol{x};S\right)+b_{y}\hat{y}\,\delta\left(\boldsymbol{x};\,\mathcal{V}\right),
\end{equation}
and since this amounts to a redefinition of $\boldsymbol{u}(\boldsymbol{x})$ by a single-valued vector field, $\boldsymbol{\Lambda}(\boldsymbol{x}) = -b_y\hat{y}\, \delta(\boldsymbol{x};\, \mathcal{V})$, as shown in Fig.~\ref{fig:slip_gauge_transformation} (see \eqref{eq:distortion_tensor_gauge_transformation} as well). Thus, transforming between $S$ and $S^\prime$ constitutes only a compatible change to the plastic strain tensor, $\varepsilon^p_{ij}(\boldsymbol{x})$. It does not alter the physical content of the dislocation, since its incompatibility is unchanged.
 
With the ``slip-gauge'' for straight edge dislocations established, we turn to computing the elastic strain fields within the helical formalism. We start from \eqref{eq:2xy_plastic_strain_source_FS}, which we repeat here as
\begin{equation}
    \varepsilon_{2xy}^{p}\left(\boldsymbol{q}\right)=4\pi\text{i}\,\delta\left(q_{z}\right)\left(\frac{b_{x}\sin\phi-b_{y}\cos\phi}{q\sin2\phi}\right), \tag{\ref{eq:2xy_plastic_strain_source_FS}}
\end{equation}
for convenience. The entire plastic strain tensor itself follows in the Cartesian basis simply as
\begin{equation}
    \varepsilon^p_{ij} = \frac{1}{2}\varepsilon^p_{2xy}\lambda_{ij}^1,
\end{equation}
meaning that it is traceless in this slip gauge, $\varepsilon_0^p = 0$, and that its Gell-Mann vectorial representation from \eqref{eq:GellMann_decomp_tensor} is $\boldsymbol{\varepsilon}^p = (0,0,0,0,\varepsilon_{2xy}^p)^{\text{T}}$. The conjugate field for the $d$-orbital nematic order parameter is then $\boldsymbol{h} = -\Gamma(\hat{q}) \cdot \boldsymbol{\varepsilon}^p$, since $\varepsilon^p_0 = 0$. For convenience, $\Gamma(\hat{q})$ is repeated here
\begin{equation}
    \Gamma=\left(1-\varrho\right)\boldsymbol{\hat{Q}}_{1}^{\phantom{\text{T}}}\boldsymbol{\hat{Q}}_{1}^{\text{T}}+\boldsymbol{\hat{Q}}_{4}^{\phantom{\text{T}}}\boldsymbol{\hat{Q}}_{4}^{\text{T}}+\boldsymbol{\hat{Q}}_{5}^{\phantom{\text{T}}}\boldsymbol{\hat{Q}}_{5}^{\text{T}}.\tag{\ref{eq:Gamma_propagator}}
\end{equation}
Due to the $\delta(q_z)$, the momentum is confined to the $xy$-plane in momentum space, where 
\begin{equation}
    \begin{aligned}\boldsymbol{\hat{Q}}_{1}^{\text{T}} & =\left(\tfrac{1}{2},\tfrac{\sqrt{3}}{2}\cos2\phi,0,0,\tfrac{\sqrt{3}}{2}\sin2\phi\right),\\
\boldsymbol{\hat{Q}}_{4}^{\text{T}} & =\left(0,0,\cos\phi,-\sin\phi,0\right),\\
\boldsymbol{\hat{Q}}_{5}^{\text{T}} & =\left(\tfrac{\sqrt{3}}{2},-\tfrac{1}{2}\cos2\phi,0,0,-\tfrac{1}{2}\sin2\phi\right).
\end{aligned}
\end{equation}
In this case, $\boldsymbol{\hat{Q}}_4 \cdot \boldsymbol{\varepsilon}^p = 0$. Likewise, it follows that  $\boldsymbol{\hat{Q}}_5 = \frac{1}{\sqrt{3}}\left(2 \boldsymbol{1}_{z^2} - \boldsymbol{\hat{Q}}_1 \right)$  where the vector $\boldsymbol{1}_{z^2} \equiv (1,0,0,0,0)^{\text{T}}$. Defining a similar vector, $\boldsymbol{1}_{2xy}\equiv(0,0,0,0,1)^{\text{T}}$, we have that
{\small
\begin{align}
    \Gamma(\hat{q};q_z=0)\cdot \boldsymbol{1}_{2xy} &= \left\{ (1-\varrho) \boldsymbol{\hat{Q}}_{1}^{\phantom{\text{T}}}\boldsymbol{\hat{Q}}_{1}^{\text{T}} + \boldsymbol{\hat{Q}}_{5}^{\phantom{\text{T}}}\boldsymbol{\hat{Q}}_{5}^{\text{T}}  \right\} \cdot \boldsymbol{1}_{2xy} \nonumber
    \\
    &= \left\{ (1-\varrho) \boldsymbol{\hat{Q}}_{1}^{\phantom{\text{T}}} - \tfrac{1}{3}\left(2\boldsymbol{1}_{z^2} - \boldsymbol{\hat{Q}}_1 \right)  \right\} \left( \boldsymbol{\hat{Q}}_{1}^{\text{T}} \cdot \boldsymbol{1}_{2xy}\right) \nonumber
    \\
    &= -\tfrac{4}{3}\left\{ \tfrac{1}{2}\boldsymbol{1}_{2xy} - \left(1- \tfrac{3}{4}\varrho\right)\boldsymbol{\hat{Q}}_1\right\}\left( \boldsymbol{\hat{Q}}_{1}^{\text{T}} \cdot \boldsymbol{1}_{2xy}\right) \nonumber
    \\
    &=-\tfrac{2}{\sqrt{3}}\sin2\phi \,\left\{ \tfrac{1}{2}\boldsymbol{1}_{2xy} - \left(1- \tfrac{3}{4}\varrho\right)\boldsymbol{\hat{Q}}_1\right\}.
\end{align}
}
Recalling that the 3D Poisson ratio is given by $\nu_3 = \lambda/2(\lambda + \mu)$ \citep{landauTheoryElasticity1970, muratoshioMicromechanicsDefectsSolids1987, kleinertGaugeFieldsSolids1989}, one can show that $1 - 3\varrho/4 = 1/2(1-\nu_3)$ using \eqref{eq:varrho_definition}. From this, it follows that
\begin{equation}
    \Gamma(\hat{q};q_z=0)\cdot \boldsymbol{1}_{2xy} = -\frac{\sin2\phi}{(1-\nu_3)\sqrt{3}} \,\left\{ (1-\nu_3)\boldsymbol{1}_{2xy} - \boldsymbol{\hat{Q}}_1\right\}. \label{eq:Gamma_times_12xy}
\end{equation}
Thus, the elastic strain follows as
\begin{align}
    \boldsymbol{h} &= -\left[\Gamma(\hat{q};q_z=0)\cdot \boldsymbol{1}_{2xy}\right]\varepsilon_{2xy}^p \nonumber
    \\
    &= \frac{\varepsilon_{2xy}^p\,\sin2\phi}{(1-\nu_3)\sqrt{3}} \,\left\{ (1-\nu_3)\boldsymbol{1}_{z^2} - \boldsymbol{\hat{Q}}_1\right\}.\label{eq:hq_disloc_compact}
\end{align}
Expanding out the expression above yields \eqref{eq:hq_disloc_expanded}, repeated here for convenience as
{\small
\begin{equation}
    \boldsymbol{h}	=\delta\left(q_{z}\right)\, \tfrac{2\pi\text{i}}{1-\nu_3}\tfrac{b_{x}\sin\phi-b_{y}\cos\phi}{q}\left(\tfrac{1-2\nu_{3}}{\sqrt{3}},-\cos2\phi,0,0,-\sin2\phi\right)^{\text{T}}.\tag{\ref{eq:hq_disloc_expanded}}
\end{equation}
}

In real-space, the elastic strain fields follow from the inverse Fourier transform. Starting with \eqref{eq:hq_disloc_expanded}, the elastic strain can be written formally as
\begin{align}
    \boldsymbol{h}\left(\boldsymbol{x}\right)	&=\tfrac{2\pi}{1-\nu_{3}}\left(b_{x}\partial_{y}-b_{y}\partial_{x}\right) \nonumber
    \\
	&\phantom{=}\times\left[-\tfrac{1-2\nu_{3}}{\sqrt{3}}\left(\partial_{x}^{2}+\partial_{y}^{2}\right),\left(\partial_{x}^{2}-\partial_{y}^{2}\right),0,0,\left(2\partial_{x}\partial_{y}\right)\right]^{\text{T}}g_{4}\left(\boldsymbol{x}\right),
\end{align}
where we have used the $n=4$  function in the family of $(n/2)$-harmonic Green's functions, $g_{n}(\boldsymbol{r})$: 
\begin{equation}
    g_{n}\left(\boldsymbol{r}\right)\equiv\int\frac{\text{d}^{3}q}{\left(2\pi\right)^{3}}\,\frac{\delta\left(q_{z}\right)\,\text{e}^{\text{i}\boldsymbol{q}\cdot\boldsymbol{r}}}{q^{n}}.\label{eq:g_n_functions}
\end{equation}
The first three functions are computed iteratively in the infinite medium in Appendix \ref{app:harmonic_Greens_functions} and $g_4(\boldsymbol{r})$ is obtained within it as \eqref{eq:g_4_fundamental_sol}. Upon differentiation, we find 
\begin{equation}
    \begin{aligned}h_{z^{2}}\left(x,y\right) & =\frac{1-2\nu_{3}}{2\pi\left(1-\nu_{3}\right)\sqrt{3}}\,\frac{b_{y}x-b_{x}y}{x^{2}+y^{2}},\\
h_{x^{2}-y^{2}}\left(x,y\right) & =-\frac{1}{2\pi\left(1-\nu_{3}\right)}\,\frac{2xy\left(b_{x}x+b_{y}y\right)}{\left(x^{2}+y^{2}\right)^{2}},\\
h_{2xy}\left(x,y\right) & =\frac{1}{2\pi\left(1-\nu_{3}\right)}\,\frac{\left(x^{2}-y^{2}\right)\left(b_{x}x+b_{y}y\right)}{\left(x^{2}+y^{2}\right)^{2}},
\end{aligned} \label{eq:eq:elastic_dislocation_fields_real-space_Cartesian}
\end{equation}
Writing $\boldsymbol{r}_\parallel \equiv (x,y,0) \equiv r_\parallel (\cos\phi_r,\sin\phi_r, 0)$, then the expressions above simplify into those given in \eqref{eq:elastic_dislocation_fields_real-space_polar}.

We now proceed to calculate the spatial correlation functions for an ensemble of uncorrelated, uniformly distributed, and isotropically oriented straight edge dislocations. Starting from the definition of the correlation function for a single realization of dislocations, \eqref{eq:definition_of_correlation_single_realization}, since each defect core can be positioned uniformly within the plane then averaging over the complex exponential term enforces that all spatial correlations must be between any given defect \textit{with itself}. That is, assuming $\llbracket\text{e}^{\text{i}\boldsymbol{q}\cdot\left(\boldsymbol{x}^{\alpha}-\boldsymbol{x}^{\beta}\right)}\rrbracket=\delta_{\alpha\beta}$, where the double brackets represent the disorder average, then 
\begin{equation}
   \begin{aligned}\left\llbracket \mathcal{C}\left(\boldsymbol{q}\right)\right\rrbracket  & =\sum_{\alpha=1}^{N}\left\llbracket \mathscr{K}^{\alpha\alpha}\left(\boldsymbol{q}\right)\right\rrbracket \\
\left\llbracket \mathscr{K}^{\alpha\alpha}\left(\boldsymbol{q}\right)\right\rrbracket  & =\Gamma\left(\hat{q}\right)\cdot\left\llbracket \boldsymbol{\varepsilon}^{p}(\boldsymbol{q};\boldsymbol{b}^{\alpha})\boldsymbol{\varepsilon}^{p}(-\boldsymbol{q};\boldsymbol{b}^{\alpha})^{\text{T}}\right\rrbracket \cdot\Gamma^{\text{T}}\left(-\hat{q}\right)
\end{aligned}
\end{equation}
To proceed, we use \eqref{eq:2xy_plastic_strain_source_FS} and the fact that $\boldsymbol{\varepsilon}^p = \varepsilon^p_{2xy}\boldsymbol{1}_{2xy}$  to compute
\begin{equation}
    \left\llbracket \varepsilon_{2xy}^{p}(\boldsymbol{q};\boldsymbol{b}^{\alpha})\varepsilon_{2xy}^{p}(-\boldsymbol{q};\boldsymbol{b}^{\alpha})^{\text{T}}\right\rrbracket 	=\frac{8\pi^{2}\delta\left(q_{z}\right)}{q^{2}\sin^{2}2\phi}\left\llbracket \left(b^{\alpha}\right)^{2}\right\rrbracket .
\end{equation}
To obtain the disorder average above, the Burgers vector was first written as $\boldsymbol{b}^\alpha \equiv b^\alpha (\cos\beta,\sin\beta)$. If we assume the Burgers vectors are isotropically distributed, then it follows that $\llbracket \cos[2(\beta -\phi)]\rrbracket = 0$. Next, we define $\llbracket(b^{\alpha})^{2}\rrbracket \equiv \llbracket b^{2}\rrbracket$  for all dislocations to obtain a disorder-averaged spatial correlation function given by 
\begin{equation}
    \left\llbracket \mathcal{C}\left(\boldsymbol{q}\right)\right\rrbracket =N\frac{8\pi^{2}\llbracket b^2 \rrbracket\delta\left(q_{z}\right)}{q^{2}\sin^{2}2\phi}\,\left\{ \Gamma\left(\hat{q}\right)\cdot\boldsymbol{1}_{2xy}^{\phantom{\text{T}}}\boldsymbol{1}_{2xy}^{\text{T}}\cdot\Gamma^{\text{T}}\left(\hat{q}\right)\right\} ,
\end{equation}
where $\boldsymbol{1}_{2xy}^{\text{T}} \equiv (0,0,0,0,1)$  and we have used the fact that $\Gamma(-\hat{q}) = \Gamma(\hat{q})$. The factor of $N$ is a consequence of the Central-Limit Theorem since the dislocations are uncorrelated and identically distributed. Despite the defects being spatially uncorrelated, the elastic strains clearly are correlated, with the incompatible projector $\Gamma(\hat{q})$ being responsible for the spatial correlations. To simplify the above, we use \eqref{eq:Gamma_times_12xy}. Defining it as the vector, $\boldsymbol{\Psi}$, yields 
\begin{equation}
    \boldsymbol{\Psi} = -\tfrac{\sin2\phi}{2(1-\nu_3)}\left(\tfrac{1-2\nu_{3}}{\sqrt{3}},-\cos2\phi,0,0,-\sin2\phi\right)^{\text{T}}.
\end{equation}
Then, the disordered averaged correlation function is
\begin{align}
   \left\llbracket \mathcal{C}\left(\boldsymbol{q}\right)\right\rrbracket &=N\frac{8\pi^{2}\llbracket b^2 \rrbracket\delta\left(q_{z}\right)}{q^{2}\sin^{2}2\phi}\,\boldsymbol{\Psi}\boldsymbol{\Psi}^{\text{T}}.
\end{align} 
The correlation function therefore will have six nonzero components (of a possible fifteen). There are three autocorrelations -- that is, correlations generated within the same orbital basis function -- and three cross-correlations. These six correlations functions are listed in \eqref{eq:dis_avg_correlation_function_FS}.

The real-space functions are obtained through inverting the Fourier transform. Recalling the following trigonometric identities,
\begin{equation}
\begin{aligned}
    q^{4}\cos\left(4\phi\right)  &=q_{x}^{4}+q_{y}^{4}-6q_{x}^{2}q_{y}^{2},
    \\
    q^{4}\sin\left(4\phi\right)&=4q_{x}q_{y}\left(q_{x}^{2}-q_{y}^{2}\right),
    \end{aligned}
\end{equation}
it follows from \eqref{eq:dis_avg_correlation_function_FS} that
\begin{equation}
    \begin{aligned}\left\llbracket \mathcal{C}_{z^2,z^2}\left(\boldsymbol{r}_{\parallel}\right)\right\rrbracket  & =2\pi^{2}N\xi^{2}\,\frac{(1-2\nu_{3})^{2}}{3}\tilde{g}_{2}\left(\tfrac{r_{\parallel}}{L}\right),\\
\left\llbracket \mathcal{C}_{x^2-y^2,x^2-y^2}\left(\boldsymbol{r}_{\parallel}\right)\right\rrbracket  & =\pi^{2}N\xi^{2}\bigg\{ \tilde{g}_{2}\left(\tfrac{r_{\parallel}}{L}\right)
\\
&\phantom{=\pi^{2}N\xi^{2}}+\left(\partial_{x}^{4}+\partial_{y}^{4}-6\partial_{x}^{2}\partial_{y}^{2}\right)g_{6}\left(\boldsymbol{r}_{\parallel}\right)\bigg\} ,\\
\left\llbracket \mathcal{C}_{2xy,2xy}\left(\boldsymbol{r}_{\parallel}\right)\right\rrbracket  & =\pi^{2}N\xi^{2}\bigg\{ \tilde{g}_{2}\left(\tfrac{r_{\parallel}}{L}\right)
\\
&\phantom{=\pi^{2}N\xi^{2}}-\left(\partial_{x}^{4}+\partial_{y}^{4}-6\partial_{x}^{2}\partial_{y}^{2}\right)g_{6}\left(\boldsymbol{r}_{\parallel}\right)\bigg\} ,
\end{aligned}
\end{equation}
\begin{equation}
    \begin{aligned}\left\llbracket \mathcal{C}_{z^2,x^2-y^23}\left(\boldsymbol{r}_{\parallel}\right)\right\rrbracket  & =2\pi^{2}N\xi^{2}\,\cdot \tfrac{1-2\nu_{3}}{\sqrt{3}}\left[\left(\partial_{x}^{2}-\partial_{y}^{2}\right)g_{4}\left(\boldsymbol{r}_{\parallel}\right)\right],\\
\left\llbracket \mathcal{C}_{z^2,2xy}\left(\boldsymbol{r}_{\parallel}\right)\right\rrbracket  & =2\pi^{2}N\xi^{2}\,\cdot\tfrac{1-2\nu_{3}}{\sqrt{3}}\left[\left(2\partial_{x}\partial_{y}\right)g_{4}\left(\boldsymbol{r}_{\parallel}\right)\right],\\
\left\llbracket \mathcal{C}_{x^2-y^2,2xy}\left(\boldsymbol{r}_{\parallel}\right)\right\rrbracket  & =\pi^{2}N\xi^{2}\,\left[4\partial_{x}\partial_{y}\left(\partial_{x}^{2}-\partial_{y}^{2}\right)g_{6}\left(\boldsymbol{r}_{\parallel}\right)\right],
\end{aligned}
\end{equation}
with $\tilde{g}_2(r_\parallel/L)$  being a formally divergent integral that needs regularization, and where $g_6(\boldsymbol{r})$ is given in \eqref{eq:g_6_fundamental_sol}.  We proceed to compute the former and save the detailed calculations of the latter for Appendix \ref{app:harmonic_Greens_functions}.

To regularize the divergent integral, we start by adding a mass, $m$, to the explicit expression for $g_2(\boldsymbol{r})$  to find 
\begin{equation}
    g_{2}\left(\boldsymbol{r};m\right)\equiv\int\frac{\text{d}^{3}q}{\left(2\pi\right)^{3}}\,\frac{\delta\left(q_{z}\right)\text{e}^{\text{i}\boldsymbol{q}\cdot\boldsymbol{r}}}{q^{2}+m^{2}}=\frac{1}{2\pi}\int_{0}^{\infty}\frac{\text{d}q}{4\pi^{2}}\,\frac{qJ_{0}\left(qr_\parallel\right)}{q^{2}+m^{2}},
\end{equation}
where the second equality was obtained by exploiting the rotational symmetry and the Jacobi-Anger expansion. When $m\rightarrow 0$, this integral is logarithmically divergent in the lower-bound, and is associated with the linear size of a finite system, $L$. Using the Laplace transform of the cosine and Bessel functions, one obtains a regularized integral given by
\begin{equation}
    g_{2}\left(r_\parallel;m\right)=\tfrac{1}{4\pi^{2}}K_{0}\left(mr_\parallel\right),
\end{equation}
where $K_\nu(x)$  is the modified Bessel function of the second kind \citep{NIST:DLMF}. The behavior of  $K_0(x)$ for small $x$ is well-known to be logarithmic: $K_0(x)\rightarrow -\log(x)$ \citep{NIST:DLMF}. In a periodic system of size $L$, the maximum separation between points is $L/2$. We therefore identify the regularizing mass, $m$, as $m^{-1} \equiv L/2$. Then, taking the limit of the Bessel function yields 
\begin{equation}
    \tilde{g}_{2}\left(\tfrac{r_{\parallel}}{L}\right)\equiv\lim_{L\rightarrow\infty}g_{2}\left(r_{\parallel};\tfrac{2}{L}\right)\rightarrow\tfrac{1}{4\pi^{2}}\log\left(\tfrac{L}{2r_{\parallel}}\right).
\end{equation}
In the limiting form, we have inverted the arguments of the logarithm by absorbing the overall negative sign. Importantly, notice that the limiting behavior of the regularized $\tilde{g}_2(r)$ yields the same fundamental Green's function in \eqref{eq:g_2_fundamental_sol}. Combining these results and differentiating $g_6(\boldsymbol{r}_\parallel)$  from \eqref{eq:g_6_fundamental_sol}  yields \eqref{eq:dis_avg_correlation_function_RS}.

\section{Multipolar Expansion of Plastic Strain \label{app:plastic_strain_defect_density_tensor}}
As discussed in Section \ref{sec:elastic_incompatibility_and_nematoplasticity} and Appendix \ref{app:elastostatic_plastic_solution}, only the incompatible part of the plastic strain tensor is physical. This defines the defect density tensor in elasticity theory in terms of the elastic strain tensor. In the main text it is given as \eqref{eq:eta_tensor_incompatibility_of_E}, and repeated here as
\begin{equation}
    \eta_{ij}(\boldsymbol{x}) \equiv \text{inc}[\mathscr{E}(\boldsymbol{x})]_{ij} = -\text{inc}[\varepsilon^p(\boldsymbol{x})]_{ij}. \tag{\ref{eq:eta_tensor_incompatibility_of_E}}
\end{equation}
The defect density tensor is related to the disclination and dislocation density tensors, both of which are defect-gauge invariant quantities \cite{kronerekkehartContinuumTheoryDefects1981,kleinertGaugeFieldsSolids1989}. Because of the manifest defect-gauge invariance of the defect density, in this appendix we compute the plastic strain tensor $\varepsilon^p_{ij}$ in terms of $\eta_{ij}$. Using \eqref{eq:elastic_strain_solution_anisotropic_medium}, we will then be able to formally compute the gauge-invariant elastic strain fields emanating from an arbitrary distribution of defects, which we will express as a multipolar expansion.

Because the plastic strain is a symmetric tensor, so is its incompatibility. Thus, $\eta_{ij}$ is a rank-2 symmetric tensor which can be written in our helical decomposition from \eqref{eq:tensor_varThetaTensor_decomposition} (also Appendix \ref{app:quadrupolar_form_factors}) as
\begin{equation}
    \eta_{ij} \equiv \frac{1}{3}\eta_0\delta_{ij} + \frac{1}{\sqrt{2}}\sum_{a=1}^5 \eta_a\vartheta^a_{ij}.\label{eq:eta_definition_varThetas}
\end{equation}
A similar decomposition follows for the plastic strain in \eqref{eq:plastic_strain_varTheta_decomp}. From \eqref{eq:incompatibility_of_A_varTheta_decomp}, applying the incompatibility operator to \eqref{eq:plastic_strain_varTheta_decomp} yields
\begin{equation}
    \text{inc}(\varepsilon^p) = \frac{1}{3}\left(\varepsilon^p_0 - \tfrac{\sqrt{3}}{2}\varepsilon^p_1\right) \text{inc}(\text{I}) + \frac{q^2}{\sqrt{2}}\sum_{a=4,5}\varepsilon^p_a\vartheta^a.
\end{equation}
Recalling from \eqref{eq:incI_fourier_space} that $
\text{inc}\left(\text{I}\right)=q^{2}\left(\mathcal{Q}^{11}-\text{I}\right)$, and using \eqref{eq:varThetaTensor_to_Qvechat_to_Qabij}, we write 
\begin{equation}
    \text{inc}\left(\text{I}\right) = \tfrac{2}{3}q^2\left( \sqrt{\tfrac{3}{2}}\vartheta^1 - \text{I}\right).
\end{equation}
Substituting this into the above, and making use of the incompatible dilatation, $\Delta_p$, from \eqref{eq:Deltap_measuring_incompatible_dilatation}, we have that 
\begin{align}
    \text{inc}(\varepsilon^p)_{ij} &= \frac{1}{3}\left(-\Delta_p\right) \cdot \frac{2}{3}q^2\left( \sqrt{\frac{3}{2}}\vartheta^1_{ij} - \delta_{ij}\right) + \frac{q^2}{\sqrt{2}}\sum_{a=4,5}\varepsilon^p_a\vartheta^a_{ij} \nonumber
    \\
    &= \frac{1}{3}\left(\frac{2}{3}q^2\Delta_p\right)\delta_{ij}+\frac{1}{\sqrt{2}}\left( -\frac{2}{3\sqrt{3}} q^2 \Delta_p \right)\vartheta^1_{ij} \nonumber
    \\
    &\phantom{=} + \frac{1}{\sqrt{2}}\sum_{a=4,5}\left(q^2\varepsilon^p_a \right)\vartheta^a_{ij}.\label{eq:incompatibility_of_plastic_strain_vartheta_decomps_for_eta}
\end{align}
Comparing this decomposition to \eqref{eq:eta_definition_varThetas}, we can express the defect density coefficients in terms of the three physical contributions to the plastic strain as 
\begin{equation}
    \begin{aligned}
        \eta_0 &= -\frac{2}{3}q^2\Delta_p, & \eta_1 &= +\frac{2}{3\sqrt{3}}q^2\Delta_p,
        \\
        \eta_2 &= \eta_3 = 0, & \eta_4 &= -q^2\varepsilon_4^p, &\eta_5 = -q^2\varepsilon^p_5.
    \end{aligned} \label{eq:eta_coeffs_plastic_coeffs}
\end{equation}
Importantly, we find that, despite the defect density tensor nominally having six elements as a rank-2 symmetric tensor, because there are only \textit{three} incompatible components of the plastic strain, there are only \textit{three} independent coefficients in $\eta_{ij}$. We choose the trace, $\eta_0$, to be the independent degree of freedom, and write $\eta_1 = -\tfrac{1}{\sqrt{3}}\eta_0$. 

With these relations, we can write the plastic strain tensor as 
\begin{align}
    \varepsilon^p_{ij}&= \frac{1}{3}\varepsilon^p_0\delta_{ij} + \frac{1}{\sqrt{2}}\left\{ \left( \tfrac{2}{\sqrt{3}}\varepsilon^p_0\right)\vartheta^1_{ij} + \sum_{a=2,3}\varepsilon^p_2\vartheta^a_{ij}\right\}\nonumber
    \\
    &\phantom{=}-\frac{1}{q^2\sqrt{2}}\left\{ \left( \sqrt{3}\eta_0 \right)\vartheta_{ij}^1 + \sum_{a=4,5}\eta_a\vartheta^a_{ij} \right\}.
\end{align}
Recognizing the first line in the second equality as a \textit{compatible} tensor from  \eqref{eq:condition_on_Trace_for_compatibility}, we can write it in terms of an unphysical defect-gauge vector field $\boldsymbol{\Lambda}$. The plastic strain then simplifies to
\begin{equation}
    \varepsilon^p_{ij} = \frac{\text{i}}{2}\left( q_j \Lambda_i + q_i \Lambda_j \right) - \frac{1}{q^2\sqrt{2}}\left\{ \left( \sqrt{3}\,\eta_0 \right)\vartheta_{ij}^1 + \sum_{a=4,5}\eta_a\vartheta^a_{ij} \right\}.
\end{equation}
Focusing on incompatible part, and using \eqref{eq:tensor_varThetaTensor_decomposition}, we can write it as 
\begin{equation}
    \frac{1}{q^2}\left\{\sqrt{\tfrac{3}{2}}\,\vartheta_{ij}^1\delta_{kl}+ \sum_{a=4,5}\vartheta^a_{ij}\vartheta^a_{kl}\right\}\eta_{kl} \equiv \mathscr{G}_{ijkl}(\boldsymbol{q})\eta_{kl}. \label{eq:generic_plastic_defect_Greens_tensor}
\end{equation}
The Green's tensor above provides the formal solution to the plastic strain induced by any distribution of defects in Fourier and momentum space as
\begin{equation}
    \begin{aligned}
        \varepsilon^p_{ij} &= \frac{\text{i}}{2}\left( q_j \Lambda_i + q_i \Lambda_j \right) - \mathscr{G}_{ijkl}(\boldsymbol{q})\eta_{kl},
        \\
        \varepsilon^p_{ij}(\boldsymbol{x}) &= \frac{1}{2}\left[ \partial_j \Lambda_i(\boldsymbol{x}) + \partial_i \Lambda_j(\boldsymbol{x}) \right] - \int_{x^\prime}\mathscr{G}_{ijkl}(\boldsymbol{x} - \boldsymbol{x}^\prime)\eta_{kl}(\boldsymbol{x}^\prime).
    \end{aligned}\label{eq:Greens_functions_plastic_strain_from_defect_density}
\end{equation}

The expression is universal for any defect density tensor in any medium as it is agnostic about the material-dependent elastic constants. Given that we have an expression for the unobservable \textit{plastic} strain, one can formally obtain the observable, gauge-invariant, \textit{elastic} strain with the use of \eqref{eq:elastic_strain_solution_anisotropic_medium}. The compatible part that depends on $\boldsymbol{\Lambda}$ will not affect the elastic strain, as proven in Appendix \ref{app:elastostatic_plastic_solution}. In the case of an isotropic medium, we can substitute \eqref{eq:eta_coeffs_plastic_coeffs} directly into \eqref{eq:elastic_strain_solution_isotropic_medium} and obtain the gauge-invariant elastic strain as
\begin{align}
    \mathscr{E}_{ij}&= \frac{1}{q^2}\left\{ \left[-\frac{1}{2}\varrho\delta_{ij}+ \frac{3}{2\sqrt{2}}(1-\varrho)\,\vartheta^1_{ij}\right]\delta_{kl} + \sum_{a=4,5}\vartheta_{ij}^{a}\vartheta^a_{kl}\right\}\,\eta_{kl}.
\end{align}
This last line defines another Green's tensor, $\mathcal{X}_{ijkl}(\boldsymbol{q})$ which explicitly depends on the elastic constants, making it non-universal. This material-dependent tensor is defined by
\begin{equation}
    \mathscr{E}_{ij}\equiv \mathcal{X}_{ijkl}(\boldsymbol{q})\eta_{kl}.
\end{equation}
By comparing this expression with the general solution for the elastic strain, \eqref{eq:elastic_strain_solution_anisotropic_medium}, and substituting in \eqref{eq:Greens_functions_plastic_strain_from_defect_density}, we find 
\begin{align}
    \mathcal{X}_{ijkl}(\boldsymbol{q}) &= -\left[\mathcal{Q}_{ij}^{1\alpha}\mathscr{S}_{\alpha\beta}\left(\hat{q}\right)\mathcal{Q}_{lk}^{1\beta}C_{klmn}-\delta_{im}\delta_{jn}\right]\mathscr{G}_{mnrs}(\boldsymbol{q}). \label{eq:X_Greens_tensor_elastic_strain_from_eta}
\end{align}

\begin{figure}
    \centering
    \includegraphics[width=1\columnwidth]{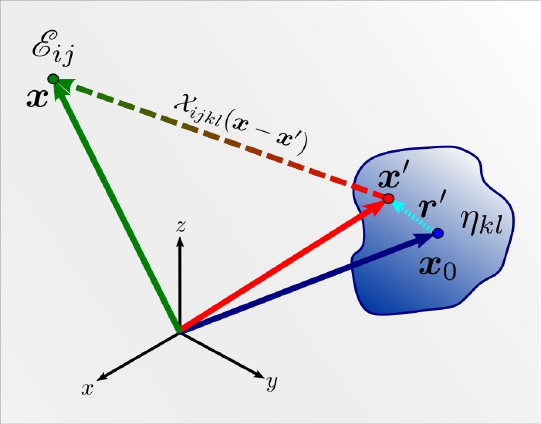}
    \caption{Geometry of the multipolar defect expansion. This expansion approximates the elastic strain field $\mathscr{E}_{ij}(\boldsymbol{x})$ from a distribution of defects $\eta_{kl}(\boldsymbol{x}^\prime)$ (blue shaded region), centered on the point $\boldsymbol{x}_0$. The propagation occurs through the Green's tensor, $\mathcal{X}_{ijkl}(\boldsymbol{x} - \boldsymbol{x}^\prime)$, defined in Fourier space as \eqref{eq:X_Greens_tensor_elastic_strain_from_eta}. The convolution in real-space, \eqref{eq:elastic_strain_eta_convolution}, runs over the defect distribution with the integration variable, $\boldsymbol{x}^\prime$, shifted relative to the center of the distribution through $\boldsymbol{r}^\prime = \boldsymbol{x}^\prime - \boldsymbol{x}_0$. In the far field limit, $\max(|\boldsymbol{r}^\prime|) \ll |\boldsymbol{x} - \boldsymbol{x}_0|$. }
    \label{fig:multipolar_defects}
\end{figure}

In real-space, the elastic strain for a medium of arbitrary symmetry is then formally written as
\begin{equation}
    \mathscr{E}_{ij}(\boldsymbol{x}) = \int_{x^\prime} \mathcal{X}_{ijkl}(\boldsymbol{x} - \boldsymbol{x}^\prime)\eta_{kl}(\boldsymbol{x}^\prime). \label{eq:elastic_strain_eta_convolution}
\end{equation}
With the elastic strain known in terms of the Green's tensor, $\mathcal{X}_{ijkl}(\boldsymbol{x})$, we can expand it formally in terms of multipolar defects. Assuming the defects are localized around some region centered on $\boldsymbol{x}_0$, then we first change the integration variable to $\boldsymbol{r}^\prime = \boldsymbol{x}^\prime - \boldsymbol{x}_0$. Assuming one measures the elastic strain in the ``far field'' regime, such that $\max(|\boldsymbol{r}^\prime|) \ll |\boldsymbol{x} - \boldsymbol{x}_0|$, then the integrand can be expanded as
\begin{align}
    \mathcal{X}_{ijkl}(\boldsymbol{x} - \boldsymbol{x}_0 - \boldsymbol{r}^\prime) &= \mathcal{X}_{ijkl}(\boldsymbol{x} - \boldsymbol{x}_0) - r^\prime_m\partial_m^\prime \mathcal{X}_{ijkl}(\boldsymbol{x} - \boldsymbol{x}_0) \nonumber
    \\
    &\phantom{=}+ \tfrac{1}{2}r^\prime_m r^\prime_n\partial_m^\prime  \partial_n^\prime \mathcal{X}_{ijkl}(\boldsymbol{x} - \boldsymbol{x}_0) +\dots
\end{align}
Using the identity $\partial_m^\prime \mathcal{X}_{ijkl}(\boldsymbol{x} - \boldsymbol{x}_0 - \boldsymbol{r}^\prime) = -\partial_m \mathcal{X}_{ijkl}(\boldsymbol{x} - \boldsymbol{x}_0 - \boldsymbol{r}^\prime)$, then the elastic strain follows as
\begin{align}
    \mathscr{E}_{ij}(\boldsymbol{x}) &= \mathcal{X}_{ijkl}(\boldsymbol{x} - \boldsymbol{x}_0)\, \overline{\eta_{kl}} + \partial_m\mathcal{X}_{ijkl}(\boldsymbol{x} - \boldsymbol{x}_0)\, \overline{\eta_{kl}^m} \nonumber
    \\
    &\phantom{=} + \tfrac{1}{2}\partial_m\partial_n\mathcal{X}_{ijkl}(\boldsymbol{x} - \boldsymbol{x}_0)\, \overline{\eta_{kl}^{mn}}+\dots \label{eq:defect_multipole_expansion}
\end{align}
with the defect monopole, dipole, and quadrupole being defined by
\begin{equation}
    \begin{aligned}
        \overline{\eta_{kl}} &\equiv \int_{x} \eta_{kl}(\boldsymbol{x}),
        \\
        \overline{\eta_{kl}^m} &\equiv \int_{x} (x_m - x_0)\eta_{kl}(\boldsymbol{x}),
        \\
        \overline{\eta_{kl}^{mn}} &\equiv \int_{x} (x_m - x_0)(x_n - x_0)\eta_{kl}(\boldsymbol{x}). 
    \end{aligned}\label{eq:defect_multipoles}
\end{equation}
Higher-order multipoles are generated in an analogous fashion. We remark that the monopolar and dipolar contributions are related to the Frank and Burgers vector of the disclination and dislocation, respectively \cite{dewitTheoryDisclinationsIII1973,dewitTheoryDisclinationsIV1973, kleinertGaugeFieldsSolids1989}. In two spatial dimensions, only $\eta_{33}$ is nonzero, and the resulting monopolar ``charge'' is the scalar-valued disclination \cite{Pretko2018_PRL, Pretko2019,gaaFractonelasticityDualityTwisted2021}. Otherwise, in three dimensions, the monopole is vector-valued, given the three independent degrees of freedom in $\eta_{ij}$ (see \eqref{eq:eta_coeffs_plastic_coeffs}).   

This multipole expansion establishes that one should expect long-ranged and potentially anisotropic elastic strain fields emanating from plastic defects. These will generically create pinning fields for the electronic nematic order parameter in the $d$-orbital basis, since the nemato-plastic conjugate field is the zero-nematic limit of the elastic strain: $\boldsymbol{h} = \boldsymbol{\mathscr{E}}|_{\varphi = 0}$. Meanwhile, because of defect-gauge invariance -- present for a medium of any symmetry -- the critical helical nematic order parameters, $(\Phi_2,\Phi_3)$, are orthogonal to the elastic strain, and do not participate in nemato-plasticity at the bilinear level.

\section{Family of Fundamental Harmonic Green's Functions\label{app:harmonic_Greens_functions}}
In this appendix, we obtain the fundamental Green's function for the two-dimensional biharmonic and triharmonic equations, defined in \eqref{eq:g_n_functions}. We first note that from the definition of $g_n(\boldsymbol{r})$, that 
\begin{equation}
\begin{aligned}\left(\partial_{x}^{2}+\partial_{y}^{2}\right)g_{2}\left(\boldsymbol{r}\right) & =-\frac{1}{2\pi}\delta\left(\boldsymbol{r}_{\parallel}\right),\\
\left(\partial_{x}^{2}+\partial_{y}^{2}\right)^{2}g_{4}\left(\boldsymbol{r}\right) & =+\frac{1}{2\pi}\delta\left(\boldsymbol{r}_{\parallel}\right),\\
\left(\partial_{x}^{2}+\partial_{y}^{2}\right)^{3}g_{6}\left(\boldsymbol{r}\right) & =-\frac{1}{2\pi}\delta\left(\boldsymbol{r}_{\parallel}\right),
\end{aligned}
\end{equation}
where $\boldsymbol{r}_\parallel\equiv(x,y,0)\equiv r_\parallel(\cos\phi_r,\sin\phi_r,0)$. Because of the $\delta(q_z)$ in \eqref{eq:g_n_functions}, there is no $z$-dependence in these partial differential equations. The fundamental Green's functions -- the solutions to the above differential equations taken on the infinite domain -- also satisfy a recursive relationship
\begin{equation}
\begin{aligned}\left(\partial_{x}^{2}+\partial_{y}^{2}\right)g_{2m}\left(\boldsymbol{r}\right) & = -g_{2\left(m-1\right)}\left(\boldsymbol{r}\right),\:m\in\mathbb{Z}^{+},\\
g_{0}\left(\boldsymbol{r}\right) & =-\tfrac{1}{2\pi}\delta\left(\boldsymbol{r}_{\parallel}\right).
\end{aligned}
\end{equation}
which can be integrated to obtain $n=2,4,6$. Starting with $n=2$, one obtains the function $g_2(\boldsymbol{r}_\parallel)$ using the same methods of electrostatics to obtain the electrostatic potential around a uniformly charged wire. Thus, we find $g_2(\boldsymbol{r}_\parallel)$ as 
\begin{equation}
    g_{2}\left(\boldsymbol{r}\right)=-\tfrac{1}{4\pi^{2}}\log\left(r_{\parallel}\right),\label{eq:g_2_fundamental_sol}
\end{equation}
which happens to source the biharmonic Green's function as
\begin{equation}
\left(\partial_{x}^{2}+\partial_{y}^{2}\right)g_{4}\left(\boldsymbol{r}\right)=-g_{2}\left(\boldsymbol{r}\right).
\end{equation}
Again, assuming infinite boundaries, one can readily obtain $g_4(r_\parallel)$, recalling that the two-dimensional cylindrically-symmetric Laplacian is 
\begin{equation}
    \left(\partial_{x}^{2}+\partial_{y}^{2}\right)g_{4}\left(r_{\parallel}\right)=\tfrac{1}{r_{\parallel}}\partial_{r_{\parallel}}\left[r_{\parallel}\partial_{r_{\parallel}}g_{4}\left(r_{\parallel}\right)\right].
\end{equation}
Integration yields 
\begin{equation}   g_{4}\left(\boldsymbol{r}\right)=\tfrac{1}{16\pi^{2}}r_{\parallel}^{2}\left[\log\left(r_{\parallel}\right)-1\right].\label{eq:g_4_fundamental_sol}
\end{equation}
Iterating one final time yields $g_6(\boldsymbol{r})$ as
\begin{equation}
g_{6}\left(\boldsymbol{r}\right)=\tfrac{1}{512\pi^{2}}r_{\parallel}^{4}\left[3-2\log\left(r_{\parallel}\right)\right].\label{eq:g_6_fundamental_sol}
\end{equation}

\bibliography{bibliography_nematics.bib, additional_references}
\end{document}